\documentstyle[rep10]{report}
\begin{document}

\parskip10pt plus 1pt minus 1pt
\setcounter{secnumdepth}{3}


\marginparwidth 20pt
\topmargin 27 pt
\headheight 12pt
\headsep 25pt
\footskip 30pt
\textheight = 43\baselineskip
\textwidth 440pt



\input feynman
%
%
\newcommand{\dirac}[1]{/ \!\!\!#1}
\newcommand{\vgl}[1]{eq.(\ref{#1})}
\newcommand{\gv}{\gamma^5}
\newcommand{\gu}[1]{\gamma^{#1}}
\newcommand{\gd}[1]{\gamma_{#1}}

\newsavebox{\uuunit}
\sbox{\uuunit}
    {\setlength{\unitlength}{0.825em}
     \begin{picture}(0.6,0.7)
        \thinlines
        \put(0,0){\line(1,0){0.5}}
        \put(0.15,0){\line(0,1){0.7}}
        \put(0.35,0){\line(0,1){0.8}}
       \multiput(0.3,0.8)(-0.04,-0.02){12}{\rule{0.5pt}{0.5pt}}
     \end {picture}}
\newcommand {\unity}{\mathord{\!\usebox{\uuunit}}}
\newcommand{\half}{{\textstyle\frac{1}{2}}}
%
\newcommand{\dr}{\raise.3ex\hbox{$\stackrel{\leftarrow}{\delta }$}}
\newcommand{\dl}{\raise.3ex\hbox{$\stackrel{\rightarrow}{\delta}$}}
%
\newcommand{\cA}{{\cal A}}
\newcommand{\cB}{{\cal B}}
\newcommand{\cC}{{\cal C}}
\newcommand{\cD}{{\cal D}}
\newcommand{\cE}{{\cal E}}
\newcommand{\cF}{{\cal F}}
\newcommand{\cG}{{\cal G}}
\newcommand{\cH}{{\cal H}}
\newcommand{\cI}{{\cal I}}
\newcommand{\cJ}{{\cal J}}
\newcommand{\cK}{{\cal K}}
\newcommand{\cL}{{\cal L}}
\newcommand{\cM}{{\cal M}}
\newcommand{\cN}{{\cal N}}
\newcommand{\cO}{{\cal O}}
\newcommand{\cP}{{\cal P}}
\newcommand{\cQ}{{\cal Q}}
\newcommand{\cR}{{\cal R}}
\newcommand{\cS}{{\cal S}}
\newcommand{\cT}{{\cal T}}
\newcommand{\cU}{{\cal U}}
\newcommand{\cV}{{\cal V}}
\newcommand{\cW}{{\cal W}}
\newcommand{\cX}{{\cal X}}
\newcommand{\cY}{{\cal Y}}
\newcommand{\cZ}{{\cal Z}}
%
\newcommand{\bA}{{\bf A}}
\newcommand{\bB}{{\bf B}}
\newcommand{\bC}{{\bf C}}
\newcommand{\bD}{{\bf D}}
\newcommand{\bE}{{\bf E}}
\newcommand{\bF}{{\bf F}}
\newcommand{\bG}{{\bf G}}
\newcommand{\bH}{{\bf H}}
\newcommand{\bI}{{\bf I}}
\newcommand{\bJ}{{\bf J}}
\newcommand{\bK}{{\bf K}}
\newcommand{\bL}{{\bf L}}
\newcommand{\bM}{{\bf M}}
\newcommand{\bN}{{\bf N}}
\newcommand{\bO}{{\bf O}}
\newcommand{\bP}{{\bf P}}
\newcommand{\bQ}{{\bf Q}}
\newcommand{\bR}{{\bf R}}
\newcommand{\bS}{{\bf S}}
\newcommand{\bT}{{\bf T}}
\newcommand{\bU}{{\bf U}}
\newcommand{\bV}{{\bf V}}
\newcommand{\bW}{{\bf W}}
\newcommand{\bX}{{\bf X}}
\newcommand{\bY}{{\bf Y}}
\newcommand{\bZ}{{\bf Z}}
\newcommand{\beq}{\begin{equation}}
\newcommand{\eeq}{\end{equation}}
\newcommand{\gras}[1]{\epsilon_{#1}}
\newcommand{\gh}[1]{\mbox{gh} \left( #1 \right)}
\newcommand{\sdet}{\mbox{sdet}}
\newcommand{\str}{\mbox{str}}
\newcommand{\tr}{\mbox{tr}}
\newcommand{\ihbar}{\frac{i}{\hbar}}
%
\newcommand{\sqrg}{\sqrt{g}}
\newcommand{\sqrabsg}{\sqrt{\vert g \vert}}
%
\newcommand{\ddr}{\raise.3ex\hbox{$\stackrel{\leftarrow}{d}$}}
\newcommand{\ddl}{\raise.3ex\hbox{$\stackrel{\rightarrow}{d}$}}
\newcommand{\wt}{\widetilde}
\def\sepand{\rule{14cm}{0pt}\and}
\def\gtwid{\raise.3ex\hbox{$>$\kern-.75em\lower1ex\hbox{$\sim$}}}
\def\ltwid{\raise.3ex\hbox{$<$\kern-.75em\lower1ex\hbox{$\sim$}}}


\evensidemargin .5in

\begin{titlepage}

\begin{center}
{\large Katholieke Universiteit Leuven}
\end{center}

\begin{center}
{\Large Faculteit Wetenschappen}
\end{center}

\begin{center}
{\Large Instituut voor Theoretische Fysica}
\end{center}

\vspace*{2cm}
\begin{center}
{\huge The Batalin-Vilkovisky \\ Lagrangian quantisation scheme} \\
{\large with applications to the study of} \\
{\huge  anomalies in gauge theories}
\end{center}

\vspace*{2cm}
\begin{flushright}
Proefschrift voorgelegd ter verkrijging \\
van de graad van Doctor in de Wetenschappen \\
door {\bf Frank De Jonghe}.
\end{flushright}

\vspace*{1cm}



{\large Leuven, december 1993.} \hspace*{2cm}
{\bf Promotor}: Dr.\ W.\ Troost

\end{titlepage}

\oddsidemargin .5in \evensidemargin 1.in
\pagenumbering{roman}
\newpage
\begin{flushright}
\vspace*{2cm}
{\it I was born not knowing,  \\
and have only had a little time \\
to change that here and there.}

{\bf Richard Feynman}
\end{flushright}

\newpage

\vspace*{10cm}

\newpage

\section*{Acknowledgements}

It is amazing how many people had to help me to accomplish so little.
Here follows an incomplete list.

First of all, there is my thesis supervisor Walter Troost, who
constantly stimulated my interest, not only in quantum field theory,
but in many other branches of physics.
Poul H. Damgaard had the burden to be responsible for me during my stay at
CERN. He provided me with many cups of coffee to keep me
awake and going. In the work of the disciple, one recognises his master's
hand. Sure enough, both will find many of their views expressed in this
dissertation. I am very grateful for everything they taught me.

Ruud, Stefan, Toine and Peter helped me to
improve my insight through many discussions during our collaborations.
I have tried to profit as much as possible from their different attitudes
towards scientific research.

It is a pleasure to acknowledge the support of Raymond Gastmans, Frans
Cerulus and Andr\'e Verbeure. Among other things, they made my visits
abroad possible, not only financially.

Speaking of money, I thank the Belgian Nationaal
Fonds voor het Wetenschappelijk Onderzoek for giving me the opportunity
to further develop my skills as a physicist through a position as Aspirant.
It will be hard to find a better employer.

The members of our Instituut voor Theoretische Fysica made me feel at home
in Leuven from the very first day. Thanks!

My entrance in the world of theoretical particle physics was facilitated by
the experimentalist Frans Verbeure. His continuing interest has always been
a stimulus for me.

Tussen het bezoek aan de opendeurdag van het RUCA en nu liggen bijna 8 jaar
vol fysica plezier. Ik dank mijn ouders omdat ze me steeds gesteund
hebben bij het verwezenlijken van mijn plannen, ook al waren zij er zich
{\em wel} van bewust dat die plannen soms vol jeugdige onbezonnenheid
zaten. Ik hoop dat ik het in mij gestelde vertrouwen niet beschaamd heb.

Lenick, bedankt voor je aansporingen om mijn grote woorden ook af en toe
eens in kleine daden om te zetten. Ik besef dat het niet gemakkelijk moet
zijn om met een theoreet getrouwd te zijn, wij brengen onze problemen
wel mee naar huis... Oh ja, ook bedankt voor je hulp bij het verluchten
van deze thesis met enkele prentjes. Angie, angie, where will it lead us
from here ?

\newpage

\vspace*{10cm}

\newpage

\begin{center}
{\bf The Batalin-Vilkovisky Lagrangian quantisation scheme, \\
with applications to the study of anomalies in gauge theories. }\\
\vspace{2cm}
{\bf Abstract}
\end{center}

{\small
\textwidth 465pt

Although gauge theories took the centre stage in theoretical physics
only in this century, they are the rule, not the exception.
The two standard examples of gauge theories are Maxwell's theory for
electromagnetic phenomena and Einstein's theory of general relativity.
Recent attempts to unify the four fundamental forces have provided new and
more complex examples of gauge theories.

We study the Batalin-Vilkovisky (BV) Lagrangian quantisation scheme. Most
of the
gauge theories known today can be quantised using this scheme. We show how
the BV scheme can be constructed by combining the gauge symmetry with the
quantum equations of motion, the so-called Schwinger-Dyson equations.
The different quantisation prescriptions that were known for the
different types of gauge theories can be reformulated in one unified
framework, using
this guiding principle. One of the most remarkable properties of the BV
scheme is that it possesses a symplectic structure that is invariant under
canonical transformations, in analogy with classical mechanics. The
possibilities offered by the canonical transformations are exploited in
some examples, e.g.\ the construction of four-dimensional topological
Yang-Mills theory.

In a second part, we show how the Lagrangian BV scheme
can be derived from the Hamiltonian description of gauge theories.

When the quantum effects destroy the gauge symmetry that was present in the
classical theory, the gauge symmetry is said to be anomalous. In the third
part, we use Pauli-Villars regularisation to determine whether a theory is
anomalous or not. We study how one can influence which symmetries will be
anomalous and a new derivation is given of the one-loop regularised BV
scheme.
}

\tableofcontents
\newpage
\vspace*{2cm}
\newpage
\pagenumbering{arabic}

\chapter*{Introduction}

{\it The number of rational hypotheses that can explain any given
phenomenon, is infinite}\footnote{R.M. Pirsig, {\it Zen, and the art of
motorcycle maintenance}, Bantam book, New Age edition, 1981, p100.}.
Physicists --and other scientists alike-- have always used two
criteria to construct an hierarchy in this vast infinity of hypotheses.
First of all, an hypothesis --or a model-- that is able to explain more
phenomena than other hypotheses, is considered to be superior.
Secondly, physicists are easily seduced by the presence of {\it
symmetry} in a model. It goes without saying, that both aspects are not
independent, as two seemingly different phenomena might be related by a
symmetry principle. In that case, symmetry allows to build one model that
describes both phenomena. After having selected some preferred
hypotheses in this way,
the final justification for the promotion of an hypothesis to
a {\it scientific fact} is of course an experimental verification.

J.C. Maxwell was one of the first to unify two different forces in one
model\footnote{Phil. Trans. R. Soc. {\bf 155} (1864) 459.}.
In 1864, he brought together the forces of electricity
and magnetism in a set of coupled differential equations, the Maxwell
equations. This Maxwell theory is also the first example of a gauge theory,
that is, a theory with an invariance with space-time dependent
transformation parameters. Even when trading the six degrees of freedom
of the electric and magnetic vector field for the four degrees of freedom
of the Lorentz vector potential field $A_{\mu}$, not all configurations
$A_{\mu}$ describe different physical systems. Indeed, the four field
equations for the $A_{\mu}$ are not
independent and the action is invariant under $A_{\mu} \rightarrow A_{\mu}
+ \partial_{\mu} \epsilon$.

\begin{figure}[t]

\begin{picture}(28000,15000)
\drawline\fermion[\E\REG](5000,11500)[3000]
\drawarrow[\E\ATTIP](\pmidx,\pmidy)
\drawline\photon[\NE\REG](\pbackx,\pbacky)[3]
\drawline\photon[\SE\REG](\pbackx,\pbacky)[3]
\global\advance\pfronty by 1050
\put(\pfrontx,\pfronty){$\gamma$}
\global\advance\pmidy by -3800
\put(\pmidx,\pmidy){(a)}
\drawline\fermion[\E\REG](\pbackx,\pbacky)[3000]
\drawarrow[\E\ATTIP](\pmidx,\pmidy)
\drawline\fermion[\W\REG](\pfrontx,\pfronty)[4500]
\drawarrow[\E\ATTIP](\pmidx,\pmidy)

\drawline\photon[\E\REG](20000,11500)[6]
\drawline\fermion[\NE\REG](\photonbackx,\photonbacky)[3600]
\drawarrow[\NE\ATTIP](\pmidx,\pmidy)
\drawline\fermion[\SE\REG](\photonbackx,\photonbacky)[3600]
\drawarrow[\NW\ATTIP](\pmidx,\pmidy)
\drawline\photon[\N\REG](\pbackx,\pbacky)[5]
\global\advance\pmidy by -3000
\global\advance\pmidx by -7000
\put(\pmidx,\pmidy){(b)}
\drawline\fermion[\NE\REG](\photonbackx,\photonbacky)[3500]
\drawarrow[\NE\ATTIP](\pmidx,\pmidy)
\drawline\fermion[\SE\REG](\photonfrontx,\photonfronty)[3500]
\drawarrow[\NW\ATTIP](\pmidx,\pmidy)
\end{picture}

\begin{picture}(28000,6000)
\put(18000,6000){\circle{2800}}
\pmidx = 18200 \pmidy = 7400
\drawarrow[\E\ATTIP](\pmidx,\pmidy)
\pmidx = 18200 \pmidy = 4600
\drawarrow[\W\ATTIP](\pmidx,\pmidy)
\pfrontx = 16500 \pfronty = 6000
\frontstemmed\drawline\photon[\W\REG](\pfrontx,\pfronty)[5]
\global\advance\pmidy by -1350
\pbackx = 19300 \pbacky = 6000
\frontstemmed\drawline\photon[\E\REG](\pbackx,\pbacky)[5]
\global\advance\pmidy by -4000
\global\advance\pmidx by -4000
\put(\pmidx,\pmidy){(c)}
\end{picture}

\begin{center}
{\it The three one loop divergences of QED. \\ The lines with an
arrow are fermions, \\ the wiggly lines photons.}
\end{center}

\end{figure}

Soon after a mathematical framework for quantum mechanics had been
developed,
people started studying the quantum theory of the interaction of electrons
and photons (the particle excitations of the electromagnetic field),
quantum electrodynamics. In the perturbative expansion, one immediately ran
into trouble. The three one loop diagrams (see
figures (a-c)) were found to lead to infinite corrections to the zeroth
order (classical) result. This problem was finally solved by R.P. Feynman,
S. Tomonaga and J. Schwinger\footnote{A collection of the early articles
can be
found in {\it Quantum Electrodynamics}, ed. J. Schwinger, Dover, New York,
1958.} by what is now known as the {\it renormalisation procedure}.
Schematically, this means that the infinities are removed by a redefinition
of the free parameters of the theory, the charge $e$ of the electron and
its mass $m$. That the three divergences depicted in (a-c) can be removed
using two parameters is the result of gauge invariance. The gauge
invariance of the classical theory also manifests itself at the quantum
level, and has as a consequence that the divergences (a) and (b)
are not independent. Hence, gauge invariance plays a crucial role
in the renormalisability of the theory.

Inspired by the example of quantum electrodynamics, C.N.~Yang and
R.~Mills\footnote{Phys. Rev. {\bf 96} (1954) 191.}
proposed a model for the strong interaction between protons and neutrons,
based on the $SU(2)$ Lie algebra. With every generator of this algebra, a
vector boson is associated. These vector bosons are the force carriers, in
analogy with the photon. This principle was generalised
to arbitrary Lie algebras,
and nowadays a whole industry of modelbuilding based
on this method exists. However, more important for our discussion is that
all these models have local symmetries, gauge symmetries. Moreover, it was
recognised that also the theory of general relativity allows for a
formulation as a gauge theory, where the Christoffel symbols play the role
of gauge fields and where arbitrary coordinate transformations are the
local gauge transformations.

It is one thing to construct a classical gauge theory, but it is a
different story to quantise the theory and to prove that it is
renormalisable. The attempts to demonstrate that Yang-Mills theories are
renormalisable have stimulated the research on the quantisation itself of
gauge theories\footnote{For a lively account, see M.J.G. Veltman, {\it The
path to renormalisability}, Third International Symposium on the History of
Particle Physics, SLAC, june 1992.}. What was needed was a means to
quantise (Yang-Mills) gauge theories in different gauges, as the
renormalisability
is more transparant in other gauges than those gauges where the physical
content of the theory is more clear. In solving this problem, the
quantisation
using path integrals proved to be superior to operator quantisation. In
1967, L.D. Faddeev and V.N. Popov\footnote{Phys. Lett. {\bf 25B}
(1967) 29.} showed how the integration over configurations that are related
by a gauge transformation can be factored out from the path integral. In
their recipe, extra fields are introduced as a technical device to
rewrite a Jacobian as resulting from fictitious particle interactions.
These extra
fields are called {\it ghost fields}, and were already foreshadowed in the
work of R.P. Feynman and B. DeWitt a few years earlier. Ever since, ghosts
wander over the battle fields of theoretical particle physics.

Gauge theories really took the centre stage after G. 't Hooft and M.
Veltman proved in 1971 that gauge theories of the Yang-Mills type are
renormalisable\footnote{Nucl. Phys. {\bf B50} (1972) 318.}. Let us stress
that all phenomenological models in particle physics, namely the gauge
theory of electroweak
interactions and QCD for the strong interactions, are of that type.
The theoretical framework of the early seventies is still sufficient to
account
for present day (1993) experimental particle physics. All subsequent
developments in theoretical
particle physics are of speculative nature and largely motivated by
the attempts to unify {\it all} four forces, including gravity, in one
(quantum) theory.

A nice reformulation of the results on quantisation and renormalisation of
gauge theories, and that later proved to
be the one amenable to generalisation, was presented by C. Becchi, A. Rouet
and R. Stora\footnote{Phys. Lett. {\bf 52B} (1974) 344.}. They discovered
that in the quantisation process the local gauge invariance is replaced
by a global invariance, which now goes under the name BRST invariance. This
transformation is encoded in a nilpotent operator $\delta$  ($\delta^2
=0$). The nilpotency of this operator has led to the introduction of
cohomological methods in the study of gauge theories.

In the second half of the seventies, it was found that in some supergravity
models a term quartic in the ghosts is needed\footnote{R.E. Kallosh,
Nucl. Phys.  {\bf B141} (1978) 141.  \newline G. Sterman, P.K.
Townsend and P. van Nieuwenhuizen, Phys. Rev. {\bf D17} (1978) 1501.} for
unitarity. Such terms can not be generated with the usual Faddeev-Popov
procedure, which only gives quadratic ghost actions. The way out was to use
BRST invariance of the quantum theory as a guiding principle. However, the
nilpotency of the BRST operator then only holds upon using the field
equations of the gauge fixed action. Theories where this is the case, are
said to have an open algebra.

Another complication that was discovered, is that for some theories the
action for the ghosts (obtained \`a la Faddeev-Popov) has itself a gauge
symmetry that needs gauge fixing. This happens when the original gauge
symmetries are not
independent, and the theory is said to be a reducible gauge theory. What is
needed in such
cases is yet another enlargement of the field spectrum, with so-called
ghosts for ghosts.

Although all these types of theories can be quantised using ad hoc rules,
there is
a quantisation scheme that encompasses all the previous developments
in one unified formalism, the Batalin-Vilkovisky (BV) scheme\footnote{
Phys. Lett. {\bf 102B} (1981) 27.}. It is the focus of our attention in
this work. The scheme uses the Lagrangian formulation of
path integral quantisation. Next to providing us with a powerful tool
for the quantisation of gauge theories of any kind known today, the BV
recipe also gives a natural environment for the study of the occurrence of
anomalies in gauge theories\footnote{W. Troost, P. van Nieuwenhuizen,
A. Van Proeyen, Nucl. Phys. {\bf B333} (1990) 727.}. A gauge theory is
anomalous, when quantum effects destroy the BRST invariance of the theory.
Here too, recent developments like string theory have led to new examples
to study.

At this moment, we should give a long list of the assets of the BV scheme,
to arouse the appetite of the reader and to motivate our effort to
investigate both the BV scheme itself and its applications. But since it is
difficult to describe the beauty of the Alps to a Dutchman who has never
left Holland, we postpone this motivation to our conclusions.

This thesis is divided in three main parts. In the
first part, we study the Batalin-Vilkovisky Lagrangian quantisation
scheme itself. In particular, we show how it naturally follows from and
encompasses the Faddeev-Popov and BRST quantisation methods. The
Lagrangian BV scheme is derived from the Hamiltonian quantisation scheme
for gauge theories of
Batalin-Fradkin-Vilkovisky in the second part of this dissertation. In the
third
part, we use the BV quantisation scheme to study some aspects of the gauge
anomalies that can appear in the quantisation process. The appendices
contain, apart from some technical rules, a discussion of
the algebraic steps that are common for all the regularised calculations of
anomalies in part III.

We start by introducing {\it gauge theories}, the subject of this
dissertation, in
chapter one. Different types of gauge theories exist, depending on the
structure functions of the gauge algebra. Whatever their type, gauge
theories lead to divergent functional integrals upon naive quantisation,
since infinitely many elements of the configuration space describe the same
physical system. The Faddeev-Popov quantisation procedure to remedy
this problem, is presented in chapter two. Following Faddeev and Popov,
one can construct a gauge fixed action that has no local gauge invariances
anymore and that can be used as a starting point for perturbative
calculations. However, this gauge fixed action has a global symmetry, the
{\it BRST symmetry}. It is argued in chapter three that it is expedient to
formulate the complete quantisation procedure from the
point of view that the local gauge invariance of the classical theory is to
be traded for BRST invariance of the quantum theory. The BRST symmetry of
the quantum theory leads to the so-called {\it Ward identities}, which
are relations between correlation functions that express the
consequences of the gauge symmetry for the full quantum theory. When
quantum effects destroy the BRST symmetry of the quantum theory, the gauge
symmetry is said to be {\it anomalous(ly broken)}.

In the fourth chapter, the equations of motion of the quantum
theory, the Schwinger-Dyson (SD) equations, are derived as Ward
identities
from a global BRST symmetry. This is done by introducing a gauge symmetry,
called the {\it SD shift symmetry}, using a collective field formalism. By
demanding that the BRST symmetry algebra of any gauge theory be enlarged
such that it includes the SD shift symmetry, we reconstruct the BV scheme
in
chapter five. One of the most remarkable properties of that scheme is that
for
every field $\phi^A$ an {\it antifield} $\phi^*_A$ is introduced. Fields
and antifields are canonically conjugated with respect to the
{\it antibracket}, in much the same way as coordinates and momenta are
conjugated in classical Hamiltonian mechanics. Most of the features of
the BRST quantisation recipe
are translated in the BV scheme using the antibracket: the
construction of the BRST invariant gauge fixed action, the Ward identities,
the condition that operators have to satisfy in order to have a gauge
invariant expectation value etc. As examples, we\footnote{F. De Jonghe and
S. Vandoren; KUL-TF-93/44, accepted for publication in Phys. Lett. {\bf
B}.} discuss the construction
of topological Yang-Mills theory and we derive a general prescription for
the construction of a BRST invariant energy-momentum tensor in the BV
scheme.

If the commutator of two infinitesimal gauge transformations is only a
linear combination of infinitesimal gauge transformations when acting on
the classical
configurations that satisfy the classical equations of motion, the algebra
is said to be {\it open}. In chapter six, we describe the BRST quantisation
of theories with an open gauge algebra. Here too, we\footnote{F. De
Jonghe; CERN-TH-6858/93, KUL-TF-93/13, accepted for publication in
J.Math.Phys.} enlarge the BRST
symmetry such that it includes the SD shift symmetry, and derive the BV
scheme
for open algebras. We\footnote{P.H. Damgaard and F. De Jonghe; Phys. Lett.
{\bf B305}
(1993) 59. \newline F. De Jonghe; KUL-TF-93/37.} construct an antifield
scheme that is invariant under both BRST and anti-BRST symmetry
in chapter seven (anti-BRST symmetry is introduced in chapter three).

Chapter eight, the final chapter of part one, contains a reformulation of
the quantisation in the BV scheme using {\it canonical transformations}.
Canonical transformations are transformations of fields and antifields that
leave the antibracket invariant, in analogy with canonical transformations
in classical mechanics that leave the Poisson bracket invariant. The
examples are the continuation of the examples of chapter five\footnote{F.
De Jonghe and S. Vandoren, Op. Cit.}.

All the developments in part I are based on the Lagrangian, Lorentz
covariant description of field theory, in casu gauge theory. In part II, we
derive the Lagrangian BV formalism from the Hamiltonian path integral
quantisation procedure for gauge theories. In chapter nine, an overview is
given of the description of gauge symmetries in the Hamiltonian canonical
formalism and of the standard quantisation procedure using an Hamiltonian
formulation of BRST symmetry. We\footnote{F. De Jonghe; Phys. Lett. {\bf
B316} (1993) 503.} then apply this recipe to derive the Schwinger-Dyson
equations as Ward identities of the Hamiltonian formalism in chapter ten.
We also show
that if we enlarge the BRST symmetry of the Hamiltonian system to include
the SD shift symmetry, we naturally obtain the Lagrangian BV formalism
after integration over the momenta.

Part III contains a detailed, one-loop regularised study of anomalies in
gauge theories. Gauge anomalies are defined and some
of their properties, in particular the Wess-Zumino consistency condition,
are discussed in chapter eleven. All these results are also reformulated
in the BV formalism.
In order to calculate an explicit expression for the anomaly in a specific
model, a regularisation scheme is required. As is explained in chapter
twelve, the functional integral can be regularised up to one loop using
Pauli-Villars regularisation. This regularisation scheme leads to
an expression for the anomaly that is of the same type as proposed by
K. Fujikawa. An important role is played by the mass term of the PV fields.
The invariances of this mass term determine which symmetries will be
anomaly free. We\footnote{F. De Jonghe, R. Siebelink and W. Troost; Phys.
Lett. {\bf B288} (1992) 47.} show how the freedom in the choice of mass
term can be exploited to calculate actions for gauge fields that are
induced by matter fields (induced gravity, Wess-Zumino-Witten model).
Chapter thirteen contains a new derivation of the regularised, one-loop
master equation of the BV scheme.
In the final chapter of the third part and of this
work, we\footnote{F. De Jonghe, R. Siebelink and W. Troost; Phys. Lett. {\bf
B306} (1993) 295.} demonstrate that one can keep preferred gauge symmetries
anomaly free by the introduction of extra (scalar) degrees of freedom.

The models in the examples are always presented without giving a raison d'
\^etre for these models. They only serve as an illustration of the points
raised in the general developments. In particular, the examples of the
third part always involve two-dimensional models. The general recipe has
been applied to four-dimensional models\footnote{F. De Jonghe,
R. Siebelink, W. Troost, S. Vandoren, P. van
Nieuwenhuizen and A. Van Proeyen; Phys. Lett. {\bf B289} (1992) 354.}
as well, although the amount of algebraic work increases drastically with
the space-time dimension.

We do not discuss the quantisation of reducible gauge theories in this
dissertation. Basically, the problem there is the construction of the
correct particle spectrum, of the BRST transformation rules and of a
suitable gauge fermion. There too, the BRST operator on the complete set of
fields is either nilpotent or on-shell nilpotent. The BV scheme can then be
developed for reducible gauge theories as well, following the steps of
the chapters 5 and 6.

\part{The Batalin-Vilkovisky Lagrangian quantisation scheme for gauge
theories}

\chapter{The definition and basic properties of gauge theories}

We start by introducing in section one the study subject of this
dissertation: {\it
gauge theories}. It is shown why gauge theories require special
quantisation
procedures. The second section contains a definition of the {\it gauge
algebra} and a classification of the different types of algebras that are
known to exist.

\section{What is a gauge theory and why does it need a special
quantisation procedure ?}

Before we start, some technical remarks.
We will use the so-called DeWitt notation. This means
that we denote the (field)degrees of freedom by $\phi^i$. Here,
the index $i$ runs over the {\it internal} degrees of freedom (e.g. the
Lorentz index of the vector potential in electromagnetism) as well as over
the space-time variable. Repeated indices are summed over, except when
explicitly indicated otherwise. The convention implies that whenever
a summation over $i$ occurs, an integral over space-time is understood.

Here and below we will loosely use notions like (in)dependence,
completeness et cetera, hoping that the context makes clear what is meant.
In the literature \cite{bv1,bv2,bv3,bv4,bv5}, these
concepts are defined by using the rank of matrices. In the proofs of various
properties, the results of finite dimensional analysis are then used.
When treating the DeWitt indices as if they only run over a finite number
of values, the difference between global and local symmetries is somewhat
obscured. Moreover, the locality of the function(al)s of the fields is not
always guaranteed.

In the following, when we use the term {\it quantisation}, we have in mind
the
quantisation method based on path integrals. This method consists of the
following steps. One starts from a configuration space with degrees of
freedom labelled by $\phi^i$,
which for the argument we take to be bosonic. On this configuration space,
an action functional $S[\phi]$ is defined, which associates with every
configuration a real number, and which specifies the dynamics. The
partition function is then defined as
\beq
  \cZ = \int [d\phi] e^{\frac{i}{\hbar} S[\phi]}   \  ;
\eeq
that is, $\cZ$ is a summation over all configurations, where every
configuration contributes a complex number of unit norm and with a phase
determined by the value of the action functional for that configuration.

The action functional (which we will henceforth call the {\it action}) is
such that the classical theory is described by the configurations
$\phi^i_0$ that extremise it. These configurations are the solutions
of the {\it field equations} (also known as {\it equations of motion} or
{\it Euler-Lagrange equations}). Denoting
\beq
  y_i(\phi^l) = \frac{\dr S}{\delta \phi^i},
\eeq
the field equations are $ y_i(\phi^l) = 0$.
The subspace of configuration space consisting of all solutions of these
field equations is called the {\it stationary surface}. We will always
assume that at least one classical solution exists \cite{bv5}, as this is
a basic requirement to set up the perturbation theory by considering
quantum fluctuations around a classical solution
\cite{Zinn-Justinboek,Ramondboek}.

Suppose now that a set of
operators $R^i_{\alpha}[\phi]$ exists, such that
\beq
   y_i (\phi)R^i_{\alpha} [\phi] \epsilon^{\alpha} = 0,
   \label{introR}
\eeq
for arbitrary values of the parameters $\epsilon^{\alpha}$. If the index
$\alpha$ does {\it not} include a space-time index, we speak of a {\it
global} or {\it rigid symmetry}. If the parameters are space-time
dependent,
the symmetry is a {\it local} or {\it gauge symmetry}. The latter are the
main subject of this work. In that case, we will sometimes refer to the
$\phi^i$ as {\it gauge fields}\footnote{This terminology is somewhat
different from the usual one. When considering models like, for instance,
QCD one distinguishes between matter fields (the fermion fields for the
quarks) that have well-defined propagators, and the vector bosons, the
gauge fields, that have ill-defined propagators. We do not make this
distinction and denote {\em all} fields present in a classical action that
has gauge symmetries by {\it gauge fields}.}.
If no such operators $R^i_{\alpha}$ exist, and hence
all field equations $y_i$ are independent, one has a theory without global
or local symmetries, and the stationary surface is really a stationary
point,
provided appropriate boundary conditions in space and time are specified.

One of the consequences of (\ref{introR}) is that the Hessian, defined by
\beq
    H_{ij} = \frac{\dl}{\delta \phi^j} \frac{\dr S[\phi]}{\delta \phi^i},
\eeq
has zeromodes when evaluated
in a point of the stationary surface\footnote{When an expression is
evaluated on the stationary surface, we
will sometimes say that {\it it is taken on shell}.}.
This is
easily proved by differentiating (\ref{introR}) with respect to
$\phi^j$: \beq
    \frac{\dl}{\delta \phi^j} \frac{\dr S[\phi^l_0]}{\delta \phi^i} \cdot
    R^i_{\alpha}[\phi^l_0] = 0 \;\; .
    \label{zeromodeHessian}
\eeq
$\phi^l_0$ denotes a field configuration that is a solution of the field
equations, and hence $y_i(\phi ^l_0) = 0$ identically.

The zeromodes\footnote{We will use the name {\it gauge generators} below
for the $R^i_{\alpha}$.} $R^i_{\alpha}[\phi^l_0]$ determine an
infinitesimal transformation
which maps a classical solution to another classical solution.
Indeed, for infinitesimal parameters $\epsilon^{\alpha}$ such that a
Taylor series expansion to linear order makes sense, we have that
\beq
   y_i( \phi^l_0 + R^l_{\alpha}[\phi^k_0] \epsilon^{\alpha} ) = 0.
\eeq
It may be necessary to impose boundary conditions on the
$\epsilon^{\alpha}$ in order for $\phi^l_0 + R^l_{\alpha}[\phi^k_0]
\epsilon^{\alpha}$ to satisfy the original boundary conditions.
This result is valid for global as well as local symmetries. The upshot is
that indeed the stationary point becomes a stationary surface. In the case
of a global symmetry, it becomes a finite dimensional space, coordinatised
by a finite set of parameters $\epsilon^{\alpha}$. For local symmetries,
the stationary surface becomes infinite dimensional: the index $\alpha$
contains a space-time point, so we are free to choose a discrete set of
parameters $\epsilon^i$ at every space-time point $x$ ($\alpha = (i,x)$).

Hence, already at the level of classical field theory, the presence of
zeromodes manifests itself. Also the difference between global and local
symmetries shows up.  Whereas global symmetries just relate a set
of solutions of the field equations parametrised by a finite set of
arbitrary parameters, the local symmetries really mean
that not all field equations are independent, and hence not all
field degrees
of freedom $\phi^i$ are fixed by the classical field equations.
Arbitrary fields appear in the classical solutions.
In order to eliminate this arbitrariness, one needs in such cases to
impose other constraints on the fields, the {\it gauge fixing conditions}.
The example everybody is familiar with, is of course classical
electromagnetism. The Maxwell equations are, even when expressed in terms
of the Lorentz vector potential $A_{\mu}$, redundant and one imposes a
gauge condition before solving them.

If one would try to quantise theories with a gauge symmetry by
ex\-po\-nen\-tia\-ting the action and summing (integrating) over all field
configurations, one would run into problems when setting up the usual
perturbation theory. Suppose that one picks a classical solution $\phi^l_0$
and one makes the saddle point approximation
\cite{Zinn-Justinboek,Ramondboek}. The quantum fluctuations around the
classical solution are composed of two contributions: fluctuations along
the stationary surface, which can be parametrised
$\epsilon^{\alpha}$, and fluctuations which take the field away from the
stationary surface, which we denote by $\delta_{\bot} \phi^i$. We have:
\beq
   \phi^i = \phi^i_0 + R^i_{\alpha} \epsilon^{\alpha} + \delta_{\bot}
   \phi^i .
   \label{fluctuations}
\eeq
Notice that only in the case of a local gauge symmetry the
$\epsilon^{\alpha}$ are field degrees of freedom. Expanding the action up
to terms quadratic in the fluctuations (which is sufficient to study the
one-loop structure of the theory, or, in other words, the first quantum
corrections of order $\hbar$) we get:
\beq
    S[\phi^i] = S[\phi^i_0] + \left(R^j_{\alpha} \epsilon^{\alpha} +
\delta_{\bot} \phi^j \right) \left[
\frac{\dr}{\delta \phi^i} \frac{\dl}{\delta
\phi^j} S[\phi^l_0] \right] .\left( R^i_{\beta} \epsilon^{\beta} +
\delta_{\bot} \phi^i \right).
\eeq
Because of (\ref{zeromodeHessian}), all terms which depend on the fields
(in the case of a local symmetry) $\epsilon^{\alpha}$ drop out, and the
previous expression reduces to
\beq
   S[\phi^i] = S[\phi^i_0] + \delta_{\bot} \phi^j
   \frac{\dr}{\delta \phi^i} \frac{\dl}{\delta \phi^j} S[\phi^i_0]
   \delta_{\bot} \phi^i.
\eeq
Owing to the decomposition (\ref{fluctuations}), the measure of the path
integral $[d \phi ]$ also splits up. We symbolically write
$[d \epsilon] [d\delta_{\bot}\phi]$.
As the integrand is independent of $\epsilon^{\alpha}$, the integration
over these degrees of freedom factorises from the path integral. This is
not really a problem for global symmetries.
Even if the integral over the parameter space of the symmetries is divergent
in that case, it can be cured by imposing boundary conditions
on the integration domain
of the path integral. For gauge symmetries however, we have field degrees
of freedom $\epsilon^{\beta}(x)$ which have no quadratic part in the
action, and hence have no propagator. The classical solution around
which one expands is often taken to be $\phi^i_0 = 0$. The Hessian is then
just the quadratic part of the original action. This leads to the common
criterion that gauge symmetries manifest themselves in the fact that the
quadratic part of the action is not invertible.

Let us finally make a small comment on the notion of a {\it regular theory}
\cite{bv5}. A theory is called regular if the non-invertibility of the
Hessian is only due to the symmetries generated by $R^i_{\alpha}$. Two
categories of irregular theories have been identified. First, it may occur
that even when the stationary surface reduces to a stationary point, the
Hessian has zeromodes in that stationary point. A pathological example is
provided by $S = \phi^3$. The classical solution is $\phi=0$, and the
second derivative evaluated for this solution is also zero. From the
previously mentioned semiclassical expansion it is clear that even when
there are no symmetries, the perturbation series can not be set up
for this kind of theories. A second type of irregular theory is found
when the notion of non-degeneracy of the Hessian itself becomes
ill-defined.

We will only consider (local) function(al)s of the fields
such that if they vanish on the stationary surface, they are
proportional to field equations everywhere in configuration space.
That is,
\beq
    X(\phi^i_0) = 0 \Rightarrow X(\phi^i) = y_j Y^j(\phi^i) \; ,
\eeq
for some functions $Y^j$. Again, the standard proof of this property
starting from the definition of regularity uses a
discrete set of degrees freedom (see e.g.\cite{fisch}).

Gauge theories are characterised by the fact that their
field equations are not independent. We have demonstrated how this leads to
divergences when setting up the perturbation theory in a naive path
integral quantisation procedure.

\section{The gauge algebra}

Let us now further study the properties of the $R^i_{\alpha}$.
We consider a set of degrees of freedom $\phi^i$, with
Grassmann parity\footnote{$\gras{i} = 0$ for a bosonic field or
monomial of fields and
$\gras{i}=1$ for a fermionic field or monomial of fields. For some basic
rules to keep in mind when working with quantities
of different Grassmann parity, we refer to the appendices.}  $\gras{i}$.
We start from a set of gauge generators $R^i_{\alpha}$ with
the following properties:
\begin{eqnarray}
   \label{ijkrel}
   y_i R^i_{\alpha } & = & 0 \nonumber \\
   \gras{R^i_{\alpha}} & = & \gras{i} + \gras{\alpha}.
\end{eqnarray}
Again, we denote $y_i = \frac{\dr S}{\delta \phi ^i}$. From now on, we
always have
the case in mind of local symmetries, i.e. $\alpha$ contains a space-time
point. Suppose that the set $R^i_{\alpha}$ is {\it complete}, in the sense
that
\beq
    \forall X^i_{\bar{\beta}} : y_i X^i_{\bar{\beta}} (\phi) = 0
      \;\; \Rightarrow  \;\;
    X^i_{\bar{\beta}}(\phi) = R^i_{\alpha} \epsilon^{\alpha}
    _{\bar{\beta}} (\phi) + y_j M^{ij}_{\bar{\beta}} (\phi),
    \label{completeR}
\eeq
where $\bar{\beta}$ is an arbitrary set of indices, and where the
$M^{ij}_{\bar{\beta}}$ have the graded antisymmetry
property:
\beq
    M^{ij}_{\bar{\beta}} (\phi) = (-1)^{\gras{i} \gras{j} +1}
    M^{ji}_{\bar{\beta}} (\phi).
\eeq
Owing to this graded antisymmetry, $y_i y_j M^{ij}_{\bar{\beta}}$
vanishes identically.

Consider now
\beq
   \frac{\dr}{\delta \phi^j} \left[ y_i R^i_{\alpha} \right] . R^j_{\beta}
   - (-1)^{\gras{\alpha} \gras{\beta}}
   \frac{\dr}{\delta \phi^j} \left[ y_i R^i_{\beta} \right].  R^j_{\alpha}
   = 0,
\eeq
where both terms vanish identically owing to (\ref{ijkrel}). This leads to
\beq
  y_i \left[
  \frac{\dr R^i_{\alpha}}{\delta \phi^j} R^j_{\beta}
  - (-1)^{\gras{\alpha} \gras{\beta}}
  \frac{\dr R^i_{\beta}}{\delta \phi^j} R^j_{\alpha}
  \right] = 0,
\eeq
since the two terms with a second derivative of the action $S$ cancel
each other.
The most general form of the {\it gauge algebra} now follows trivially from
(\ref{completeR}), which states that $T^{\alpha}_{\beta \gamma}[\phi]$
and $E^{ij}_{\alpha \beta} [\phi]$ exist, such that
\beq
  \frac{\dr R^i_{\alpha}}{\delta \phi^j} R^j_{\beta}
  - (-1)^{\gras{\alpha} \gras{\beta}}
  \frac{\dr R^i_{\beta}}{\delta \phi^j} R^j_{\alpha}
  = 2 R^i_{\gamma} T^{\gamma}_{\alpha \beta}  (-1)^{\gras{\alpha}}
  - 4 y_j E^{ji}_{\alpha \beta} (-1)^{\gras{i}} (-1)^{\gras{\alpha}}.
  \label{algebra}
\eeq
We introduced some numerical and sign factors for later convenience.
The different types of gauge algebras are classified as follows.
When $E^{ji}_{\alpha \beta} = 0$, one says that the algebra is {\it
closed}. Closed algebras can be divided in several categories, depending
on the type of structure functions $T^{\gamma}_{\alpha \beta}$.
If the structure functions are just arbitrary functions of the fields, the
algebra is called {\it soft}. When they reduce to field independent
constants, the {\it structure constants}, the algebra becomes an ordinary
{\it Lie algebra}, or their infinite dimensional generalisations.
When the structure constants all vanish, i.e. when
$T^{\gamma}_{\alpha \beta}=0 $, the algebra is {\it abelian}.
If the $E^{ji}_{\alpha \beta} \neq 0$, they are called the {\it
non-closure functions}, and the algebra is an {\it open algebra}.
Theories with this type of gauge algebra will require special attention.
Examples of all types will be given below when discussing their
quantisation.

Two more remarks are in order.
In \cite{bv4}, I.A. Batalin and G.A. Vilkovisky showed that it is always
possible to
choose a different set of generators for the gauge algebra, such that it is
closed. Once this is achieved, they even proved that any closed algebra is
a disguised form of an abelian one. However, in rewriting an (open) algebra
in its abelian form, other required features of the theory might get lost,
like a Lorentz covariant formulation or locality in some preferred set of
variables. Hence, the necessity remains to have a prescription for
quantising theories with an arbitrary algebra.

Furthermore, the following properties of the structure functions $T$ and
$E$ are useful for later developments. We find the Grassmann parities:
\begin{eqnarray}
      \gras{T^{\gamma}_{\alpha \beta}} & = & \gras{\alpha} + \gras{\beta} +
\gras{\gamma} \nonumber \\
      \gras{E^{ji}_{\alpha \beta}} & = &  \gras{\alpha}  +  \gras{\beta}
      +\gras{i}+\gras{j}.
\end{eqnarray}
Under the exchange of $\alpha$ and $\beta$, we see that
\beq
    T^{\gamma}_{\beta \alpha} = (-1)^{(\gras{\alpha} + 1)(\gras{\beta} +
1)} T^{\gamma}_{\alpha \beta},
\eeq
and an analogous property for $E^{ji}_{\beta \alpha}$.
The structure functions also satisfy the {\it Jacobi identity}. This can be
seen as follows. Define first the structure functions $t^{\gamma}_{\alpha
\beta} = 2 T^{\gamma} _{\alpha \beta} (-1)^{\gras{\alpha}}$,
for which under the exchange of
$\alpha$ and $\beta$ the property $t^{\gamma}_{\beta \alpha }
=(-1)^{\gras{\alpha}\gras{\beta} +1 }t^{\gamma}_{\alpha \beta}$ holds.
Define now
\begin{eqnarray}
   Y^i_{\alpha \beta \gamma} & = & \frac{\dr}{\delta \phi^k}
 \left[ \frac{\dr R^i_{\alpha}}{\delta \phi^j} R^j_{\beta}
  - (-1)^{\gras{\alpha} \gras{\beta}}
  \frac{\dr R^i_{\beta}}{\delta \phi^j} R^j_{\alpha} \right] R^k_{\gamma}
  \nonumber \\
  &  & - (-1)^{(\gras{\alpha}+\gras{\beta})\gras{\gamma}} \frac{\dr
R^i_{\gamma}}{\delta \phi^j}.
 \left[ \frac{\dr R^j_{\alpha}}{\delta \phi^k} R^k_{\beta}
  - (-1)^{\gras{\alpha} \gras{\beta}}
  \frac{\dr R^j_{\beta}}{\delta \phi^k} R^k_{\alpha} \right] \; ,
\end{eqnarray}
which is just the generalisation of $[A,[B,C]]$ to the case of graded
commutators. The Jacobi identity then follows from the fact that
\beq
   Y^i_{\alpha \beta \gamma} +
(-1)^{\gras{\alpha}.(\gras{\beta}+\gras{\gamma})} Y^i_{\beta \gamma \alpha}
(-1)^{\gras{\gamma}.(\gras{\beta}+\gras{\alpha})} Y^i_{\gamma \alpha \beta}
=0
\eeq
For a closed algebra, the Jacobi identity becomes
\begin{eqnarray}
     \label{jacobi}
     0 & = & (-1)^{\gras{\alpha}\gras{\gamma}} t^{\delta}_{\alpha \beta,k}
  R^k_{\gamma} + (-1)^{\gras{\alpha}\gras{\beta}} t^{\delta}_{\beta \gamma ,k}
  R^k_{\alpha} + (-1)^{\gras{\beta}\gras{\gamma}} t^{\delta}_{\gamma \alpha,k}
     R^k_{\beta}  \\
  &  & + (-1)^{\gras{\gamma}(\gras{\beta} + \gras{\mu})} t^{\delta}_{\mu
\gamma } t^{\mu}_{\alpha \beta} +
 (-1)^{\gras{\alpha}(\gras{\gamma} + \gras{\mu})} t^{\delta}_{\mu
\alpha} t^{\mu}_{\beta \gamma}  +
 (-1)^{\gras{\beta}(\gras{\alpha} + \gras{\mu})} t^{\delta}_{\mu
\beta } t^{\mu}_{\gamma \alpha } \nonumber \; .
\end{eqnarray}
Notice the terms with right derivatives of the structure functions
with respect to the fields $\phi^k$, denoted by $_{,k}$.

Now that we have classified the different types of gauge theories depending
on the properties of their gauge algebra, one final concept has to be
introduced, i.e. {\it reducible gauge theories}. As we have shown above,
gauge theories are characterised by the fact that the Hessian has zeromodes
on-shell, that is
\beq
         H_{ij}[\phi^l_0] . R^i_{\alpha}[\phi^l_0] = 0.
\eeq
Although we assumed completeness of the set of gauge generators, we did not
consider the fact that they might be dependent with respect to the index
$\alpha$. Indeed,
it can happen that operators $Z^{\alpha}_{\alpha_1}[\phi]$ exist, such
that
\beq
    R^i_{\alpha} [ \phi^l_0] Z^{\alpha}_{\alpha_1} [ \phi^l_0] = 0
    \label{redrel}  .
\eeq
Notice that this relation only has to hold on the stationary surface.
Again, a distinction could be made depending on whether $\alpha_1$
contains a space time index or not, but we will only consider the case
where it does. If the $Z^{\alpha}_{\alpha_1}[\phi^l_0]$
themselves are an independent set labelled by $\alpha_1$, one speaks of a
{\it first order reducible gauge theory}.
It is clear that (\ref{redrel}) expresses the fact that not
all zeromodes of the Hessian are independent. If no such
$Z^{\alpha}_{\alpha_1}$ exist, and hence all gauge generators
$R^i_{\alpha}$ are independent, one speaks of an {\it irreducible gauge
theory}.

Using the regularity of the theory, (\ref{redrel}) can be generalised
to a relation between the gauge  generators which is valid everywhere in
configuration space:
\beq
    R^i_{\alpha} [\phi] Z^{\alpha}_{\alpha_1} [\phi] = 2 y_j B^{ji}_{\alpha
_1} (-1)^{\gras{i}}.
\eeq
Some factors were again introduced for later convenience. Like for the
gauge generators themselves, we demand that the reducibility relations are
complete. That is,
\beq
   \forall X^{\alpha}_{\bar{\beta}} [\phi] : R^i_{\alpha}[\phi^l_0]
X^{\alpha}_{\bar{\beta}}[\phi^l_0] = 0 \;\; \Rightarrow \;\;
X^{\alpha}_{\bar{\beta}}[\phi^l_0] = Z^{\alpha}_{\alpha_1}[\phi^l_0]
N^{\alpha_1}_{\bar{\beta}} [\phi^l_0],
\eeq
or, in an off-shell notation,
\beq
X^{\alpha}_{\bar{\beta}}[\phi^l] = Z^{\alpha}_{\alpha_1}[\phi^l]
N^{\alpha_1}_{\bar{\beta}} [\phi^l] + y_i D^{\alpha i}_{\bar{\beta}}
[\phi^l].
\eeq

Of course, the $Z^{\alpha}_{\alpha_1}[\phi ^l_0]$ may not all be
independent. Some relations may exist among them
\beq
     Z^{\alpha}_{\alpha_1}[\phi^l_0]   Z^{\alpha_1}_{\alpha_2}[\phi^l_0]
     = 0,
\eeq
again on the stationary surface. If the $Z^{\alpha_1}_{\alpha_2}$ are
independent with respect to $\alpha_2$,
we have a {\it second order reducible gauge theory}.
As the reader has understood by now, this can go on to lead to reducible
theories of an arbitrary order. Even theories of infinite
order reducibility have been encountered (see e.g.\cite{Brink}).

Owing to the regularity condition, one could again start eliminating the
reducibility relations and construct a set of independent gauge generators.
However, the same objections as raised against the
abelianisation
procedure also apply here: Lorentz covariance and space-time locality may
be lost when rewriting the theory this way. We will not discuss the
quantisation of reducible gauge theories.

This finishes our overview of the different types of gauge algebras and
their basic properties. In the
following chapters, we first quantise the simplest cases, gradually
attacking the more difficult ones. We show how all the procedures we
will develop, can be incorporated into one, the Batalin-Vilkovisky scheme
for Lagrangian BRST quantisation. The latter is at present the most general
quantisation prescription at our disposal.

\chapter{Finite gauge transformations and the Faddeev-Popov quantisation
procedure}

In this chapter, we present a procedure which allows the quantisation of
theories with a closed, irreducible gauge algebra. In studying this
problem, the path integral formulation of quantum field theory proved
to be of major importance. The quantisation recipe described below, was
first developed by L.D. Faddeev and V.N. Popov \cite{FadPop} using path
integrals.
Only much later, an operator quantisation prescription was given by T. Kugo
and I. Ojima \cite{Kugo1}, based on BRST symmetry. We discuss BRST symmetry
in extenso in the next chapter. Whatever the point of view,
one is led to enlarging the set of field degrees of
freedom by the introduction of {\it ghost fields}. This latter fact was
already foreshadowed in \cite{Feynman,DeWitt}.

The basic rationale behind the definition of the path integral for gauge
theories, is that one should not integrate over the whole
configuration space, but only over the space of so-called {\it gauge
orbits}. These
are subsets of the configuration space consisting of configurations that
are related by (finite) gauge transformations, and therefore have the same
action.
Every gauge orbit contributes one term to the path integral summation. In
order to select one configuration on every gauge orbit, {\it gauge fixing
conditions} are introduced. The action of such a selected configuration
determines the contribution to the path integral of the gauge orbit it
belongs to. Of course,
the whole procedure should be independent of the gauge fixing conditions.

In the exposition, we will restrict ourselves again to bosonic degrees of
freedom and bosonic gauge symmetries, in order not to obscure the structure
of the derivation. In the first section, we describe the gauge orbits in a
bit more detail. There we already see one important difference between
theories with a closed or an open algebra. The Faddeev-Popov quantisation
prescription for closed algebras is developed in the second section.
The final section of this chapter contains an example: Yang-Mills theory.
We will treat this type of theory using the Faddeev-Popov
procedure.

\section{Gauge orbits}

Let us first define the gauge
orbits more carefully. Every gauge orbit is coordinatised by a set of
parameters $\theta^{\alpha}$, where the index $\alpha$ runs over all gauge
symmetries determined by the generators $R^i_{\alpha}$. Hence, loosely
speaking, one can say that the dimension of the gauge orbits is equal to
the number of gauge generators. The basic idea that we want to implement is
that two configurations on a gauge orbit that have infinitesimally
differing coordinates $\theta^{\alpha}$, are connected by an
infinitesimal gauge transformation. Following \cite{batalin1}, we introduce
a function $\phi^i(\theta)$, which satisfies the {\it Lie equation}
\footnote{We import terminology from the theory of Lie groups and Lie
algebras. If the structure functions of the closed algebra are structure
constants, then our considerations reduce to that case.
A point on a gauge orbit is then coordinatised by specifying a point on the
group manifold for every space-time point.} :
\beq
   \frac{\delta \phi^i(\theta)}{\delta \theta^{\beta}} =
R^i_{\alpha}[\phi^l(\theta)] \lambda^{\alpha}_{\beta}(\theta) \; .
   \label{LieEquation}
\eeq
The unspecified functions $\lambda^{\alpha}_{\beta}$ express the fact
that we can choose different
sets of gauge generators $R^i_{\alpha}$. On the next page, we show
that doing a coordinate transformation on the gauge orbits leads to other
functions $\lambda^{\alpha}_{\beta}$. We can choose any configuration
$\Phi ^i_0$ as a boundary condition for solving the Lie equation,
$\phi^i(\theta =0) = \Phi ^i_0$. All
configurations one gets by taking all values for $\theta^{\alpha}$ for a
fixed solution of the Lie equation are said to form a gauge orbit.
It is clear that all configurations on a gauge orbit have indeed the same
action:
\beq
       \frac{\delta S[\phi^i(\theta)]}{\delta \theta^{\alpha}} =
       y_i R^i_{\alpha} \lambda^{\alpha}_{\beta} = 0.
       \label{dSdtheta}
\eeq

The functions $\lambda^{\alpha}_{\beta}$ are not completely arbitrary.
They
have to satisfy the analogue of the {\it Maurer-Cartan} equation, which
follows from the requirement that (\ref{LieEquation}) be integrable. This
means that
\beq
    \frac{\delta^2 \phi^i(\theta)}{\delta \theta^{\gamma} \delta
\theta^{\beta}} -
    \frac{\delta^2 \phi^i(\theta)}{\delta \theta^{\beta} \delta
\theta^{\gamma}} = 0,
\eeq
which leads to
\beq
    \frac{\delta \lambda^{\alpha}_{\beta}}{\delta \theta^{\gamma}} -
    \frac{\delta \lambda^{\alpha}_{\gamma}}{\delta \theta^{\beta}}
    + t^{\alpha}_{\mu \nu} \lambda^{\mu}_{\beta} \lambda^{\nu}_{\gamma}
    = 0                .
    \label{MaurerCartan}
\eeq
To arrive at this result one has to use (\ref{LieEquation}) and the
commutation relation for a closed algebra ($E^{ij}_{\alpha \beta} =0$ in
(\ref{algebra})). For the Maurer-Cartan equation, one chooses the boundary
condition $\lambda^{\alpha}_{\beta}(\theta =0) = \delta^{\alpha}_{\beta}$.
The equation
(\ref{MaurerCartan}) itself is integrable for closed algebras due to the
Jacobi identity (\ref{jacobi}). This only works for closed algebras,
because for open
algebras an extra term, proportional to the field equations, appears in
(\ref{MaurerCartan}). The final result is that the Lie equation
(\ref{LieEquation}) as it stands above is not integrable for open algebras
\cite{bv4}, except of course on the
stationary surface where both the Maurer-Cartan equation and the Jacobi
identity have the same form for open as for closed algebras. This does
however
not mean that one can not define finite gauge transformations for open
algebras off the stationary surface.
We will point out the appropriate starting point below (\ref{xequations}).

The gauge orbits can be reparametrised by an invertible coordinate
transformation, specified by some functions
$\theta^{\alpha}(\xi^{\beta})$. It is then easy to see that
$\tilde{\phi}^i(\xi)=\phi^i(\theta(\xi))$ satisfies the Lie equation
\beq
       \frac{\delta \tilde{\phi}^i(\xi)}{\delta \xi^{\alpha}} =
       R^i_{\gamma}(\tilde{\phi}) \tilde{\lambda}^{\gamma}_{\alpha}(\xi),
\eeq
with $\tilde{\lambda}^{\gamma}_{\alpha}(\xi) =
\lambda^{\gamma}_{\beta}(\theta (\xi)) \frac{\delta \theta^{\beta}}{\delta
\xi^{\alpha}}$. It is as straightforward to show that if the Maurer-Cartan
equation is satisfied in one set of coordinates, it also is satisfied in
another. Hence, it is clear that different choices of functions
$\lambda^{\alpha }_{\beta}$ when writing down the Lie equation, give rise
to different coordinatisations of the gauge orbits. When describing the
Faddeev-Popov procedure, we will at every step keep invariance under
coordinate transformations of the gauge orbits. In section 4 of chapter 8,
changes of the set of generators are discussed from the point of view of
canonical transformations of the BV scheme.

Let us now restrict the freedom in the choice of the functions
$\lambda^{\alpha}_{\beta}$ by imposing the extra condition
\beq
    \lambda^{\alpha}_{\beta} \theta^{\beta} = \theta^{\alpha}.
    \label{restrictionlambda}
\eeq
This restriction allows the derivation of a different
differential equation that defines the gauge orbits and finite gauge
transformations. Define the functions
\begin{eqnarray}
      \varphi^i(x,\theta) & = & \phi^i(x\theta), \nonumber \\
      \Lambda^{\alpha}_{\beta}(x,\theta) & = & x \lambda^{\alpha}_{\beta}
      (x\theta).
\end{eqnarray}
The functions determining the gauge orbits are
$\varphi^i(1,\theta)$. Using (\ref{restrictionlambda}), the Lie equation
(\ref{LieEquation}) and the Maurer-Cartan equation (\ref{MaurerCartan})
lead to the following two equations for $\varphi^i$ and
$\Lambda^{\alpha}_{\beta}$:
\begin{eqnarray}
    \label{xequations}
    \frac{d \varphi^i(x)}{dx} & = & R^i_{\alpha}[\varphi^l] \theta^{\alpha}
     , \nonumber \\
    \frac{d \Lambda^{\alpha}_{\beta}(x)}{dx} & = & \delta^{\alpha}_{\beta}
    + t^{\alpha}_{\mu \nu} \theta^{\mu} \Lambda^{\nu}_{\beta},
\end{eqnarray}
with the boundary conditions $\varphi^i(x=0, \theta) = \Phi^i_0$ and
$\Lambda^{\alpha}_{\beta}(x=0, \theta) = 0$.

The latter equations (\ref{xequations}) can be taken as the starting point
for the definition of finite gauge transformations for closed as well
as for open algebras \cite{batalin1,bv4}. For closed algebras, the Lie
equation, the Maurer-Cartan equation and (\ref{restrictionlambda}) can be
rederived. For open algebras however, the Lie equation that
one obtains, contains an extra term proportional to field equations, the
hallmark of open algebras:
\beq
    \frac{\delta \phi^i}{\delta \theta^{\beta }} = R^i_{\alpha}
\lambda^{\alpha}_{\beta} + y_j M^{ji}_{\beta}.
\eeq
Instead of just two equations, one for $\phi ^i$ and one for the functions
$\lambda^{\alpha}_{\beta}$, one gets an infinite sequence of differential
equations. Every equation in this sequence is
obtained by imposing integrability on the previous equation.
For more details we refer again to \cite{batalin1,bv4}.

\section{The Faddeev-Popov procedure}

Having been through all this trouble to define gauge orbits, the derivation
of the Faddeev-Popov procedure \cite{FadPop} becomes quite straightforward.
The measure of the path integral can be decomposed into two pieces, an
integration over the different gauge orbits and an integration over all
configurations on a fixed orbit:
\beq
    [d\phi] = [d \Phi^i_0] . \prod_l [d \theta_l^{\alpha}]. \det \lambda
    .
\eeq
Symbolically, $[d \Phi_0^i]$ denotes the integration over the different
gauge orbits. We introduced an index $l$,
labelling the different gauge orbits. The coordinates of the $l$-th orbit
are $\theta^{\alpha}_l$, so $[d\theta^{\alpha}_l]$ denotes the integration
over the configurations that lie on the $l$-th orbit. We coordinatise all
orbits in
the same way, using the same functions $\lambda^{\alpha}_{\beta}$, although
this is not necessary. The $\det \lambda$ is introduced in order to define
the integration over the gauge orbits in a coordinate invariant way.

The culprit for the infinities of the naive path integral is of course the
integration over the coordinates of the gauge orbits, since the
classical action, and hence also the contribution to the naive path
integral,
does not change under variation of $\theta^{\alpha}$ (\ref{dSdtheta}). An
obvious way to
cure this problem, is to introduce $\delta$-functions that select a
specific set of values for the $\theta^{\alpha}$, i.e. that select one
configuration
on every orbit. Again maintaining coordinate invariance, we {\em replace}
$[d\phi]$ by
\beq
  [d\Phi^i_0] \prod_l [d\theta_l] \det \lambda .
  \frac{\delta \left( \theta^{\alpha}_l - \Theta^{\alpha}_l \right)}{\det
  \lambda} = [d\phi]. \prod_l
  \frac{\delta \left( \theta^{\alpha}_l - \Theta^{\alpha}_l \right)}{\det
  \lambda}.
\eeq
The $\Theta^{\alpha}_l$ are the coordinates of the configuration that is
selected on gauge orbit $l$. Notice that we do not need to fix the same
values for the coordinates on the different orbits.

This procedure is however not a very practical way of selecting one
configuration on every orbit. A more useful, but less explicit, way is by
choosing a set of {\it gauge fixing functions} $F^{\alpha}(\phi^i)$. There
are as many gauge fixing functions as there are coordinates for the gauge
orbit. Instead of introducing $\delta$-functions that select specific
values for the coordinates, we will introduce
$\delta(F^{\alpha}(\phi) - f^{\alpha}(x))$
in the measure, for some space time dependent functions $f^{\alpha}$. We
will not dwell here on the questions whether $F^{\alpha}(\phi) - f^{\alpha}
= 0$ has one solution on every gauge orbit or possibly more than one.
If on every orbit there is not exactly one solution,
there is omission or double counting of certain gauge orbits in
the path integral. These possible problems go under the name {\it Gribov
ambiguities}. Some details can be found in
\cite{Ramondboek,Gribov,Govaertsbook}.

For every gauge orbit, we can relate the $\delta$-function for the
coordinates to the $\delta$-function of the implicit gauge fixing
conditions in the usual way:
\beq
    \delta \left( F^{\alpha}(\phi(\theta_l)) - f^{\alpha} \right) =
    \frac{1}{\det M} \delta \left( \theta ^{\alpha}_l - \Theta^{\alpha}_l
    \right) \; .
\eeq
The $\Theta^{\alpha}_l$ are such that
$F^{\alpha}(\phi(\Theta^{\alpha}_l)) - f^{\alpha} = 0$.
The matrix $M$ is defined by
\beq
     M^{\alpha}_{\beta} = \frac{\delta F^{\alpha}(\phi (\theta))}{\delta
     \theta^{\beta}} \; .
\eeq
Using the Lie equation for {\it closed} algebras (\ref{LieEquation}), this
can be rewritten as
\beq
     M^{\alpha}_{\beta} = \frac{\delta F^{\alpha}(\phi (\theta))}{\delta
    \phi^i} R^i_{\gamma} \lambda^{\gamma}_{\beta}.
\eeq
Hence, the measure of the path integral is taken to be
\beq
  [d\phi] \rightarrow [d \phi] \frac{1}{\det \lambda}. \det M .
\delta(F^{\alpha} - f^{\alpha}) \;  ,
\eeq
and the complete path integral becomes
\beq
  \tilde{\cZ} = \int [d \phi] \frac{1}{\det \lambda}. \det M .
\delta(F^{\alpha} - f^{\alpha}) . e^{\frac{i}{\hbar} S} \; .
   \label{Zwelldefined}
\eeq

Of course, $\tilde\cZ$ is not a good starting point for diagrammatic
calculations.
Fortunately, both factors which appeared in the measure owing to the
definition of the path integral, can be rewritten using extra fields.
Enlarge the set of field degrees of freedom with one pair of fields
$(b_{\alpha},c^{\alpha})$ of
odd Grassmann parity for every gauge generator $R^i_{\alpha}$.
Then we can write
\beq
   \det M = \int [db][dc] \exp \left[ \frac{i}{\hbar} b_{\alpha}
    M^{\alpha}_{\beta} c^{\beta} \right]  = \int [db][dc]
e^{\frac{i}{\hbar} S_{ghost}}.
\eeq
These extra fields are called {\it ghost fields} or {\it ghosts}.
The $c^{\beta}$ are the {\it ghosts}, the $b_{\alpha}$ the {\it
antighosts}. In the literature, the antighosts are often denoted by
$\bar{c}_{\alpha}$ for historical reasons. However, in order to prevent the
misleading interpretations this might cause (see e.g. \cite{Kugo1} in this
context), we will not follow this tradition.

The dependence of $S_{ghost}$ on the coordinatisation of the gauge orbit,
via $M$, can be removed, by redefining $c^{\gamma}$ as
$\lambda^{\gamma}_{\beta} c^{\beta}$, leading to
\beq
   S_{ghost} = b_{\alpha} \frac{\delta F^{\alpha}}{\delta \phi^i}
R^i_{\beta} c^{\beta}.
\label{ghostaction}
\eeq
The Jacobian of this redefinition, $\det \lambda$, cancels with its inverse
in the path integral.

As could already be seen from (\ref{Zwelldefined}), the gauge fixing
conditions $F^{\alpha}$ are only {\it admissible} if $\det M \neq 0$. As
$M^{-1}$ is the propagator for the ghost, we see that this condition is
equivalent to saying that the ghosts have a well-defined propagator.

Finally, we can also rewrite the $\delta$-function in a more tractable
way. By construction, $\tilde{\cZ}$ does not depend on the specific
choice of $f^{\alpha}$ in the gauge fixing. This implies that we are
allowed to integrate over $f^{\alpha}$ with a suitable weight factor $W[f]$
such that
\beq
   \int [df] W[f] = 1.
\eeq
A very popular choice for $W[f]$ is a Gaussian damping factor, $W[f]=N
e^{\frac{i}{2\hbar} f^2}$. We then define
\begin{eqnarray}
      \label{hatcZ}
      \widehat{\cZ} & = & \int [df] \; \tilde{\cZ} \; W[f] \nonumber \\
      & = & \int [d\phi][db][dc] \; e^{\ihbar S_{com}}.
\end{eqnarray}
Here, the complete action is given by
\begin{eqnarray}
     S_{com} & = & S + S_{ghost} + S_{gf} \nonumber \\
     & =  & S + b_{\alpha} \frac{\delta F^{\alpha}}{\delta \phi^i}
     R^i_{\beta} c^{\beta}+ \frac{1}{2}F^2 .
     \label{Seff}
\end{eqnarray}
This way of gauge fixing is called {\it Gaussian gauge fixing}. The path
integral (\ref{hatcZ}) can be used as a starting point for perturbative,
diagrammatic calculations.
Before applying the Faddeev-Popov recipe to an example, let us finish this
section with some comments.

\begin{itemize}
\item
First of all, the generalisation to the case where the fields and
the gauge symmetries can have either Grassmann parities is
straightforward. The general rule is that for bosonic gauge symmetries
fermionic ghost fields are added to the configuration space and vice versa.
\item
In our derivation, we used the Lie equation for closed algebras, so we
should not apply this formalism to open algebras. However, it also fails
for reducible gauge theories. This can be seen as follows. In the
semiclassical approximation, the propagator for the quantum fluctuations of
the ghosts is the inverse of
\beq
     \frac{\delta F^{\alpha}(\phi^i_0)}{\delta \phi^i}
     R^i_{\beta} [\phi^i_0] \; .
\eeq
But for reducible gauge theories, this matrix has a zeromode owing to the
on-shell reducibility relations (\ref{redrel}). Notice that although the
symptoms are the same as for non admissible gauge choices, i.e. $\det
M=0$, the rationale is of course different. For reducible gauge theories
the problem exists for whatever gauge fixing function one chooses, while as
far as admissibility is concerned, the problem can be solved by choosing
the gauge fixing functions judiciously.
\item
A tacit assumption in the derivation is also that the final action
$S_{com}$
describes a local field theory. This restricts the choice of gauge fixing
functions $F^{\alpha}$, which are normally polynomials in the fields and a
finite order of their derivatives.
\item
Another weighing procedure that is often used is
\beq
   \int [d\lambda][df] \;\; e^{\ihbar \lambda_{\alpha} f^{\alpha}} = 1
   ,
\eeq
leading to the gauge fixed action $S_{com} = S + S_{ghost} +
\lambda_{\alpha} F^{\alpha}$. The {\it auxiliary field} $\lambda_{\alpha}$
is sometimes called the {\it Nakanishi-Lautrup} field. This way of gauge
fixing is called {\it delta function gauge fixing}.
\end{itemize}

This concludes the description of the first quantisation procedure that was
developed for gauge theories. To make the abstract construction more
concrete, we now turn to the example of Yang-Mills gauge theory.

\section{Example : non-abelian Yang-Mills theory}

Historically, the first gauge theories that were encountered are
electromagnetism and the theory of general relativity. The former is the
abelian case of the more general type of gauge theories which go under the
name {\it Yang-Mills theories}, as C.N. Yang and R.L.Mills were the first
to consider them \cite{YM}. All the experimentally confirmed
models fall in this class. The Faddeev-Popov quantisation
is hence sufficient to construct a quantum theory for any phenomenological
model in particle physics.

These models are based on a Lie
algebra, in the true mathematical sense, defined by a set of structure
constants $f^c_{ab}$:
\beq
     [ T_a , T_b ] = T_c f^c_{ab}.
\eeq
Here, the $T_a$ form a set of matrices that satisfy this commutation
rule\footnote{For a space-time dependent field which takes
its values in the algebra, we use the boldface style. For example,
$ \bA_{\mu}(x) = A^a_{\mu}(x) T_a$, $\bB(x) = B^a(x) T_a$.}.
Introduce the {\it covariant derivative}
\beq
    \bD_{\mu} \bB = \partial_{\mu} \bB + [ \bA_{\mu} , \bB].
\eeq
The curvature tensor of the gauge field is then defined by
\beq
   \bF_{\mu \nu} = \partial_{\mu} \bA_{\nu} -\partial_{\nu} \bA_{\mu}
                   + [\bA_{\mu} , \bA_{\nu}].
\eeq
Notice that the curvature tensor is antisymmetric in its two Lorentz
indices, $ \bF_{\mu \nu} = - \bF_{\nu \mu}$.

The action for pure Yang-Mills theory is
\beq
    S^{YM} = -\frac{1}{4} \int d^4 x \; \tr \bF_{\mu \nu} \bF^{\mu \nu},
\eeq
We work in the 4 space-time dimensions of experimental physics and
the trace of two generators of the algebra is normalised to $\tr T_a
T_b = \delta_{ab}$ in this representation. $S^{YM}$ is not an action of
free fields, as cubic and
quartic terms in the fields $A^a_{\mu}$ occur, except when $f^a_{bc}=0$,
in which case everything reduces to electromagnetism. By varying the
action $S^{YM}$ with respect to the field variable $A^a_{\mu}$, we get
\beq
   \delta S^{YM} \sim \int d^4x \; \tr [ \bD_{\mu}\bF^{\mu \nu} T_a ]
\delta A^a_{\mu},
\eeq
which gives the field equations
\beq
    \tr[ \bD_{\mu}\bF^{\mu \nu}(x) T_a ] = 0.
\eeq

Comparing with the general notation for field equations, we see that
$i=(\nu,a,x)$. The index $\alpha$ of the gauge generators runs over a
space-time point $y$ (local symmetry) and an index of the algebra $c$. The
gauge generators are then given by
\beq
    R^{\nu,a,x}_{y,c} = \left[\partial^x_{\nu} \delta^a_c + A^b_{\nu}
f^a_{bc} \right] \delta(x-y).
\eeq
$\partial^x$ denotes a derivative with respect to
$x$. The $R^i_{\alpha}$ do indeed satisfy
\begin{eqnarray}
y_i R^i_{\alpha} & = & \int d^4 x \; \tr[ \bD_{\mu}\bF^{\mu \nu}(x) T_a ]
   . \left[\partial_{\nu} \delta^a_c + A^b_{\nu} f^a_{bc} \right] \delta(x-y)
   \nonumber \\
   & = &  - \tr[ \bD_{\nu} \bD_{\mu}\bF^{\mu \nu}(y) T_c ]  \nonumber
   \\
   & = & + \frac{1}{2} \tr [ [\bF_{\mu \nu} , \bF^{\mu \nu}] T_c ]
\nonumber \\
   &=& 0,
\end{eqnarray}
owing to the antisymmetry of $\bF^{\mu \nu}$. It is easy to verify
that owing to the Jacobi identity, the gauge generators satisfy
the algebra commutation relations (\ref{algebra}) with
the same structure constants as the Lie algebra we started from and with
$E^{ij}_{\alpha \beta} = 0$.

A popular, and Lorentz covariant, gauge fixing condition for Yang-Mills
theories is given by the functions $F^{\alpha}$:
\beq
     F^a(x) = \partial^{\mu} A_{\mu}^a(x) .
\eeq
The ghost fields are $b_a(x)$ and $c^a(x)$ in this case. Following the
general prescription (\ref{ghostaction}), we get the ghost action
\beq
    S_{ghost} = \int d^4 x \; b_a \partial^{\mu} \left[ \partial_{\mu}
\delta_c^a + A_{\mu}^b f^a_{bc} \right] c^c.
\eeq
We rewrite this using the trace in any representation of the
algebra as
\begin{eqnarray}
   S_{ghost} & = & \int d^4 x \; \tr \left[ {\bf b} \partial^{\mu} (
\partial_{\mu} {\bf c} + [ \bA_{\mu} , {\bf c} ] ) \right] \nonumber \\
    & = & \int d^4 x \; \tr \left[ {\bf b} \partial^{\mu} \bD_{\mu} {\bf c}
\right].
\end{eqnarray}
If the algebra is abelian, the case of electromagnetism, then the ghosts
and the antighosts are not coupled to the gauge fields $A_{\mu}$ and are
free fields, since the covariant derivative in the ghostaction reduces to
an ordinary one. This is why quantum electrodynamics could be quantised
without
the need for (anti)ghosts. If the structure constants are not zero, a three
point vertex is present in the theory, where a gauge field couples to both
the ghost fields. The complete gauge fixed action becomes
\beq
  S_{com} = -\frac{1}{4} \int d^4 x \; \tr \bF_{\mu \nu} \bF^{\mu \nu}
    + \int d^4 x \; \tr \left[ {\bf b} \partial^{\mu} \bD_{\mu} {\bf c}
\right] + \frac{1}{2} \int d^4 x \; \tr (\partial_{\mu} \bA^{\mu})^2 \; ,
\eeq
if we use Gaussian gauge fixing.

\chapter{BRST quantisation}

The gauge fixed action we have obtained (\ref{Seff}) using the
Faddeev-Popov construction,
has no local gauge invariances anymore for admissible gauge choices.
This is of course the goal we wanted to achieve. However, the
gauge invariance of the classical theory is expected to manifest itself
in the quantum theory as well. This is indeed the case. The classical,
local
gauge invariance is traded for a global invariance, the so-called {\it
BRST-invariance}, which is present during the complete quantisation
process.
This acronym stands for C. Becchi, A. Rouet, R. Stora and I.V. Tyutin who
were the first to notice and use the fact that the gauge fixed action
(\ref{Seff}) has a global invariance \cite{BRST}.

We start by constructing the BRST transformation rules and by pointing
out how the gauge fixed action $S_{com}$ can be obtained from them. This
way, it will be immediately clear that $S_{com}$ is indeed invariant under
the transformation. This fact will then be used to derive the most general
Ward identity, which is at the heart of the proofs of
perturbative renormalisability and unitarity of gauge theories.
We point out the possible need for quantum corrections in order to
guarantee that the naive Ward identities are valid. We also
introduce the concept of BRST-cohomology. In a final subsection,
BRST--anti-BRST symmetry is briefly discussed.

\section{The BRST operator}

Starting from a classical action $S_0[\phi^i]$ with gauge invariances
determined by $R^i_{\alpha}[\phi]$ , the BRST transformation rules can be
constructed as follows. First of all, the BRST operator, denoted by
$\delta$, is a
fermionic, linear differential operator, acting from the right. This means
that
\beq
   \delta (X.Y) = X. \delta Y + (-1)^{\gras{Y}} \delta X . Y,
   \label{Leibniks}
\eeq
and that
\beq
     \gras{\delta X} = \gras{X} + 1.
\eeq
$\delta$ is completely known if we specify its action on the fields, as the
Leibnitz rule (\ref{Leibniks}) allows to work out how it acts on any
polynomial. With
every gauge generator $R^i_{\alpha}$ we associate a ghost field
$c^{\alpha}$ with statistics {\em opposite} to $\alpha$,
i.e. $\gras{c^{\alpha}} = \gras{\alpha} + 1$.
The BRST transformations of the classical fields $\phi^i$ are
defined to be
\beq
    \delta \phi^i = R^i_{\alpha}[\phi] c^{\alpha}.
\eeq
Notice that for function(al)s that only depend on these classical fields,
gauge invariance is equivalent to BRST invariance. Especially, the
classical
action is BRST invariant : $\delta S_0 = y_i R^i_{\alpha} c^{\alpha} = 0$.

A crucial property that we impose on $\delta$, is that it be a {\em
nilpotent} operator, i.e. that $\delta^2 = 0$. One easily verifies that if
$\delta$ is nilpotent on a set of fields --$\delta^2 A^l= 0$--, it is also
nilpotent when acting on any function(al) of these fields,
$\delta^2 F(A^l) = 0$. For {\sl closed} algebras, the nilpotency of the
BRST operator can be guaranteed easily by choosing the BRST transformation
of the ghost field. Imposing nilpotency on the classical fields $\phi^i$,
we get
\begin{eqnarray}
   0 & = & \delta^2 \phi^i \nonumber \\
     & = & R^i_{\alpha} \delta c^{\alpha} +
\frac{\dr R^i_{\alpha} c^{\alpha}}{\delta \phi^j} R^j_{\beta} c^{\beta}.
    \label{nilphi}
\end{eqnarray}
If we choose
\beq
    \delta c^{\gamma} = T^{\gamma}_{\alpha \beta}[\phi] c^{\beta}
c^{\alpha} \; ,
\eeq
the condition (\ref{nilphi}) is satisfied. Indeed, consider the closed
gauge algebra ($E^{ji}_{\alpha \beta} = 0$ in (\ref{algebra})) and multiply
both sides from the right with $(-1)^{\gras{\alpha} \gras{\beta} +
\gras{\beta}} c^{\alpha} c^{\beta}$. We obtain
\beq
   \frac{\dr R^i_{\alpha} c^{\alpha}}{\delta \phi^j} R^j_{\beta} c^{\beta}
   + R^i_{\gamma} T^{\gamma}_{\alpha \beta} c^{\beta} c^{\alpha} = 0.
   \label{newalgebra}
\eeq
The numerical and signfactors were precisely introduced in
(\ref{algebra}) to have this simple form here. Hence, we have that $\delta$
is nilpotent when acting on $\phi^i$.

It remains to verify that with the two definitions above, $\delta$ is also
nilpotent when acting on the ghost $c^{\gamma}$, i.e. that $\delta^2
c^{\gamma} =0$. To see that for closed algebras this is indeed the case, it
suffices to use the Jacobi identity (\ref{jacobi}).
We multiply all terms of that
identity from the right with $(-1)^{\gras{\alpha} \gras{\gamma}
+\gras{\beta}} c^{\gamma} c^{\beta} c^{\alpha}$, which gives
\beq
    \frac{\dr T^{\gamma}_{\alpha \beta}[\phi] c^{\beta} c^{\alpha}}{\delta
\phi^i} R^i_{\mu} c^{\mu} + 2 T^{\gamma}_{\alpha \beta} c^{\beta}
T^{\alpha}_{\mu \nu} c^{\nu} c^{\mu} = 0.
   \label{newjacobi}
\eeq
This result straightforwardly implies $\delta^2 c^{\gamma} = 0$.

So far, the BRST operator acts on fields and ghosts. One can however
always enlarge the set of fields by pairs of fields $A^l$ and $B^l$,
for an arbitrary set of indices $l$,
with the following set of BRST transformation
rules:
\begin{eqnarray}
       \delta A^l & = & B^l \nonumber \\
       \delta B^l & = & 0 \; .
\end{eqnarray}
The interpretation of this goes as follows. The classical action we started
from, does not depend on the fields $A^l$. Therefore, shifting
the $A^l$ over an arbitrary amount is a symmetry of that action. It is a
local symmetry and the $B^l$ are the ghosts associated with this shift
symmetry. The nilpotency of the BRST operator is clearly maintained by this
way of extending the set of fields.
Such pairs of fields form together a {\it trivial system}.
Although this is at first sight a rather
trivial construction, hence the name, it has many applications. For
instance,
the antighosts which were used in the Faddeev-Popov procedure are exactly
introduced in this way:
\begin{eqnarray}
       \delta b_{\alpha} & = & \lambda_{\alpha} \nonumber \\
       \delta \lambda_{\alpha} & = & 0
\end{eqnarray}

In the configuration space, one can define {\it gradings}. Every field can
be assigned a sort of {\sl charge}, and, as is typical for the $U(1)$
charge that is known from electromagnetism, the {\sl charge} of a monomial
in the fields is the sum of the {\sl charges} of the fields of which it is
the product. We already have been using an example of such a grading, the
Grassmann parity of every field. In that case, the {\sl charge}, i.e. the
parity,
takes the values $0$ or $1$. A new grading that we will use very often is
called {\it ghost number}, and the two basic assignments are:
\begin{eqnarray}
               \gh{\phi^i} & = & 0 \nonumber \\
               \gh{c^{\alpha}} & = & 1 .
\end{eqnarray}
The classical action satisfies $\gh{S_0}=0$, and we impose the same
requirement on the gauge fixed action (\ref{Seff}) which we obtained using
the Faddeev-Popov prescription. As a result, we have that $\gh{b_{\alpha}}
= -1$ and $\gh{\lambda_{\alpha}} = 0$.

The BRST operator also carries a ghostnumber, $\gh{\delta} = 1$. This means
that for any functional $F$ with a specific ghostnumber,
\beq
    \gh{\delta  F} = \gh{F} + 1.
\eeq
All the examples above, were the $F$ are just fields, satisfy this rule.
The ghostnumber serves many times as a good bookkeeping device.

We come to the main purpose of the construction of the nilpotent
BRST operator: it can be used to construct the gauge fixed
Faddeev-Popov action of the previous chapter (\ref{Seff}). The claim is
that
\beq
     S_{com} = S_0 + \delta \Psi .
     \label{SEFF}
\eeq
$\Psi$ is called the {\it gauge fermion}, as it obviously has to have
odd Grassmann parity.
Since we want the gauge fixed action to have ghostnumber zero, and
since the BRST operator raises the ghostnumber with one unit, $\gh{\Psi } =
-1$. The $\phi^i$ and $c^{\alpha}$
have respectively ghostnumber zero and one, which does not allow the
construction of a suitable $\Psi $. Therefore, we introduce a trivial pair
$(b_{\alpha},\lambda^{\alpha})$ with $\gh{b_{\alpha}} = -1$. Given any set
of admissible gauge fixing functions $F^{\alpha}$, we can consider $\Psi =
b_{\alpha} (F^{\alpha} - \lambda^{\alpha} a )$, which leads to the terms
\begin{eqnarray}
    \label{SEFFP}
\delta \Psi  & = & \delta \left[ b_{\alpha}
  (F^{\alpha} - \lambda^{\alpha} a ) \right] \nonumber \\
  & = & b_{\alpha} \frac{\dr F^{\alpha}(\phi)}{\delta \phi^i} R^i_{\beta }
  c^{\beta } + F^{\alpha} \lambda_{\alpha} - a \lambda^2
(-1)^{\gras{\alpha}}.
\end{eqnarray}
These are precisely the terms we got in the previous chapter (\ref{Seff})
by rewriting the determinant and the gauge fixing $\delta$-function. A
commuting parameter $a$ was introduced. Notice that the
$a$-dependent
term is only present for bosonic gauge symmetries, because for fermionic
gauge symmetries $\lambda$ is fermionic and hence its square is zero. For
bosonic gauge symmetries, $a = 1$ gives the Gaussian gauge fixing after
integrating over $\lambda$, while $a=0$ leads to the $\delta$-function
gauge.

Now that we have established that the gauge fixed action that one obtains
using the Faddeev-Popov procedure is of the form $S_{com} = S_0 + \delta
\Psi $, it is clear that $\delta S_{com} = 0$, since the classical action
is BRST invariant. Our main conclusion is then that {\it after the gauge
fixing, the original gauge invariance manifests itself as the global
BRST invariance of the gauge fixed action}. In section 3, we
argue that also for operators other than the action, gauge invariance is to
be replaced by BRST invariance (\ref{delom}).

In the BRST quantisation scheme we are not obliged to take for
the gauge fermion the one taken above (\ref{SEFFP}) in order to reproduce
the Faddeev-Popov expression for the gauge fixed action. More general
choices are allowed, leading e.g. to four ghost interactions. In the next
section we prove --formally-- that as long as $\Psi$ leads to path
integrals that
are well-defined, meaning that they do not have gauge invariances, the
partition function is indeed independent of the gauge fixing fermion.

The quantisation based on BRST symmetry also needs to be modified in order
to handle gauge theories with an open algebra. With the
two basic definitions of $\delta \phi^i$ and $\delta c^{\gamma}$ of above,
the BRST operator is not nilpotent when the algebra is open. Indeed,
$\delta^2
\phi^i$ becomes proportional to field equations, as such a term appears in
(\ref{newalgebra}) for open algebras. As the nilpotency of $\delta$ is
crucial for having a BRST invariant action after gauge fixing, we see
that this latter invariance is not present for open algebras, when
quantised as described above. In chapter 6, we show how the BRST
transformations and the gauge fixing procedure have to be modified in the
case of an open algebra, in order to end up with a BRST invariant gauge
fixed action.

\section{Ward identities}

In the previous section, we have seen that the classical gauge symmetry
gives rise to the BRST invariance of the gauge fixed action. In this
section, we derive the Ward identity $\langle \delta  X \rangle = 0$.

Given an action\footnote{The $\phi^A$ denote all the fields present in
the gauge fixed action:$\phi^i$,$c^{\alpha}$,trivial pairs.}
$S[\phi^A]$ that is invariant under the BRST transformation
rules $\delta \phi^A$, we have that for any $X(\phi)$
\beq
    \langle \delta  X(\phi) \rangle = \int [d\phi] . \delta X . e^{\ihbar
S[\phi]} = 0   \; .
    \label{WaTaSlaTa}
\eeq
The simple proof goes as follows. Consider the expectation value of an
operator $X(\phi)$:
\beq
   \label{William1}
   \chi = \int [d\phi] . X(\phi) . e^{\ihbar S[\phi]}  \; .
\eeq
We are always allowed to redefine the integration variables
\footnote{We use a global parameter $\mu$ of Grassmann parity one
(a fermionic parameter) to construct $\tilde{\delta} A = \delta A
.\mu$. Then $\gras{\tilde{\delta} A} = \gras{A}$.}
$\phi \rightarrow \phi + \delta \phi . \mu$. If we assume for the moment
that this redefinition does not give a Jacobian, we have
\beq
   \label{William2}
   \chi = \int [d\phi] . [ X(\phi) + \delta X.\mu] . e^{\ihbar
S[\phi+\delta \phi .\mu ]} \; .
\eeq
Since $S[\phi]$ is BRST invariant, we can subtract (\ref{William1}) from
(\ref{William2}) to find
\beq
     \int [d\phi] . \delta X . e^{\ihbar S[\phi]} = 0   \; ,
\eeq
which is the desired result. We will come back to the Jacobian that can
appear in the redefinition at the end of this section. Notice that once a
BRST invariant action is given, nowhere in this proof the nilpotency of
$\delta$ is required.

{}From this simple property, all Ward-Takahashi-Slavnov-Taylor
identities\footnote{Below we will use the name Ward identities.} \cite{WTST}
can be
derived by choosing $X$. The use of these Ward identities is manifold.
In the first place, they allow us to prove that the gauge fixed actions
of
the form $S_{com} = S_0 +\delta \Psi$ do indeed lead to partition functions
which are independent of the gauge fixing, that is, of $\Psi$. Consider
two partition functions, one calculated with the gauge fermion $\Psi$ and
one with an infinitesimally different gauge fermion $\Psi +d \Psi $.
In an obvious notation we have that
\begin{eqnarray}
    \label{dpsiward}
   \cZ_{\Psi +d \Psi} - \cZ_{\Psi} & = & \int [d\phi] \exp \left[
\ihbar(S_0 + \delta [\Psi +d \Psi ]) \right]
- \int [d\phi] \exp \left[ \ihbar(S_0 + \delta \Psi
) \right] \nonumber \\
     & = & \ihbar \langle \delta d \Psi \rangle_{\Psi} \\
     & = & 0. \nonumber
\end{eqnarray}
This shows that the quantisation based on BRST symmetry is internally
consistent. No reference is needed to the results obtained with the
Faddeev-Popov procedure.

Perturvative proofs of unitarity and renormalisability are also
heavily based on the Ward identities.
For instance, the different divergences that can occur when doing loop
calculations are related as a consequence of these Ward identities. This
retricts the number of independent divergences that have to be absorbed in
the parameters of the theory and thus aids in proving renormalisability.
For examples in practical calculations, see e.g. chapter 7 of
\cite{ItzyZuber}.
A treatment which strongly stresses the perturbative, diagrammatic point of
view can be found in \cite{THV}. More formal discussions, e.g. on the use
of BRST invariance to constrain the renormalised effective action, can be
found in \cite{Zinn-Justinboek,Baulieu1}.

Let us now correct for our carelessness in the proof of the Ward identity
and investigate what effect a possible Jacobian has on the derivation of
the Ward identity. The Jacobian is given by\footnote{We follow the
conventions of \cite{superBryce}. When doing a redefinition of integration
variables $y^i = y^i(x)$, the measure changes by
\beq
   dy = \sdet \left[ \frac{\dl y^i(x)}{\delta x^j} \right] dx \; .
\eeq
The supertrace and the superdeterminant are related by
$ \delta \ln \sdet M = \str (M^{-1} \delta M)$. The supertrace is defined
here by
\beq
     \str \left[ \frac{\dl y^i(x)}{\delta x^j} \right] = (-1)^{\gras{i}}
     \frac{\dl y^i(x)}{\delta x^i} \; .
\eeq}
\beq
 \label{BRSTJacob}
 \sdet \left[ \delta^A_B + \frac{\dl (\delta \phi^A)}{\delta  \phi^B}
\mu \right] = \exp \left[ \str
 \frac{\dl (\delta \phi^A)}{\delta \phi^B} \mu \right]
 \stackrel{not.}{\equiv} e^{{\cal A} \mu}.
\eeq
Instead of (\ref{William2}), we get
\beq
   \chi = \int [d\phi] . [ X(\phi) + \delta X.\mu] . e^{\ihbar
S[\phi]}. [ 1 + \cA.\mu] \; .
    \label{William2p}
\eeq
If we subtract (\ref{William1}) from (\ref{William2p}) we find
\beq
    \label{AX}
    \langle \delta  X(\phi) \rangle + \langle X \cA \rangle = 0.
\eeq

So, this Jacobian really makes a difference, it gives a correction
to the Ward identity. Suppose now that a functional
$M(\phi)$ --a {\it local} function of the fields-- exists such that $
\delta M
= i {\cal A} $. Then, by considering as weight of the path integral $S + \hbar
M$ instead of $S$, one can still derive the result that
\beq
 \int [d\phi] e^{\frac{i}{\hbar}( S + \hbar M) } . \delta X = 0,
\eeq
since the possible non-invariance of the measure is then cancelled
by the non-invariance of this $M$ when repeating the derivation above.
Hence, we see that
possible Jacobians are quite harmless if they are the BRST variation of
something, because adding this something as so-called {\it counterterm} enables
us to derive the so desired Ward identity. When such a local $M$ can not be
found, and consequently we can not get our naive result, we speak of a {\it
genuine anomaly}. The Ward identities are then modified in the quantum
theory, and one says that the gauge symmetry is {\it anomalously broken}.
Notice also that $M$ is imaginary as $\cA$ is real,
at least when we work in Minkowski space.

Let us stress once
again that the manipulations above are at a formal level. As soon as one
tries to evaluate such a Jacobian, one runs into the usual infinities of
quantum field theory. The Jacobians are often neglected on formal grounds
(see e.g. the argumentation in section 4 of \cite{Baulieu1}, or in the
introduction of \cite{ik2}). K. Fujikawa was the first to point out that
genuine anomalies can be understood in the functional integral framework
for quantisation as coming from Jacobians \cite{Fujikawa}. He also proposed
a regularisation method to obtain finite expressions for these Jacobians.
The third part of this work is devoted to a one-loop
regularised treatment of these counterterms and of genuine anomalies.

\section{BRST cohomology}

Using the Ward identity, we can also see that the classical gauge
invariance is replaced by BRST invariance for the full quantum theory.
In classical electrodynamics, for instance, the quantities that one can
measure experimentally, the electric and magnetic vector fields $\vec{E}$
and $\vec{B}$, are independent of the chosen gauge. Analogously, we can
look for operators that have a quantum expectation value that is
invariant
under (infinitesimal) changes of the gauge fixing fermion. Consider any
local operator $\Omega[\phi^A]$, i.e. a function(al) of the fields
$\phi^A$
and a finite number of their derivatives. Adopting the same notation as
above (\ref{dpsiward}), we
express this gauge invariance condition as
\begin{eqnarray}
   0 & = & \langle \Omega[\phi] \rangle_{\Psi + d \Psi}
           -\langle \Omega[\phi] \rangle_{\Psi}  \nonumber \\
     &= & \ihbar \int [d\phi] \;\;  \delta d \Psi . \Omega .
     e^{\ihbar(S_0 +\delta \Psi + \hbar M)}     \\
     &= & \ihbar \int [d\phi] \;\; \delta \Omega . d \Psi .
     e^{\ihbar(S_0 +\delta \Psi + \hbar M)} \nonumber.
\end{eqnarray}
The last step is obtained by using the Ward identity, which allows to {\it
integrate by parts} the BRST operator. As this should be zero for any
infinitesimal deformation $d \Psi$ of the gauge fermion, we see that
{\it gauge invariant operators} are characterised by the condition
\beq
    \label{delom}
    \delta \Omega[\phi] = 0,
\eeq
that is, they {\em have to be BRST invariant}. Hence, the classical gauge
invariance, which is a property of functions of the original fields
$\phi^i$, is replaced by BRST
invariance, which is a property of functions depending on all
variables $\phi^A$, including the ghosts. This generalises the conclusion
of the previous section that gauge invariance of the classical action is
traded for BRST invariance of the gauge fixed action.

Owing to the Ward identity, one sees that two operators that
differ by the BRST variation of something, i.e. $\Omega_2 = \Omega_1 +
\delta \Theta$ have the same expectation value. Indeed,
\beq
    \langle \Omega_2 \rangle_{\Psi} = \langle \Omega_1 \rangle_{\Psi} +
    \langle \delta \Theta  \rangle_{\Psi} = \langle \Omega_1
\rangle_{\Psi}.
\eeq

Since the BRST operator is nilpotent for closed algebras, we can combine
the latter two results in the following {\bf theorem}. {\it The gauge
invariant operators are the non-trivial cohomology classes
of the BRST operator $\delta$ at ghostnumber zero}.

What do we
mean by this? Whenever one has a nilpotent operator acting on some space,
one can define equivalence classes by
\beq
     X \sim Y \Leftrightarrow \exists Z : X = Y + \delta Z.
\eeq
Two elements of the same equivalence class lead to the same result when
$\delta$ acts on them because of the nilpotency, $X \sim Y
\Rightarrow \delta X=\delta Y$ . Because of the condition
of BRST invariance, we are only interested in those equivalence classes
which are in the kernel of $\delta$.
We adopt terminology familiar from differential geometry, where the
exterior derivative $d$ which acts on forms is also nilpotent.
A quantity $Z$ is {\it BRST exact} if it is the BRST variation
of something, that is if $Z = \delta X$. A quantity $Y$ is {\it BRST
closed}, if it has a vanishing BRST variation, $\delta Y = 0$.
The equivalence classes are called {\it cohomology classes}. Of course, we
can construct operators of any ghostnumber. Hence the addition {\sl at
ghostnumber zero} for the physical observables, which means that we only
consider the cohomology classes of operators with ghostnumber zero. The
cohomology classes at ghostnumber one are important for the
nonperturbative study of anomalies, as will be pointed out below (see the
discussion following (\ref{anosom})).

One final comment. Above we based the equivalence of operators on
whether they had the same quantum expectation value or not. We can replace
in the above discussion {\it expectation value} by {\it  correlation
functions with BRST invariant operators}.
However, two operators which differ by a BRST exact term need
not have the same correlation functions with non BRST invariant operators.

\section{BRST--anti-BRST symmetry}

In this section, we give a short discussion of an extension of the BRST
construction of above, which goes under the name {\it extended BRST
symmetry} or {\it BRST--anti-BRST symmetry}\footnote{We will use the name
BRST--anti-BRST symmetry when we refer to the construction presented in
this section. In the rest of this work, we will sometimes use {\em extended
BRST transformation rules} which will {\em not} refer to this section.}. In
the literature
also the name $Sp(2)$ {\it invariant quantisation} is sometimes used.
This anti-BRST symmetry  was first discovered
as {\sl yet another symmetry} of gauge fixed actions, where instead of the
ghost field, the antighost field plays a more important role \cite{CurOj}.
Although this BRST--anti-BRST symmetry pops up from time to
time, it does not seem to add anything substantial to the BRST quantisation
procedure. However, a superfield formulation of Yang-Mills theory has
been constructed using it \cite{Baulieu1,BoTo}.

Instead of introducing one BRST operator, two operators are defined,
$\delta_1$ and $\delta_2$, also sometimes denoted $\delta$ and
$\bar{\delta}$. Both are fermionic, linear differential operators acting
from the right. $\delta_1$ is called BRST-operator, $\delta_2$ anti-BRST
operator. We construct them such that the BRST operator
raises the ghostnumber by one, while the anti-BRST operators lowers it with
the same amount.
The configuration space is extended by the introduction of two
ghost fields, $c^{\alpha}_1$ and $c^{\alpha}_2$, with $\gh{c^{\alpha}_a} =
- (-1)^a$. So, using the terminology of above, $c^{\alpha}_1$ is a ghost
and $c^{\alpha}_2$ an antighost, as far as ghostnumber is concerned. The
basic transformation rules are given by:
\begin{eqnarray}
     \delta_a \phi^i & = & R^i_{\alpha} c^{\alpha}_a \nonumber \\
     \delta_a c^{\alpha}_a & = & T^{\alpha}_{\beta \gamma} c^{\gamma}_a
c^{\beta}_a,
\end{eqnarray}
where in the last line there is {\sl no} summation over $a$. Further,
\beq
   \label{som}
   \delta_1 c^{\alpha}_2 + \delta_2 c^{\alpha}_1 =
   2 T^{\alpha}_{\beta \gamma} c^{\gamma}_1 c^{\beta}_2.
\eeq
This requirement does not yet fix $\delta_1 c^{\alpha}_2$ and
$\delta_2 c^{\alpha}_1$. This is done by introducing an extra field
$b^{\alpha}$ with $\gh{b}=0$, and with transformation properties
\begin{eqnarray}
        \delta_1 c^{\alpha}_2 & = & b^{\alpha} \nonumber \\
        \delta_1 b^{\alpha} & = & 0.
\end{eqnarray}
The transformation of the ghost under the anti-BRST operator is then
\beq
   \delta_2 c^{\alpha}_1  =  - b^{\alpha} +
   2 T^{\alpha}_{\beta \gamma} c^{\gamma}_1 c^{\beta}_2,
\eeq
as a consequence of (\ref{som}).

Finally, $\delta_2 b^{\alpha}$ is determined by imposing nilpotency
of the anti-BRST operator on $c^{\alpha}_1$. We find:
\beq
   \delta_2 b^{\alpha} = - 2 T^{\alpha}_{\gamma \beta} c^{\beta}_2
b^{\gamma} - \frac{\dr T^{\alpha}_{\mu \nu} c^{\nu}_1 c^{\mu}_2}{\delta
\phi^k} R^k_{\gamma} c^{\gamma}_2.
\eeq
We used the equality that one obtains by multiplying the Jacobi identity
with $(-1)^{\gras{\alpha}\gras{\gamma} + \gras{\beta}} c^{\gamma}_2
c^{\beta}_1 c^{\alpha}_2$, in analogy with (\ref{newjacobi}).
With these transformation rules, both the BRST
and the anti-BRST operator are nilpotent. Moreover,
\beq
    \delta_1 \delta_2  + \delta_2 \delta_1 = 0.
\eeq
This can again be checked with the by now familiar tricks based on the
closed gauge algebra and the Jacobi identity.

Gauge fixing in a BRST--anti-BRST invariant way is done by adding
\beq
    S_{gf} = \frac{1}{2} \epsilon^{a b} \delta_a \delta_b \Psi
\eeq
to the classical action \footnote{$\epsilon^{a b} = - \epsilon^{b a}$ is
the antisymmetric tensor, which is typical for $Sp(2)$. Our convention is
such that $\epsilon^{12}=-1$.}. Since the classical action is invariant
under both BRST and anti-BRST transformations and since
$\delta_1 +\delta_2$ is nilpotent, this leads indeed to gauge
fixed actions which are invariant under BRST--anti-BRST symmetry. In
contrast to ordinary BRST quantisation, $\gh{\Psi} = 0$.

As the gauge fixed action is now invariant under both the BRST and the
anti-BRST symmetry, there are two general Ward identities.
We have
\begin{eqnarray}
     \langle \delta_1 X \rangle & = & 0 \nonumber \\
     \langle \delta_2 X \rangle & = & 0.
\end{eqnarray}
Again, these Ward identities guarantee that the partition functions are
gauge independent.

We will come back to BRST--anti-BRST symmetry only once below. We
construct an antifield scheme for BRST--anti-BRST invariant quantisation
in chapter 7.

\section{Overview}

In this chapter, we have shown that in the process of constructing a gauge
fixed action, the local gauge invariance is replaced by a global
invariance, the BRST invariance. The BRST invariance of the quantum theory
implies Ward identities $\langle \delta X \rangle = 0$ if the theory
is anomaly free. These Ward identities are relations between correlation
functions, encoding the consequences of gauge symmetry for the full quantum
theory.

We have argued that gauge invariant operators are the cohomology classes of
the BRST operator at ghostnumber zero. Finally, we have discussed a
generalisation of the BRST symmetry, that treats ghosts and antighosts on
an equal footing. In the next chapter, we show how Schwinger-Dyson
equations can be obtained as Ward identities of a BRST symmetry.

\chapter{The Schwinger-Dyson BRST-symmetry}

The Schwinger-Dyson equations \cite{SDEq} are equations satisfied by the
Green's functions of any theory, with or without gauge symmetries.
In principle, they determine the quantum theory completely. The
Schwinger-Dyson equations (SD equations) are the quantum equations of
motion. In the standard textbook arguments
\cite{ItzyZuber,Zinn-Justinboek}, they are derived as a consequence of the
generalisation to path integrals of the invariance of an integral under a
redefinition of the integration variable from $x$ to $x+a$. They are used
in
several domains of quantum field theory, like the study of bound states and
the study of theories with spontaneously or dynamically broken symmetries,
to name only
two. In this brief chapter, we show how Schwinger-Dyson equations can
be obtained as Ward identities of an extra symmetry following
\cite{AD1,AD2}. This serves both as an illustration of the
previous section in a simple setting and as one of the cornerstones of the
subsequent developments.

Start from any action $S_0[\phi^i]$ which leads to a well defined path
integral. This may be an action without gauge symmetries, or an action
obtained after gauge fixing. In the latter case the $\phi^i$ contain the
gauge fields, the ghosts and antighosts and possible auxiliary fields.
Double now the configuration space, by copying every field $\phi^i$ with a
{\it collective field} $\varphi^i$, and consider the action
$S_0[\phi^i-\varphi^i]$. This action now has a gauge symmetry, it is
invariant under $\delta \phi^i = \epsilon^i$, $\delta \varphi^i =
\epsilon^i$. Introducing a ghost field for this {\it shift symmetry}, we
have the following BRST transformation rules:
\begin{eqnarray}
        \delta \phi^i & = & c^i \nonumber \\
        \delta \varphi^i & = & c^i \nonumber \\
        \delta c^i & = & 0,
\end{eqnarray}
which is obviously nilpotent. We want to remove the gauge symmetry by gauge
fixing the collective field to zero. For that purpose, we introduce a
trivial pair consisting of an antighost $\phi^*_i$ and an auxiliary field
$b_i$. They
have the BRST transformations $\delta \phi^*_i = b_i$ and $\delta b_i = 0$.
The gauge fixing functions are taken to be $F^i = \varphi^i$, i.e. we fix
the collective field $\varphi^i$ to zero. The gauge fixing
is done by adding to the classical action $S_0$ the term $\delta [ \phi^*_k
\varphi^k]$. We obtain
\begin{eqnarray}
    \label{fisterc}
       S_{com} & = & S_0[\phi -\varphi] + \delta [\phi^*_k \varphi^k ]
\nonumber
\\ & = & S_0 [ \phi -\varphi ] + \phi^*_k c^k + (-1)^{\gras{k}} b_k
\varphi^k.
\end{eqnarray}
The second term is the ghost action (\ref{ghostaction}), the third the
delta function gauge fixing\footnote{Notice the notation. The $\phi^*_i$ is
now the antighost ($b_{\alpha}$ in (\ref{ghostaction})) while $b_i$ is now
the Nakanishi-Lautrup field ($\lambda^{\alpha}$ of (\ref{SEFFP})).}. The
ghostnumber assignments here are \begin{eqnarray} \label{ghnranti}
        \gh{\phi^i} & = & \gh{\varphi^i} = x^i \nonumber ,\\
        \gh{c^i} & = & x^i + 1 \nonumber, \\
        \gh{\phi^*_i} & = & - x^i -1 = -\gh{\phi^i} - 1.
\end{eqnarray}
Notice especially the last line, which we will encounter again in the next
chapter.

Owing to the BRST invariance of this gauge fixed action, we have the Ward
identity (\ref{WaTaSlaTa}):
\beq
   0 = \langle \delta \left[ \phi^*_i X(\phi) \right] \rangle
   = \int [d\phi] [d\varphi] [dc] [d\phi^*] [db] \left[ \phi^*_i \frac{\dr
X}{\delta \phi^j} c^j + (-1)^{\gras{X}} b_i X \right] e^{\ihbar S_{gf}} \;
{}.
\eeq
Notice that, at least at the formal
level, no counterterms are needed to maintain the BRST invariance, as the
measure of the functional integral is assumed to be translational
invariant. We will now integrate over the
Nakanishi-Lautrup field $b_i$, the collective field $\varphi^i$ and over
the ghost and antighost field. As a result, we see that this Ward identity
is nothing
but the Schwinger-Dyson equation satisfied by $X(\phi)$.

In the first term the integral over $b$ and $\varphi$ can be done
trivially, fixing the collective field $\varphi$ to zero. This leads to
\beq
    T_1 = \int [d\phi] [dc] [d\phi^*]  \left(
    \phi^*_i \frac{\dr X}{\delta \phi^j} c^j  \right)
  e^{\ihbar \left( S_0[\phi] + \phi^*_k c^k \right)}.
\eeq
Integrating over the ghosts is a bit more subtle, as they also occur
outside the exponentiated action. We rewrite
\beq
    T_1 = \int [d\phi] [dc] [d\phi^*] \;\; . \; \phi^*_i \frac{\dr
X}{\delta \phi^j} e^{\ihbar  S_0[\phi]} . \;\;
\frac{\hbar}{i} \frac{\dl}{\delta \phi^*_j} e^{\ihbar \phi^*_k c^k}.
\eeq
Integrating by parts, and integrating over the ghostfields, we get
\beq
    T_1 = - (-1)^{(\gras{X}+1)(\gras{i}+1)} \int [d\phi] \;\; \frac{\dr
X}{\delta \phi^i} . \frac{\hbar}{i} e^{\ihbar S_0}.
\eeq

For the second term, we do an analogous manipulation to integrate out the
collective and auxiliary field:
\beq
    T_2 = \int [d\phi] [d\varphi] [dc] [d\phi^*] [db](-1)^{\gras{X}}
\frac{\hbar}{i} \frac{\dl}{\delta \varphi^i} \left[ e^{\ihbar \varphi^k
b_k} \right] . X(\phi) . e^{\ihbar ( S_0[\phi -\varphi] + \phi^*_k c^k )}.
\eeq
Again integrating by parts, using that $\frac{df(x-y)}{dx} =
-\frac{df(x-y)}{dy}$, and integrating out all the fields of the BRST shift
symmetry leads to
\beq
   T_2 = - \int [d\phi] (-1)^{(\gras{X}+1)(\gras{i}+1)} X(\phi) \frac{\dr
S_0[\phi]}{\delta \phi^i} e^{\ihbar S_0}.
\eeq

Adding up both terms, we find that
\beq
    \langle X(\phi) \frac{\dr S_0}{\delta \phi^i} + \frac{\hbar}{i}
\frac{\dr X(\phi)}{\delta \phi^i} \rangle = \int [d\phi] \left[
 X(\phi) \frac{\dr S_0}{\delta \phi^i} + \frac{\hbar}{i}
\frac{\dr X(\phi)}{\delta \phi^i} \right] e^{\ihbar S_0}= 0 .
\eeq
This is the general form of the Schwinger-Dyson equations. We derived it
here using the {\it collective field formalism} of J. Alfaro and P.H.
Damgaard. The BRST symmetry which
implies these Schwinger-Dyson equations as Ward identities, will below be
referred to as {\it Schwinger-Dyson BRST symmetry}.

In the next chapter we will study the interplay between this
Schwinger-Dyson BRST symmetry and gauge symmetries which may originally
be present in the theory in more detail. This will lead to
the Batalin-Vilkovisky scheme for the quantisation of gauge theories.
We will show that the Batalin-Vilkovisky scheme combines BRST symmetry and
the quantum equations of motion (the Schwinger-Dyson equations) in one
formalism \cite{AD3}.

\chapter{The BV antifield formalism for closed, irreducible gauge
algebras}

The Batalin-Vilkovisky scheme \cite{bv1,bv2,bv3,bv4,bv5} combines, loosely
speaking, the BRST symmetry associated with gauge
symmetries and the Schwinger-Dyson shift symmetry. This was first noticed
by J. Alfaro and P.H. Damgaard recently \cite{AD3}. In this and the
following chapter, we present the BV scheme from this point of view.
This way, the key features of the BV scheme (classical and quantum master
equation, quantum BRST operator) are linked with their counterparts in BRST
quantisation.

In the first two sections of this chapter, we develop the BV scheme
fol\-lowing the idea of \cite{AD3}. The BRST symmetry of the theory is
enlarged
such that the Schwinger-Dyson equations are Ward identities of the theory.
We can do this by using the collective field formalism of the previous
chapter. In contrast to \cite{AD3}, we argue that the collective field
formalism is more than a technical means to implement the Schwinger-Dyson
BRST symmetry. This becomes clear in section 3, where a slight
generalisation of \cite{AD3} is discussed\footnote{For further arguments
that emphasize the importance of the collective field, see the next
chapter on open algebras and chapter 7 where an antifield formalism for
BRST--anti-BRST invariant quantisation is developed.}. Section 4 contains
some examples, mainly from \cite{ik8}.

\section{Classical BV}

Whatever the gauge structure of the theory that we want to quantise, the
basic requirement will be that the BRST symmetry of the theory is extended
in such a way that the
Schwinger-Dyson equations are included in the BRST Ward identities of the
theory \cite{AD3}. It is clear that the collective field formalism
of the previous chapter is the appropriate tool to implement this.

We start from a classical action $S_0[\phi^i]$, depending on a set of
fields $\phi ^i$. Suppose that this classical action has gauge invariances
which are irreducible and form a closed algebra. Then one can construct a
nilpotent BRST operator, acting on an extended set of fields $\phi^A$ as is
described in section 1 of chapter 3. The $\phi ^A$ include the original
gauge fields $\phi ^i$,
the ghosts $c^{\alpha}$ and the pairs of trivial systems used for the
construction of the gauge fermion and for the gauge fixing. We summarise
all their BRST transformation rules by
\beq
    \delta \phi^A = \cR^A[\phi^B].
\eeq
The nilpotency of $\delta$,
\beq
    \delta^2 \phi^A = \frac{\dr \cR^A[\phi]}{\delta \phi^B} \cR^B[\phi] =
0,
\eeq
contains both the commutation relations of the algebra (for $\phi^A =
\phi^i$) and the Jacobi identity (for $\phi^A = c^{\alpha}$). With this
BRST operator, we can in principle construct the gauge fixed action.

Instead, we first enlarge the set of fields by replacing the field
$\phi^A$ wherever it occurs, by $\phi^A-\varphi^A$. Again, $\varphi^A$ is
called the {\it collective field}. Like in the previous chapter,
this leads
to a new symmetry, the shift symmetry, for which we introduce a ghost field
$c^A$, and a trivial pair consisting of an antighost field $\phi^*_A$ and
an auxiliary field $B_A$. The BRST transformation rules are taken to be:
\begin{eqnarray}
  \label{extBRST}
      \delta \phi^A & = & c^A \nonumber \\
      \delta \varphi^A & = & c^A - \cR^A[\phi -\varphi] \nonumber \\
      \delta c^A & = & 0 \\
      \delta \phi^*_A & = & B_A \nonumber \\
      \delta B_A & = & 0. \nonumber
\end{eqnarray}
These rules are constructed such that $\delta[\phi^A - \varphi^A] =
\cR^A[\phi -\varphi]$, as the classical action $S_0[\phi -\varphi]$ should
be BRST invariant. This leaves of course the arbitrariness to take
\begin{eqnarray}
       \delta \phi^A & = & c^A +\alpha \cR^A[\phi -\varphi] \nonumber \\
       \delta \varphi^A & = & c^A -( 1- \alpha) \cR^A[\phi -\varphi].
\end{eqnarray}
We have chosen the parameter $\alpha $ to be zero, as it is this choice
which leads to what is known as the Batalin-Vilkovisky quantisation
scheme. In the next chapter, when describing the quantisation of open
algebras, we will give a more compelling reason for this choice
(\ref{deltaN}). Possible other choices of $\alpha$ lead also to
well-defined
path integrals, but will not be considered. Notice that the BRST operator
(\ref{extBRST}) is still nilpotent.

Instead of one, we now have two gauge symmetries to fix, the original
symmetries and the shift symmetry. Like in the previous chapter,
we will
impose as gauge fixing condition $\varphi^A = 0$. So, we subtract $\delta
[\phi^*_A \varphi^A]$ from the classical action, which gives the terms
\beq
 - \delta [\phi^*_A \varphi^A] = - \phi^*_A \left( c^A - \cR^A[\phi
-\varphi ] \right) - (-1)^{\gras{A}} B_A \varphi^A.
\eeq
The original gauge symmetry is fixed by constructing a gauge fermion
$\Psi$ that we take to be a function of the fields $\phi^A$ only. Remember
that the $\phi^A$ include the trivial systems which are introduced in the
usual BRST quantisation precisely for the purpose of gauge fixing. We add
\beq
    \delta \Psi[\phi] = \frac{\dr \Psi}{\delta \phi^A} c^A
\eeq
to $S_0[\phi -\varphi]$.
The complete gauge fixed action that we obtain is then given by
\begin{eqnarray}
    \label{SGF}
    S_{com} & = & S_0[\phi^i-\varphi^i] - \delta [\phi^*_A \varphi^A]
                 + \delta \Psi[\phi ^A] \nonumber \\
           & = & S_0[\phi^i-\varphi^i] + \phi^*_A \cR^A[\phi -\varphi]
           - \phi^*_A c^A + \frac{\dr \Psi}{\delta \phi^A} c^A
           -\varphi^A B_A.
\end{eqnarray}
Define now $S_{BV}(\phi^A,\phi^*_A) = S_0[\phi^i] + \phi^*_A \cR^A[\phi]$,
which allows us to rewrite the gauge fixed action as
\beq
   S_{com} = S_{BV}(\phi -\varphi, \phi ^*) - \left( \phi^*_A - \frac{\dr
\Psi[\phi]}{\delta \phi^A} \right) c^A - \varphi^A B_A. \label{gfbv}
\eeq
Below we will study the properties of $S_{BV}$, the so-called {\it extended
action} of the Batalin-Vilkovisky scheme, in more detail.

The partition function constructed with the gauge fixed action becomes
\beq
   \cZ_{\Psi} = \int [d\phi^A] [d\varphi^A] [d\phi^*_A] \;\;
   e^{\ihbar S_{BV}(\phi -\varphi , \phi^*)} \;\; \delta \left(\phi ^*_A -
      \frac{\dr \Psi[\phi]}{\delta \phi^A} \right)  \delta (\varphi^A).
\eeq
The two delta-functions appear because we have integrated over the ghost of
the shift-symmetry $c^A$ and over the auxiliary field $B_A$. Integrating
out the collective field --it is to put to zero owing to
the $\delta$-function-- brings the partition function in the form which is
typical for the BV scheme:
\beq
   \cZ_{\Psi} = \int [d\phi^A] [d\phi^*_A] e^{\ihbar S_{BV}(\phi, \phi^*)}
   \delta \left(\phi ^*_A -
      \frac{\dr \Psi[\phi]}{\delta \phi^A} \right) .
    \label{BVint}
\eeq
Gauge fixed path integrals are obtained by exponentiating the
extended action, which is a function of fields and antifields, and
afterwards replacing the antifields $\phi^*_A$ by a derivative of an
admissible gauge fermion with respect to the corresponding field.

{}From section 1 of chapter 3 on BRST quantisation, we know that
for $\phi^A=\phi^i$ one has
\beq
     \cR^A[\phi^B] = R^i_{\alpha}[\phi^j] c^{\alpha},
\eeq
and that for $\phi^A = c^{\gamma}$,
\beq
     \cR^A[\phi^B] = T^{\gamma}_{\alpha \beta}[\phi^i] c^{\beta}
c^{\alpha}.
\eeq
This gives the following terms in the extended action
\beq
   \phi^*_A \cR^A[\phi] = \phi^*_i R^i_{\alpha}[\phi] c^{\alpha} +
c^*_{\gamma} T^{\gamma}_{\alpha \beta}[\phi] c^{\beta} c^{\alpha}.
\eeq
Analogously, the trivial system $\delta b_{\alpha} = \lambda_{\alpha}$,
$\delta \lambda_{\alpha} = 0$ leads to the extra term $ b^{* \alpha}
\lambda_{\alpha} $. This is the form trivial systems take in the BV scheme.
The complete extended action for a theory with a closed, irreducible gauge
algebra plus some trivial systems, is
\beq
    S_{BV}(\phi^A, \phi^*_A ) = S_0[\phi^i] +
    \phi^*_i R^i_{\alpha}[\phi] c^{\alpha} + c^*_{\gamma}
 T^{\gamma}_{\alpha \beta}[\phi] c^{\beta} c^{\alpha} + b^{* \alpha}
\lambda_{\alpha} .
   \label{SolCM}
\eeq

If the gauge fermion is of the simple form
\beq
    \Psi [\phi^A] = b_{\alpha} F^{\alpha}(\phi^i),
\eeq
we have that
\begin{eqnarray}
     \frac{\dr\Psi}{\delta \phi^i} & = & b_{\alpha} \frac{\dr F^{\alpha}}
     {\delta \phi^i} \nonumber \\
     \frac{\dr\Psi}{\delta c^{\gamma}} & = & 0 \\
     \frac{\dr\Psi}{\delta b_{\alpha}} & = & F^{\alpha}(\phi^i) \nonumber,
\end{eqnarray}
which leads to the by now familiar expression for the gauge fixed action
(cfr. $a=0$ in (\ref{SEFFP})):
\beq
    S_{BV} \left(\phi^A, \phi^*_A = \frac{\dr \Psi[\phi]}{\delta \phi^A}
   \right) = S_0[\phi^i] + b_{\alpha} \frac{\dr F^{\alpha}} {\delta \phi^i}
   R^i_{\beta} c^{\beta} + F^{\alpha} \lambda_{\alpha}.
\eeq
So, the Faddeev-Popov procedure is still contained in the BV scheme.
Notice that the choice of the gauge fermion may be more
complicated than the one used above to rederive the old expressions,
as was already the case for BRST quantisation.

So far, we have verified that upon integration over all the fields that
were introduced in the collective field formalism, i.e. $\varphi^A$, $c^A$,
$\phi^*_A$ and $B_A$, the previously derived expressions for well-defined
path integrals are found back. Although this is necessary for consistency,
we have of course not gone
through all the trouble of introducing the extra shift symmetry and extra
fields just to integrate them out again immediately.

The BV formalism corresponds to the stage where we have integrated over all
extra fields, except over the antighosts of the shift symmetry, $\phi^*_A$.
In the context of BV, one calls $\phi^*_A$ the {\it antifield} of $\phi^A$.
Like before (\ref{ghnranti}), we have the following important relations:
\begin{eqnarray}
    \gras{\phi^*_A} & = & \gras{\phi^A} + 1 \nonumber \\
    \gh{\phi^*_A} & = & - \gh{\phi^A} - 1 .
\end{eqnarray}
Notice that the antifields have an indexstructure opposite to that of their
associated field. For example, with a covariant vectorfield $A_{\mu}$ a
contravariant antifield vector $A^{*\mu}$ is to be associated.
The reason for keeping precisely the antifield in the scheme, is that it
leads to an elegant formulation of BRST invariance using the {\it
antibracket}. This we discuss below (\ref{antibracket}).

Before turning to the Ward identities, let us derive the
most important property of the extended action. $S_{BV} (\phi ,\phi^*)$
satisfies the so-called {\it classical master equation}. It is precisely
this equation which generalises to the cases of open and
reducible gauge algebras. It follows from the fact that $S_{com}$
(\ref{SGF}) is BRST invariant under the transformation rules
(\ref{extBRST}).
We have that
\begin{eqnarray}
    0 & = & \delta S_{com} \nonumber \\
      & = & \delta S_{BV} (\phi -\varphi, \phi^*) - \delta \left[
   \left( \phi^*_A - \frac{\dr \Psi[\phi]}{\delta \phi^A} \right) c^A
\right]  - \delta  \left[ \varphi^A B_A \right] \\
      & = & \frac{\dr S_{BV}(\phi -\varphi, \phi ^*)}{\delta \phi^A}
\cR^A[{\phi -\varphi}]. \nonumber
\end{eqnarray}
Using the explicit expression for $S_{BV}$, which allows to rewrite $\delta
[\phi^A -\varphi^A] = \frac{\dl S_{BV}(\phi -\varphi,\phi ^*)}{\delta
\phi^*_A}$,  this can be cast in the form
\beq
 \frac{\dr S_{BV}(\phi,\phi ^*)}{\delta \phi^A}
       \frac{\dl S_{BV}(\phi, \phi ^*)}{\delta \phi^*_A} = 0.
 \label{ClaMas}
\eeq
This is the {\it classical master equation} of the BV formalism. It is a
consequence of the BRST invariance of the gauge fixed action $S_{gf}$ and
hence of the nilpotency of the BRST operator
(\ref{extBRST}). In the next section we will rederive the classical
master equation as the $\hbar=0$ stage of an infinite tower of equations,
which are all gathered in the {\it quantum master equation}.

Let us finish this section with a brief summary of the BV recipe for the
quantisation of gauge theories as it has emerged so far. One first has to
find a solution of the classical master equation (\ref{ClaMas}). For
closed, irreducible gauge algebras, the standard solution is of the
form (\ref{SolCM}). In fact, this is not the only solution. Also the
classical action $S_0[\phi]$ itself, for instance, satisfies the classical
master equation, as it does not depend on antifields. Hence, to find the
solution of the desired form when solving the master equation,
one has to impose the extra condition that it be of the form
$ S_{BV} = S_0[\phi^i] + \phi^*_i R^i_{\alpha} c^{\alpha} + \ldots$
Here, the $R^i_{\alpha}$ have to be a complete set of gauge generators
(\ref{completeR}). If the $R^i_{\alpha}$ form a complete set, we say that
$S_{BV}$ is
{\it proper}\footnote{Basically, this condition expresses that we have to
introduce a gauge fixing and ghost action for {\em all} gauge symmetries of
the classical action. If we do not do that, we can not start the
perturbative expansion (cfr. chapter 1).}.
The solution obtained in this way is called the {\it minimal
proper solution}.
Before one is able to construct the  gauge fermion, which has
ghost number $-1$, we have to enlarge the set of fields with trivial
systems. This we can do by taking the non minimal solution of the master
equation
\beq
   \label{SNM}
   S_{n.m.} = S_{BV} + b^{*X} \lambda_X,
\eeq
for an arbitrary set of indices $X$. The non minimal solution satisfies
the classical master equation if the minimal one does. Owing to the
interpretation of the added term as a set of fields $b_X$ with their
arbitrary shift symmetries with ghosts $\lambda_X$, we still have a proper
solution\footnote{A different way to construct a non-minimal proper
solution is by adding trivial systems of the form \cite{VT}
\beq
   S_{n.m.} = S_{min} + b^{* X}b^{* Y} M_{XY}  \;
\eeq
for an invertible matrix $M_{XY}$. The fields $b_X$ and $b^{* X}$ form a
trivial pair for the classical antibracket cohomology (\ref{BracBRST}).}.
Once this is done, the gauge
fixed action is obtained by replacing the antifields by the derivative of
an admissible gauge fermion with respect to the associated field.

\section{Ward identities, the quantum master equation and the quantum
cohomology}

\subsection{Ward identities}

As was argued above (section 2 of chapter 3), the Ward
identities are crucial properties of any gauge theory. For instance, they
guarantee that the partition function is independent of the gauge fixing.
We also showed that it may be necessary to add quantum corrections, or
counterterms, to the action in order for the identities to be valid. Here,
we will study how all this is translated in the BV scheme. This will
lead us to the introduction of the antibracket and a second order
differential operator. With these two ingredients, we can define the
quantum BRST operator.

{}From the expressions for the BRST transformations (\ref{extBRST}), we see
that the possible variation of the measure under a BRST transformation
gives a Jacobian that is a function of $\phi^A -\varphi^A$. As
$\delta(\phi^A-\varphi^A) = \cR^A(\phi -\varphi)$, we are led to add a
counterterm $\hbar M(\phi^A-\varphi^A)$. However, we will also allow
dependence on the antifields. We consider the Ward identity
\begin{eqnarray}
   0 & = & \langle \delta X(\phi^A,\phi^*_A) \rangle \nonumber \\
     & = &\int [d\phi^A] [d\varphi^A] [d\phi^*_A] [dc^A] [dB_A]
     \left[ \frac{\dr X}{\delta \phi^B} c^B + \frac{\dr X}{\delta \phi^*_B
     } B_B \right] \\
     &  & \times \exp \left[ \ihbar \left(W[\phi^i-\varphi^i,\phi ^*]
           - \phi^*_A c^A + \frac{\dr \Psi}{\delta \phi^A} c^A
           -\varphi^A B_A \right) \right].  \nonumber
\end{eqnarray}
The {\it quantum action} is defined by $W(\phi,\phi^*)=
S_{BV}(\phi,\phi^*) + \hbar M(\phi ,\phi^*)$. We restricted ourselves here
to quantities $X(\phi,\phi^*)$, as all the other fields are integrated out
to get the BV formalism. We could have considered quantities $X(\phi
-\varphi,\phi^*)$, but this would not change our final result, as the
collective field $\varphi$ is fixed to zero anyway. The integration over
$B_A$, $\varphi^A$ and $c^A$ can be done following the same steps that
are described in detail in the previous chapter on the
Schwinger-Dyson
equations. We find that the Ward identity becomes
\begin{eqnarray}
   0 & = & \int [d\phi ][d\phi^*] \left[ (X,W) - i \hbar \Delta X \right]
          e^{\ihbar W(\phi ,\phi^*)} \delta \left(\phi ^*_A -
      \frac{\dr \Psi[\phi]}{\delta \phi^A} \right)  \nonumber \\
      & \stackrel{not.}{=} & \langle \sigma X \rangle_{\Psi} .
\end{eqnarray}
Let us first explain the multitude of new notations introduced here.
The {\it antibracket} between two quantities $F$ and $G$ of arbitrary
Grassmann parity is defined by
\beq
   (F , G) = \frac{\dr F}{\delta \phi^A} \frac{\dl G}{\delta \phi^*_A} -
 \frac{\dr F}{\delta \phi^*_A} \frac{\dl G}{\delta \phi^A}.
 \label{antibracket}
\eeq
It plays a crucial part in the whole BV formalism. Many of its
properties are listed in the appendices. The antibracket ressembles a lot
the Poisson bracket used in classical Hamiltonian mechanics. Fields and
antifields are {\sl canonically conjugated} with respect to the
antibracket,
$(\phi^A,\phi^*_B) = \delta^A_B$, in much the same way as coordinates and
momenta are conjugated in classical mechanics with respect to the
Poisson bracket. Below, in
chapter 8, we present the BV formalism from a more algebraic point of
view, based on the properties of {\it canonical transformations}, which are
transformations of the fields and antifields that leave the antibracket
invariant. This is in analogy with the canonical transformations which are
used in Hamiltonian mechanics and which leave the Poisson bracket
invariant.

Furthermore, there is the fermionic second order differential operator
which we will often refer to with the name {\it delta operator} or {\it
box}. It is defined by
\beq
   \label{DELTA}
   \Delta X = (-1)^{\gras{A}+1} \frac{\dr}{\delta \phi^*_A}
\frac{\dr}{\delta \phi^A} X  = (-1)^{\gras{X}}
   (-1)^{\gras{A}} \frac{\dl}{\delta \phi^*_A}
\frac{\dl}{\delta \phi^A} X      \;\;   .
\eeq
This operator $\Delta$ has two nasty properties. Firstly, it is a
non-linear
differential operator, which means that $\Delta (XY) \neq X\Delta Y +
(-1)^{\gras{Y}}\Delta X.Y$. An extra term, $(X,Y)$ is present. The correct
formula, and a few more, can again be found in the appendices.
Secondly, when acting on local functionals of fields and antifields,
$\Delta$
leads to expressions proportional to $\delta (0)$. The third part of
this work is devoted completely to a one-loop regularisation prescription
to handle
this difficulty. For the moment, we will neglect this problem and all the
manipulations with path integrals are understood to be {\it formal}.
Finally, the operator $\sigma X$ is defined to be
\beq
   \sigma X = (X,W) -i\hbar\Delta X.
\eeq
Below we will argue that $\sigma$ is the {\it quantum BRST operator}.

Let us present two applications of this Ward identity. Consider first
the partition function constructed with the gauge fermion $\Psi[\phi]$:
\beq
    \cZ_{\Psi } = \int [d\phi][d\phi^*] e^{\ihbar W(\phi ,\phi^*)}
\delta(\phi^*_A - \Psi _A).
   \label{BVPsi}
\eeq
The new notation is $\Psi _A = \frac{\dr \Psi}{\delta \phi^A}$. An
infinitesimal change of the gauge fermion from
$\Psi[\phi]$ to $\Psi[\phi] +d\Psi[\phi]$ gives the partition function
\beq
    \cZ_{\Psi+d\Psi  } = \int [d\phi][d\phi^*] e^{\ihbar W(\phi ,\phi^*)}
\delta(\phi^*_A - \Psi _A - d\Psi _A),
\eeq
in an obvious notation. Redefine now the integration variable $\phi^{*'}_A
= \phi^*_A - d\Psi_A$, which formally gives Jacobian $1$. Using the
fact that $d\Psi$ is infinitesimal, we can expand to linear order to get
(dropping the primes)
\beq
    \cZ_{\Psi+d\Psi  } = \int [d\phi][d\phi^*] e^{\ihbar W(\phi ,\phi^*)}
\left( 1 + \frac{\dr W}{\delta \phi^*_B} \frac{\dl d\Psi}{\delta \phi^B}
\right) \delta(\phi^*_A - \Psi _A). \label{BVdPsi}
\eeq
Subtracting (\ref{BVPsi}) from (\ref{BVdPsi}) and using that $d\Psi$ does
not depend on antifields, we get
\beq
    \cZ_{\Psi +d\Psi} - \cZ_{\Psi } = \langle \sigma d\Psi \rangle_{\Psi} =
0.
\eeq
Hence, the Ward identity still implies gauge independence.

As a second application, consider $X(\phi ,\phi^*) = \phi^*_A
F(\phi ,\phi^*)$. Using the properties of the antibracket and of the box
operator as listed in the appendix, we easily see that
\beq
\sigma [ \phi^*_A F ] = \phi^*_A \sigma  F + (-1)^{(\gras{A}
+1)(\gras{F}+1)} \left[ F \frac{\dr W}{\delta \phi^A}+ \frac{\hbar}{i}
\frac{\dr F}{\delta \phi^A} \right].
\eeq
If $\sigma F=0$, we see that we find back
the Schwinger-Dyson equation as a Ward identity. This was already pointed
out in \cite{NucHenneaux}.

\subsection{Quantum master equation}

Of course, turning the argument around, the fact that the Ward identity is
valid for all $X(\phi,\phi^*)$ implies that $W(\phi ,\phi^*)$ has to
satisfy certain conditions. We now derive the {\it quantum master
equation} by removing, by means of partial integrations, all derivatives
acting on $X$. We start from
\begin{eqnarray}
0 & = & \int [d\phi][d\phi^*] \left[ \frac{\dr X}{\delta \phi^A}
\frac{\dl W}{\delta
 \phi^*_A} - \frac{\dr X}{\delta \phi^*_A} \frac{\dl W}{\delta
\phi^A} - i\hbar (-1)^{\gras{A}+1} \frac{\dr}{\delta \phi^*_A}
\frac{\dr}{\delta \phi^A} X \right] \nonumber \\
 &  & \times  e^{\ihbar W(\phi ,\phi^*)} \;\; \delta \left(\phi ^*_A -
      \frac{\dr \Psi[\phi]}{\delta \phi^A} \right).
\end{eqnarray}
We will do a field redefinition of the variables in the path integral:
$\phi^{*}_A = \phi^{*'}_A  + \dr \Psi[\phi] / \delta \phi^A$ and the
fields themselves are not altered. The Jacobian of this transformation is,
at least formally, $1$. We use the notation $\Psi_A$ as defined on the
previous page. We will make use of the fact that
\beq
   \frac{\dr Y(\phi ,\phi^*)}{\delta \phi^A}\vert_{
\phi^{*}_A \rightarrow  \phi^{*}_A  + \Psi_A}
   = \frac{\dr Y(\phi ,\phi^*_A +\Psi_A)}{\delta \phi^A} -
 \frac{\dr Y(\phi ,\phi^*_A +\Psi_A)}{\delta \phi^*_B}.
\frac{\dr}{\delta \phi^A} \frac{\dr}{\delta \phi^B} \Psi .
\label{shifttrick}
\eeq
This allows us to rewrite the first contribution to the Ward identity as
\begin{eqnarray}
   &  & i\hbar \int [d\phi][d\phi^*] \;\; X(\phi,\phi^*_A +\Psi_A).
\delta(\phi^*) . \Delta \exp \left[ \ihbar W(\phi ,\phi^*_A +\Psi_A)
\right]  \\
  & - & \int [d\phi][d\phi^*] \;\; \frac{\dr X(\phi,\phi^*_A
+\Psi_A)}{\delta
\phi^*_B} .\frac{\dr}{\delta \phi^A} \frac{\dr}{\delta \phi^B} \Psi .
\frac{\hbar}{i} \frac{\dl e^{\ihbar W(\phi,\phi^*_A+\Psi _A)}}{\delta
\phi^*_A}  \delta(\phi^*)  \nonumber  .
    \label{moetweg}
\end{eqnarray}
Before doing the field redefinition in
the second and third term of the Ward identity, we combine them by
isolating the derivative with respect to $\phi^A$. Then doing the
field redefinition, applying the shift trick
(\ref{shifttrick}) and dropping a total divergence, leads to
\beq
\begin{array}{c}
   -\frac{\hbar}{i} \int [d\phi][d\phi^*] \;\; (-1)^{\gras{A}+1}
\frac{\dr}{\delta \phi^*_B} \left[ e^{\ihbar W(\phi ,\phi^*_A+\Psi_A)}
\frac{\dr X(\phi ,\phi^*_A +\Psi_A)}{\delta \phi^*_A} \right] \\
\times \left( \frac{\dr}{\delta \phi^A} \frac{\dr}{\delta \phi^B} \Psi
\right) . \delta(\phi^*) .
\end{array}
\eeq
Working out the derivative acting on the square brackets gives two terms,
one which vanishes identically and a second one which cancels the second
term of (\ref{moetweg}). Thus the first term of (\ref{moetweg}) is zero for
all possible choices of $X$, which gives the {\it quantum master
equation}
\beq
   0 = \Delta \exp \left[\ihbar W(\phi^A,\phi^*_A + \Psi_A) \right] .
\label{QME}
\eeq
This master equation can be rewritten in the more tractable form
\beq
    \label{QME2}
    (W,W) - 2i\hbar \Delta W = 0 \ ,
\eeq
by using the explicit expression for $\Delta$.
The natural Ansatz for solving this equation, is by expanding $W$ in a
powerseries in $\hbar$:
\beq
   W(\phi ,\phi^*_A +\Psi_A ) = S_{BV} (\phi ,\phi^*_A +\Psi_A )
    + \sum _{n=1} \hbar^n M_n.
    \label{Planckexpan}
\eeq
Grouping the terms order by order in $\hbar$, we get the set of equations
\beq
  \begin{array}{lcl}
    \hbar^0 & \mbox{\hspace{1cm}} & (S_{BV},S_{BV} )  =  0 \\
    \hbar^1 &  & (S_{BV},M_1) =  i \Delta S_{BV} \\
    \hbar^2 &  & (S_{BV},M_2) +\frac{1}{2} (M_1,M_1) = i\Delta M_1 \\
            &  & \ldots
  \end{array}
\eeq
The $\cO(\hbar^0)$ is the only one where $\Delta$ does not appear, and
hence it is the only one that can be studied without having to introduce a
regularisation scheme.
Although the set of equations that one has to solve is possibly infinite,
one very
rarely has to worry about more than the first two equations. In fact, the
regularisation prescription that we will discuss in the third part of this
work, is only capable to give a regularised expression for $\cO(\hbar)$
equation. No regularised computations have been done in the BV scheme for
the $\cO(\hbar^2)$ equation\footnote{It is known that $W_3$ gravity
(the model is presented as an example in the next chapter) has a two loop
anomaly \cite{Matsuo}. It would hence be interesting to study this model
in a two loop regularised BV formalism.}. The equation at $\cO(\hbar^0)$
is just the classical master equation we encountered before. The extra term
generated by the fact that we here (\ref{Planckexpan}) have the gauge fixed
action rather than just $S_{BV}(\phi ,\phi^*)$, is identically zero,
as an explicit calculation shows. In chapter 8,
we will see that this is due to the fact that gauge fixing is
a canonical transformation.

Instead of expanding in integer powers of $\hbar$, one can also take an
expansion in halfinteger powers as Ansatz: $W = S_{BV} + \sqrt{\hbar}
M_{1/2} + \hbar M_1 + \ldots$ This leads to the set of equations
\beq
  \begin{array}{lcl}
     \hbar^0 & \mbox{\hspace{1cm}} &(S_{BV},S_{BV} )  =  0 \\
     \hbar^{1/2} &  & (S_{BV}, M_{1/2}) = 0 \\
     \hbar^1 &  & (S_{BV},M_1) + \frac{1}{2} ( M_{1/2} , M_{1/2} )=  i
\Delta S_{BV} \\ \ldots
  \end{array}
\eeq
It has been found (\cite{ik3}, see also \cite{Stefan}) that the method
of introducing {\it background charges}  (in conformal field theory) to
cancel the anomalies, can be incorporated in BV in this way. We
come back to this issue in the last chapter of this dissertation.

\subsection{Classical and quantum cohomology}

The form of the Ward identity, $\langle \sigma X \rangle = 0$,
suggests that we should consider
\beq
   \sigma X = ( X , W ) - i\hbar \Delta X
\eeq
as the BRST operator in the BV formalism. Moreover, it is easy to show that
$\sigma$ is a nilpotent operator if the quantum action $W$ satisfies the
quantum master equation. Owing to the non-linearity of the delta operator
and the fact that a regularisation scheme is needed to calculate $\sigma
X$, the {\it quantum BRST operator} and its cohomology have not been
studied in detail yet. In the examples in section 4, we give a formal
derivation of an operator satisfying $\sigma X = 0$.

The classical part of $\sigma$, $\cS X = (X , S_{BV})$ is easier to handle.
As $S_{BV}$ satisfies the classical master equation, it is easy to see that
this is also a nilpotent operator: $\cS^2 X = 0$. In contrast to the full
quantum BRST operator, it is a linear differential operator acting from the
right:
\beq
    \cS X(\phi ,\phi^*) = \frac{\dr X}{\delta \phi^A} \cS \phi^A
                       + \frac{\dr X}{\delta \phi^*_A}  \cS \phi^*_A.
   \label{BracBRST}
\eeq
This differential operator acts on the space of (local) function(al)s
of the fields and antifields. For an extensive study of the
antibracket cohomology, see \cite{fisch,FH,Stefan}.

A third BRST operator can be defined that only acts on function(al)s of the
fields, not of the antifields. It is defined by
\beq
    Q F(\phi) = ( F(\phi) , S_{BV})\vert_{\phi^*=0}.
\eeq
If $S_{gf} = S_{BV}(\phi , \Psi_A)$ denotes the gauge fixed action, it
follows from eva\-lua\-ting the master equation for $\phi^*=0$ that
$QS_{gf} =
0$. For closed algebras, one has that $Q^2 F(\phi) = 0$. For open algebras,
which we discuss in the next chapter, it is easy to show that
the nilpotency of $Q$ holds only using the field equations.

\section{Room for generalisation}

We close the theoretical developments of this chapter with two comments,
both concerning the choice of the gauge fermion $\Psi$ (\ref{SGF}). They
will provide the link
with chapter 8 on canonical transformations. First off all, consider a
specific configuration for the antifields $\phi^*_A$, say $\theta^*_A$, and
take as gauge fermion:
\beq
    \Psi[\phi] = \theta^*_A \phi^A + \Psi_0[\phi].
\eeq
This leads to the partition function
\begin{eqnarray}
    \cZ(\theta^*) & = & \int [d\phi][d\phi^*] e^{\ihbar W(\phi ,\phi^*)}
\delta (\phi^*_A - \theta^*_A - \Psi_{0A} ) \nonumber \\
     & = & \int [d\phi][d\phi^*] e^{\ihbar W(\phi ,\phi^*_A + \Psi _{0A})}
\delta (\phi^*_A - \theta^*_A).
\end{eqnarray}
Hence, we do not have to fix the antifields to zero, but we can keep them
as arbitrary, external sources for the BRST transformations.

Secondly, we can lift the restriction that the gauge fermion only
depends
on the fields $\phi^A$. We can consider $\Psi [\phi ,\phi^*]$, and still we
have that $\delta^2 \Psi = 0$\footnote{The property that $\delta^2 \Psi=0$
is of crucial importance in the quantisation of open algebras, as we will
see in the next chapter. There too, this property is valid for gauge
fermions depending
on both fields and antifields.}. Gauge fixing the original symmetries with
such a gauge fermion, leads instead of (\ref{gfbv}) to
\beq
   S_{com} = S_{BV}(\phi -\varphi, \phi ^*) - \left( \phi^*_A - \frac{\dr
\Psi[\phi,\phi^*]}{\delta \phi^A} \right) c^A -  \left( \varphi^A -
\frac{\dr \Psi[\phi ,\phi^*] }{\delta \phi^*_A} \right) B_A.
\eeq
For such a gauge fermion, the collective field is not fixed to zero. We
obtain the partition function
\beq
   \cZ = \int [d\phi][d\phi^*] \exp \left[ \ihbar S_{BV}(\phi -
\frac{\delta \Psi[\phi ,\phi^*]}{\delta \phi^*}, \phi^*)
\right] \delta(\phi^*_A - \frac{\delta \Psi[\phi ,\phi^*]}{\delta \phi^A}),
   \label{More}
\eeq
If we make a specific choice for the gauge fermion of the form
$ \Psi[\phi,\phi^*] = \theta^*_A \phi^A + \Psi_0[\phi,\phi^*]$ for an
infinitesimal $\Psi_0$, we can solve the condition in the delta-function
for $\phi^*$, up to terms quadratic in $\Psi_0$. The antifields can then be
integrated out. Below we will see that this expression can be
reinterpreted
elegantly using canonical transformations. The derivation of the Ward
indentity and hence of the quantum master equation as sketched above
becomes more difficult. We discuss these generalisations from
the more algebraic point of view of canonical transformations in section 2
of chapter 8.

\section{Some hopefully clarifying examples}

\subsection{Extended action of $W_2$-gravity}

Let us consider a first example: $W_2$-gravity \cite{Polyakov1}.
The classical action is given by
\beq
   \label{SOW2}
   S_0 = \frac{1}{2\pi} \int d^2x \left[ \partial \phi \bar \partial \phi -
h (\partial \phi)^2 \right].
\eeq
The fields in the configuration space, $\phi(z,\bar z)$ and $h(z,\bar
z)$, are both of even Grassmann parity
and we used the notation $\partial = \frac{d}{dz}$ and $ \bar \partial =
\frac{d}{d\bar z}$. The action is invariant under the transformations
\begin{eqnarray}
        \delta_{\epsilon} \phi  & = & \epsilon \partial \phi \nonumber \\
        \delta_{\epsilon} h  & = & \bar \partial \epsilon - h \partial
\epsilon + \partial  h .\epsilon .
\end{eqnarray}
{}From this we can derive the $R^i_{\alpha}$. They are given by
\begin{eqnarray}
    R^{\phi(x)}_y & = & \partial_x \phi(x) . \delta(x-y) \\
    R^{h(x)}_y & = & \partial_x h(x) . \delta(x-y) - \bar \partial_y
    \delta(x-y) + \partial_y [ h(x) \delta(x-y) ] \nonumber \\
\end{eqnarray}
Here, $x$ and $y$ are complex coordinates. We can calculate the structure
functions $T^{\alpha}_{\beta \gamma} = T^z_{y \tilde y}$ explicitly by
evaluating the commutator of the gauge generators (\ref{algebra}) for any
choice of
$i$, by using $i=\phi(x)$ or $i=h(x)$. We find
\beq
  2 T^z_{y \tilde y} = \delta(z-y) \partial_z \delta(z-\tilde y)
                       - \delta(z-\tilde y) \partial_z \delta(z-y) .
\eeq
Notice that in order to verify that the algebra is closed, one has to
calculate the commutator for every value of $i$. Using all this, we find
the extended action
\begin{eqnarray}
    \label{W2extac}
S & = & S_0 + \phi^*_i R^i_{\alpha} c^{\alpha} + c^*_{\alpha}
T^{\alpha}_{\beta \gamma} c^{\gamma} c^{\beta} \nonumber \\
  & = & \frac{1}{2\pi} \left[ \partial \phi \bar \partial \phi -
h (z,\bar z) (\partial \phi)^2 \right] \\
  & & + \phi^* c \partial \phi + h^* (\bar \partial c - h \partial c +
\partial h . c ) + c^* (\partial c) c \nonumber .
\end{eqnarray}
In practice, we do not evaluate the algebra and its possible non-closure
functions. One takes $S = S_0 + \phi^*_i R^i_{\alpha} c^{\alpha}
+\ldots $ and one tries to find the dots such that $(S,S)=0$.

To construct a gauge fixed action, we first
introduce a trivial system, as up to now, all fields have positive
ghostnumber. Take the non-minimal solution of
the classical master equation (\ref{SNM})
\beq
      S_{n.m.} = S + b^* \lambda.
\eeq
Hence, we have introduced an antighost $b$ and an auxiliary field
$\lambda$. The gauge field $h$ can for instance be fixed to a background
field $\hat h$ by taking as gauge fermion $\Psi = b (h - \hat h )$.
This gives
\beq
  S_{n.m.}(\phi , \phi^*_A + \Psi_A ) = S + (b^*+h-\hat h) \lambda + b
(\bar \partial c - h \partial c + \partial h . c ).
\eeq
Of course, the first extra term is the gauge fixing while the second extra
term is the ghost action. Notice that upon integration over the auxiliary
field $\lambda$, we are left with the delta-function $\delta (b^*+h-\hat
h)$. If we integrate over $h$, imposing this gauge fixing, $h$ is
replaced everywhere by $\hat h - b^*$. Normally, the gauge fixing of $h$ on
this background field is done in order to prevent the accidental vanishing
of the anomaly. Now we see that $\hat h$ occurs always together with $b^*$.
{}From this we can conclude that if we do not put $b^*$ to zero, it may play
the part of the background field. This is a first reason to keep the
antifield dependence after gauge fixing \cite{ik2}.

\subsection{Construction of topological Yang-Mills theory}

In this second example, we \cite{ik8} construct the
extended action for topological Yang-Mills (YM) theory
\cite{Witten4,TFTA,TFTB}.
This model is presented as it exemplifies a non-standard gauge fixing
procedure where the antighosts that are introduced, have a different
(tensor) structure than the ghosts.

We start from a compact, four-dimensional manifold, endowed with a metric
$g_{\alpha \beta }$ which may be of Euclidean or Minkowski signature. On
this manifold we define the Yang-Mills fields $A_\mu =A_\mu ^aT_a$.
The $T_a$ are the generators of a Lie
algebra. The classical action is
the topological invariant known as the {\it Pontryagin index} or
{\it winding number}. So we have
\begin{equation}
S_0=\int _{\cal M}d^4x \sqrabsg F_{\mu \nu }\tilde {F}^{\mu \nu }\ .
\end{equation}
The {\it dual} of an antisymmetric tensor $G_{\mu \nu }$ is defined by
\begin{equation}
\tilde {G}_{\mu \nu }=\frac{1}{2}[\epsilon ]_{\mu \nu \sigma \tau
}G_{\alpha \beta }g^{\alpha \sigma }g^{\beta \tau }\ .
\end{equation}
The Levi-Civita tensor
tensor is defined by $[\epsilon ]_{\mu \nu \sigma \tau }=\sqrt{g}\epsilon
_{\mu \nu \sigma \tau }$, where $\epsilon _{\mu \nu \sigma \tau }$ is the
permutation symbol and $g=\det g_{\alpha \beta }$. Remark that it is
complex for a Minkowski metric. We normalise our representation for the
algebra such that $Tr [T_a T_b] = \delta _{ab}$, and a trace over the
Yang-Mills indices is always understood. The classical action
is invariant under continuous deformations of the gauge fields that do not
change the winding number:
\begin{equation}
\delta A_\mu =\epsilon _\mu \ .  \label{SS}
\end{equation}
We will not specify the conditions to be imposed on $\epsilon_{\mu}$. We
associate the ghosts $\psi _\mu $ with the shift parameters
$\epsilon_{\mu}$. Then we immediately obtain the BV extended action
\begin{equation}
S=S_0+A^{*\mu}\psi _\mu \ . \label{S1}
\end{equation}
In the literature \cite{Witten4,TFTA,TFTB,TFTRep}, one works with
the reducible set of gauge
symmetries consisting of the shift symmetries $\delta A_\mu =\epsilon _\mu$
and the usual
Yang-Mills symmetry $\delta A_{\mu} = D_{\mu} \epsilon$. We will not do
this, as this is merely a complicated disguise of the construction that we
will describe here. We can always reintroduce this reducible set of
symmetries by adding a trivial system and doing two canonical
transformations. We refer to section 4 of chapter 8,
where we discuss this in detail as an example of the use of
canonical transformations to enlarge the set of fields.

Let us now gauge-fix the shift symmetry (\ref{SS}) in order to obtain the
topological
field theory that is related to the moduli space of self dual YM instantons
\cite{Witten4}. We take the gauge fixing conditions
\begin{eqnarray}
\label{GF}
F^+_{\mu \nu }&=&0\nonumber\\
\partial _\mu A^\mu &=&0\ ,
\end{eqnarray}
where $G_{\mu \nu }^{\pm}=\frac{1}{2}(G_{\mu \nu }\pm \tilde {G}_{\mu \nu
})$. Fields $G_{\mu \nu}$ that satisfy $G_{\mu \nu} = G^+_{\mu \nu}$ are
{\it selfdual}, while $G_{\mu \nu} = G^-_{\mu \nu}$ is the {\it
anti-selfduality} condition. These projectors are orthogonal to eachother,
so that we have
for general $X$ and $Y$ that $X^+_{\alpha \beta }Y^{-\alpha \beta }=0$. The
above gauge choice does not fix all the gauge freedom because there may not
be a unique solution of (\ref{GF})
for a given winding number. To use the picture of the gauge orbits, the
gauge fixing may not select one configuration on every orbit. If that
is the case,
then the moduli space would consist out of one single point for every
winding number.
However, this gauge choice is admissible in the sense that the gauge fixed
action will have well defined propagators. Moreover, the degrees of freedom
that are left (the space of solutions of (\ref{GF})) form exactly the
moduli space of the instantons that
one wants to explore. We now
introduce auxiliary fields in order to construct a gauge
fermion. Obviously we should add
\begin{equation}
S_{nm}=S+\chi _{0\alpha \beta }^*\lambda _0^{\alpha \beta }+b^*\lambda \ ,
\label{S2}
\end{equation}
and consider the gauge fermion
\begin{equation}
\Psi _1=\chi _0^{\alpha \beta }(F^+_{\alpha
\beta }-x\lambda _{0\alpha \beta })+b(\partial _\mu A^\mu-y\lambda )\ ,
\end{equation}
where $x$ and $y$ are some arbitrary gauge
parameters. We introduced here an
antisymmetric field $\chi _0^{\alpha \beta }$\footnote{In the literature
\cite{Witten4,TFTA,TFTB,TFTRep}, one usually introduces a selfdual
two-tensor as antighost. However, in taking the variation of the action, to
obtain the field equations or to calculate the energy-momentum tensor, ad
hoc rules are then needed to maintain this selfduality. This becomes
particularly cumbersome when the projection operators on the selfdual or
anti-selfdual pieces are dependent of other fields in the theory, as is the
case for topological $\sigma$-models.}. This field has
six components, which are used to impose three gauge conditions.
After the gauge fixing, the action has the gauge
symmetry $\chi _0^{\alpha \beta }\rightarrow \chi _0^{\alpha \beta
}+\epsilon
_0^{-\alpha \beta }$. So we fix this symmetry by imposing the condition
$\chi _0^-=0$.
This can be done by adding an extra trivial system $(\chi _{1\alpha
\beta
},\lambda _{1\alpha \beta })$ and with the extra gauge fermion $F=\chi
_{1\alpha \beta }\chi _0^{-\alpha \beta }$. But then we have again
introduced too much fields,
and this leads to a new symmetry $\chi _{1\alpha \beta }\rightarrow
\chi _{1\alpha \beta }+\epsilon^+_{1\alpha \beta }$ which we have to
gauge fix. One easily sees that this procedure repeats itself ad infinitum.
We could, in principle, also solve this problem by only
introducing $\chi _0^{+\alpha \beta }$ as a field. Then we have
to integrate over the space of self dual fields.
To construct the measure on this space, we have to solve the constraint
$\chi =\chi ^+$.
Since this in general can
be complicated (as e.g. in the topological $\sigma
$--model) we will keep the $\chi _{\alpha \beta }$ as the
fundamental
fields. The path integral is with the measure $[d\chi _0^{\alpha
\beta}]$ and we do not split this into the measures in the spaces of
self and anti--selfdual fields. The price we have to pay is an infinite
tower of auxiliary fields. These we denote by
$(\chi _n^{\alpha \beta },\lambda _n^{\alpha \beta
})$\footnote{One remark has to be made here concerning the place of the
indices. We choose the indices of $\chi _n$ and $\lambda _n$ to be upper
resp. lower indices when $n$ is even resp. odd. Their antifields have
the opposite property, as usual.}
with Grassmann parities $\gras{\lambda _n}=n$,$\gras{\chi _n}=n+1$
(modulo $2$) and ghostnumbers
$\gh{\lambda _n}$ equal to zero for $n$ even and one for $n$ odd.
Similarly, $\gh{\chi _n}$ equals $-1$ for $n$ even and zero for $n$ odd.
We then add to the action (\ref{S2}) the term $\sum _{n=1}^{\infty}\chi
^*_{n,\alpha
\beta }\lambda _n^{\alpha \beta }$ and take as gauge fixing fermion
\begin{equation}
\Psi _2=\sum _{n=1}^{\infty}\chi _n^{\alpha \beta }\chi _{n-1,\alpha \beta
}^{(-)^{n}}\ + \Psi _1, \label{IJK}
\end{equation}
where $G_{\alpha \beta }^{(-)^n}$ is the selfdual part of
$G_{\alpha
\beta }$ if $n$ is even and the anti-selfdual part if $n$ is odd.
After doing the gauge fixing we end up with
the following non--minimal solution of the classical master equation
\footnote{Note that from $(\chi ,\chi ^*)=1$, it follows that
$(\chi ^{\pm},\chi ^{*\pm})=P^{\pm}$ and $(\chi ^+,\chi ^{*-})=0$, where
$P^{\pm}$ are the projectors onto the (anti)-selfdual sectors.}~:
\begin{eqnarray}
S_{nm}&=&S_0+A^{*\mu }\psi _\mu +(\partial _\mu A^\mu + b^*)\lambda
+( F^+_{\alpha \beta } + \chi^-_{1\alpha \beta } + \chi^*_{0\alpha \beta })
\lambda _0^{\alpha \beta }  \nonumber\\&&
- y\lambda ^2  -x\lambda _0^{\alpha \beta }
\lambda _{0\alpha \beta }+\chi _0^{+\alpha \beta }D_{[\alpha }\psi _{\beta
] }+b\partial _\mu \psi ^\mu \nonumber\\&&
+\sum _{n=1}^{\infty}(\chi ^*_{n\alpha \beta }+\chi
_{n+1,\alpha \beta }^{(-)^{(n+1)}}+\chi _{n-1,\alpha \beta
}^{(-)^{n}})\lambda _n^{\alpha \beta }\ .
\end{eqnarray}
Performing the $\lambda _n,
n\geq 1$ integrals would give the gauge fixing
delta functions $\delta (\chi _{n+1}^{(-)^{n+1}}+\chi
_{n-1}^{(-)^n}+\chi _n^*)$ .
Doing only the Gaussian $\lambda _0$ and $\lambda$ integral, we arrive
at \begin{eqnarray}
S&=&S_0+\frac{1}{4x}(\partial _\mu A^\mu
+b^*)^2+\frac{1}{4y}(F^++\chi^-_1 +\chi
_0^*)^2\nonumber\\&&+b\partial _\mu \psi ^\mu +\chi _0^{+\alpha \beta
}D_{[\alpha }\psi _{\beta ]}+A_\mu ^*\psi ^\mu \nonumber \\
& &
+\sum _{n=1}^{\infty}(\chi ^*_{n\alpha \beta }+\chi
_{n+1,\alpha \beta }^{(-)^{(n+1)}}+\chi _{n-1,\alpha \beta
}^{(-)^{n}})\lambda _n^{\alpha \beta }\ .
\end{eqnarray}
Notice that we now have terms quadratic in the antifields. This means that
the BRST operator defined by $Q\phi ^A=(\phi ^A,S)\vert_{\phi ^*=0}$ is
only
nilpotent using field equations. Indeed, $Q^2b=\frac{1}{2x}\partial _\mu
\psi ^\mu \approx 0$, using the field equation of the field $b$. The fact
that we have to use the field equations to prove the nilpotency of the BRST
operator is the hallmark of an open algebra. In the next chapter, we will
show that open algebras manifest themselves in the BV scheme by non-linear
terms in the antifields in the extended action.

\subsection{The energy-momentum tensor as BRST invariant operator in BV}

We \cite{ik8} construct formally the energy-momentum tensor
$T^q_{\alpha \beta}$ that
satisfies the condition $\sigma T^q_{\alpha \beta} = (T^q_{\alpha \beta}
,W)-i\hbar \Delta T^q_{\alpha \beta} = 0$
or, classically only, $(T_{\alpha \beta},S)=0$, provided $W$($S$)
satisfies the quantum (classical) master equation. In a first subsection,
we derive
expressions for the derivation of the antibracket and $\Delta$-operator
with respect to the metric. We
then define an energy-momentum tensor that is classical or quantum
BRST invariant.
In the chapter on canonical transformations, we show that the energy
momentum tensor as we define it here is canonically invariant
(\ref{there}).

\subsubsection{Metric dependence of the antibracket and $\Delta$}
We
have to be precise on the occurences of the metric in all our expressions,
and specify a consistent set of conventions. All
integrations are with the volume element $dx\sqrabsg $. The functional
derivative is then defined as
\beq
   \frac{\delta \phi^A}{\delta \phi^B} = \frac{1}{\sqrabsg_B}
\frac{d\phi^A}{d\phi^B} = \frac{1}{\sqrabsg_B} \delta_{AB}\ ,
\label{FD}
\eeq
and the same for the antifields. The notation is that $A$ and $B$
contain both the discrete
and space-time indices, such that $\delta_{AB}$ contains both space-time
$\delta$-functions (without $\sqrabsg$) and Kronecker deltas ($1$ or
zero) for the discrete indices . $g$ is
$\det g_{\alpha \beta}$, and its subscript $B$ denotes that we evaluate it
in the space-time index contained in $B$.  Using this, the antibracket and
box operator are defined by\footnote{We then have that $(\phi ,\phi
^*)=\frac{1}{\sqrabsg}$. In this convention the extended action takes the
form $S=\int dx\sqrabsg
[{\cal L}_0+\phi ^*_i\delta \phi ^i+\phi ^*\phi ^*...]$. Demanding that
the total Lagrangian is a scalar
amounts to taking the antifield of a scalar to be a
scalar, the antifield of a covariant vector to be a contravariant vector,
etc. One could also use the following set of conventions. We integrate
with the volume element $dx$ without metric, and define the functional
derivative (\ref{FD}) without $\sqrabsg$. Also the antibracket is defined
having no metric in the integration. Therefore, $(\phi ,\phi
^*)'=1$. With this
bracket the extended action takes the form $S'=\int dx[\sqrabsg{\cal L}_0+
\phi ^*_i\delta \phi ^i+\frac{1}{\sqrabsg}\phi ^*\phi ^*...]$. The
relation between the two sets of conventions is a transformation
that scales the antifields with the metric, i.e. $\phi ^*\rightarrow
\sqrabsg \phi ^*$. In these variables, general
covariance
is not explicit and requires a good book--keeping of the $\sqrabsg$ 's in
the extended action and in other computations. Therefore, we will not use
this convention.}
\begin{eqnarray}
 ( A, B ) & = & \sum_i \int dx \sqrabsg_X \left( \frac{\dr A}{\delta
\phi^X} \frac{\dl B}{\delta \phi^*_X} - \frac{\dr A}{\delta \phi^*_X}
\frac{\dl B}{\delta \phi^X} \right) \nonumber \\
  \Delta A & = & \sum_i \int dx \sqrabsg_X (-1)^{\gras{X}+1}
\frac{\dr}{\delta \phi ^X}\frac{\dr}{\delta \phi^*_X} A \; .
\end{eqnarray}
For once, we made the summation that is hidden in the DeWitt summation
more explicit. $X$ contains the discrete indices $i$ and the space-time
index $x$. These definitions guarantee that the antibracket of two
functionals is again a functional.
Using the notation introduced above, we have that
\begin{eqnarray}
 ( A, B ) & = & \sum_i \int dx \frac{1}{\sqrabsg_X} \left( \frac{\ddr
A}{d \phi^X} \frac{\ddl B}{d \phi^*_X} - \frac{\ddr A}{d \phi^*_X}
\frac{\ddl B}{d \phi^X} \right) \nonumber \\
  \Delta A & = & \sum_i \int dx \frac{1}{\sqrabsg_X} (-1)^{\gras{X}+1}
\frac{\ddr}{d \phi ^X}\frac{\ddr}{d \phi^*_X} A\ .
\end{eqnarray}
It is now simple to differentiate with respect to the metric. We use the
following rule~:
\begin{equation}
\frac{\delta g^{\alpha \beta }(x)}{\delta g^{\rho \gamma }(y)}=\frac{1}{2}
(\delta ^\alpha _\rho \delta ^\beta _\gamma +\delta ^\alpha _\gamma \delta
^\beta _\rho )\delta (x-y)\ ,
\end{equation}
where the $\delta $--function does {\em not} contain any metric, i.e.
$\int dx \delta (x-y)f(x)=f(y)$. This we do in order to agree with the
familiar recipe to calculate the energy-momentum tensor. Then we find that
\beq \label{metricbracket}
\frac{\delta (A,B)}{\delta g^{\alpha \beta}(y)}=\left( \frac{\delta
A}{\delta g^{\alpha \beta}(y)} , B \right) + \left( A , \frac{\delta
B}{\delta g^{\alpha \beta}(y)} \right)
+\frac{1}{2}g_{\alpha \beta }(y)\sqrabsg (y)[A,B](y)\ ,
\eeq
and
\beq
\frac{\delta \Delta A}{\delta g^{\alpha \beta}(y)}=\Delta \frac{\delta
A}{\delta g^{\alpha \beta}} +\frac{1}{2}g_{\alpha \beta }(y)\sqrabsg
(y)[\Delta A](y)\ ,
\label{metricdelta}
\eeq
with the notation that $(A,B)=\int dx\sqrabsg [A,B]$ and $\Delta A=\int
dx\sqrabsg [\Delta A]$. Notice that in $[A,B]$ and $[\Delta A]$ a
summation over the discrete indices is understood, but no integration over
space-time. Before applying this to define the energy momentum tensor
in the BV scheme, consider the following properties.
For any two operators $A$ and $B$, we have that
\beq
   \label{lemma1}
   \sum_i \phi^*_X \frac{\dl}{\delta \phi ^*_X}
   (A,B)=( \sum_i \phi^*_X \frac{\dl A}{\delta \phi ^*_X}, B)
   +( A , \sum_i \phi^*_X \frac{\dl B}{\delta \phi ^*_X})
   -[A,B](x)\ ,
\eeq
and
\beq
   \label{lemma2}
   \sum_i \phi^*_X \frac{\dl \Delta  A}{\delta \phi ^*_X} = \Delta \left(
   \sum_i \phi^*_X \frac{\dl A}{\delta \phi ^*_X}  \right)
   -[\Delta A](x)\ .
\eeq
In both expressions, $X=(i,x)$ with discrete indices $i$ and continuous
indices $x$. There is no integration over $x$ understood, only a summation
over $i$, which is explicitised.

Let us define the differential operator
\beq
     D_{\alpha \beta} = \frac{2}{\sqrabsg} \frac{\delta}{\delta g^{\alpha
\beta}} + g_{\alpha \beta} \sum_i \phi^*_X \frac{\dl}{\delta \phi^*_X}\ .
\eeq
Then it follows from (\ref{metricbracket}) and (\ref{lemma1}) that this
operator satisfies
\beq
    D_{\alpha\beta} (A,B) = (D_{\alpha\beta} A, B) + (A , D_{\alpha\beta}
B)\ .
\eeq
Owing to (\ref{metricdelta}) and (\ref{lemma2}), $D_{\alpha\beta}$ is seen
to commute with the $\Delta$-operator:
\beq
      D_{\alpha\beta} \Delta  A = \Delta D_{\alpha\beta} A\ .
\eeq

\subsubsection{Definition of the energy-momentum tensor}

Let us now apply all these results to define an expression which can be
interpreted as being the BRST invariant energy-momentum tensor and that is
invariant under the BRST transformations in the antibracket sense. Define
\beq
      \theta_{\alpha \beta} = \frac{2}{\sqrabsg} \frac{\delta
S}{\delta g^{\alpha \beta}}\ .
\eeq
By differentiating the classical master equation $(S,S)=0$ with
respect to the metric $g^{\alpha \beta}(y)$, and by multiplying with
$2/\sqrabsg$, we find from (\ref{metricbracket}) that
\beq
   0 = 2 ( \theta_{\alpha \beta}(y) , S ) + 2 g_{\alpha \beta}(y)
\sum_i  \frac{\dr S}{\delta \phi^X} \frac{\dl S}{\delta
\phi^*_X}\ .
\eeq
In the second term, $X=(y,i)$ and there is only a summation over $i$.
Hence, we see that $\theta_{\alpha \beta}$ is not BRST invariant in the
antibracket sense.

However, if we define the energy-momentum tensor by
\beq
    T_{\alpha \beta} =  D_{\alpha\beta}  S,
\eeq
then it follows immediately that
\beq
    D_{\alpha\beta} (S,S) = 0 \;\; \Leftrightarrow \;\; ( T_{\alpha \beta}
, S ) =0. \eeq
It is then clear that $T_{\alpha \beta}$ is a BRST invariant
energy-momentum tensor\footnote{
Notice that this quantity is the energy momentum tensor that one
immediately obtains when using the variables mentioned in the
previous footnote, i.e.
after scaling the antifields. One can then check that $T_{\alpha \beta
}=\frac{2}{\sqrabsg}\frac{\delta S'}{\delta g^{\alpha
\beta }}$. In this sense the modification of $\theta _{\alpha \beta
}$ is an artifact of the used conventions.}.
Moreover, $\theta_{\alpha \beta}\vert_{\phi^*=0} =
T_{\alpha \beta}\vert_{\phi^*=0}$.
Whether this
is a trivial element of the cohomology, i.e. equivalent to zero, can of
course not be derived on general grounds. By adding to this expression for
$T_{\alpha \beta}$ terms of the form $(X_{\alpha \beta},S)$, one can
obtain
cohomologically equivalent expressions. For example, by subtracting the
term $(\frac{1}{2} g_{\alpha \beta} \sum_i \phi^*_X \phi^X, S)$, the
terms that have to be added to $\theta_{\alpha \beta}$ to obtain $T_{\alpha
\beta}$ take a form that is symmetric in fields and antifields.

We can generalise this result and define an energy-momentum tensor that
is quantum BRST invariant. Consider
the quantum extended action $W$ that satisfies the quantum master equation
$ (W,W) - 2i\hbar \Delta W =0$. Define the quantum analogue of
$\theta_{\alpha \beta}$, i.e.
\beq
    \theta^q_{\alpha \beta} = \frac{2}{\sqrabsg}  \frac{\delta W}{\delta
g^{\alpha \beta}}\ .
\eeq
Again, one easily sees that this is not a quantum BRST invariant quantity.
Define however
\beq
     T^q_{\alpha \beta} =D_{\alpha\beta} W,
\eeq
then it follows by letting $D_{\alpha\beta}$ act on the quantum master
equation that
\beq
\begin{array}{c}
  D_{\alpha \beta} [ (W,W) - 2i\hbar \Delta W] = 0 \\ \Updownarrow \\
  \sigma T^q_{\alpha \beta} = (T^q_{\alpha \beta} , W ) - i\hbar \Delta
T^q_{\alpha \beta} = 0\ .
\end{array}
\eeq

\section{Overview}

Since this is a rather long chapter where we have introduced many new
concepts, a short recapitulation is in place. Imposing that the quantum
equations of motion (the Schwinger-Dyson equations) are included in the
Ward identities of any theory, we have constructed the BV antifield scheme
for closed, irreducible algebras. One can do this by enlarging the symmetry
algebra to include the shift symmetries. The BRST invariance of the gauge
fixed action under the enlarged symmetry implies that the extended action
of the BV scheme satisfies the classical master equation. We have defined
the quantum BRST operator $\sigma$, and have derived the quantum master
equation as the condition that guarantees that the Ward identities $\langle
\sigma X \rangle = 0$ hold. After a short discussion of different
cohomologies, we have given some examples. Let us stress that most of the
developments above generalise to open algebras, as we show
in the next chapter.

\chapter{Open gauge algebras}

In this chapter, we give a quantisation recipe for theories
with an open gauge algebra. First, we describe in detail an inductive
approach \cite{ik6}, combining the recipe of B. de Wit and J.W.
van Holten \cite{Open} for the BRST quantisation of gauge
theories with an open algebra with the requirement of J. Alfaro and P.H.
Damgaard that the BRST Ward identities must include the
Schwinger-Dyson equations. This way, we show that also for open algebras
the central object is the extended action that is a solution
of the BV classical master equation. Secondly, we turn the argument
around, and point out that the collective field formalism leads in a
straightforward fashion to a quantisation recipe for open algebras.

\section{From BRST to BV quantisation for open algebras}

Whatever the quantisation procedure for closed algebras that we have
discussed, be
it the Faddeev-Popov recipe, the quantisation based on BRST symmetry or the
reformulation of the latter in the BV scheme, an essential step in the
quantisation process is the choice of gauge.
Let us assume that the gauge choice is encoded in the gauge conditions
$F^{\alpha}= 0$. Any quantisation scheme should at least satisfy the
following
two requirements. First of all, the functional integrals for the partition
function and for the correlation functions can be made
well-defined. This means that all fields can be given a propagator
by a careful choice of the functions $F^{\alpha}$. This
has previously been called the {\it admissability of the gauge choice}. No
general prescription can be given for this, {\it gauge fixing is an art}.
Moreover, although the partition function is defined using a specific
choice for the $F^{\alpha}$, it should nevertheless be invariant
under (infinitesimal) deformations of these functions.
This means that the partition function should be gauge independent.

Let us recapitulate how the BRST quantisation for closed algebras that
was described in chapter 3, satisfies these requirements.
Having a nilpotent BRST operator $\delta$, the gauge fixed action is
constructed by adding the BRST variation of a gauge fermion --which encodes
the gauge functions $F^{\alpha}$-- to the classical action: $S_{com} = S_0
+ \delta \Psi$. Owing to the nilpotency of $\delta$, we have that $\delta
S_{com} = 0$, which allows the derivation of the Ward identities $\langle
\delta X \rangle =0$ (\ref{WaTaSlaTa}). These Ward identities then
imply invariance of the partition under infinitesimal deformation of the
gauge fermion (\ref{dpsiward}). One may have to add quantum counterterms to
the action to cancel the BRST non-invariance of the measure in the
derivation of the Ward identity (section 2 of chapter 3).

This recipe fails for open algebras. Remember that by {\it open algebra},
we mean that a term proportional to the classical field equations appears
in the commutation relations of the algebra\footnote{This term proportional
to field equations is not arbitrary. $E^{ji}_{\alpha \beta}
(-1)^{\gras{i}}$
is graded antisymmetric in $i$ and $j$ (\ref{completeR}).} (\ref{algebra}):
\beq
\frac{\dr R^i_{\alpha}}{\delta \phi^j} R^j_{\beta}
  - (-1)^{\gras{\alpha} \gras{\beta}}
  \frac{\dr R^i_{\beta}}{\delta \phi^j} R^j_{\alpha}
  = 2 R^i_{\gamma} T^{\gamma}_{\alpha \beta}  (-1)^{\gras{\alpha}}
  - 4 y_j E^{ji}_{\alpha \beta} (-1)^{\gras{i}} (-1)^{\gras{\alpha}}.
\eeq
Multiplying both sides with $(-1)^{(\gras{\alpha} + 1)\gras{\beta}}
c^{\alpha} c^{\beta}$, we get
\beq
   \frac{\dr R^i_{\alpha} c^{\alpha}}{\delta \phi^j} R^j_{\beta} c^{\beta}
   + R^i_{\gamma} T^{\gamma}_{\alpha \beta} c^{\beta} c^{\alpha} =
   2 (-1)^{\gras{i}} y_j E^{ji}_{\alpha \beta} c^{\beta} c^{\alpha}.
   \label{Openghost}
\eeq
If we take the naive BRST transformation rules of section 1 of chapter 3,
i.e. $ \delta \phi^i = R^i_{\alpha}c^{\alpha}$ and $\delta c^{\gamma} =
T^{\gamma}_{\alpha \beta} c^{\beta} c^{\alpha}$, we see that
(\ref{Openghost}) implies that the BRST operator $\delta$ is not
nilpotent when acting on $\phi^i$, instead we have that
\beq
    \delta^2 \phi^i =
   2 (-1)^{\gras{i}} y_j E^{ji}_{\alpha \beta} c^{\beta} c^{\alpha}.
\eeq
Hence, the BRST operator is only nilpotent on the stationary surface .
Similar terms, proportional to field equations, may appear when calculating
$\delta^2 c^{\alpha}$. If we again adopt the condensed notation $\phi^A$
for $\phi^i$, $c^{\alpha}$ and trivial systems and $\delta \phi^A =
\cR^A[\phi]$, we can sum up the situation by
\beq
   \delta^2 \phi^A = \frac{\dr \cR^A}{\delta \phi^B} \cR^B = y_j M^{jA}.
\eeq
It is then clear that if we would construct the gauge fixed action like for
closed algebras, it would no longer be BRST invariant off the stationary
surface. Consequently, we can not prove gauge independence in the way
we have described in section 2 of chapter 3, using Ward identities.

Gauge theories with an open algebra were first encountered in the study of
supergravity theories in the second half of the seventies. It was found
that terms quartic in the ghost fields are needed, which can of course
{\em not} be obtained from the Faddeev-Popov procedure like we presented
it in chapter 2. The action with the quartic ghost term is still
invariant under modified BRST transformations \cite{fourghost}.
Inspired by these results, B. de Wit and J.W. van Holten
gave a general recipe for BRST quantisation of theories with an open
algebra \cite{Open}. The basic
observation is that one can drop the nilpotency requirement of the BRST
transformation and just demand that one constructs a gauge fixed action
that is invariant under a set of BRST transformation rules. If the theory
is anomaly free, a BRST invariant action allows the derivation of Ward
identities, as we described in section 2 of chapter 3.
Furthermore, the gauge fixed action
should be such that changing the gauge (fermion) infinitesimally, amounts
to adding the BRST variation of an infinitesimal quantity. The Ward
identities then guarantee gauge independence of the partition function.

The recipe of \cite{Open} is to consider the gauge fermion
$F=b_{\alpha} F^{\alpha}(\phi^i)$ and to define $F_i = \dr F/\delta
\phi^i$. The gauge fixed action $S_{com}$ is obtained by adding an
expansion
in powers of these $F_i$ to the classical action, where the linear term is
taken to be the Faddeev-Popov quadratic ghost action. The
modified BRST transformation rules for the gauge fields and the ghosts
are also obtained by adding an expansion in these $F_i$ to the BRST
transformation rules for closed algebras. All field dependent coefficients
in these expansions are fixed by requiring $\delta S_{com}=0$.

In accordance with our collective field approach to enlarge the BRST
algebra in such a way that the Ward identities include the
Schwinger-Dyson equations,
we introduce again collective fields $\varphi^A$ for $\phi^A$, shift ghosts
$c^A$ and the trivial pair $(\phi^*_A,B_A)$. The naive BRST
transformation rules are taken to be
\begin{eqnarray}
  \label{deltaN}
    \delta_N \phi^A & = & c^A \nonumber \\
    \delta_N \varphi^A & = & c^A - \cR^A[\phi -\varphi] \nonumber \\
    \delta_N c^A & = & 0 \\
    \delta_N \phi^*_A & = & B_A \nonumber \\
    \delta_N B_A & = & 0. \nonumber
\end{eqnarray}
Notice that these transformation rules are still only nilpotent
using the field equations, as can be seen from $\delta^2 (\phi^A -
\varphi^A)$. More importantly however, the BRST transformation as we have
constructed it, is now nilpotent when acting on $\phi^A$ : $\delta_N^2
\phi^A = 0$. This implies that the original gauge symmetry can be gauge
fixed in a BRST invariant way by adding $\delta_N \Psi[\phi]$.
This is the real reason for using the freedom to shift the $\cR^A[\phi
-\varphi]$ between the BRST transformation of the field and the collective
field the way we do. We shift the off-shell non-nilpotency in the
transformation rules of the collective field.

If we impose the same gauge fixing conditions as in chapter 5 (\ref{SGF}),
the complete gauge fermion $F$ is given by
\beq
   F = - \phi^*_A \varphi^A + \Psi [\phi].
\eeq
The prescription of de Wit and van Holten then implies that we should add
expansions in
\beq
         \frac{\dr F}{\delta \phi^A} = \frac{\dr \Psi}{\delta \phi^A}
\stackrel{not.}{=} \Psi_A,
\eeq
and
\beq
    \frac{\dr F}{\delta \varphi^A} = - \phi^*_A
\eeq
to the classical action and to the naive BRST transformation rules.
Owing to the nilpotency of $\delta_N$ when acting on $\phi^A$, we only get
a linear term in $\Psi_A$ and an expansion in the antighost (or
antifield) $\phi^*_A$ remains. We then look for quantities $M_n^{A_1 \ldots
A_n}(\phi)$, with the Grassmann parities
\beq
   \epsilon(M_n^{A_1 \ldots A_n}) = \sum_{i=1}^{n} (\gras{A_i}+1),
\eeq
and the antisymmetry property
\beq
  M_n^{A_1 \ldots A_i A_{i+1} \ldots A_n} =
(-1)^{(\gras{A_i}+1)(\gras{A_{i+1}}+1)} M_n^{A_1 \ldots A_{i+1} A_i \ldots
A_n},
\eeq
such that we can construct the gauge fixed action
\begin{eqnarray}
   \label{label}
  S_{com} & = & S_0[\phi -\varphi] -\varphi^A B_A - (\phi^*_A - \Psi_A)c^A
\nonumber \\
    &  & + \phi^*_A \cR^A[\phi -\varphi] - \sum_{n \geq 2} \frac{1}{n}
\phi^*_{A_1} \ldots \phi^*_{A_n} M_n^{A_1 \ldots A_n} (\phi -\varphi)
\end{eqnarray}
to be invariant under the BRST transformation rules
\begin{eqnarray}
    \delta \phi^A & = & c^A \nonumber \\
    \delta \varphi^A & = & c^A - \cR^A[\phi -\varphi] + \sum_{n \geq 2}
\phi^*_{A_2} \ldots \phi^*_{A_n} M_n^{A A_2 \ldots A_n} (\phi -\varphi)
    \nonumber \\
    \delta c^A & = & 0 \\
    \delta \phi^*_A & = & B_A \nonumber \\
    \delta B_A & = & 0. \nonumber
\end{eqnarray}
We introduced a factor $\frac{1}{n}$ in the last term of the gauge fixed
action. Therefore, we have
\begin{eqnarray}
 \delta \left[ \frac{1}{n} \phi^*_{A_1} \ldots \phi^*_{A_n}
 M_n^{A_1 \ldots A_n} \right] & = & \frac{1}{n}
\phi^*_{A_1} \ldots \phi^*_{A_n} \delta M_n^{A_1 \ldots A_n} \\
&  & + (-1)^{\gras{A_1}+1} B_{A_1}
\phi^*_{A_2} \ldots \phi^*_{A_n} M_n^{A_1 \ldots A_n} \nonumber .
\end{eqnarray}
This makes all $B$-dependent terms cancel in the expression for $\delta
S_{com}$.

Calculating $\delta S_{com}$ and equating it order by order in $\phi^*_A$
to zero, we get a set of equations. At the {\bf order $(\phi^*)^0$} we have
\beq
  \frac{\dr S_0(\phi -\varphi )}{\delta \phi^A}
    \cR^A(\phi - \varphi ) = 0 \; .
\eeq
The first condition, the term in $\delta S_{com}$ that is independent of
antifields, is just that the classical action has invariances.
{}From the {\bf order $(\phi^*)^1$} in $\delta S_{com}=0$ we get the
condition
\beq
   \frac{\dr \cR^A [\phi  - \varphi ]}
   {\delta \phi^B } \cR^B (\phi -\varphi ) -
   (-1)^{(\gras{A} + 1)\gras{B}} \frac{\dr S_0( \phi -\varphi )}
   {\delta \phi^B} M_2^{BA} ( \phi -\varphi )
   =  0  \;
\eeq
If we take for $\phi^A$ the original gauge fields $\phi^i$, we have that
\beq
   \frac{\dr R^i_{\alpha} c^{\alpha}}{\delta \phi^j} R^j_{\beta} c^{\beta}
+ R^i_{\gamma} T^{\gamma}_{\alpha \beta} c^{\beta} c^{\alpha} -
(-1)^{(\gras{i}+1)\gras{j}} y_j M_2^{ji} = 0.
\eeq
Comparing this with (\ref{Openghost}), we find that we can take
\beq
     M_2^{ij} = - 2 E^{ji}_{\alpha \beta} c^{\beta} c^{\alpha},
\eeq
leading to a contribution $\phi^*_i \phi^*_j E^{ji}_{\alpha \beta}
c^{\beta} c^{\alpha}$ in $S_{com}$. Analogously, for $\phi^A=c^{\alpha}$,
we find
\beq
   \frac{\dr T^{\alpha}_{\beta \gamma} c^{\gamma} c^{\beta}}{\delta \phi^j}
   R^j_{\mu} c^{\mu} + 2 T^{\alpha}_{\beta \gamma} c^{\gamma}
T^{\beta}_{\mu \nu} c^{\nu} c^{\mu} - (-1)^{\gras{\alpha} \gras{j}} y_j
M^{j\alpha}_2 = 0.
\eeq
Provided that the first two terms equal an expression proportional to
a field equation, we can conclude that we can take
\beq
    M_2^{i\alpha} = - D^{i\alpha}_{\mu \nu \sigma} c^{\sigma} c^{\nu}
c^{\mu},
\eeq
for some function $D^{i\alpha}_{\mu \nu \sigma}$. This gives an extra term
in $S_{com}$ of the form \newline $\frac{1}{2} \phi^*_i c^*_{\alpha} D^{i
\alpha}_{\mu \nu \sigma} c^{\sigma} c^{\nu} c^{\mu}.$
No further terms come from this condition. All the other
fields that are included in $\phi^A$ belong to trivial systems and hence
have that
\beq
    \frac{\dr \cR^A}{\delta \phi^B} \cR^B = 0.
\eeq
Thus, we see that for all these fields $M_2^{jA}=0$.

In principle, the set of equations coming from the condition $\delta
S_{com} = 0$ may be infinite. It
is clear that every $M_n$ is to be determined from an equation of the form
\beq
   - \frac{\dr S_0}{\delta \phi^A} \phi^*_{A_2} \ldots \phi^*_{A_n} M_n^{A
A_2 \ldots A_n} + G(\cR^A,M_2,\ldots,M_{n-1}) = 0.
\eeq
Although knowing $M_2$,\ldots,$M_{n-1}$ allows in principle to
determine $M_n$, it is not a priori guaranteed that $ G(\cR^A,M_2,
\ldots,M_{n-1})$ is indeed of the required form proportional to a field
equation. It can be proven however that this is indeed the case, i.e. that
a BRST invariant action of the form (\ref{label}) exists. Moreover, the
solution is not unique. It has however been shown that different
solution are related by a canonical transformation of fields and
antifields\footnote{Canonical transformations are discussed in chapter 8.}.
For more details on the existence proof, we refer to
\cite{Open,fisch,FH,Stefan,Henneauxbook,HenLoc}.

Like for closed algebras (\ref{gfbv}), we decompose
\beq
   S_{com} = S_{BV}(\phi -\varphi ,\phi^*) - (\phi^*_A - \Psi_A) c^A -
\varphi^A B_A.
\eeq
{}From all the above, we then conclude that
\beq
  \label{Sopen}
  S_{BV}(\phi ,\phi^*) = S_0[\phi] + \phi^*_A \cR^A[\phi] +
\phi^*_i \phi^*_j E^{ji}_{\alpha \beta} c^{\beta} c^{\alpha} +
\frac{1}{2} \phi^*_i c^*_{\alpha} D^{i \alpha}_{\mu \nu \sigma}
c^{\sigma} c^{\nu} c^{\mu} + \ldots,
\eeq
where the dots stand for terms cubic and higher order in the antifields.
Notice that in the terms that are non-linear in the antifields, at least
one antifield $\phi^*_i$ of the original gauge fields $\phi^i$ is present,
as $S_0$ only depends on these.

As was the case for closed algebras (\ref{ClaMas}), we have that
\beq
\delta(\phi^A -
\varphi^A) = \frac{\dl S_{BV}(\phi -\varphi ,\phi^*)}{\delta \phi^*_A}.
\eeq
Therefore, BRST invariance of $S_{com}$ implies that also for open
algebras, the extended action $S_{BV}$ satisfies the classical master
equation \beq
    \frac{\dr S_{BV}}{\delta \phi^A} \frac{\dl S_{BV}}{\delta \phi^*_A} =
0.
\eeq
We now have a unifying principle for quantising theories with an
algebra that may be closed or open. In both cases we have to solve the
classical
master equation $(S_{BV},S_{BV})=0$, with the boundary condition that
$S_{BV} = S_0 + \phi^*_i R^i_{\alpha} c^{\alpha} +\ldots$ As for the rest
of the recipe to obtain a well-defined partition function, everything goes
through as for closed algebras (\ref{SNM}).
Especially, for a gauge fermion of the form
$\Psi = b_{\alpha} F^{\alpha}(\phi)$, we see that the replacement in
$S_{BV}$ (\ref{Sopen}) of antifields
by derivatives with respect to the fields, gives the four ghost interaction
mentioned above:
\beq
    \phi^*_i \phi^*_j E^{ji}_{\alpha \beta} c^{\beta} c^{\alpha}
    \rightarrow
    b_{\alpha} \frac{\dr F^{\alpha}}{\delta \phi^i}
    b_{\beta} \frac{\dr F^{\beta}}{\delta \phi^j} E^{ji}_{\mu \nu} c^{\nu}
    c^{\mu} \; .
\eeq

Hence, we see that for theories with an open gauge algebra, the combination
of the BRST quantisation recipe with the SD BRST symmetry, gives rise to
the same (BV) scheme as was derived for closed algebras in the previous
chapter.

\section{The role of the collective field}

A posteriori, we can see that precisely by introducing the collective
fields, we can gauge fix the original gauge symmetries in the
same way for open algebras as for closed algebras and
prove that the functional integral is formally gauge independent. We
discuss this point of view here in some detail, as it will be our starting
point for the derivation of an antifield scheme for BRST--anti-BRST
invariant quantisation of theories with an open gauge algebras in the next
chapter.

The introduction of collective fields allows us to construct
the BRST transformation rules such that $\delta^2 \phi^A = 0$, since we
can shift the
off-shell nilpotency problem of open algebras to the transformation rules
of the collective field by defining $\delta \phi^A = c^A$. We can then fix
the originally present gauge symmetry in a manifest BRST invariant way by
adding $\delta \Psi[\phi]$. As $(\phi^*_A,B_A)$ form a trivial system,
here
too we can generalise the choice of gauge fermion to $\Psi[\phi ,\phi^*]$
without spoiling the fact that $\delta^2 \Psi = 0$.

The gauge fixed action can be decomposed as $S_{com} = S_{inv}
+ \delta \Psi$, and we consider the gauge fixed partition function
\beq
   \cZ_{\Psi} = \int [d\phi][d\phi^*][d\varphi][dc][dB] \;\; e^{\ihbar
S_{com}} \ .
\eeq
It now remains to make sure that the
Ward identities are valid. Then the partition function $\cZ_{\Psi}$ is
invariant
under infinitesimal deformations of the gauge fermion $\Psi$. For that
purpose, we construct $S_{inv}$ to be BRST invariant.
If we make the decomposition,
\beq
    \label{decompoS}
    S_{inv} = S_{BV} (\phi -\varphi ,\phi^*) - \phi^*_A c^A -\varphi^A B_A,
\eeq
we know that $\delta S_{inv} = 0$, under the BRST transformation rules
\begin{eqnarray}
    \label{THEBRST}
    \delta \phi^A & = & c^A \nonumber \\
    \delta \varphi^A & = & c^A - \frac{\dl S_{BV}(\phi -\varphi ,\phi
^*)}{\delta \phi ^*_A} \nonumber \\
    \delta c^A & = & 0 \\
    \delta \phi^*_A & = & B_A \nonumber \\
    \delta B_A & = & 0 ,  \nonumber
\end{eqnarray}
if $S_{BV}$ satisfies the classical master equation of the BV scheme;
$(S_{BV},S_{BV})=0$. The question whether open
algebras can be quantised then amounts to proving that the classical
master equation of the BV scheme can be solved for open algebras
\cite{Henneauxbook,Stefan}. The reason for the decomposition
(\ref{decompoS}) is that the auxiliary collective field $\varphi^A$ is
fixed to zero to remove it from the final functional integrals. The
$\phi^*_A c^A$ is the Faddeev-Popov ghost-antighost action associated with
this gauge fixing. Moreover, we know that this way the SD equations with
the gauge fixed action are included in the Ward identities of the theory.

Things get only slightly more complicated when quantum counterterms have to
be used to cancel the BRST variation of the measure in the derivation of
the Ward identity. In that case, we
decompose $S_{com} = S_q + \delta \Psi$ after the introduction of the
collective field.
Now $S_q$ is {\em not} BRST invariant. Its BRST variation has to
cancel the Jacobian of the measure under BRST transformations.
We decompose
\beq
   S_q = W(\phi -\varphi,\phi^*) - \phi^*_A c^A -\varphi^A B_A \; .
\eeq
With the BRST transformation rules of (\ref{THEBRST}), except for
the collective field for which we take
\beq
    \label{colquant}
    \delta \varphi^A  =  c^A - \frac{\dl W(\phi -\varphi ,\phi
^*)}{\delta \phi ^*_A}  ,
\eeq
we find that
\beq
    \delta S_q .\mu = \frac{1}{2} (W , W) . \mu \; ,
\eeq
where the expression of the antibracket is evaluated in $(\phi -\varphi
,\phi ^*)$. With these BRST transformation rules
(\ref{THEBRST},\ref{colquant}),
the Jacobian from the measure in the derivation of the Ward identity
(\ref{BRSTJacob}) is
\beq
   J = \exp [ \ihbar ( -i\hbar \Delta W ).\mu ] .
\eeq
Then we see that the product of the measure and $e^{\ihbar S_q}$ is BRST
invariant if $W$ satisfies the quantum master equation
\beq
    ( W , W ) - 2i\hbar \Delta W = 0 .
\eeq
We are now back at the result derived in previous chapter for closed
algebras (\ref{QME2}). If the quantum master equation is satisfied, we can
reverse steps of section 2 of chapter 5 to prove that the Ward identity
$\langle \sigma X \rangle =0$. This then implies gauge independence of the
partition function $\cZ_{\Psi}$.

In this chapter, we have shown how the quantisation of gauge theories with
an open algebra leads to the same classical and quantum master equation of
the BV scheme. We have also argued that the design of the collective
field formalism itself plays a crucial role in the
construction of a quantisation scheme for open algebras. We close this
chapter by giving an example of a theory with an open algebra.

\section{Example: $W_3$ gravity}

We give here $W_3$ gravity \cite{Hull} (again a two dimensional model) as
an example of a theory with an open algebra and
give its extended action \cite{Stefan}. Consider the classical action
\beq
   S_0 =  \int d^2x \left[ \frac{1}{2} \partial \phi \bar \partial \phi
-\frac{1}{2} h (\partial \phi)^2 - \frac{1}{3} B (\partial \phi)^3 \right].
 \label{SOW3}
\eeq
The fields $\phi$,$h$ and $B$ all have even Grassmann parity. Notice that,
up to a scale factor $\frac{1}{\pi}$, the first two terms are the classical
action of $W_2$ gravity (\ref{SOW2}). The invariances of this action are
\begin{eqnarray}
   \delta \phi  & = & \epsilon \partial \phi  + \lambda (\partial
\phi)^2 \nonumber \\
   \delta h  & = & ( \bar \partial \epsilon - h \partial \epsilon
+ \partial  h .\epsilon ) + (\lambda .\partial B - B . \partial \lambda )
(\partial \phi)^2 \nonumber \\
   \delta B & = & (\epsilon .\partial B - 2 B.\partial \epsilon) + (\bar
\partial \lambda - h .\partial \lambda + 2 \lambda .\partial h ) .
\end{eqnarray}
The transformation parameters are $\epsilon$ and $\lambda$, for which we
introduce respectively the ghosts $c$ and $l$.

A long but straightforward calculation allows to verify that
\begin{eqnarray}
S_{BV} & = & \frac{1}{2} \partial \phi \bar \partial \phi
-\frac{1}{2} h (\partial \phi)^2 - \frac{1}{3} B (\partial \phi)^3
\nonumber \\
    & & + \phi^* \left[ c .\partial \phi + l (\partial \phi)^2 \right]
    \nonumber \\
    & & + h^* \left[ ( \bar \partial c - h. \partial c + \partial  h .c)
+ (l .\partial B - B . \partial l) (\partial \phi)^2 \right] \nonumber \\
    & & + B^* \left[ (c .\partial B - 2 B.\partial c) + (\bar
\partial l - h .\partial l + 2 l .\partial h ) \right] \nonumber \\
    & & + c^* . \partial c.c + c^* . \partial l.l. (\partial \phi)^2
    + 2 l^* . \partial c . l - l^*.c.\partial l \nonumber \\
    & & + 2 \phi^* h^* .\partial l.l.\partial \phi \; ,
\end{eqnarray}
satisfies the classical masterequation. Moreover, it clearly is a proper
solution. An integration over two dimensional space time is understood.
As is discussed in \cite{Stefan}, this particular solution involves a
choice: what are $T^{\alpha}_{\beta \gamma}$ and
$E^{ji}_{\alpha \beta}$. In particular, the term $c^*.\partial l.l.
(\partial \phi)^2$ is due to taking a non-vanishing structure function
$T^{\epsilon}_{\lambda \lambda}$. However, $(\partial \phi)^2\sim y_h$,
and one can also take $T^{\epsilon}_{\lambda \lambda} = 0$ and modify the
non-closure functions. The price to pay is that then also terms
proportional $h^*h^*$ and $h^*B^*$ are needed to construct a solution of
the master equation. All these different solutions are related by canonical
transformations. For some examples of the change of the gauge generators
and the structure functions under canonical transformations, we refer to
section 4 of chapter 8.

\chapter{An antifield scheme for BRST--anti-BRST invariant quantisation}

In this chapter, we derive an antifield scheme for quantisation in a
BRST--anti-BRST invariant way. Instead of only one antifield for every
field, our construction will lead to
three antifields: one acting as a source term for BRST
transformations, one as a source term for anti-BRST transformations and one
as a source term for mixed transformations. The whole structure of BV is
doubled. There are two master equations, one of ghostnumber one and
one of ghostnumber minus one. They correspond to two BRST operators, one
that
raises the ghostnumber by one and one that lowers the ghostnumber by the
same amount. For closed algebras, the scheme was derived using the usual
collective field formalism in \cite{ik4}. However, in order to obtain a
better agreement, especially for the gauge fixing, with the earlier
algebraic derivation of \cite{BLT}, an improved collective field formalism
was set up in \cite{ik7}. There, we introduced {\it two} collective fields
for every field. We first describe how the Schwinger-Dyson equations
can be derived using this formalism with two collective fields. Then we
derive the antifield scheme.
For some alternative formulations of BRST--anti-BRST symmetry using
antifields, we refer to \cite{antiHull}.

\section{Schwinger-Dyson Equations from two collective fields}

In this section, we present the collective field formalism with {\em two}
collective
fields. We derive the SD equation as a Ward identity using this formalism
and postpone the complication of possible gauge symmetries of the
classical action to the next section.

We start from an action $S_0[\phi]$, depending on bosonic
degrees of freedom $\phi^i$ and that has no gauge symmetries. The index $i$
is suppressed in this section. We introduce {\em two} copies of the
original field, the two
so-called {\it collective fields}, $\varphi_1$ and $\varphi_2$ and consider
the action $S_0[\phi -\varphi_1 -\varphi_2]$. There now are two gauge
symmetries for which we introduce two ghostfields $\pi_1$ and $\phi^*_2$
and two antighost fields $\phi^*_1$ and $\pi_2$. The BRST--anti-BRST
transformation rules are taken to be (we follow the construction of section
4 of chapter 3)
\beq
  \begin{array}{lcl}
     \delta_1 \phi = \pi_1 & \mbox{\hspace{2cm}} & \delta_2 \phi = \pi_2 \\
     \delta_1 \varphi_1 = \pi_1 - \phi^*_2 &  & \delta_2 \varphi_1 = -
\phi^*_1 \\
     \delta_1 \varphi_2 = \phi^*_2 & & \delta_2\varphi_2 = \pi_2 + \phi^*_1
\\
     \delta_1 \pi_1 = 0 &  & \delta_2 \pi_2 = 0 \\
     \delta_1 \phi^*_2 = 0 & & \delta_2 \phi^*_1 = 0.
\end{array}
\eeq

Imposing $(\delta_2 \delta_1 +\delta_1 \delta_2)\phi =0$ gives the  extra
condition $\delta_2 \pi_1 + \delta_1 \pi_2 =0$, while analogously
$(\delta_2 \delta_1 +\delta_1 \delta_2)\varphi_1 =0$ gives $\delta_1
\phi^*_1 +\delta_2 \phi^*_2 = \delta_2 \pi_1$, and
$(\delta_2 \delta_1 +\delta_1 \delta_2)\varphi_2 =0$ leads to no new
condition. We introduce two extra bosonic fields $B$ and $\lambda$ and the
BRST transformation rules:
\beq
  \begin{array}{lcl}
      \delta_1 \pi_2 = B & \mbox{\hspace{2cm}} & \delta_2 \pi _1 = -B \\
      \delta_1 B = 0 &  & \delta_2 B = 0 \\
      \delta_1 \phi^*_1 = \lambda - \frac{B}{2} &  & \delta_2 \phi^*_2 =
-\lambda - \frac{B}{2} \\
      \delta_1 \lambda = 0 &  & \delta_2 \lambda = 0.
\end{array}
\eeq
All these rules together guarantee that $\delta_1^2 = \delta_2^2 = \delta
_1\delta_2 +\delta_2\delta_1 = 0$ on the complete set of fields.

With all these BRST transformation rules at hand, we can construct a gauge
fixed action that is invariant under BRST--anti-BRST symmetry. We will fix
both the collective fields to be zero. To that end, we add
\begin{eqnarray}
   S_{col} & = & \frac{1}{2} \delta_1 \delta_2 [ \varphi_1^2 - \varphi_2^2
] \nonumber \\
    & = & - (\varphi_1 + \varphi_2) \lambda+\frac{B}{2}(\varphi_1 - \varphi_2)
    + (-1)^a \phi^*_a \pi_a .
\end{eqnarray}
In the last term, there is a summation over $a=1,2$. Denoting
$\varphi_{\pm} = \varphi_1 \pm \varphi_2$, we have the gauge fixed action
\beq
   S_{com} = S_0[\phi -\varphi_+]
 - \varphi_+\lambda+\frac{B}{2}\varphi_- + (-1)^a \phi^*_a \pi_a .
\eeq
$S_{com}$ has both BRST and anti-BRST symmetry.

The Schwinger-Dyson equations can be derived as Ward identities in the
following way.
\begin{eqnarray}
  0 & = & \langle \delta_1 [ \phi^*_1 F(\phi) ] \rangle \\
    & = & \int d\mu  \left[ \phi^*_1 \frac{\dr F}{\delta \phi} \pi_1 +
(\lambda - \frac{B}{2}) F(\phi) \right] e^{\ihbar S_{com}}. \nonumber
\end{eqnarray}
$d\mu$ denotes the integration measure over all fields. The term
$\langle B F(\phi) \rangle$ is zero. This can be seen by noticing that
$B = \delta_1\delta_2 \varphi_+$. The Ward identities themselves allow to
{\sl integrate by parts} to get
\beq
\langle B F(\phi) \rangle = - \langle \varphi_+ \delta_2 \delta_1 F(\phi)
\rangle,
\eeq
which drops out as $\varphi_+$ is fixed to zero.

The SD equation then results as in chapter 4 or in \cite{AD1,AD2}, by
integrating out
$\pi_a$,$\phi^*_a$,$\lambda$,$B$,$\varphi_+$ and $\varphi_-$. Of course,
the SD equations can also be derived as Ward identities of the anti-BRST
transformation $\delta_2$.

\section{Closed algebras}

Given any classical
action $S_0[\phi^i]$ with a closed and irreducible gauge algebra, the
configuration space is enlarged by introducing the
necessary ghosts, antighosts and auxiliary fields, needed for the
construction of BRST--anti-BRST transformation rules as is described in
section 4 of chapter 3. The complete set
of fields is denoted by $\phi_A$ and their
BRST--anti-BRST transformation rules are all summarised by $\delta_a \phi_A
= \cR_{Aa} (\phi)$. For $a=1$, we have the BRST transformation rules, for
$a=2$ the anti-BRST transformation. Since the algebra is closed, we have
that ($\delta_a^2 \phi_A =0$)
\beq
       \frac{\dr \cR_{Aa}(\phi)}{\delta \phi_B} \cR_{Ba}(\phi)  =  0
\eeq
and that ($ (\delta_1\delta_2 + \delta_2\delta_1) \phi_A =0$)
\beq
       \frac{\dr \cR_{A1}(\phi)}{\delta \phi_B} \cR_{B2}(\phi) +
       \frac{\dr \cR_{A2}(\phi)}{\delta \phi_B} \cR_{B1}(\phi)  =  0.
\eeq
In the first formula, there is no summation over $a$.

Instead of constructing a gauge fixed action that is invariant under the
BRST--anti-BRST symmetry, we introduce collective fields and
associated extra shift symmetries. We
introduce {\it two} collective fields $\varphi_{A1}$ and $\varphi_{A2}$,
collectively denoted by $\varphi_{Aa}$, and replace everywhere $\phi_A$ by
$\phi_A - \varphi_{A1} - \varphi_{A2}$.
There now are two shift symmetries for which we
introduce the ghosts $\pi_{A1}$ and $\phi^{*2}_A$ with ghostnumber
$\gh{\pi_{A1}} = \gh{\phi^{*2}_A} = \gh{\phi_A} + 1$ and the antighosts
$\phi^{*1}_A$ and $\pi_{A2}$ with ghostnumber
$\gh{\pi_{A2}} = \gh{\phi^{*1}_A} = \gh{\phi_A} - 1$.
Again, we will use $\pi_{Aa}$ and $\phi^{*a}_A$ as compact notation.
Of course, one has to keep in mind that for $a=1$, $\pi_{Aa}$ is a ghost,
while for $a=2$, $\pi_{Aa}$ is an antighost and vice versa for $\phi^{*a}_A$.

We construct the BRST--anti-BRST transformations as follows:
\begin{eqnarray}
   \label{colanti}
             \delta_a \phi_A & = & \pi_{Aa}  \\
      \delta_a \varphi_{Ab} & = & \delta_{ab} \left[ \pi_{Aa} - \epsilon
      _{ac} \phi^{*c}_{A} - \cR_{Aa}(\phi -\varphi_1 -\varphi_2) \right]
       + (1 - \delta_{ab}) \epsilon_{ac} \phi^{*c}_A \nonumber ,
\end{eqnarray}
with no summation over $a$ in the second line\footnote{
Our convention: $\epsilon_{12} = 1$,$\epsilon^{12} = -1$.}.
These rules are constructed such that
\beq
      \delta_a(\phi_A - \varphi_{A1} -  \varphi_{A2}) =
      \cR_{Aa}(\phi -\varphi_1 -\varphi_2).
\eeq
The two collective fields lead to even more freedom to shift the
$\cR_{Aa}$ in the transformation rules, than the one in the
collective field formalism for BV (\ref{extBRST}).
The choice above incorporates
the antifield formalism for BRST--anti-BRST symmetry \cite{BLT}.
Furthermore, the discussion of open algebras in the previous
chapter (\ref{deltaN}) also indicates that it is useful to construct the
rules such that
$\delta_a^2 \phi_A = 0$ and $(\delta_1 \delta_2 + \delta_2 \delta_1)\phi_A
= 0 $, independently of the closure of the algebra.
We can make sure that $\delta_a^2 = 0$ ($a=1,2$) and that
$\delta _1\delta _2 +\delta _2 \delta_1 = 0 $ when acting on any field,
by the introduction of two extra fields $B_A$ and $\lambda_A$ and
the new transformationrules:
\begin{eqnarray}
   \label{ccolanti}
       \delta _a \pi_{Ab} & = & \epsilon_{ab} B_A \nonumber \\
       \delta_a B_A & = & 0 \\
       \delta_a \phi^{*b}_A &= & -\delta_a^b \left[ (-1)^a \lambda_A
        + \frac{1}{2} \left( B_A +
       \frac{\dr \cR_{A1}(\phi-\varphi_1 - \varphi_2)}{\delta \phi_B}
\cR_{B2}(\phi-\varphi_1 - \varphi_2 )  \right) \right] \nonumber \\
       \delta _a\lambda _A & = & 0  .\nonumber
\end{eqnarray}

We gauge fix both the collective
fields to zero in a BRST--anti-BRST invariant way. For that purpose, we
need a matrix $M^{AB}$, with constant c-number entries and which is
invertible. Moreover, it has to have the symmetry property $M^{AB} =
(-1)^{\gras{A} \gras{B}} M^{BA}$ and all the entries of $M$ between
Grassmann odd and Grassmann even sectors have to vanish. It has to be
such that $\phi_A M^{AB} \phi_B$ has over all ghostnumber zero and has even
Grassmann parity. Except for the constraints above, the precise form of $M$
is of no concern. It will drop out completely in the end.
The collective fields are then gauge fixed to zero in a BRST--anti-BRST
invariant way by adding the term
\begin{eqnarray}
S_{col} & = & - \frac{1}{4} \epsilon^{ab} \delta_a \delta_b \left[
          \varphi_{A1} M^{AB} \varphi_{B1} - \varphi_{A2} M^{AB}
          \varphi_{B2} \right] \nonumber \\
       & = & - (\varphi_{A1} + \varphi_{A2}) M^{AB} \lambda_B
          + \frac{1}{2} ( \varphi_{A1} - \varphi_{A2}) M^{AB} B_B\\
      & +& (-1)^{\gras{B}+1} \phi^{*1}_A M^{AB} \pi_{B1}
          + (-1)^{\gras{B}} \phi^{*2}_A M^{AB} \pi_{B2}  \nonumber \\
     & + &  (-1)^{\gras{B}} \phi^{*1}_A M^{AB} R_{B1} (\phi
-\varphi_1-\varphi_2) + (-1)^{\gras{B}+1} \phi^{*2}_A M^{AB} R_{B2} (\phi
-\varphi_1-\varphi_2)  \nonumber \\
     & + & \frac{1}{2} (\varphi_{A1} - \varphi_{A2}) M^{AB}
       \frac{\dr \cR_{B1}(\phi-\varphi_1 - \varphi_2)}{\delta \phi_C}
\cR_{C2}(\phi-\varphi_1 - \varphi_2 ) \nonumber.
\end{eqnarray}
The relative sign between the two contributions of the gauge fixing is
needed to make two terms containing the product $\phi^{*1}_A M^{AB}
\phi^{*2}_B$, cancel. Redefine now $\varphi_{A\pm} = \varphi_{A1}
\pm \varphi_{A2}$, which allows us to rewrite the gauge fixing terms
in a more compact and suggestive form:
\begin{eqnarray}
     \label{nogeenlabel}
S_{col} & = & - \varphi_{A+} M^{AB} \lambda_B + \frac{1}{2} \varphi_{A-}
         M^{AB} B_B + (-1)^a (-1)^{\gras{B}} \phi^{*a}_{A} M^{AB} \pi_{Ba}
         \nonumber \\
     &  & +\frac{1}{2} \varphi_{A-}  M^{AB}
       \frac{\dr \cR_{B1}(\phi-\varphi_+)}{\delta \phi_C}
\cR_{C2}(\phi-\varphi_+)  \\
    &   & + (-1)^{a+1}
      (-1)^{\gras{B}} \phi^{*a}_A M^{AB} R_{Ba} (\phi -\varphi_+).
      \nonumber
\end{eqnarray}
Notice that this time a summation over $a$ {\it is} understood in the third
and fifth term. The $\phi^{*a}$ have indeed become source terms for the
BRST and anti-BRST transformation rules of the fields $\phi-\varphi_+$,
while the difference
of the two collective fields $\varphi_-$ acts as a source for mixed
transformations. The sum of the two collective fields is fixed to
zero.

The original gauge symmetry can be fixed in a BRST--anti-BRST invariant way
by adding the variation of a gauge {\em boson} $\Psi(\phi)$, of ghostnumber
zero.
We take it to be only a function of the original fields $\phi_A$. This
gives the extra terms
\begin{eqnarray}
S_{\Psi} & = & \frac{1}{2} \epsilon^{ab} \delta_a \delta_b \Psi(\phi)
           \nonumber \\
         & = & - \frac{\dr \Psi}{\delta \phi_A} B_A + \frac{1}{2}
\epsilon^{ab} (-1)^{\gras{B}+1} \left[ \frac{\dr}{\delta \phi_A}
\frac{\dr}{\delta \phi _B} \Psi \right] . \pi_{Aa} \pi_{Bb}.
\end{eqnarray}

In order to make contact with the antifield formalism that was derived
on algebraic grounds in \cite{BLT}, we first have to make the following
(re)definitions.
We incorporate the matrix $M^{AB}$ introduced above in the antifields:
\begin{eqnarray}
\label{rredef}
\phi ^{*Aa'} &=& (-1)^{\gras{A}}\phi ^{*\ a}_B
M^{BA}(-1)^{a+1}\hspace*{1cm}a = 1,2\nonumber\\
\mbox{}\\
\bar \phi ^A &=& \frac{1}{2} \varphi _{B-} M^{BA} . \nonumber
\end{eqnarray}
Owing to the properties of the matrix $M^{AB}$ above, the ghostnumber
assignments {\it after the redefinition} are given by
\begin{eqnarray}
\gh{\phi ^{*Aa'}} &=& (-1)^a - \gh{\phi _A} \nonumber \\
\gh{\bar \phi ^A} &=& -\gh{\phi _A},
\end{eqnarray}
while the Grassmann parities are of course
\beq
\varepsilon _{\phi ^{*Aa'}} = \varepsilon _{\phi _A} + 1\ \ ;\ \
\varepsilon _{\bar \phi ^A} = \varepsilon _{\phi _A}\,.
\eeq
In \cite{BLT}, I.A. Batalin, P.M. Lavrov and I.V. Tyutin
introduced the so-called {\it extended action}, which we denote by
$S_{BLT}$.  Using the new variables and dropping the primes,
it is defined by
\beq
S_{BLT}(\phi _A,\phi ^{*Aa},\bar \phi ^A) = S_0[\phi _A]
+ \phi ^{*Aa}R_{Aa}(\phi ) + \bar \phi ^A \frac{\dr
 R_{A1}(\phi )}{\delta \phi _B}R_{B2}(\phi )\,.
\label{SBLT}
\eeq
$S_{BLT}$ is the sum of the classical action, plus the last two terms of
(\ref{nogeenlabel}), up to a substitution of $\phi$ by $\phi-\varphi_+$.
The remaining terms of $S_{col}$, are denoted by $S_\delta $, hence
\beq
S_\delta = -\varphi _{A+}M^{AB}\lambda _B
   + \bar \phi ^AB_A - \phi ^{*Aa}\pi _{Aa}\,.
\eeq
Integrating over $\pi _{Aa}$, $B_A$ and $\lambda _B$,
$S_{\delta}$ leads to a set of $\delta $-functions removing all the
fields of the collective field formalism.  The situation is then
analogous to the BV scheme.  Before the gauge fixing term $S_\Psi $ is
added, all antifields are fixed to zero.

With all these definitions at hand, we have that
\begin{eqnarray}
\label{Sgf}
S_{com} &=& S_0[\phi - \varphi _+] + S_{col} + S_\Psi\\
&=& S_{BLT}[\phi - \varphi _+,\phi ^{*a},\bar \phi ] + S_\delta + S_\Psi
\,,\nonumber
\end{eqnarray}
which gives the gauge fixed partition function
\beq
\cZ = \int [d\phi ][d\phi ^{*a}][d\bar \phi ][d\pi_a][dB]
 e^{\frac{i}{\hbar }S_{BLT}[\phi ,\phi ^{*a},\bar \phi
]}e^{\frac{i}{\hbar }S_\Psi }e^{\frac{i}{\hbar }\tilde S_\delta }\,.
\eeq
We already integrated out $\lambda $ and $\varphi _+$, and $\tilde
S_\delta $ is $S_\delta $ with the term $-\varphi _{A+}M^{AB}\lambda _B$
omitted.  The gauge fixing term $\exp(\frac{i}{\hbar }S_\Psi )$ can be
obtained by acting with an operator $\hat V$ on $\exp(\frac{i}{\hbar
}\tilde S_\delta )$, i.e.
\beq
e^{\frac{i}{\hbar }S_\Psi }e^{\frac{i}{\hbar }\tilde S_\delta } = \hat
V e^{\frac{i}{\hbar }\tilde S_\delta }\,.
\eeq
{}From the explicit form of $\tilde S_\delta $ and $S_\Psi $, and using
that $e^{a(y)\frac{\delta }{\delta x}}f(x) = f(x + a(y))$,
we see that
$ \hat V(\Psi) = e^{-T_1(\Psi )- T_2(\Psi)}$ with
\begin{eqnarray}
T_1 (\Psi ) & = & \frac{\dr \Psi (\phi )}{\delta \phi
_A}\,\cdot\,\frac{\dl }{\delta \bar \phi ^A} \nonumber \\
T_2(\Psi)  & = & \frac{i\hbar
}{2}\varepsilon ^{ab}\frac{\dl }{\delta \phi ^{*Bb}} \left[
\frac{\dr }{\delta \phi_A}\frac{\dr}{\delta \phi_B}\Psi \right]
\frac{\dl }{\delta \phi ^{*Aa}}\,.
\end{eqnarray}
The convention is that the derivatives with respect to the antifields
$\phi^*$ and $\bar \phi$ act on everything standing to the
right of them.  The operator $\hat V$ can be integrated by parts, such that
\beq
\cZ = \int [d\phi ][d\phi ^{*a}][d\bar \phi ] \delta (\phi ^{*A1})
\delta(\phi ^{*A2}) \delta (\bar \phi ^A)
\left[\hat U(\Psi ) e^{\frac{i}{\hbar }S_{BLT}}\right] \, ,
\eeq
with the operator $\hat U = e^{ + T_1 - T_2 }$.
This form of the path integral agrees with \cite{BLT}.

Let us finally derive the {\it classical master equations} which are
satisfied by $S_{BLT}$.  They follow from the fact that $S_{com}$
(\ref{Sgf}) is invariant under both the BRST and the anti-BRST
transformation.  Furthermore, one has to use the fact that the matrix
$M^{AB}$ only has non-zero entries for $\varepsilon _A = \varepsilon _B$,
and hence that $M^{AB} = (-1)^{\varepsilon_ A} M^{BA} = (-1)^{\varepsilon_
B} M^{BA}$.  Also, in the collective field BRST--anti-BRST transformation
rules, we may replace $R_{Aa}(\phi - \varphi _+)$ by
$\dl S_{BLT}/\delta \phi ^{*Aa'}$.
Since $\delta _aS_\Psi = 0$, we have that
\begin{eqnarray}
0 &=& \delta _a S_{com}  \nonumber \\
&=& \delta _a S_{BLT} + \delta _a S_\delta \\
&=& \frac{\dr S_{BLT}}{\delta \phi _A}\,\cdot\,\frac{\dl
S_{BLT}}{\delta \phi ^{*Aa'}} + \varepsilon _{ab} \phi ^{*Ab}
\frac{\dl S_{BLT}}{\delta \bar \phi ^A} \nonumber
\end{eqnarray}
We introduce two antibrackets, one for every $\phi ^{*Aa}$, defined by
\beq
(F,G)_a = \frac{\dr F}{\delta \phi
_A}\,\cdot\,\frac{\dl G}{\delta \phi ^{*Aa}} - \frac{\dr
F}{\delta \phi ^{*Aa}}\,\cdot\,\frac{\dl G}{\delta \phi _A}\,.
\eeq
Of course, they have the same properties as the antibrackets from the usual
BV scheme, so that we finally can write the classical master equations as
\beq
\frac{1}{2}(S_{BLT},S_{BLT})_a + \varepsilon _{ab} \phi ^{*Ab}
\frac{\dl S_{BLT}}{\delta \bar \phi ^A} = 0\,.
\label{ClasBLT}
\eeq
For closed, irreducible algebras, we know that the proper solution is of
the form (\ref{SBLT}), if a complete set of gauge generators
$R^i_{\alpha}$ is used.

The quantisation prescription is then to construct $S_{BLT}$,
function of fields and antifields, by solving the classical master
equations. The gauge fixing is done by acting
with the operator $\hat U(\Psi )$.  Then the antifields $\phi ^{*Aa}$ and
$\bar \phi ^A$ are removed by the $\delta$-functions which fix them to
zero. Notice however that instead of acting with
$\hat U$ on $e^{\ihbar S_{BLT}}$, it is a lot easier to take as realisation
of the gauge fixing $S_{\Psi} + \tilde S_{\delta}$, especially when
$S_{BLT}$ becomes non-linear in the antifields.

\section{Ward identities and quantum master equations}

In this section, we first derive the Ward identities for the
BRST--anti-BRST symmetry and then we take these identities
as a starting point to derive the quantum master equation.
This is in analogy with section 2 of chapter 5.

\subsection{Ward identities}

Since the gauge fixed action we constructed (\ref{Sgf}) is invariant under
both the BRST and anti-BRST transformation rules, the standard procedure
of section 2 of chapter 3 allows the derivation of 2 types
of Ward identities.  For any $X$, we have that
\begin{eqnarray}
\langle \delta _1 X \rangle &=& 0 \nonumber \\
\langle \delta _2 X \rangle &=& 0\,,
\end{eqnarray}
where $\langle {\cal O} \rangle$ denotes the quantum expectation value
using
the gauge fixed action (\ref{Sgf}) of an operator $\cO$.  As we are only
interested in the
theory after having integrated out $\varphi_+$, we will restrict ourselves
to quantities $X(\phi _A,\phi ^{*Aa},\bar \phi )$.
For a closed algebra, the Jacobian of the measure of the functional
integral under either a BRST or an anti-BRST transformation is a function
of
$\phi -\varphi_+$ (\ref{colanti}). Since $\delta_a (\phi_A - \varphi_{A+})
= \cR_{Aa}(\phi -\varphi_+)$, we consider quantum counterterms of the form
$M(\phi - \varphi _+)$. The quantum extended action is defined by
\beq
W_{BLT}(\phi ,\phi ^{*Aa},\bar \phi ) = S_{BLT}(\phi ,\phi ^{*Aa},\bar
\phi ) + \hbar M(\phi) \,.
\eeq
The Ward identities become
\begin{eqnarray}
0 &=& \langle \delta_a X \rangle \nonumber \\
&=& \int [d\phi ][d\phi ^{*a}][d\bar \phi ][d\varphi _+][d\pi
_a][dB][d\lambda ] \;\; \delta _a X\,\cdot\,e^{\frac{i}{\hbar
}W_{BLT}(\phi - \varphi _+,\phi ^{*Aa},\bar \phi )} \nonumber \\
& & \hspace*{1cm}\cdot\, e^{\ihbar S_{\Psi}} e^{\ihbar
S_\delta }\,.
\end{eqnarray}
Let us take $a = 1$.  Then
\begin{eqnarray}
\label{del1X}
\delta _1 X &=& \frac{\dr X}{\delta \phi _A}\,\cdot\,\pi _{A1} +
\frac{\dr X}{\delta \phi ^{*A1'}}(-1)^{\gras{A}} M^{BA} \left[\lambda _B -
\frac{1}{2}\left(B_B + \frac{\dr R_{B1}}{\delta \phi
_C}R_{C2}\right)\right] \nonumber \\
 & & + \frac{\dr X}{\delta \bar \phi^A}\,\cdot\,\frac{1}{2}
M^{BA} [-2\phi _B^{*2} + \pi _{B1} - R_{B1}(\phi - \varphi _+)]\,.
\end{eqnarray}
We reintroduced the primes for the $\phi ^{*Aa'}$ in order to distinguish
the antifields before and after the redefinition (\ref{rredef}).
In the second term of (\ref{del1X}), we can replace
\beq
B_B + \frac{\dr  R_{B1}(\phi - \varphi _+)}{\delta
\phi _C } R_{C2}(\phi  - \varphi _+)\, ,
\eeq
by
\beq
  \frac{\dl}{\delta \bar \phi^B} ( S_{\delta} +
  W_{BLT}(\phi - \varphi _+,\phi ^{*Aa},\bar \phi ) ) \; .
\eeq
In the third term of (\ref{del1X}), $\pi_{B1} - R_{B1}(\phi - \varphi _+)$
equals
\beq
    - \frac{\dl}{\delta \phi^{*B1'}} ( S_{\delta} +
  W_{BLT}(\phi - \varphi _+,\phi ^{*Aa},\bar \phi ) ) \; .
\eeq
Under the path integral, $(\dl S_{\delta} / \delta Q ). e^{\ihbar
S_{\delta}}$ can be replaced by $ (\hbar/i) \dl e^{\ihbar S_{\delta}} /
\delta Q$ ($Q = \bar \phi^B$ or $Q = \phi^{*B1'}$), and analogously for the
derivatives on $W_{BLT}$. By partial integrations,
one sees that these two contributions cancel.

$d\mu$ denotes the complete measure of the path integral.
The remaining Ward identity is
\begin{eqnarray}
0 &=& \int d\mu \left[\frac{\dr X}{\delta \phi _A} \pi _{A1} +
\frac{\dr X}{\delta \phi ^{*A1'}} (-1)^{\gras{A}} M^{BA} \lambda _B + \phi
^{*A2'} (-1)^{\gras{X}} \frac{\dl X}{\delta \bar \phi^A}\right]
\nonumber \\
& & . e^{\frac{i}{\hbar }W_{BLT}(\phi - \varphi _+,\phi ^{*Aa},\bar
\phi )} .\left[ \hat V e^{\frac{i}{\hbar }\tilde S_\delta }\right].
e^{-\frac{i}{\hbar }\varphi _{A+} M^{AB} \lambda _B}\,.
\end{eqnarray}
In the first term, the $\varphi _{A+}$ can trivially be integrated out.
Then, considering the expressions for $\hat V$ and $\tilde S_\delta $, we
see that $\pi _{A1}$ can be replaced by $-\frac{\hbar
}{i}\,\dl (\exp \ihbar \tilde S_{\delta}) /\delta \phi ^{*A1'}$.
Integrating by parts over $\phi ^{*A1'}$, gives
\begin{eqnarray}
&&\frac{\hbar }{i} (-1)^{\varepsilon _X(\varepsilon _A + 1)}
\frac{\dl}{\delta \phi ^{*A1'}} \left[\frac{\dr X}{\delta
\phi _A} e^{\frac{i}{\hbar }W_{BLT}}\right] .\left[ \hat V
e^{\frac{i}{\hbar }\tilde S_\delta }\right]\\
&&= \left[-i\hbar \Delta _1 X + \frac{\dr X}{\delta \phi
_A}\,\cdot\,\frac{\dl W_{BLT}}{\delta \phi ^{*A1'}}\right]
 e^{\frac{i}{\hbar }W_{BLT}} .\left[ \hat V e^{\frac{i}{\hbar }\tilde
S_\delta }\right] \nonumber
\end{eqnarray}
in the path integral.  Here, we generalised that other operator
well-known from BV (\ref{DELTA}) :
\beq
\Delta_a X = (-1)^{\varepsilon _A + 1} \frac{\dr }{\delta \phi
^{*Aa'}}\ \frac{\dr}{\delta \phi _A} X\,.
\eeq
For the second term we can proceed analogously by replacing $M^{AB} \lambda
_B e^{- \frac{i}{\hbar }\phi _{A+}M^{AB}\lambda _B}$ by
$\left(-\frac{\hbar }{i}\right) \frac{\dl }{\delta \varphi
_{A+}}e^{-\frac{i}{\hbar }\varphi _{A+}M^{AB}\lambda _B}$.
Integrating by parts over $\varphi _{A+}$, we see that the derivative can
only act on $W_{BLT}(\phi - \varphi _+,\phi ^{*a},\bar \phi )$, and we get
under the path integral
\beq
\frac{\dr X}{\delta \phi ^{*A1'}}\ \frac{\hbar }{i}\
\frac{\dl}{\delta \varphi _{A+}}\ e^{\frac{i}{\hbar } W_{BLT} (\phi
- \varphi _+,\phi ^{*a},\bar \phi )}.\left[ \hat V e^{\frac{i}{\hbar
}\tilde S_\delta }\right]. \delta (\varphi _+)\,.
\eeq
The derivative with respect to $\varphi_+$ can be replaced by a derivative
with respect to $\phi$. This leads finally to
\beq
- \int d\mu \;\;\; \frac{\dr X}{\delta \phi ^{*A1'}}\
\frac{\dl W_{BLT}}{\delta \phi _A}\ . \left[\hat
V e^{\frac{i}{\hbar }\tilde S_\delta} \right]\,.
\eeq
The complete Ward identity hence becomes, dropping the primes again,
\begin{eqnarray}
0 &=& \left\langle (X,W_{BLT})_1 - i\hbar \Delta _1 X + (-1)^{\varepsilon
_X} \phi ^{*A2} \frac{\dl X}{\delta \bar \phi^A}\right\rangle\\
&=& \int [d\phi ][d\phi ^{*a}][d\bar \phi ]\left[(X,W_{BLT})_1 -
i\hbar \Delta _1 X + (-1)^{\varepsilon
_X} \phi ^{*A2} \frac{\dl X}{\delta \bar \phi^A}\right]
\nonumber
\end{eqnarray}
\[e^{\frac{i}{\hbar }W_{BLT}} .\left[ \hat V e^{\frac{i}{\hbar }\tilde
S_\delta }\right]\,.\]
An analogous property is obtained by going through the same steps
for the Ward identities $\langle \delta _2 X \rangle = 0$.

\subsection{Quantum Master Equation}

Analogous to the case of the BV formalism (section 2 of chapter 5), the
fact that these Ward identities
are valid for all $X(\phi ,\phi ^{*a},\bar \phi )$, leads to two conditions
on
$W_{BLT}$, the so-called {\it quantum master equations}. Starting from the
most general Ward identity, the purpose is to remove all
derivative operators acting on $X$ by partial integrations. Again, $d\mu$
denotes the measure of the path integral. We start from
\begin{eqnarray}
   0 & = & \int d\mu \left[ \frac{\dr X}{\delta \phi_A} \frac{\dl
W_{BLT}}{\delta \phi^{*Aa}} - \frac{\dr X}{\delta \phi^{*Aa}}\frac{\dl
W_{BLT}}
{\delta \phi_A} -i\hbar (-1)^{\gras{A}+1} \frac{\dr}{\delta \phi^{*Aa}}
\frac{\dr}{\delta \phi_A} X \right. \nonumber \\
 & & \left. + (-1)^{\gras{X}} \epsilon_{ab} \phi^{*Ab} \frac{\dl X}{\delta
\bar \phi^A} \right] e^{\ihbar(W_{BLT} + S_{\Psi} + \tilde S_{\delta})}.
   \label{Ward}
\end{eqnarray}
Notice that the operator $\hat V$ was explicitised again as
$e^{\ihbar S_{\Psi}}$.

By integrating by parts over $\phi_A$ in the first term, we get the
following two terms:
\begin{eqnarray}
 & &   \int d\mu \;\; i\hbar. X. \Delta_a e^{\ihbar W_{BLT}} . e^{\ihbar(
S_{\Psi} + \tilde S_{\delta})} \nonumber \\
 &  & + \int d\mu \;\; i\hbar. X. (-1)^{\gras{A}+1} \frac{\dr e^{\ihbar
W_{BLT}}} {\delta \phi^{*Aa}}. \frac{\dr}{\delta \phi_A} \left[
 e^{\ihbar( S_{\Psi} + \tilde S_{\delta})} \right].
 \label{Term1}
\end{eqnarray}
The second and third contribution to the Ward identity (\ref{Ward}) can be
combined to give
\beq
   \int d\mu  \;\; (-i\hbar) (-1)^{\gras{A}+1} \frac{\dr}{\delta \phi_A}
\left[ \frac{\dr X}{\delta \phi^{*Aa}} e^{\ihbar W_{BLT}} \right] .
 e^{\ihbar( S_{\Psi} + \tilde S_{\delta})} .
\eeq
Integrating by parts twice, first over $\phi_A$, then over $\phi^{*Aa}$
gives us the terms:
\begin{eqnarray}
 &  & \int d\mu \;\; i\hbar (-1)^{\gras{A}}. X. e^{\ihbar W_{BLT}}.
\frac{\dr}{\delta \phi^{*Aa}} \frac{\dr}{\delta \phi_A} \left[
 e^{\ihbar( S_{\Psi} + \tilde S_{\delta})} \right]  \nonumber \\
 & + & \int d\mu \;\; i\hbar (-1)^{\gras{A}}. X. \frac{\dr e^{\ihbar
W_{BLT}}}{\delta \phi^{*Aa}} . \frac{\dr}{\delta \phi_A}
\left[ e^{\ihbar( S_{\Psi} + \tilde S_{\delta})} \right].
\label{Term2}
\end{eqnarray}
Notice that the second term of (\ref{Term1}) cancels the second term of
(\ref{Term2}).

Also in the fourth term of (\ref{Ward}), we have to integrate by parts,
over $\bar \phi^A$. This gives us again two terms:
\begin{eqnarray}
 & - &\int  d\mu \;\; X . \epsilon_{ab} \phi^{*Ab}
  \frac{\dl e^{\ihbar W_{BLT}}}{\delta \bar \phi^A}
   e^{\ihbar( S_{\Psi} + \tilde S_{\delta})} \nonumber \\
 & - & \int d\mu \;\; X . \epsilon_{ab} \phi^{*Ab} e^{\ihbar W_{BLT}}
\frac{\dl }{\delta \bar\phi^A}
\left[ e^{\ihbar( S_{\Psi} + \tilde S_{\delta})} \right].
\label{Term3}
\end{eqnarray}
We now show that the first term in (\ref{Term2}) and the
second term in (\ref{Term3}) cancel. Working out the two derivatives and
using the explicit form of $\tilde S_{\delta}$, we rewrite the first term
of (\ref{Term2}) as
\beq
   \int d\mu  \;\; (i\hbar) (\ihbar)^2 .X. e^{\ihbar W_{BLT}}
 e^{\ihbar( S_{\Psi} + \tilde S_{\delta})}
 \frac{\dr S_{\Psi}}{\delta \phi_A} \pi_{Aa}.
\eeq
Now, we know that $\delta_a S_{\Psi} = 0$, which allows us to replace
$ \frac{\dr S_{\Psi}}{\delta \phi_A} \pi_{Aa}$ by $- \frac{\dr
S_{\Psi}}{\delta \pi_{Ab}} \epsilon_{ab} B_A$. Using the explicit form of
$ \tilde S_{\delta}$ again, this is
\beq
   - \int d\mu  \;\; (i\hbar). X. e^{\ihbar W_{BLT}}. \frac{\dr e^{\ihbar
S_{\Psi}}}{\delta \pi_{Ab}} .\frac{\dl e^{\ihbar \tilde S_{\delta}}}{\delta
\bar \phi^A} \epsilon_{ab}.
\eeq
One more partial integration, over $\pi_{Ab}$, is needed to see that the
terms do cancel as mentioned above.

Summing up (\ref{Term1},\ref{Term2},\ref{Term3}), we see that the Ward
identities (\ref{Ward}) are equivalent to
\beq
  0 = \int d\mu  \;\; X \left [\hat V
e^{\ihbar \tilde S_{\delta}} \right]
  \left[ i\hbar \Delta_a - \epsilon_{ab} \phi^{*Ab}
\frac{\dl}{\delta \bar\phi^A} \right] e^{\ihbar W_{BLT}} .
\eeq
As this is valid for all possible choices for $X(\phi,\phi^{*a},\bar\phi)$,
we see that $W_{BLT}$ has to satisfy the {\it quantum master equation}
\beq
   \left[ i\hbar \Delta_a - \epsilon_{ab} \phi^{*Ab}
\frac{\dl}{\delta \bar\phi^A} \right] e^{\ihbar W_{BLT}} = 0.
\eeq
This is equivalent to
\beq
   \frac{1}{2} (W_{BLT},W_{BLT})_a + \epsilon_{ab} \phi^{*Ab} \frac{\dl
W_{BLT}}{\delta \bar \phi^A} = i\hbar \Delta_a W_{BLT}.
\eeq
Remember that these are two equations, $a=1,2$. By doing the usual
expansion $W_{BLT} = S_{BLT} + \hbar M_1 + \hbar^2 M_2 + \ldots$, we
find the classical master equation (\ref{ClasBLT}) for $S_{BLT}$ back.

\section{Open Algebras}

In the previous chapter, we pointed out how combining the collective field
approach
and the recipe of \cite{Open}, one is naturally led to the construction of
an extended action that contains terms of quadratic and higher order in the
antifields for BRST invariant quantisation. As we do not have a principle
analogous to the one described in \cite{Open} and chapter 6,
for constructing a gauge fixed action that is invariant under
BRST--anti-BRST
symmetry for the case of an open algebra, we will have to take the other
point of view advocated in section 2 of chapter 6.

The collective field method is
{\em a method sometimes employed in French
cuisine : a piece of pheasant meat is cooked between two slices of veal,
which are then discarded} \cite{GM}. Nevertheless,
like in the case of ordinary BRST collective field quantisation (see
chapter 5 and 6), the introduction of the collective fields allows
to shift the problem of the off-shell non-nilpotency to the (anti-)BRST
transformations of the collective fields. Indeed, $\delta_a \phi_A =
\pi_{Aa}$,
$\delta_a \pi_{Ab} = \epsilon_{ab} B_A$ and $\delta_a B_A = 0$ guarantee
that $\delta_a^2 \phi_A = 0 $  and that $(\delta_1 \delta_2 + \delta_2
\delta_1)\phi_A = 0$. Therefore, the originally present gauge symmery can
be fixed in a BRST--anti-BRST invariant way
like for closed algebras, i.e.\ by adding $S_{\Psi} = \frac{1}{2}
\epsilon^{ab} \delta_a \delta_b \Psi$ to a BRST--anti-BRST invariant
action,
$S_{inv}$. This way, the BRST and anti-BRST Ward identities guarantee that
whatever way we
choose to construct $S_{inv}$, the partition function will be independent
of the gauge choice if $S_{inv}$ is BRST--anti-BRST invariant.

We decompose again
\begin{eqnarray}
   S_{inv} & = & S_{BLT}(\phi -\varphi_+,\phi^{*a'},\bar \phi)\\
   &  & - \varphi_{A+} M^{AB} \lambda_B + \bar \phi^A B_A - \phi^{*Aa'}
\pi_{Aa} \nonumber.
\end{eqnarray}
It is useful to keep the redefinitions (\ref{rredef}) in mind in the
following. $S_{inv}$ will be BRST--anti-BRST invariant, that is $\delta_a
S_{inv} = 0$, under the transformation rules
(\ref{colanti},\ref{ccolanti}), except for the generalisations
\begin{eqnarray}
 \delta_a\varphi_{Ab} & = & \delta_{ab} \left[ \pi_{Aa} - \epsilon_{ac}
\phi^{*c}_A - \frac{\dl S_{BLT}(\phi - \varphi_+)}{\delta \phi^{*Aa'}}
\right] + (1 - \delta_{ab}) \epsilon_{ac} \phi^{*c}_A \nonumber \\
  \delta_a \phi^{*b}_A & = & - \delta_a^b \left[ (-1)^a \lambda_A +
\frac{1}{2} (B_A + \frac{\dl S_{BLT}(\phi -\varphi_+)}{\delta \bar
\phi^A})\right] \; ,
\end{eqnarray}
if $S_{BLT}$ satisfies the two classical master equations of the antifield
scheme. Hence, we see that the question whether open algebras can be
quantised in
a BRST--anti-BRST invariant way, reduces to the fact whether a solution to
(\ref{ClasBLT}) can be found for open algebras with the extra condition
that $S_{BLT} = S_0 + \phi^{*Aa} \cR_{Aa} + \ldots$ It has been proved that
such solutions exist \cite{BLT,Gregoire,Brussel}.

When quantum counterterms are needed to derive the Ward identities, we
obtain the two quantum
master equations, following the same steps as in section 2 of chapter 6,
as the conditions that guarantee the validity of the Ward identities.
These identities then imply gauge independence of the partition function.

\chapter{Canonical transformations}

After the inductive approach of the previous chapters, where we have shown
how the antifield scheme for BRST invariant quantisation can be constructed
from the BRST quantisation
recipes, we present here a more algebraic approach. As was already pointed
out above (\ref{antibracket}), the antibracket with fields and antifields
ressembles
the Poisson bracket of classical mechanics in its Hamiltonian formulation.
Inspired by this analogy, we will look for {\it canonical transformations}
of the fields and the antifields that leave the antibracket of two
function(al)s invariant.

A large class of canonical transformations, although not all, are those
that are obtained from a generating function, which has to be fermionic
here.
We will first show that such transformations leave both the classical and
quantum cohomology invariant. For the former, this follows trivially from
the definition of canonical transformations itself. For the latter
however, we have to study carefully how $\Delta$ transforms, which is
related to the transformation of the measure of the path integral.

The gauge fixing procedure as defined above (\ref{BVint}), is a canonical
transformation
generated by $F = \unity + \Psi$ \cite{Siegel}, where $\unity$ is a
symbolic notation for the identity tranformation. We replace
the condition of gauge invariance of the expectation value of an
arbitrary operator $X$ by invariance under arbitrary
canonical transformations. This way we rederive the quantum master
equation and the condition that the operator $X(\phi ,\phi^*)$ has to
satisfy in order to have the same expectation value in two different sets
of canonical coordinates.

Besides gauge fixing, the use of canonical transformations is manifold.
First of all, they can be used to construct other realisations as a field
theory of the same physical degrees of freedom, i.e. to construct
cohomologically equivalent theories with a different field content. This
will be demonstrated in the examples below and used in
chapter 14 on the hiding of anomalies. Conversely, auxiliary fields can be
removed in a consistent way by doing canonical transformations that bring
them under the form of a trivial system, which can then be discarded
\cite{MH,TheBible}. Taking a different set of gauge generators
$R^i_{\alpha}$ can also be seen as a canonical transformation \cite{bv4}.

Before studying the canonical transformations in more detail, let us make
another comment. The analogy with the Poisson bracket has also served as a
starting point for recent investigations on the geometrical structure of BV
\cite{geomBV}. Like for Poisson brackets, antibrackets have been defined
using a general Grassmann odd symplectic 2-form which has to be closed
(Jacobi identity). The form of the antibracket that we always use in this
work corresponds to working with the Darboux coordinates. The same 2-form
is then used to define a second order differential operator $\Delta$, that
has all the properties listed in the appendix. It is believed that these
constructions will serve in the attempts at a construction of
string field theories \cite{Witten1,Zwiebach}.

\section{Canonical transformations and the cohomologies}

Two sets of canonical variables (=fields and antifields) are respectively
denoted by $\{\phi^A ,\phi^*_A \}$ and $\{ \phi^{A'} , \phi^{*'}_A \}$. A
transformation from the unprimed to the primed indices is then said to be
{\it canonical} if for any two function(al)s $A(\phi ,\phi^*)$ and $B(\phi
,\phi^*)$ calculating the antibracket in the unprimed variables and
transforming the result gives the same expression as first transforming $A$
and $B$ to the primed variables and then calculating the antibracket with
respect to the primed variables.

A large class of canonical transformations consists of those
transformations for which
\beq
     \frac{\dr \phi^B(\phi',\phi^{*'})}{\delta \phi^{'A}}\vert_{\phi^{*'}}
\eeq
is invertible. It is possible to show \cite{WPT,TheBible} that they can be
obtained
from a fermionic generating function $F(\phi ,\phi^{*'})$ of ghostnumber
$-1$. The transformation rules are then given by
\begin{eqnarray}
   \label{CT}
       \phi^{A'} & = & \frac{\delta F(\phi,\phi^{*'}) }{\delta
\phi^{*'}_A} \nonumber \\
\phi^*_A & = & \frac{\delta F( \phi,\phi^{*'} )}{\delta \phi^A}.
\end{eqnarray}
The other way around, if $F(\phi,\phi^{*'})$ is such that
\beq
     \frac{\dl \delta}{\delta \phi^{*'}_A \delta \phi^B} F(\phi,\phi^{*'})
\eeq
is invertible, then the transformation given by (\ref{CT}) is canonical.

We study here infinitesimal canonical transformations generated by
\beq
    F(\phi,\phi^{*'}) = \unity + f(\phi,\phi^{*'}) = \phi^A \phi^{*'}_A
    +f(\phi,\phi^{*'}),
\eeq
with $f$ small. Below we will use the name {\it generating fermion} for
$f$. The transformation rules become
\begin{eqnarray}
      \phi^{A'} & = &
      \phi^A + \frac{\delta f(\phi ,\phi^*)}{\delta \phi^*_A}
\nonumber \\
      \phi^{*'}_A & = &
      \phi^*_A - \frac{\delta f(\phi,\phi^*)}{\delta \phi^A}.
\end{eqnarray}
We replaced $\phi^{*'}$ on the RHS by $\phi^*$ since we are making an
infinitesimal transformation.

The expression in the primed coordinates for any function(al) given in the
unprimed coordinates can be obtained by direct substitution of the
transformation rules. Owing to the infinitesimal nature of the
transformation, we can expand in a Taylor series to linear order in $f$ and
we find
\begin{eqnarray}
   X'(\phi ',\phi^{*'}) & = & X \left(
      \phi^{A'} - \frac{\delta f(\phi' ,\phi^{*'})}{\delta \phi^{*'}_A},
      \phi^{*'}_A + \frac{\delta f(\phi',\phi^{*'})}{\delta
\phi^{A'}}\right) \nonumber \\
      & = & X (\phi^{'} , \phi^{*'}) - (X,f) (\phi',\phi ^{*'}).
\end{eqnarray}
By $(\phi^{'},\phi^{*'})$ we do of course not mean the antibracket of a
field and an antifield but only that the preceding expression is a function
of fields and antifields. We can
drop the primes of the arguments and denote the transformed functional by
\beq
    X'(\phi ,\phi^*) = X(\phi ,\phi^*) - (X,f)(\phi ,\phi^*)
    \label{ClasTraf}.
\eeq
With this result, it becomes easy to show that (infinitesimal)
transformations of the form (\ref{CT}) are indeed canonical. Consider
\begin{eqnarray}
  ( A',B' ) & = & ( A - (A,f), B - (B,f)) \nonumber \\
     & = & (A,B) - ((A,B),f) + \cO(f^2) \\
     & = & (A,B)' \nonumber,
\end{eqnarray}
where in the second step the Jacobi identity for the antibracket is used.
Remember that $\cO(f^2)$ is neglected.
As a corollary of this, it is straightforward to see that the classical
cohomology is invariant under (infinitesimal) canonical transformations.
Also, a solution of the classical master equation is transformed in a
solution of the classical master equation. We discuss two examples in
section 4 of canonical transformations that are merely redefinitions of the
gauge generators $R^i_{\alpha}$. The structure functions and non-closure
functions of the algebra (\ref{algebra}) of the new gauge generators can
then be determined from the transformed extended action.

Let us now turn to the box operator. Using again the expressions of the
appendix, we have that
\begin{eqnarray}
     \label{TranDX}
      \Delta X' & = & \Delta X - \Delta (X,f) \nonumber \\
      & = & (\Delta X)' - (X, \Delta f).
\end{eqnarray}
Hence, we see that acting with $\Delta$ is not a canonical invariant
operation. An extra term $- (X,\Delta f)$ appears. We show in the next
section that $\Delta f=\ln J$, with $J$ the Jacobian of the change of
integration variables in the path integral.

We want to define the quantum BRST operator in the transformed
coordinates,
$\sigma' Y = ( Y, \tilde W ) - i\hbar \Delta Y$. In order for this
operator to be nilpotent, we know that $\tilde W$ has to satisfy the
quantum master equation. In contrast with the classical master equation,
the transformation of a solution $W$ of the
quantum master equation does {\it not} give a
solution in the new variables. Indeed, we have that
\begin{eqnarray}
   ( W',W' ) - 2i\hbar \Delta W' & = & [ (W,W) - 2 i \hbar \Delta W ]' + 2
i \hbar ( W , \Delta f ) \nonumber \\
  & = & 2 i \hbar (W,\Delta f),
\end{eqnarray}
if $W$ satisfies the quantum master equation (\ref{QME2}) in the original
variables. Instead,
\beq
    \tilde W = W' - i\hbar \Delta f = W + \sigma f
    \label{Wtilde}
\eeq
does satisfy the quantum master equation in the transformed coordinates, as
follows from the nilpotency of $\Delta$.
Notice that (\ref{Wtilde}) generalises (\ref{ClasTraf})
in the sense that both the classical and quantum extended action transform
under an infinitesimal canonical transformation by the addition of
respectively the classical and quantum BRST transformation of the
generating fermion $f$.
Then we have that
\begin{eqnarray}
    \sigma'X' & = & ( X' , \tilde W ) - i\hbar \Delta X' \nonumber \\
       & = & (\sigma X)' .
\end{eqnarray}
Hence, we see that also the quantum cohomology is invariant under
infinitesimal transformations. If $\sigma X = 0$, then $\sigma'X'=0$ and if
$Y=\sigma X$ then $Y' = \sigma 'X'$.

Although we have only shown that the antibracket and both the classical and
quantum cohomology are invariant under {\em infinitesimal} canonical
transformations, these results also hold for finite transformations. For
proofs of this statement, see \cite{TheBible}.

\section{From gauge invariance to invariance under canonical
transformations}

As was pointed out in section 3 of chapter 5, we can choose more general
gauge fermions, depending on both the fields and the antifields, leading to
a gauge fixed action (\ref{More})
$S(\phi-\frac{\delta \Psi}{\delta \phi^*},\phi^* + \frac{\delta
\Psi}{\delta \phi})$.
This gauge fixed action is obtained from $S(\phi ,\phi^*)$ by doing a
canonical transformation. This observation will be our starting
point here. In chapter 13 we copy the derivation of the master equation and
of the quantum BRST operator that we give in this section, using one-loop
regularised path integrals.

We construct the expression for the expectation value of an
arbitrary operator $X(\phi,\phi^*)$ in two
sets of coordinates, related by an infinitesimal canonical transformation
generated by a fermion $f$ and we calculate their difference in function of
the fermion $f$. So we have
\beq
    \chi(\phi^*) = \int [d\phi] X(\phi ,\phi^*) e^{\ihbar W(\phi ,\phi^*)}
\eeq
and
\beq
    \chi'(\phi^*) = \int [d\phi] \left[ X - (X,f) \right] e^{\ihbar
\left( W - (W,f) - i\hbar \Delta f \right) }.
\eeq
Expanding in $\chi'(\phi^*)$ to linear order in $f$, and using the
notation $\cW = \ihbar W$, we find that
\begin{eqnarray}
   \delta \chi & = & \chi'(\phi^*) - \chi(\phi^*) \nonumber \\
               & = & \int [d\phi] \left[ - X (\cW,f) + X\Delta f - (X,f)
\right] e^{\cW} \\
               & = & \int [d\phi] \left[ - X \frac{\dr e^{\cW}}{\delta
\phi^A} \frac{\dl f}{\delta \phi^*_A} + X \frac{\dr e^{\cW}}{\delta
\phi^*_A} \frac{\dl f}{\delta \phi^A} + X .\Delta f. e^{\cW} \right.
\nonumber \\
            &  & \left. - \frac{\dr X}{\delta \phi^A} \frac{\dl f}{\delta
\phi^*_A} e^{\cW} + \frac{\dr X}{\delta \phi^*_A} \frac{\dl f}{\delta
\phi^A} e^{\cW} \right] .
\end{eqnarray}
The first, third and fourth term can be combined to a total derivative with
respect to $\phi^A$, and can hence be discarded. In the two remaining
terms, we integrate by parts over $\phi^A$, which leads to:
\beq
   \label{deltachi}
   \delta \chi = \int [d\phi] \left[ X.\Delta e^{\cW}.f + \left[ (X,\cW) +
\Delta X \right] . e^{\cW} . f \right].
\eeq
Imposing that $\delta \chi=0$ for all $f$, i.e. that the
expectation value of the operator $X$ is invariant under infinitesimal
canonical transformations (=infinitesimal deformations of the gauge
choice), we have the sufficient conditions\footnote{In fact, it seems that
from $\delta \chi = 0$ for all $f$ it only follows that $\Delta [ X e^{\cW}
] = 0$. However, as we want that for $X=1$ the partition function is
independent of $f$, we separate this one condition in two.}
\begin{eqnarray}
\Delta e^{\cW} & = & 0 \nonumber \\ (X,\cW) + \Delta X & = & 0.
\end{eqnarray}
These are easily seen to be equivalent to the quantum master equation
\beq
    (W,W) - 2 i \hbar \Delta W = 0,
\eeq
and the condition that a gauge invariant operator has a vanishing quantum
BRST variation
\beq
     \sigma X = (X,W) - i\hbar \Delta X = 0.
\eeq
We can also impose gauge independence \footnote{We will use equivalently
{\it independent of the set of canonical coordinates} and {\it gauge
independent}.}
on correlation functions. Consider for $X = Y e^{J(\phi ,\phi^*)}$, where
the $J(\phi ,\phi^*)$ can be interpreted as source terms. From the
expression
for $\sigma [AB]$ in the appendix, we conclude:
\begin{eqnarray}
         \sigma e^J  & = & 0 \nonumber \\
         (Y , e^J ) & = & 0 \\
         \sigma Y & = & 0 \nonumber .
\end{eqnarray}
The first line is the gauge independence condition for the sources $J$.
If the sources $J$ do not satisfy this requirement, the correlation
functions become gauge dependent with a dependence given by
(\ref{deltachi}).

It is straightforward to show, using again
partial integrations, that for all functions $X(\phi,\phi^*)$,
\beq
 \int [d\phi]\, \sigma X(\phi ,\phi ^*). e^{\ihbar W(\phi ,\phi^*)} = 0,
\eeq
provided $W$ satisfies the quantum master equation.
This is the form of the Ward identities in the BV scheme (see section 2 of
chapter 5).
Here too, by taking $X=Ye^J$, Ward identities for the correlation functions
of $\sigma Y$ are obtained.

Notice that demanding that the partition function is the same in both sets
of coordinates is a non-trivial condition. Of course, the fact that we can
redefine $\phi^A = \phi^{A'} - \delta f/\delta \phi^{*'}_A$ is a
mere consequence of the freedom to redefine the variables in a
(path) integral. It is the invariance under redefinition of the antifields
that leads to conditions on the integrand. The extra term $-i\hbar \Delta
f$ in the transformation the quantum extended action $W$ is exactly the
Jacobian that one expects from the change of the integration variables:
\begin{eqnarray}
    \label{deltaf}
     J & = & \sdet \left[ \delta^A_B - \frac{\dl}{\delta \phi^{B'}}
\frac{\dr f}{\delta \phi^{*'}_A} \right] \nonumber \\
       & = & e^{\ihbar (-i \hbar \Delta f)} \; ,
\end{eqnarray}
again neglecting $\cO(f^2)$ corrections.

\section{The Zinn-Justin equation}

Let us here mention a nice result for the sake of completeness, but which
we will not explicitely use below \cite{NucHenneaux,TheBible}.
After introducing a (non gauge invariant) sourceterm for all
fields, the generating functional $\cW_c$ for connected diagrams is
defined as:
\beq
   e^{- \ihbar \cW_c(J,\phi^*)} = \int [d\phi] e^{\ihbar W(\phi ,\phi^*) +
  \ihbar J_A \phi^A}.
\eeq
As usual, $\cW_c$ depends on the sources of the fields, but it now also
depends on the sources $\phi^*$ of the BRST transformations. Using a
Legendre transform, we can pass from $\cW_c$ to the {\it effective action}
$\Gamma$. This goes as follows. The {\it classical field} is defined by
\beq
    \phi^A_{cl} = - \frac{\dl \cW_c}{\delta J_A} \; .
\eeq
We assume that this relation is invertible to give
$J(\phi_{cl},\phi^*)$. The effective action is then defined by
\beq
   \Gamma(\phi_{cl},\phi^*) = \cW_c(J(\phi_{cl},\phi^*),\phi^*) +
   J_B (\phi_{cl},\phi^*) \phi^B_{cl}.
\eeq
If we further define the antibracket
\beq
   (\Gamma , \Gamma) = \frac{\dr \Gamma}{\delta \phi^A_{cl}} \frac{\dl
\Gamma}{\delta \phi^*_A}
- \frac{\dr \Gamma}{\delta \phi^*_A} \frac{\dl \Gamma}{\delta
\phi^A_{cl}} \; \; ,
\eeq
it is easy to show that
\beq
\begin{array}{c}
\frac{1}{2} (\Gamma , \Gamma) e^{-\ihbar \cW_c(J(\phi_{cl},\phi^*),\phi^*)}
\\
= \left[  \int [d\phi] [\frac{1}{2} (W,W) - i \hbar \Delta W] e^{\ihbar
W(\phi ,\phi^*) + \ihbar J_A \phi^A} \right]_{J(\phi_{cl},\phi^*)}.
\label{ZJeq}
\end{array}
\eeq
If the quantum extended action $W$ satisfies the quantum master equation
(\ref{QME2}), we clearly have that
\beq
    ( \Gamma ,\Gamma )=0.
\eeq
This equation goes under the name {\it Zinn-Justin equation}. It is the
generalisation to all types of algebras of the result discussed in chapter
19 of \cite{Zinn-Justinboek}. {\it The effective action satisfies the
classical master equation if the theory is anomaly free}.

\section{Examples}

We give here some examples of the use of canonical transformations.
First we show how a change of basis of gauge generators can be realised by
a canonical transformation. Then we give two more examples that
are continuations \cite{ik8} of the examples given in chapter 5, the
topological Yang-Mills theory and the BRST invariant energy-momentum
tensor.

\subsection{Changing gauge generators using canonical transformations}

The form of the minimal proper extended action for a specific choice of a
complete set gauge generators $R^i_{\alpha}$ is given by (see chapters 5
and 6):
\beq
    S(\phi^A, \phi^*_A ) = S_0[\phi^i] +
    \phi^*_i R^i_{\alpha}[\phi] c^{\alpha} + c^*_{\gamma}
 T^{\gamma}_{\alpha \beta}[\phi] c^{\beta} c^{\alpha} +
\phi^*_i \phi^*_j E^{ji}_{\alpha \beta} c^{\beta} c^{\alpha} + \ldots
\eeq
$T^{\gamma}_{\alpha \beta}$ and $E^{ji}_{\alpha \beta}$ are determined by
(\ref{algebra}).

Consider now the canonical transformation generated by
\beq
    F_1 = \phi^{*'}_i \phi^i + c^{*'}_{\alpha} M^{\alpha}_{\beta}[\phi]
c^{\beta},
\eeq
where $M^{\alpha}_{\beta}[\phi]$ is invertible. From (\ref{CT}), we have
that
\begin{eqnarray}
   \phi^{i'} & = & \phi^i \nonumber \\
   c^{\alpha'} & = & M^{\alpha}_{\beta} c^{\beta} \nonumber \\
   \phi^*_i & = & \phi^{*'}_i + c^{*'}_{\alpha} \frac{\dr
M^{\alpha}_{\beta}[\phi] c^{\beta}}{\delta \phi^i} \nonumber \\
   c^*_{\beta} & = & c^{*'}_{\alpha} M^{\alpha}_{\beta}[\phi] \; .
\end{eqnarray}
If we only consider the terms linear in the antifields, we see that the
gauge generators $R^i_{\alpha}$ and the structure functions
$T^{\gamma}_{\alpha \beta}$ of the algebra change. In particular,
\beq
  R^{'i}_{\alpha} =  \left[ \frac{\dl}{\delta \phi^{*'}_i}
\frac{\dr}{\delta c^{\alpha'}} S'
\right]_{\phi^*_A =0} = R^i_{\beta}[\phi] M^{-1}{}^{\beta}_{\alpha} [\phi]
\; \eeq
We see that this canonical transformation transforms
a complete set of generators (\ref{completeR}) $R^i_{\alpha}$ in a
different complete set $R^{'i}_{\alpha}$ if $M^{\alpha}_{\beta}[\phi]$ is
invertible.

On the other hand, the fermion
\beq
   F_2 = \unity + \frac{1}{2} \phi^{*'}_i \phi^{*'}_j M^{ji}_{\alpha}
c^{\alpha}  \; ,
\eeq
where $M^{ij}_{\alpha} = (-1)^{(\gras{i}+1)(\gras{j}+1)} M^{ji}_{\alpha}$,
and where $M^{ij}_{\alpha}$ is field independent,
generates the transformation rules
\begin{eqnarray}
   \phi^{i'} & = & \phi^i + \phi^{*'}_j M^{ji}_{\alpha} c^{\alpha}
\nonumber \\
c^{\alpha'} & = & c^{\alpha} \nonumber \\
   \phi^*_i & = & \phi^{*'}_i \nonumber \\
   c^*_{\alpha} & = & c^{*'}_{\alpha} + \frac{1}{2} \phi^{*'}_i \phi^{*'}_j
 M^{ji}_{\alpha}  \; .
\end{eqnarray}
The transformed gauge generators are
\beq
R^{'i}_{\alpha} =
\left[\frac{\dl}{\delta \phi^*_i} \frac{\dr}{\delta c^{\alpha}} S'
\right]_{\phi^*_A =0} = R^i_{\alpha} + (-1)^{\gras{i}+1} \frac{\dr
S_0}{\delta \phi^j} M^{ji}_{\alpha} \; .
\eeq
The non-closure functions are also altered by this transformation, owing to
the transformation of $c^*_{\alpha}$. Clearly, using this type of canonical
transformation we can open a closed algebra.

\subsection{Enlarging the set of fields}

Canonical transformations combined with the introduction of trivial
systems, can be used to enlarge the set of fields and
the set of gauge symmetries without changing the cohomology of the theory.
In other words, they allow us to construct other field realisations of the
same physics.

Take an extended action
$S(\phi^A,\phi^*_A)$ that satisfies the classical master equation. Suppose
that we want to enlarge the set of fields by a set $\alpha_k$. As $S$ does
not depend on $\alpha_k$, the action is invariant under arbitrary shifts
of these fields. The properness condition implies that we
have to introduce ghosts $\beta_k$ for that symmetry and consider the new
extended
action $S' = S + \alpha^{*k} \beta_k$. It is trivial to see that
these extra fields do not change the (classical) cohomology. We are now
allowed to disguise these extra fields and extra symmetries by doing
whatever canonical transformation we want.

A nice application of this procedure is
for instance smooth bosonisation \cite{DaNiS}. There, one starts from
fermions (in 2d) coupled to an external source.  An extra scalar field is
introduced via a trivial system, and the transformation that is done to
disguise this trivial system is a chiral rotation, where the extra scalar
field gives the space-time dependent rotation angle. The Jacobian
of the transformation plays an important part, it provides the coupling of
this extra scalar field to the sources. Afterwards, the fermionic
degrees of freedom one started from, can be decoupled from the source
using a gauge fixing, and one is left with the bosonised theory. An
analogous scenario
has been used to extract mesonic degrees of freedom from the QCD field
theory \cite{DaNiS2}. We describe the first step of this bosonisation
procedure using a one loop regularised BV scheme as an example in the
third chapter of the third part. Another application,
which we will discuss in extenso later on, is the hiding of anomalies
\cite{ik3}. We will now briefly discuss how in
the model of topological Yang-Mills a realisation with a reducible set of
gauge symmetries can be obtained along these lines.

We had (\ref{SS}) that $\delta A_{\mu} = \epsilon_{\mu}$, but
usually \cite{TFTA,TFTB,TFTRep} the Yang--Mills gauge symmetry $\delta
A_\mu = D_\mu \epsilon
$ is included in the $R^i_{\alpha}$ and one starts from $ \delta A_{\mu} =
\epsilon_{\mu}+D_\mu \epsilon$. This is clearly a reducible set of gauge
generators as for $\epsilon_{\mu} = D_{\mu} \eta$ and $\epsilon = -\eta$,
we have $\delta A_{\mu} = 0$. We can go over to this reducible set
following the general lines sketched above.
First, we enlarge the configuration space by introducing a fermionic
ghost field $c$. As it does not appear in
the extended action so far, the extended action is invariant under
arbitrary
shifts of $c$, for which we introduce a {\it ghost for ghost} $\phi $. The
new extended action then becomes
\begin{equation}
S=S_0+A^{*\mu}\psi _\mu+c^*\phi \ .
\end{equation}
Now we do a canonical transformation,
generated by the fermion
\begin{equation}
F=\unity-\psi '^{*\mu }D_{\mu }c\ .
\end{equation}
This gives the transformation rules
\begin{eqnarray}
\psi _\mu &=&\psi '_\mu +D_\mu c\nonumber\\
c^*&=&c'^*+\partial _\mu \psi '^{*\mu }-\psi '^{*\mu }[A_\mu ,\cdot
]\nonumber\\
A^{*\mu }&=&A'^{*\mu }-\psi '^{*\mu }[\cdot ,c] \ .
\end{eqnarray}
The transformed extended action
is then (dropping the primes)~:
\begin{equation}
S=S_0+c^*\phi -\psi ^{*\mu }D_\mu (\phi -cc)+\psi ^{*\mu }(\psi _\mu
c+c\psi _\mu )+A^{*\mu }(\psi _\mu +D_\mu c)\ .
\end{equation}
Notice that the antifields of the ghosts $c$ and $\psi_{\mu}$ now act as
sources for the reducibility transformations: $c^* \phi$ and $-\psi^{\mu *}
D_{\mu}\phi$.
Also, $A_\mu $ transforms under the shifts as well as under the Yang--Mills
symmetry.
These are two typical properties of the solution of the classical master
equation in the case of reducible gauge symmetries. In order to make the
connection to the description with reducible symmetries
complete, we do yet another canonical transformation that
makes the familiar $c^*cc$ term of the Yang-Mills symmetry appear. This
transformation is generated by
\begin{equation}
G=\unity-\phi '^*cc \ .
\end{equation}
This gives $\phi '=\phi -cc$ and $c^*=c'^*-\phi '^*[c,\cdot ]$. After doing
these two canonical transformations, we have that
\begin{eqnarray}
S&=&S_0+A^{*\mu }(\psi _\mu +D_\mu c)+\psi ^{*\mu }(\psi _\mu c+c\psi _\mu
-D_\mu \phi )\nonumber\\&&+ c^*(\phi +cc)-\phi ^*[c,\phi ] \ .
\end{eqnarray}
Of course, this extra symmetry with ghost $\phi$, has to be gauge fixed
too. This is done by introducing a Lagrange multiplier and
antighost (sometimes called $\eta $ and $\bar \phi$).

\subsection{Canonical invariance of the energy-momentum tensor}

We now show that the definition of the energy-momentum tensor that we
have given in section 4 of chapter 5,
is invariant under (infinitesimal) canonical transformations, up to a
BRST exact term. Under an infinitesimal canonical transformation
generated by the fermion $F = \unity + f$, the classical
action and the energy-momentum tensor transform as follows:
\begin{eqnarray}
      S^{'} & = & S - (S,f) \nonumber \\
      T^{'}_{\alpha \beta} & = & T_{\alpha \beta} - (T_{\alpha \beta}, f)
      .
\end{eqnarray}
Here, $T_{\alpha \beta}$ is the energy-momentum tensor that is obtained
following the recipe given in chapter 5 starting from the extended action
$S$ \footnote{Let us briefly recapitulate this recipe. Define the operator
\beq
     D_{\alpha \beta} = \frac{2}{\sqrabsg} \frac{\delta}{\delta g^{\alpha
\beta}} + g_{\alpha \beta} \sum_i \phi^*_X \frac{\dl}{\delta \phi^*_X}\ .
\eeq
If $S$ satisfies the classical master equation, then $T_{\alpha \beta} =
D_{\alpha \beta} S$ is a classical gauge invariant operator: $(T_{\alpha
\beta},S)=0$. If $W$ satisfies the quantum master equation, then $T^q
_{\alpha \beta} = D_{\alpha \beta} W$ is quantum BRST invariant (at least
formally): $\sigma T^q_{\alpha \beta} = 0$.}.
Analogously, we can apply the recipe to the transformed action $S^{'}$,
which leads to an energy-momentum tensor $\tilde T_{\alpha \beta}$. Using
(\ref{metricbracket}) and (\ref{lemma1}), it is easy to show that
\begin{eqnarray}
  \label{there}
  \tilde T_{\alpha \beta} &=&
\frac{2}{\sqrabsg} \frac{\delta S'}{\delta g^{\alpha \beta}} + g_{\alpha
\beta} \sum_i \phi^*_X \frac{\dl S'}{\delta \phi ^*_X} \nonumber \\
  & = & T^{'}_{\alpha \beta} - (S ,
\frac{2}{\sqrabsg} \frac{\delta f}{\delta g^{\alpha \beta}} + g_{\alpha
\beta} \sum_i \phi^*_X \frac{\dl f}{\delta \phi ^*_X}) \nonumber \\
  & = & T^{'}_{\alpha \beta} + (D_{\alpha \beta} f , S'),
\end{eqnarray}
as for infinitesimal transformations terms of order $f^2$ can be neglected.

We finally verify that also $T^q_{\alpha \beta}$ is canonically
invariant. Under an infinitesimal canonical transformation, we have the
following transformation properties:
\begin{eqnarray}
          \tilde W & = & W + \sigma  f = W + (f,W) -i\hbar \Delta f
\nonumber \\ T^{q'}_{\alpha \beta} & = & T^q_{\alpha \beta} - (T^q_{\alpha
\beta},f),
\end{eqnarray}
with the same definition of $f$ as above. Let $\tilde T^q_{\alpha \beta}$
denote the energy-momentum tensor that we obtain by applying the recipe to
the transformed action $\tilde W$. We then easily see that
\begin{eqnarray}
    \tilde T^q_{\alpha \beta} & = & \frac{2}{\sqrabsg} \frac{\delta
\tilde W}{\delta g^{\alpha \beta}} + g_{\alpha \beta} \sum_i \phi^*_X
\frac{\dl \tilde W}{\delta \phi ^*_X} \nonumber \\
 & = & T^{q'}_{\alpha \beta} + \sigma \left[ D_{\alpha \beta} f \right].
\end{eqnarray}
Here too, rewriting the last term using $\sigma'$, the quantum BRST
operator in the transformed basis, only involves $f^2$ corrections.

\part{The equivalence of the Hamiltonian Batalin-Fradkin-Vilkovisky
and the Lagrangian Batalin-Vilkovisky quantisation schemes}

\chapter{Hamiltonian quantisation of gauge theories}

In this chapter, the Hamiltonian approach to the quantisation of gauge
theories is outlined, without any justification or proof. Since we
believe that most of the introduced concepts and procedures are evident
from the preceding discussion of the Lagrangian quantisation procedure,
this chapter is somewhat less selfcontained. In the next chapter, we
apply the recipe given here, to describe the Schwinger-Dyson BRST
symmetry in the
Hamiltonian formalism. This will allow us to prove the equivalence of the
Hamiltonian scheme given in this chapter and the Lagrangian approach on
which the rest of this work is focussed. We again restrict ourselves
to irreducible gauge symmetries.

In the Hamiltonian approach, two main parts can be distinguished. First,
one has to perform an analysis of the classical system, using what is often
denoted by {\it Dirac's constraints analysis} \cite{Dirac}. With the
results of this analysis at hand, one formulates the
classical system in terms of the global BRST symmetry, if gauge symmetries
are present. We discuss
the method of I.A. Batalin, E.S. Fradkin and G.A. Vilkovisky \cite{BFV}.
This prepares the stage for the quantisation of the model, using either
operator or path integral methods. In the spirit of
this work, we choose the latter.

A detailed account of both steps can be found in
\cite{Govaertsbook,Henneauxbook}.
A discussion of the second step, the BFV Hamiltonian formalism,
is given in \cite{HenneauxRep}.

\section{Dirac's constraints analysis}

Consider a classical dynamical system, described by a
Lagrangian function $L(q_n(t),\dot q_n(t))$. We use the notation of a
system with a discrete set of degrees of freedom, although $n$ may
represent a possibly continuous set of indices (=classical field theory).
The $q_n(t)$ may be of either Grassmann parity and as usual, $\dot q_n$
denotes the time derivative of $q_n$. The Hamiltonian formalism is obtained
by trading the velocities $\dot q_n$ for momenta $p_n$ via a Legendre
transformation. The momenta are defined by
\beq
       p_n(q,\dot q) = \frac{\dl L}{\delta \dot q_n}.
       \label{defmom}
\eeq
If the matrix $\dl p_n/\delta \dot q_m$ is of maximal rank, then
the relation (\ref{defmom}) between the momenta and velocities can be
inverted to $\dot q_n(p,q)$ and the Hamiltonian is defined by
\beq
     H(p,q) = \dot q_n(p,q) p_n - L (q, \dot q(p,q)).
\eeq
Theories for which this procedure can be followed, are not of interest as
gauge theories.

Much more exciting is the case where the equations (\ref{defmom}) can not
be solved for the velocities. A carefull analysis \cite{Dirac} leads to the
following results. There exists a set of {\it constraints}, that is, a set
of
functions $\phi_j (p,q)$ such that for the classical system $\phi_j(p,q) =
0$. Functions $F(p,q)$ that vanish on the subspace of phase space
that is defined by $\phi_j = 0$, are said to be {\it weakly vanishing}, and
we write $F(p,q) \approx 0$. Especially, $\phi_j \approx 0$. We restrict
the functions $F(p,q)$ to these functions such that $F \approx 0
\Leftrightarrow F = \sum_j c^j(p,q) \phi_j(p,q)$ for some phase space
functions $c^j(p,q)$ (regularity condition).

Functions $F$ defined on the phase space can now be divided into two
categories, {\it first class} and {\it second class} functions. A function
$F$ is said to be first class, if it has a weakly vanishing Poisson
bracket\footnote{For the definition and a list of properties of
Poisson brackets, see the appendix at the end of {\it this chapter}.}
with {\em all} constraints. That is, $F$ is first class if
\beq
    \forall j : [ F , \phi_j ] \approx 0 \;\; \Leftrightarrow \;\; \forall
j : [F,\phi_j] = \sum_k C_j^k (p,q) \phi_k \; .
\eeq
All other functions defined on phase space are called second class.

Of course, this divides also the set of constraints themselves into two
classes, first class constraints and second class constraints.
A constraint is first class if it has a weakly vanishing Poisson bracket
with all other constraints. A constraint is second class if there exists at
least one other constraint with which it does not have a weakly vanishing
Poisson bracket. Second class constraints can be accounted for
by replacing the Poisson brackets by {\it Dirac brackets}.
These Dirac brackets have the same properties as
Poisson brackets (distributivity, Jacobi-identities),
but they are constructed using a symplectic two
form that is determined from the Poisson bracket of the second class
constraints. We will in this and the following chapter assume that no
second class constraints are present.
If they were, we only have to put in the
associated symplectic form at the right places.

After the elimination of the second class constraints, we are left with a
set of first class constraints, $\phi_{\alpha}$; i.e.
\beq
 [ \phi_{\alpha} , \phi_{\beta}] \approx  0 \;\; \Leftrightarrow  \;\;
    [ \phi_{\alpha} , \phi_{\beta}] = C^{\gamma}_{\alpha \beta} (p,q)
\phi_{\gamma} \; .
\eeq
The phase space functions $C^{\gamma}_{\alpha \beta}$ are called {\it
(first order) structure functions}. It was conjectured, by P.A.M. Dirac,
that all first class constraints generate gauge symmetries.

Besides constraints, the second ingredient of the Hamiltonian formalism is
a first class Hamiltonian $H_0$:
\beq
   [ H_0 , \phi_{\alpha} ] \approx 0 \Leftrightarrow [H_0,\phi_{\alpha}] =
V_{\alpha}^{\beta} (p,q) \phi_{\beta}.
\eeq
With this Hamiltonian and this set of constraints a first order action
(linear in the velocities) can be constructed:
\beq
   S = \int dt \, \left[ \dot q_n p_n - H_0 - \sum_{\alpha}
\lambda^{\alpha}(t) \phi_{\alpha}  \right].
  \label{FOA}
 \eeq
Here, $\lambda^{\alpha}(t)$ is a Lagrange multiplier, imposing the
constraints.
This action is invariant under gauge transformations. The transformation of
an arbitrary phase space function $F$ is generated by the constraints:
\beq
\delta_{\epsilon} F(p,q) = [ F , \epsilon^{\alpha}(t) \phi_{\alpha} ] \; .
\eeq
The $\epsilon^{\alpha}$ are the parameters of the transformation.
The Lagrange multiplier $\lambda^{\alpha}$ has to be given the
transformation rule
\beq
   \delta_{\epsilon} \lambda^{\alpha} (t) = \dot \epsilon^{\alpha} -
\epsilon^{\beta}
V_{\beta}^{\alpha} + \lambda^{\gamma} \epsilon^{\beta} C^{\alpha}_{\beta
\gamma} \; ,
\eeq
in order for $S$ (\ref{FOA}) to be invariant.
These are the main results of Dirac's constraints analysis, the first step
of the Hamiltonian quantisation of (gauge) theories.

\section{The BFV formalism}

We now turn to the Batalin-Fradkin-Vilkovisky (BFV) scheme \cite{BFV}. A
first step is to promote the Lagrange multiplier $\lambda^{\alpha}$  in
(\ref{FOA}) to a full dynamical variable. When one considers (\ref{FOA})
as a Lagrangian action and one starts the constraints
analysis, an extra set of constraints appears. The momenta that are
canonically conjugated to the $\lambda^{\alpha}$ have to vanish:
$\pi_{\lambda^{\alpha}} = 0$. The set of constraints $(\pi_{\lambda
^{\alpha}}, \phi_{\alpha}) = G_a$ is the set we will use. Notice that we
could now introduce Lagrange multipliers for the constraints $G_a$, and
repeat the procedure. This goes on ad infinitum. A discussion of this
possibly infinitely nested structure and of the reason to treat the
Lagrange multipliers as dynamical degrees of freedom can be found in
respectively chapter 2 and 3 of \cite{Govaertsbook}. With the extra pair of
conjugate variables $\lambda^{\alpha}$ and $\pi_{\lambda^{\alpha}}$ and the
extended set of constraints $G_a$, we still have
\beq
    \begin{array}{lcl}
          [ G_a , G_b ] & = & C^c_{ab} G_c  \cr
          [ H_0 , G_a ] & = & V_a^b G_b .
    \end{array}
\eeq

The second step in the BFV procedure is the construction of the {\it
extended phase space}. For every constraint $G_a$ a ghost field $\eta^a$ is
introduced. Its Grassmann parity is opposite to the Grassmann parity of
the constraint
it is associated with: $\gras{\eta^a} = \gras{G_a} + 1= \gras{a} + 1$.
The momenta $\cP_a$ canonically conjugated to $\eta^a$ (i.e.
$[\cP_a,\eta^b]=-\delta_a^b$) have the same Grassmann parity as $\eta^a$.
Furthermore,
$\gh{\eta} = 1$ and $\gh{\cP} = -1$.

In this extended phase space, one constructs the {\it generator of the BRST
transformations} $\Omega$,
which acts on functions of the phase space variables. This $\Omega$ is also
called the {\it BRST charge}. $\Omega$ is a functional of odd Grassmann
parity and has ghostnumber $1$. It has to satisfy
\beq
          [ \Omega ,\Omega ] = 0 ,
          \label{nilpotom}
\eeq
with the extra condition that
\beq
      \Omega = \eta^a G_a + \ldots
\eeq
Depending on the symmetry algebra, more terms of higher order in the ghosts
and their momenta may appear where the dots are.
For any function $F(p,q)$, we have $\delta F = [F,\Omega]
= [F , \eta^a G_a ] + \dots$  In analogy with the extended action of the
Lagrangian BV formalism, $\Omega$ is only specified by these two conditions
up to canonical transformations in the extended phase space
\cite{HenneauxRep}.

The last important ingredient of the BFV formalism is the Hamiltonian, $H$.
We impose that $H$ is BRST invariant, i.e. we look for a solution of
\beq
      [ H , \Omega ] = 0,
\eeq
with the extra condition that $H = H_0 + \ldots$ and with $\gh{H} =0$.
Again, terms that are a function of the ghosts and their momenta may appear
where the dots are.

If the structure functions $C^a_{bc}$ and $V_a^b$ are independent of the
phase space variables, we have that:
\begin{eqnarray}
     \Omega & = & \eta^a G_a - \frac{1}{2} (-1)^{\gras{b}} \eta^b \eta^c
C^a_{cb} \cP_a \nonumber \\
      H & = & H_0 + \eta^a V_a^b \cP_b.
\end{eqnarray}
Owing to (\ref{nilpotom}), $H$ is only determined up to a term $ - [\Psi ,
\Omega ]$. We define
\beq
    H_{eff} = H - [\Psi ,\Omega ],
\eeq
and $\Psi$ is called the {\it gauge fermion}.

So far, the classical gauge system has been reformulated using BRST
technology. The quantisation using functional integrals goes as
follows. Consider the action
\beq
S_{eff} = \int dt \, \left[ \dot q_n p_n + \dot \lambda^{\alpha}
\pi_{\lambda^{\alpha}} + \dot \eta^a \cP_a - H_{eff} \right].
   \label{SBFV}
\eeq
The gauge fermion that is contained in $H_{eff}$ is to be chosen such that
all fields in this functional integral have propagators.
The first three terms are determined by the symplectic form of the extended
phase space. A general
theorem, the Fradkin-Vilkovisky theorem, states that the path integral
\beq
    \cZ_{\Psi} = \int [d\mu] e^{\ihbar S_{eff}}
\eeq
is independent of the specific choice of the gauge fermion $\Psi$. The
measure is the product over time of the Liouville measure that is defined
on the phase space. As a corrollary, it is easy to derive the form of the
Ward identities in the Hamiltonian formalism. By considering $\Psi
\rightarrow \Psi + \epsilon X$ for an infinitesimal parameter $\epsilon$,
we find that
\beq
    \int [d\mu] [ X,\Omega ] e^{\ihbar S_{eff}} = 0.
    \label{WardBFV}
\eeq
This expression is the Hamiltonian version of the Ward identities
(\ref{WaTaSlaTa}).

Two final remarks are in order before applying the recipe in the next
chapter. Although we introduced the $\eta^a$ as ghost fields and
the $\cP_a$ as their momenta, the following reinterpretation will be made:
\beq
\label{split}
\begin{array}{|c||c|c|} \hline
     G_a &  \pi_{\lambda^{\alpha}} & \phi_{\alpha} \\  \hline
     \eta^a &  - i^{\gras{\alpha}+1} \cP^{\alpha} & c^{\alpha} \\
     \cP_a &  i^{\gras{\alpha}+1} b_{\alpha} & \bar \cP_{\alpha} \\
 \hline
\end{array}
\eeq
Now the fields are taken to be the ghost $c^{\alpha}$ and the antighost
$b_{\alpha}$. The $\bar \cP_{\alpha}$ and $\cP^{\alpha}$ are considered to
be momenta. The canonical Poisson
brackets are $[\bar \cP_{\alpha}, c^{\beta} ] = [ \cP^{\beta} , b_{\alpha}]
= - \delta_{\alpha}^{\beta}$.

One often chooses a gauge fermion $\Psi$ of the form
\beq
   \Psi = i^{\gras{\alpha}+1} b_{\alpha} X^{\alpha} + \bar{\cP}_{\alpha}
\lambda^{\alpha}.
  \label{alweerlabel}
\eeq
Here, the $X^{\alpha}$ are gauge fixing functions that do not depend on the
ghosts, antighosts nor on the momenta of both.

This concludes our outline of the Hamiltonian treatment of gauge theories.
In the next chapter, we first demonstrate this recipe by applying it
to a Lagrangian with a shift symmetry. This serves as a preparation for the
incorporation of the shift symmetry in a general Hamiltonian system with
first class constraints. Thus we will prove the equivalence of the
Hamiltonian BFV and Lagrangian BV formalism.

\section*{Appendix: Poisson brackets}

Suppose that we can divide the coordinates $q_n$ and momenta $p_n$ in
bosonic degrees of freedom $(q^i,p_i)$ and fermionic degrees of freedom
$(\theta^{\alpha}, \pi_{\alpha})$. Here, $q_i$ and $\theta_{\alpha}$ are
the
coordinates, $p_i$ and $\pi_{\alpha}$ the momenta. The canonical Poisson
bracket of two phase space function $F$ and $G$ is then defined by:
\begin{eqnarray}
    [ F , G ] & = & \frac{\dr F}{\delta q^i} \frac{\dl G}{\delta p_i}
   - \frac{\dr F}{\delta p_i} \frac{\dl G}{\delta q^i} \nonumber \\
     &  &
   - \frac{\dr F}{\delta \theta^{\alpha} } \frac{\dl G}{\delta
\pi_{\alpha}}
   - \frac{\dr F}{\delta \pi_{\alpha} } \frac{\dl G}{\delta
\theta^{\alpha}}.
\end{eqnarray}
The Poisson bracket is a bosonic bracket, $\gras{[F,G]} = \gras{F} +
\gras{G}$, and $\gh{[F,G]} = \gh{F} + \gh{G}$. Furthermore, we have the
following list of properties: \beq
   \begin{array}{lc}
      1. & [ F , G ] = (-1)^{\gras{F} \gras{G} +1}  [ G , F ] \\
      2. & [ F , G H ] = [ F , G ] H + (-1)^{\gras{F} \gras{G}} G [ F , H ]
\\
      3. & [[F,G],H] + (-1)^{\gras{F} ( \gras{G} +\gras{H})} [[G,H],F] +
      (-1)^{\gras{H} ( \gras{F} + \gras{G} )} [[H,F],G] = 0.
\end{array}
\eeq
As a consequence of the definition of the Poisson brackets, the following
derivative rules hold. Suppose that
$x^k(p_n,q_n)$ are some phase space functions and suppose $F(x^k)$. Then
\beq
    \begin{array}{lcl}
        [ F , G ] & = & \frac{\dr F}{\delta x^k} [x^k , G ] \cr
        [ G , F ] & = & [ G , x^k ] \frac{\dl F}{\delta x^k}.
    \end{array}
\eeq

\chapter{The equivalence of Hamiltonian BFV and Lagrangian BV}

We now show \cite{ik5} how the Lagrangian antifield formalism can be
derived from the Hamiltonian BFV formalism that is described in the
previous
chapter. The transition from the Hamiltonian to the Lagrangian description
of systems with gauge symmetries comprises two aspects. The first, and this
is not restricted to gauge theories, concerns the question how the
Hamiltonian description, with its special treatment of
the time coordinate, leads to the Lorentz covariant Lagrangian formalism.
We will not address this question. Let us only mention that the Lagrange
multipliers of the Hamiltonian formalism (\ref{FOA}) are considered
dynamical fields for exactly the purpose of Lorentz covariance in the
Lagrangian formulation. For instance, in Maxwell's theory for
electromagnetism, the fourth component $A_0$ of the Lorentz vector is a
Lagrange multiplier that imposes the constraint $\partial_i F^{0i} =0$
(Gauss' law). A detailed discussion of this example can be found in
\cite{Govaertsbook}. The second aspect is the relation
between the Hamiltonian BFV and the Lagrangian BV formalisms themselves.
Our approach will be to enlarge the BRST symmetry of the
Hamiltonian system with the Schwinger-Dyson shift symmetry. This way, we
introduce the antifields (antighosts of the shift symmetry) in the
Hamiltonian path integral. Integrating out the momenta of the Hamiltonian
formalism, the gauge fixed action of the BV formalism is obtained. The
Fradkin-Vilkovisky theorem, which guarantees that the Hamiltonian path
integral is independent of the gauge
fermion, is shown to imply the BV quantum master equation.

Given the importance of this equivalence, a large effort has already been
devoted to its study. In \cite{Siegel2,FischHen}, an approach different
from ours, is followed.
There, the starting point is the first order action (\ref{FOA})
of the previous chapter, which is treated as any other Lagrangian using the
BV antifield scheme. The $\phi^i$ of the BV formalism are the phase space
variables
of the Hamiltonian scheme, i.e. the fields {\it and} their momenta.
The basic observation is that an extended action that satisfies the
classical master equation can be obtained by taking as antifield dependent
terms $\phi^*_i [ \phi^i , \Omega]$ if $[\Omega ,\Omega]=0$. By gauge
fixing, i.e. by replacing the antifields of the fields and the momenta of
the Hamiltonian formalism by derivatives with respect to a gauge fermion,
the action (\ref{SBFV}) as prescribed by the BFV formalism is reobtained
\cite{FischHen}. An analogous, although less general, result is described
in \cite{Jordi1}.

The approach we follow is closer to that of \cite{GriGriT}, where the
transition from the Hamiltonian to Lagrangian formalism was made directly
at the path integral level. However, in \cite{GriGriT}, the antifields are
introduced in a rather ad hoc way as sources for the BRST transformations
and the Lagrangian quantum master equation requires a seperate proof,
whereas one would expect it to follow from the Fradkin-Vilkovisky theorem.
Our presentation below clarifies exactly these two points.

As a preparation to the equivalence proof,
we first derive the Schwinger-Dyson equation as Ward identity
in the Hamiltonian formalism. In the second section of this chapter we
present our proof of the equivalence of the Hamiltonian and Lagrangian
formalism.

\section{Schwinger-Dyson equation in the BFV formalism}

Consider a Lagrangian depending on fields and their time-derivatives and
which describes a system without gauge symmetries:
$L(t) = L (\phi^a, \dot{\phi}^a) $. Introducing collective fields
amounts here to considering $ L(\phi^a- \varphi^a, \dot{\phi}
^a - \dot{\varphi}^a ) $.
The momenta conjugate to $\phi^a$ are denoted by $\pi_a$ and to $\varphi^a$
by $\varpi_a$. The first class constraints are
\beq
\chi_a = \pi_a + \varpi_a = 0 .
\eeq
The structure constants $C^a_{bc}$ of the constraints algebra vanish,
 $[ \chi_b , \chi_c ] = 0$, since $ [ \pi_a , \phi^b ] = [\varpi_a ,
\varphi ^b ] = - \delta_a^b $ are the only non-vanishing Poisson brackets.
The constraints are clearly first class.
These constraints also have a vanishing Poisson
bracket with the Hamiltonian, as the latter only depends on the difference
$\phi^a - \varphi^a$. This is equivalent to saying that the structure
constants $V^b_a$ vanish.
With every constraint $\chi_a$, we associate one Lagrange multiplier
$\lambda^a$ and its canonical momentum $\pi_{\lambda a}$
 ($[ \pi_{\lambda a} , \lambda^b ] = - \delta_a^b $).
The complete set of constraints $G_a = ( \pi_{\lambda a} , \chi_a )$ still
has vanishing structure constants. We construct the extended phase
space, following the prescription of the previous chapter. The ghost
and antighost fields are introduced as in (\ref{split}):
\beq
\begin{array}{|c|| c | c | c |}
    \hline
    G_a &  \pi_{\lambda^a} & \chi_a & {\mbox{ghnr.}} \\  \hline
    \eta^a & -i^{\gras{a}+1} {\cal P}^a  & c^a & 1 \\
    P_a    & i^{\gras{a}+1} \phi^*_a &  \bar{{\cal P}}_a & -1 \\
    \hline
\end{array}
\eeq
with the only non-vanishing brackets
$ [\bar{{\cal P}}_a,c^b ] = [ {\cal P}^b , \phi^*_a ] = - \delta^b_a $.
The Grassmann parities are as
follows:$\;\;
\gras{\phi^a}=\gras{\varphi^a}=\gras{\pi_a}=\gras{\varpi_a}=
\gras{\lambda^a}=\gras{\pi_{\lambda a}} = a$ and $\gras{\cP^a}=\gras{\bar
\cP_a}=\gras{c^a}=\gras{\phi^*_a}=a+1$.
Notice that $\phi^*_a$ denotes again the antighost.
In the extended phase space, one can straightforwardly construct the BRST
generator of the
shift symmetries. All structure constants vanish, so we have $\Omega_s =
\eta^a G_a$, which gives
\beq
\label{OmegaS}
\Omega_s = -i^{\gras{a}+1}{\cal P}^a \pi_{\lambda a} + c^a ( \pi_a +
\varpi_a).
\eeq
Indeed, $[\phi^a,\Omega_s] = [\varphi^a,\Omega_s] = (-1)^{\gras{a}} c^a$.
Since $[H_0 , \Omega_s ] = 0$, we have that $H=H_0$.
Gauge fixing the collective field to zero can be done by taking as
gauge fermion
\beq
 \Psi_s = \bar{{\cal P}}_a \lambda^a + \frac{i^{\gras{a}+1}}{\beta}
\phi^*_a \varphi^a \label{phizero} ,
\eeq
where $\beta$ is an arbitrary parameter. We will take the $\beta \rightarrow
0$ limit later. This is a standard procedure in the Hamiltonian formalism.
See e.g.\ \cite{HenneauxRep}, see also \cite{Govaertsbook} for a critical
examination of this procedure.
The Poisson bracket of this gauge fermion with the BRST-charge gives \beq
[ \Psi_s , \Omega_s ] = i^{\gras{a}+1} {\cal P}^a \bar{{\cal P}}_a
        - \frac{1}{\beta} \varphi^a \pi_{\lambda a} - \lambda^a (\pi_a +
\varpi_a ) + \frac{i^{\gras{a}+1} (-1)^{\gras{a}}}{\beta} \phi^*_a c^a .
\eeq
The action to be used in the path integral,
is then given by (\ref{SBFV})
\beq
S = \dot{\phi}^a \pi_a + \dot{\varphi}^a \varpi_a
+ \dot{\lambda}^a \pi_{\lambda a}  + \dot{\eta}^a P_a
 - H ( \pi , \varpi , \phi - \varphi ) + [ \Psi , \Omega_s ]
\label{actie}
.\eeq
An integration over time is understood. We want to construct a Ward
identity like (\ref{WardBFV}), with $ X = \frac{i^{\gras{a}+1}
\phi^*_a}{\beta} F(\phi)$. We calculate
\beq
   [ X , \Omega_s ] = \frac{i^{\gras{a}+1} \phi^*_a}{\beta} \frac{\dr
F}{\delta \phi^b} (-1)^{\gras{b}} c^b - \frac{1}{\beta} (-1)^{\gras{F}}
\pi_{\lambda a} F.
\eeq

We now redefine\footnote{Formally, this redefinition
leads to a $\beta$ independent Jacobian in the path integral measure as the
two fields that are rescaled by the inverse of $\beta$ have opposite
Grassmann parity. However, these formal
manipulations may require a more careful treatment on a case by case
basis. See \cite{Govaerts1}.}
\begin{eqnarray}
\label{redef}
      \frac{1}{\beta} \pi_{\lambda a} & \rightarrow & \pi_{\lambda a}
      \nonumber \\
      \frac{i^{\gras{a}+1}}{\beta} \phi^*_a & \rightarrow & \phi^*_a
        \\
      (-1)^{\gras{a}} c^a & \rightarrow & c^a  .
  \nonumber
\end{eqnarray}
After this rescaling, we take the limit $\beta \rightarrow 0$.
Thereafter, the momenta of the ghosts,
 ${\cal P}^a$ and $\bar{{\cal P}}_a$, can be integrated out trivially,
leading to
\beq
 S = \dot{\phi}^a \pi_a + \dot{\varphi}^a \varpi_a - H (\pi , \varpi , \phi
-\varphi) - \varphi^a \pi_{\lambda a}  - \lambda^a ( \pi_a + \varpi_a )
 + \phi^*_a c^a .
\eeq
The first three terms are grouped in $S_1$, the other three in $S_2$. $S_2$
clearly removes all the extra fields of the collective field formalism.
Indeed, integration over $\pi_{\lambda a}$ gives a delta-function fixing
the collective field to zero, integration over the Lagrange multiplier
$\lambda^a$ leads to a delta-function imposing the constraint $\chi_a$ and
integration over $c^a$ gives a delta-function fixing the antifield to zero.
However, in the path integral for the expectation value of $[X ,\Omega_s]$,
these integrations can not be done immediately. After the
rescaling (\ref{redef}), we have
\beq
[ X , \Omega_s ] = \phi^*_a \frac{\dr F}{\delta \phi^b} c^b
- (-1)^{\gras{F}} \pi_{\lambda a} F.
\eeq
The Ward identity hence becomes
\begin{eqnarray}
   0 & = & \int [d\phi][d\varphi][d\pi][d\varpi][dc][d\phi^*][d\lambda]
   [d\pi_{\lambda}] e^{\ihbar S_1} \nonumber \\
     &  & \times \frac{\hbar}{i} \left[
 \phi^*_a \frac{\dr F}{\delta \phi^b} \frac{\dl e^{\ihbar S_2}}{\delta
\phi^*_b} + (-1)^{\gras{F}} \frac{\dl e^{\ihbar S_2}}{\delta \varphi^a} F
\right].
\end{eqnarray}
In the first term, we integrate by parts over $\phi^*$. Thereafter, the
integrations contained in $S_2$ can be done. In the second term, we
integrate by parts over $\varphi$, which again allows to do the
integrations leading to $\delta(\varphi)\delta(\pi +
\varpi)\delta(\phi^*)$. We integrate out the momentum $\varpi$ of the
collective field, after which we can replace the derivative with respect to
$\varphi$ by minus a derivative over $\phi$. If we then integrate out the
momenta $\pi_a$ of the original fields and define
\beq
\exp[\frac{i}{\hbar} {\cal S}]= \int [d \pi_a ] \exp \left[\frac{i}{\hbar}
\left( \int dt \, \dot{\phi}^a \pi_a -
H (\pi , -\pi , \phi ) \right) \right] ,
\eeq
we find back the Schwinger-Dyson equation
\beq
   \int [d\phi] \;\; e^{\ihbar \cS} \left[ F \frac{\dr \cS}{\delta \phi^a}
+ \frac{\hbar}{i} \frac{\dr F}{\delta \phi^a} \right] = 0.
\eeq
This parallels the result of chapter 4.

\section{From Hamiltonian BFV to Lagrangian BV via Schwinger-Dyson
symmetry}

Now we turn to the case of an Hamiltonian system with gauge symmetries. We
start from an extended phase space, on which a nilpotent BRST
charge $\Omega_0(\Pi_A, \phi^A)$ and a BRST invariant Hamiltonian
$H_0(\Pi_A, \phi^A)$ are defined. More specifically, the
extended phase space has coordinates ($\phi^i, c^{\alpha},
 b_{\alpha}, \lambda^{\alpha}$
for the case of first class irreducible theories) which we collectively
denote by $\phi^A$, and conjugate momenta $(\pi_i, \bar{\cP}_{\alpha},
\cP^{\alpha} , \pi_{\lambda {\alpha}})$ denoted by $\Pi_A$.
The fundamental Poisson bracket is $[\Pi_A, \phi^B  ] = - \delta
_A^B$. We use the specific field content of $\phi^A$
and $\Pi_A$ only in the end.
The BRST-charge and the the BRST-invariant Hamiltonian
satisfy the standard conditions  $ [ \Omega_0 , \Omega_0 ] =
[ H_0 , \Omega_0 ] = 0 $.

We double the phase space by introducing for every field $\phi^A$ a
collective field $\varphi^A$ which has the conjugate momentum $\Upsilon_A$.
The fundamental Poisson brackets are also copied. We take
$ [ \Upsilon_A, \varphi^B] = - \delta_A^B$.
With every function $F (\Pi_A,\phi^A)$ defined on the original
extended phase space, we associate a function $ \wt{F}
= F (- \Upsilon_A, \phi^A - \varphi^A)$, defined on the doubled extended
phase space. The following two properties are then easily seen to hold:
\beq
[ \wt{F} , \wt{G} ] = \wt{[F,G]}
\label{FtGt=FGt} ,
\eeq
and
\beq
[ \wt{F} , \chi_A ] = 0  \label{FtC} .
\eeq
Again, $\chi_A = \Pi_A + \Upsilon_A$ is the constraint that generates the
shift symmetry.
In the case that the original extended phase space has an arbitrary
symplectic form $ [ \Pi_A , \phi^B ] = \omega_A^B(\Pi,\phi)$ that depends
on the phase space variables, the brackets in the doubled extended phase
space have to be defined as $ [ \Pi_A , \phi^B ] = \tilde \omega_A^B$ and
$[\Upsilon_A ,\varphi^B] =\tilde \omega_A^B$. This way, the two crucial
properties (\ref{FtGt=FGt},\ref{FtC}) on which all following developments
are based, can be generalised. Notice that the bracket on the RHS of
(\ref{FtGt=FGt}) is the one defined in the original extended phase space.

The next step is to see how the BRST charge and the Hamiltonian have to be
modified to take the new symmetry (the shift symmetry) into account. We
look for
\begin{eqnarray}
        \Omega = \wt{\Omega_0} + \Omega_s \\
        H = \wt{H_0} + \Delta H
\end{eqnarray}
and demand that $[ \Omega , \Omega ] = [H ,\Omega ] = 0$, with the extra
condition that $\Omega$ generates {\em all} BRST symmetries, also the BRST
shift symmetry. For that purpose,
we construct the extended phase space\footnote{In fact we mean the
extended phase space associated with the shift symmetry in the
doubled extended phase space, but for obvious linguistic reasons we speak
of {\it the extended phase space}.}
as in the case of no internal symmetries discussed above:
\beq
\begin{array}{|c|| c | c | c |}
    \hline
    G_A &  \pi_{\lambda^A} & \chi_A & {\mbox{ghnr.}} \\  \hline
    \eta^A & -i^{\gras{A}+1} \cP^A  & c^A & 1 \\
    P_A    & i^{\gras{A}+1} \phi^*_A  &  \bar{\cP}_A & -1 \\
    \hline
\end{array}
\eeq
This table summarises the following, by now familiar, steps (\ref{split}).
For every shift constraint $\chi_A$
we have introduced a Lagrange multiplier $\lambda^A$ and its conjugate
momentum $\pi_{\lambda^A}$. The latter is constrained to zero. For the set
of constraints $G_A = ( \pi_{\lambda^A} , \chi_A)$ we have added ghosts,
antighosts and their momenta. As usual, the fundamental Poisson brackets
are defined $[\cP^B,\phi^*_A] = [ \bar \cP_A , c^B ] = - \delta_A^B$.

The total BRST-charge is obtained by taking for $\Omega_s$ the expression
of the case without gauge symmetries (\ref{OmegaS}):
\beq
 \Omega = \wt{\Omega_0} -i^{\gras{A}+1}\cP^A \pi_{\lambda^A}
 + c^A ( \Pi_A + \Upsilon_A) .
 \eeq
This quantity satisfies $ [ \Omega , \Omega ] = 0 $ owing to
(\ref{FtGt=FGt}) and (\ref{FtC}). It is also
clear that $ \Delta H = 0 $, i.e. $H = \wt{H_0}$.
In fact, these two results simply reflect the vanishing of
the structure constants associated with the Poisson
brackets of the original constraints and of the original Hamiltonian with
the constraints of the shift symmetry, when the former are evaluated in
$(-\Upsilon_A, \phi^A - \varphi^A)$.

Let us calculate the BRST transformations of the fields $\phi^A$ and of
the collective fields $\varphi^A$.  This clarifies the meaning of
the abstract construction above. Moreover, it {\sl explains}
why the $\tilde{F}$ operation also involves a substitution
of the momenta by minus the momenta of the collective fields.
The fields only transform under the shift symmetry
\beq
[ \phi^A , \Omega ]  =  (-1)^{\gras{A}} c^A,
\eeq
while the original gauge transformations have shifted to the collective
fields
\beq
[ \varphi^A ,\Omega ] =  (-1)^{\gras{A}} c^A + [ \varphi^A, \wt{\Omega_0}
]. \eeq
It is precisely by our momentum substitution
rule that the BRST transformations of the originally present gauge
symmetries end up entirely in the collective field transformation.
This is analogous to what we did in the construction of the BV scheme from
BRST quantisation (\ref{extBRST}). Here, we have no a priori reason for
doing this, contrary to our discussion in the chapters 5,6 and 7. There the
BRST (--anti-BRST) transformation rules were organised this way in order to
be able to gauge fix theories with an open gauge algebra in the same way as
theories with a closed algebra. Making the same choice here, we will obtain
terms where the antifields act as sources for the BRST transformation rules
of their associated field.

To gauge fix the collective field to zero, we use again (\ref{phizero})
\beq
 \Psi_s = \bar{\cP}_A \lambda^A + \frac{i^{\gras{A}+1}}{\beta}
    \phi^*_A \varphi^A \ ,
\eeq
leading to
\begin{eqnarray}
  [ \Psi_s , \Omega ] & = & i^{\gras{A}+1} \cP^A \bar{\cP}_A -
\frac{1}{\beta} \varphi^A  \pi_{\lambda^A} - \lambda^A (\Pi_A + \Upsilon_A)
 \nonumber \\
  &  & +\frac{i^{\gras{A}+1} (-1)^{\gras{A}}}{\beta} \phi^*_A c^A
+ \frac{i^{\gras{A}+1}}{\beta} \phi^*_A [ \varphi^A , \wt{\Omega_0} ] .
\end{eqnarray}
We can rewrite the last term as
\beq
[ \varphi^A , \wt{\Omega_0} ] =  - [ \wt{\phi^A} , \wt
{\Omega_0} ] = - \wt{ [ \phi^A , \Omega_0 ] } ,
\eeq
since $\wt{\Omega_0}$ does not depend on the momenta conjugate to
$\phi^A$.

It only remains to gauge fix the original symmetries. This can be done by
taking a fermion $ \Psi(\Pi_A , \phi^A)$ and adding
\beq
[ \wt{\Psi} , \Omega ] = \wt{ [ \Psi , \Omega_0 ] } \label{gf}.
\eeq
Before taking the $\beta \rightarrow 0$ limit, we scale like
in (\ref{redef}), with a replacement of the indices $a$ by $A$.
This leads to the standard BFV action (\ref{SBFV})
\begin{eqnarray}
   S & = & \dot{\phi}^A \Pi_A + \dot{\varphi}^A \Upsilon_A
     + \beta \dot{\lambda}^A \pi_{\lambda^A} - \beta i^{\gras{A}+1}
     \dot{\cP}^A \phi^*_A + (-1)^A \dot{c}^A \bar{\cP}_A
     \nonumber \\
    &  & - \wt{H_0} + i^{\gras{A}+1} \cP^A \bar{\cP}_A
   - \varphi^A \pi_{\lambda^A} - \lambda^A ( \Pi_A + \Upsilon_A)
      \nonumber \\
    &  & + \phi^*_A c^A - \phi^*_A \wt { [ \phi^A , \Omega_0 ] }
    + \wt {[ \Psi , \Omega_0 ] } .
\end{eqnarray}
Taking the limit $\beta \rightarrow 0 $, the integrations over the ghost
momenta $\cP$ and $\bar{\cP}$  become trivial.
Integrating over the momenta $ \pi_{\lambda^A}$ of the Lagrange
multipliers, we obtain a delta-function $\delta(\varphi)$
while integrating out the Lagrange multipliers $\lambda^A$ themselves
leads to $\delta(\Pi_A+\Upsilon_A)$, imposing the constraint. It is trivial
to see that these two delta-function constraints allow us to drop the
tildes upon integration over the collective field and its momentum.
We obtain in this way
\beq
 S = \dot{\phi}^A \Pi_A - H_0 - \phi^*_A [ \phi^A, \Omega_0 ]
  + \phi^*_A c^A + [\Psi , \Omega_0  ] .
  \label{thisone}
\eeq
Again, the antighosts (antifields) start acting as a source term
for the BRST transformations of $\phi^A$.

We will now integrate out the momenta of the Hamiltonian formalism to
obtain an action that satisfies the Lagrangian quantum master equation.
In order to do that, we rewrite (\ref{thisone}) in a more useful form.
We make the most popular choice for the gauge
fermion\footnote{The Lagrange multipliers, ghosts
and antighosts that appear below, must not be confused
with the analogous fields for the shift
symmetries, which have been integrated out above. The former have
indices $\alpha$, while the latter had $A$. In order to proceed
further we have for the first time to specify in detail what fields
are contained in $\phi^A$ and $\Pi_A$.} (\ref{alweerlabel})
\beq
\Psi = \Psi_0( \phi^A,\pi_{\lambda \alpha} ) + \bar{\cP}_{\alpha}
\lambda^{\alpha} .
\eeq
$\Omega_0$ is generally of the form
\beq
\Omega_0 = -i^{\gras{\alpha}+1} \cP^{\alpha} \pi_{\lambda \alpha} +
\Omega_{min},
\eeq
where $\Omega_{min}$ does not depend on the Lagrange multiplier
$\lambda^{\alpha}$ nor on its momentum $\pi_{\lambda \alpha}$
\cite{HenneauxRep}.
Taking these two facts into account, the terms for the gauge fixing of the
original gauge symmetries are of the form:
\beq
[\Psi , \Omega_0 ] = \frac{\dr \Psi_0}{\delta \phi^A}
[ \phi^A, \Omega_0 ] + i^{\gras{\alpha}+1} \cP^{\alpha} \bar{\cP}_{\alpha}
+ [ \bar{\cP}_{\alpha}\lambda^{\alpha} , \Omega _{min} ].
\eeq
The important term here is the first one.
It is now convenient to define $\hat{S}$ by
\beq
S =  \widehat{S} (\phi^A,\phi^*_A, \pi_{\lambda \alpha} ,
 \cP^{\alpha}, \bar{\cP}_{\alpha}  , \pi_i)
+ \phi^*_A c^A - \frac{\dr \Psi_0}{\delta \phi^A}
\frac{\dl \widehat{S}}{\delta \phi^*_A}.
\eeq
Let us stress that $\widehat{S}$ is linear in the antifields
(\ref{thisone}), which allows us to write the last term the way we do.
The partition function, obtained by Hamiltonian methods
is then seen to be given by
\beq
\cZ = \int [d\phi^A][d\pi_{\lambda \alpha}] [d \phi^*_A] [d c^A]
[d\cP^{\alpha}] [d\bar{\cP}_{\alpha}] [d\pi_i]
  \exp\left[\frac{i}{\hbar} \phi^*_A c^A \right] \hat{U}
\left(  \exp \left[ \frac{i}{\hbar} \widehat{S} \right]  \right) .
\eeq
The differential operator $\hat{U}$ produces the gauge fixing term
and is defined by
\beq
\hat{U} = \exp \left[ - \frac{\dr \Psi_0}{\delta \phi^A}
     \frac{\dl}{\delta \phi^*_A} \right] .
\eeq
Define the quantum extended action $W$ by
\beq
e^{\frac{i}{\hbar} W(\phi ,\phi^*)} = \int
[d\cP^{\alpha}] [d\bar{\cP}_{\alpha}][d\pi_i]  \exp [ \ihbar \widehat{S} ]
 \label{WBV} .
\eeq
The operator $\hat{U}$ commutes with the integrations over the momenta that
define $W$. Integrating out the momenta, the partition function becomes
\beq
\cZ = \int [d\phi^A][d\pi_{\lambda \alpha}] [d \phi ^*_A] [d c^A]
 \exp \left[ \frac{i}{\hbar} \left( \phi^*_A c^A + W
( \phi^A , \phi^*_A - \frac{\dr \Psi_0}{\delta \phi^A}
)\right) \right].
\eeq
The shift in the antifields is obtained by using the well-known relation
\newline
$ \exp \left[ a(y) \frac{\delta} {\delta x} \right] f(x) = f(x + a(y)) $.
We have now recovered the form of the BV path integral (\ref{BVint}),
as the integration over the ghost $c^A$ leads to a
$\delta(\phi^*_A)$, removing the antifields.
In contrast with \cite{GriGriT}, we do not integrate over the Lagrange
multipliers, but include them together with their momenta in the set of
degrees of freedom of the obtained Lagrangian system.
Notice that $W$ need not be
linear in the antifields. This suggests that the non-linear terms in the
antifields that are typical for open algebras in the BV scheme, appear when
integrating out the momenta.

It is now trivial to show that $W$ satisfies the
BV quantum master equation. The
Fradkin-Vilkovisky theorem states that changing the gauge fermion
$\Psi_0$ to $\Psi_0 + d\Psi$ leaves the partition
function $\cZ$ invariant. For an infinitesimal change in the
gauge fermion we have
\beq
\int [d\phi^A] [d\phi^*_A] [d c^A] [d \pi_{\lambda \alpha }] \;\;
 e^{\ihbar \phi^*_A c^A}.
\frac{\dr d \Psi}{\delta \phi^A} . \frac{\dl}{\delta \phi^*_A}
\exp[\frac{i}{\hbar} \hat{W}]= 0 ,
\eeq
for any choice of $d \Psi$.
We denoted $W (\phi^A , \phi^*_A - \frac{\dr \Psi_0 }{\delta \phi^A})
= \hat{W} $. As $d\Psi$ is completely arbitrary,
a partial integration gives us the by now
well-known quantum master equation:
\beq
       \Delta e^{\ihbar \hat{W}} = 0.
\eeq

Let us finally try to get a better understanding of $W$
defined in (\ref{WBV}).
We consider the familiar expansion in $\hbar$ (\ref{Planckexpan})
\beq
\hat W = S_0 + \sum_{i=1}^{+\infty} \hbar^i M_i .
\eeq
We will only discuss $S_0$, which satisfies the classical master equation
$(S_0 , S_0) = 0 $. It can be calculated by applying the saddle-point
approximation to the momentum integrals in (\ref{WBV}).
Solving the field equations (the equations that determine the extrema of
the integrand)
\beq
\frac{\delta \hat{S}}{\delta \cP^{\alpha}}=
\frac{\delta \hat{S}}{\delta \bar{\cP}_{\alpha}} = \frac{\delta \hat{S}}
{\delta \pi_i} = 0
\eeq
leads to functions
$ \cP^{\alpha} (\phi^A , \phi^*_A , \pi_{\lambda \alpha})$,
$ \bar{\cP}_{\alpha} (\phi^A , \phi^*_A , \pi_{\lambda \alpha}) $ and
$\pi_i (\phi^A , \phi^*_A , \pi_{\lambda \alpha})$. When we plug in these
solutions, we denote this by $\vert_{\Sigma}$.
Clearly,
\beq
S_0 (\phi, \phi^*, \pi_{\lambda \alpha} ) = \hat{S} \vert_{\Sigma}.
\eeq
Finally, we have that the BRST transformations in the Lagrangian formalism
(\ref{extBRST}) are
\beq
   \cR^A[\phi] =  \left[ \frac{\dl S_0}{\delta \phi^*_A} \right]_{\phi^*=0}
= - [ \phi^A , \Omega_0 ]_{\Sigma,\phi^*=0} + \ldots
\eeq

This finishes our proof of the equivalence of the Hamiltonian BFV
and the Lagrangian BV formalism. Our guiding principle was that the
Schwinger-Dyson shift symmetry allows for a natural introduction
of antifields. Using the prescriptions of the BFV scheme to
implement the Schwinger-Dyson BRST symmetry, we see that the presence of the
antifields need not be restricted to Lagrangian BV. However,
integrating out the momenta
leads straightforwardly to an interpretation like that of
the Lagrangian scheme of BV.
The Lagrangian action we have in the end, satisfies the BV
quantum master equation, as a result of the Fradkin-Vilkovisky theorem.
We thus have linked the two principles which assure that the Hamiltonian and
Lagrangian method can be used for quantising gauge theories at all, namely,
that the partition functions constructed following their prescription, are
independent of the chosen gauge fixing.

\part{A regularised study of anomalies in BV}

\chapter{Gauge anomalies and BV}

The third part of this dissertation is devoted entirely to one-loop
aspects of the quantisation of gauge theories. We will especially
focus on the occurrence of gauge anomalies (for a first definition, see the
discussion following (\ref{AX})) and on their description in the BV
scheme. In this chapter, we first highlight some aspects of gauge anomalies
in the case of a closed algebra. In a second part of this chapter, we
rephrase gauge anomalies in the BV terminology, which generalises the
preceding results to all types of algebras. In this chapter, we only work
on the formal level, that is, without using a regularisation scheme.
This will be remedied in extenso in the following chapters.
Regularisation should be one of the cornerstones of any discussion on
anomalies. In the next chapter, we
introduce a one-loop regularisation scheme for the path integral, which
is then used to derive a regularised version of the BV quantum master
equation. In the final chapter, the
previously discussed regularisation techniques are applied to study
how extra fields can be introduced to keep some preferred symmetries
anomaly free.

\section{Gauge anomalies in quantum field theory}

In general, when a symmetry of the classical action, be it a local or a
global one, is not a symmetry of the effective action owing to quantum
corrections, then this symmetry is said to be {\it anomalous} or
{\it anomalously broken}. Ideally, one would like to turn
this procedure around and define a quantum theory with a certain set of
symmetries. This set may then enlarge upon taking the classical limit.
This way, the negative consequences of gauge anomalies, which we will point
out below, maybe are to be seen as an artifact of our as yet incomplete
understanding of quantisation of gauge theories \cite{Kowalski}.

We study {\it perturbative gauge anomalies}\footnote{{\it
Perturbative} here is in contrast with non-perturbative anomalies,
so-called {\it global anomalies} \cite{Witten2}.}, i.e. anomalies in local
symmetries.
Usually, the classical gauge symmetry leads to BRST symmetry of the gauge
fixed action and of the quantum theory (see chapter 3). When the product of
the measure
of the path integral and the exponential of the (quantum) action can not be
made BRST invariant, we have a gauge anomaly, see (\ref{AX}).
A crucial ingredient is the Jacobian of the path integral measure under
BRST transformations.
When the Jacobian is different from 1, we can distinguish two cases. Either
the Jacobian can be cancelled by the addition of a {\it local} quantum
counterterm to the action, or it can not be cancelled this way.
We will loosely speak of {\it anomaly} when the Jacobian is different
from $1$ and denominate a Jacobian that can not be countered by adding a
{\it local} quantum counterterm, a {\it genuine anomaly}.
Genuine anomalies lead to a quantum correction to the Ward
indentity. The consequences are that the partition function becomes gauge
dependent and that the customary proofs of renormalisability and unitarity
of gauge theories are jeopardized. The other way around, imposing the
absence of anomalies has served as a criterion in the selection of
{\sl healthy}
theories. Here, the famous example is of course the structure of the
matter families in the standard model of electroweak interactions.
Within one generation of matter fields, the contributions of the different
particles to the anomaly cancel. This requires a careful choice of the
representations of the gauge group under which both chiralities of the
fermions transform
and gives evidence for the existence of quarks in three colours \cite{BIM}.
Also, the interest in string
theory was triggered by the observation that the anomalies in these models
can be cancelled by working in specific space-time dimensions
\cite{SchwaGre}.

The most important consequence of a gauge anomaly, which manifests itself
time and again, is the fact that degrees of freedom that can classically be
fixed to zero by a choice of gauge, start propagating in the quantum
theory. This can be seen as follows. Suppose that we start from a classical
action $S_0[\phi^i]$, with gauge generators $R^i_{\alpha}$ that form a
closed algebra. We take both
the $\phi^i$ and the gauge generators to be bosonic for the sake of the
argument. Enlarge the configuration space, as usual, with ghosts
$c^{\alpha}$ and the trivial systems consisting of the antighosts
$b_{\alpha}$ and the Lagrange multipliers $\lambda^{\alpha}$. On this
enlarged configuration space, the nilpotent BRST operator
$\delta$ is defined (see the chapters 2 and 3). In order to gauge fix, one
chooses as many gauge fixing functions,
denoted e.g. by $F^{\alpha}$, as there are gauge symmetries.
The choice of the $F^{\alpha}$ can be interpreted as the selection of
degrees of freedom of the system that will be fixed to zero.
By taking as gauge fermion $\Psi= b_{\alpha} F^{\alpha}$, the gauge fixed
action is of the form (\ref{SEFF}), with $a=0$ in (\ref{SEFFP}). Instead of
fixing these degrees of freedom $F^{\alpha}$ to zero, we could fix them on
any configuration $\theta^{\alpha}$ with the gauge fermion
\beq
        \Psi_{\theta} = b_{\alpha} ( F^{\alpha} - \theta^{\alpha}).
\eeq
The gauge fixed path integral then becomes
\beq
    \cZ_{\theta} = \int [d\phi^A] e^{\ihbar ( S_0 + \delta \Psi _{\theta})}
    .
\eeq
By taking all possible configurations for $\theta^{\alpha}$, we cover
a range of gauge fixings. When there is no anomaly, one has that
\beq
          \frac{\delta  \cZ_{\theta}}{\delta \theta^{\alpha}} = 0.
\eeq
However, in the case of an anomaly, it follows from
(\ref{dpsiward},\ref{AX}), that
\beq
     \label{exprop}
          \frac{\delta \cZ_{\theta}}{\delta \theta^{\alpha}} =
\ihbar \langle \cA b_{\alpha} \rangle .
\eeq
The different gauge choices, i.e. the different choices for the
configurations $\theta^{\alpha}$ give now different partition functions.
It is then natural to consider the $\theta^{\alpha}$ as new degrees of
freedom and to include an integration over them in the functional integral.
We will come back to these extra degrees of freedom in the
last chapter of this dissertation.

Let us now derive the {\it Wess-Zumino consistency condition} for anomalies
\cite{WZcon}. Consider a theory where the fields can be divided in matter
fields $\phi^i$ and external gauge fields $A^a$, and where the classical
action
$S[\phi ,A]$ has a gauge invariance. This classical gauge invariance is
traded for BRST invariance. $\delta$ denotes the BRST transformation.
Define
\beq
    \exp [\ihbar W[A] ] = \int [d\phi] e^{\ihbar S[\phi ,A] + i M},
    \label{geinduc}
\eeq
where we have included a possible {\it local} counterterm $M$.
We suppose that the matter fields have well-defined propagators and
therefore, we do not need to gauge fix. Despite all these restrictions,
in many interesting examples these requirements are satisfied (see next
chapter for an example). Consider now
\begin{eqnarray}
    \delta e^{\ihbar W[A]} .\mu & = & \int [d\phi] \exp \left[ \ihbar
    S[\phi ,A+ \delta A.\mu] + i M[\phi, A + \delta A.\mu]  \right]
    \nonumber \\
    &  & - \int [d\phi] \exp \left[ \ihbar S[\phi ,A] + i M \right] ,
\end{eqnarray}
where $\mu$ is again a space-time independent, Grassmann odd parameter. As
in the derivation of the Ward identities (section 2 of chapter 3),
we can now redefine the integration
variables in the first integral, and use that $S$ is BRST invariant to
obtain
\beq
    e^{\ihbar W[A]}. \ihbar \delta W[A] . \mu  = \int [d\phi] e^{\ihbar S +
i M} ( \cA + i \delta M )\mu,
\eeq
where $\cA$ is defined as in (\ref{BRSTJacob}). If neither $\cA$
nor $M$ depend on the matter fields $\phi^i$, we are led to
\beq
     \ihbar \delta W[A] = \cA + i \delta M.   \label{dw}
\eeq
Again, if we can find a {\it local} $M$ such that the RHS of (\ref{dw}) is
zero,
the naively expected result $\delta W[A]= 0$ is obtained. Whether one can
or can not find such an $M$, we always have that
\beq
     \delta \cA = 0  \label{WZ},
\eeq
owing to the nilpotency of the BRST operator for closed algebras. This
condition is the {\it Wess-Zumino consistency condition}. The logarithm of
the Jacobian of the measure of the path integral under a BRST
transformation is BRST invariant. $\cA$ is called a {\it consistent
anomaly}. A different form of the anomaly has been introduced and used in
\cite{BarZu}, the so-called {\it covariant anomaly}. We restrict our
attention to consistent anomalies and we will introduce a regularisation
scheme in next chapter that gives consistent anomalies.

Notice that $\gh{\cA} = 1$, since it is the BRST variation of $W[A]$ and
$M$, both of ghostnumber zero.
In a theory with an irreducible gauge algebra, the ghosts
$c^{\alpha}$ are the only fields with ghostnumber $1$. In such cases, the
general form of $\cA$ is given by
\beq
\cA = c^{\alpha} \cA_{\alpha},
\label{anosom}
\eeq
with
$\gh{\cA_{\alpha}} = 0$.

As a result of all the above, it is sometimes possible
to determine whether a theory is anomaly free or is
possibly anomalous without doing perturbative
calculations. Indeed, if all BRST invariant functions of ghostnumber 1
are BRST exact, then there can be no consistent genuine anomaly as any
possible Jacobian can then be neutralised by choosing a counterterm. On the
other hand, if there exist BRST invariant functions of ghostnumber 1 that
are not BRST exact, one only knows that the
theory is possibly anomalous. The mathematical structure to investigate is
clearly the cohomology of the BRST operator at ghostnumber 1. Notice
however, that the actual expression for the local counterterm or the
anomaly in a specific regularised calculation, can not be obtained from
cohomological arguments. As we show on a few examples in the next chapters,
these actual expressions depend on the regularisation scheme that one uses.

The consistency condition has been used as a starting point for geometric
approaches to anomalies \cite{BarZu,Zumino}. This has led to an input of
results from mathematics, centered around the Atiyah-Singer index theorems.
The interested reader is referred to \cite{AGG} for a physicist's
introduction.

In the next section, we first indicate how genuine anomalies manifest
themselves in the context of BV \cite{WPT}. The Wess-Zumino condition for a
consistent anomaly naturally appears there.

\section{Gauge anomalies in BV}

In this section, we translate the results of the previous section in the
language of BV \cite{WPT}. This generalises them at the same time to all
types of gauge algebras.

As was pointed out in the previous section, if the theory has a
genuine gauge anomaly,
the partition function becomes gauge dependent. One has that
(\ref{AX}),
\beq
    \cZ_{\Psi + d\Psi} - \cZ_{\Psi} = \ihbar \langle \cA d\Psi \rangle.
\eeq
This is easily compared with (\ref{deltachi}) for $X=1$, where the change
of the partition function under infinitesimal canonical transformations is
considered. In analogy, we define the genuine anomaly in the BV scheme as
\beq
   ( W , W ) - 2 i \hbar \Delta W = -2i\hbar \cA  \label{defan}.
\eeq
Hence, we see that the genuine anomaly expresses the failure to construct
a quantum extended action $W$ that satisfies the quantum master equation.
Notice that a genuine anomaly also implies that the Zinn-Justin equation
changes (\ref{ZJeq}). The effective action $\Gamma(\phi_{cl},\phi^*)$ does
not satisfy the classical master equation in that case.

If we take for $W$ the usual expansion in $\hbar$: $W= S + \hbar M_1 +
\ldots$ and for the anomaly $\cA = \cA_0 + \hbar \cA_1 + \hbar^2 \cA_2 +
\ldots$, we get from (\ref{defan}), order in order in $\hbar$,
\begin{eqnarray}
                (S,S) & = & 0  \\
 \label{1LAN}  (S,M_1) - i \Delta S & = & - i \cA_0  \\
                 \ldots &  &  \nonumber
\end{eqnarray}
The dots denote the infinite tower of equations, one for every order in
$\hbar$. Since we will use a regularisation
prescription that is only capable to handle one loop
in the next chapters, we will not discuss them.

It was already pointed out in chapter 6 that the classical master
equation can always be solved \cite{fisch,FH,Stefan,Henneauxbook}, starting
from
a given classical action $S_0$ and a given complete set of gauge generators
$R^i_{\alpha}$. With the extended action $S$ at hand, one then has to
calculate $\Delta S$. Remember that in the antifield scheme, the BRST
transformation in the antibracket sense is defined by (\ref{BracBRST}) $\cS
\phi^A = (\phi^A ,S) = \dl S/\delta \phi^*_A$. It is then not difficult to
see that formally, $\Delta S$ is indeed proportional to the
logarithm of the Jacobian of the path integral measure $[d\phi^A]$
under BRST transformation\footnote{Notice the slight change of notation
with respect to the previous section and section 2 of chapter 3. What we
there denoted by $\cA$ corresponds to $\Delta S$ here. The notation $\cA$
is now reserved for Jacobians that can not be countered by an $M_1$
(\ref{defan}).}. Since $S$ is a local functional of fields and
antifields, one has that $ \Delta S \sim \delta(0)$. To remedy this
situation, a regularisation scheme is required. We postpone a discussion of
this to the next chapter.

If $\Delta S=0$, then that is all there is to it. The $M_i$ are then only
determined by the renormalisation process. However, if $\Delta
S \neq 0$, we have what we called an {\it anomaly} in the previous section.
In that case, one has to look for a {\it local} $M_1$, function of fields
and antifields, such that $(S,M_1) = i \Delta S$. For a specific $ \Delta
S$,
the {\it counterterm} $M_1$ is not uniquely defined. Indeed, we can always
consider $\tilde M_1 =M_1 + (S,X)$, because $( S,\tilde M_1) = (S,M_1)$,
owing to the fact that $(S,S)=0$. This $X$ may contain divergent terms of
order $\hbar$ that are needed in the renormalisation process etc. We will
not study this in detail and refer to \cite{TLV,Anselmi}. If no $M_1$,
local in a preferred set of variables, can be found such that $\cA_0 = 0$,
then there is a {\it genuine anomaly}. A specific
expression for $\cA_0$ is obtained by choosing a counterterm $M_1$.
As we showed above (\ref{anosom}), $\cA_0 = c^{\alpha} \cA_{0 \alpha}$.
By different choices of the counterterm $M_1$, it may be possible to make
$ \cA_{0 \alpha} = 0$ for certain values of $\alpha$. This way, one can
specify which gauge symmetries are kept anomaly free by a carefull choice
of $M_1$.

We can also see how $\cA_0$ of (\ref{1LAN}) transforms under
infinitesimal canonical transformations generated by $F = \unity + f$. We
find,
\beq
    \tilde \cA_0 = i (S',M'_1) + \Delta S' = \cA_0 ' - ( S , \Delta f).
\eeq
Here we used the notation (\ref{ClasTraf}). The extra term comes from the
transformation of $\Delta S$, given in (\ref{TranDX}). Hence, we see that
in a different set of coordinates, an extra counterterm $i\Delta f$
appears. In a formal reasoning, which needs justification in a
regularised treatment, we can even go further. For a closed algebra, $S$ is
linear in the antifields. Therefore, it is expected that $\cA_0$ does not
depend on antifields (for $M_1=0$). If one then only considers $f(\phi)$,
independent of the antifields, one has that $\cA_0' = \cA_0$ and one sees
that changing gauge only results in a change of counterterm. Consequently,
in different gauges, different gauge symmetries may be anomalous.

Let us finally reformulate the Wess-Zumino consistency condition in the BV
scheme \cite{WPT,White}. From the definition of the genuine anomaly
(\ref{defan}),
and from the properties of the antibracket and the box operator as listed
in the appendix, we easily arrive at
\beq
   \sigma \cA = ( \cA , W ) - i \hbar \Delta \cA = 0.
\eeq
The full anomaly is quantum BRST invariant.
The usual expansion in $\hbar$ gives
\begin{eqnarray}
           ( \cA_0 , S ) & = & 0  \nonumber \\
           (\cA_0 , M_1 ) + (\cA_1 , S ) - i\Delta \cA_0 & = & 0 \\
           \ldots &  & \nonumber
\end{eqnarray}
The first condition at $\cO(\hbar^0)$ is that the one loop anomaly is
classical BRST
invariant. This is the condition that we will meet below in the examples
and we will denote it by {\it Wess-Zumino consistency condition}. In fact,
since $S$ satisfies the classical master equation, the
$\cO(\hbar^0)$ consistency condition becomes $(\Delta S,S)=0$.

This finishes our overview of the definition and the major properties of
the description of gauge anomalies in BV. It is important to remember that
we have so far two related ways of changing the explicit expression of the
genuine anomaly, viz the choice of the {\it local} counterterm $M_1$ and
the choice of
gauge. This way, we obtain other representants of the same cohomology class
that determines the anomaly. In the second chapter of this part,
we will discuss the regularisation of the formal expressions of section 1.
Only in the following chapter we will return to the
regularised treatment of anomalies in BV.

\chapter{Pauli-Villars regularisation for consistent anomalies}

In this and the following chapters, a
one-loop regularised study of anomalies is presented.
In the first section, we sketch the core idea of the procedure proposed by
K. Fujikawa \cite{Fujikawa} to obtain a regularised expression for the
Jacobian associated with a transformation of the fields. Typically, the
explicit regularised expression that one gets for such a Jacobian is
determined by the
transformation itself {\em and} by the way one chooses to regularise the
determinant by means of what we will call below a {\it regulator}. Not all
choices for the regulator that are allowed in the Fujikawa scheme,
give consistent anomalies. In the second section, we follow A. Diaz, M.
Hatsuda, W.
Troost, P. van Nieuwenhuizen and A. Van Proeyen \cite{Diaz,Hatsuda,WPT} and
use Pauli-Villars regularisation \cite{PV} to
obtain regulators that give consistent anomalies. In the final section, the
freedom one has in choosing a mass term for the Pauli-Villars fields, is
exploited \cite{ik1} to calculate the induced action for $W_2$ gravity.

\section{Fujikawa's proposal for regularised Jacobians}

Following \cite{Fujikawa}, we give here --schematically-- a procedure to
calculate a regularised expression for anomalies. Although a few steps are
rather ad hoc, this procedure provides
a first contact with the type of regularised expressions that we will meet
below. Typically, one starts from a path integral (cfr. (\ref{geinduc}))
\beq
    e^{- W[A]} = \int [d\phi] \exp [- \phi^{\dagger} D[A] \phi ] \ .
\eeq
We have put $\hbar=1$, as in the rest of this chapter.
We work now in Euclidean space, and consider bosonic fields $\phi$. A
space-time integration is understood in the exponent on the RHS. $D[A]$
is a first or second order differential operator, depending on an external
field $A$. Internal indices are understood, i.e. $D[A]$ may actually be
a matrix and $\phi^{\dagger}$ and $\phi$ respectively a row and a column.
We assume that $D[A]$ has a complete set of orthonormal
eigenfunctions, denoted by $\phi_n$:
\begin{eqnarray}
          D[A] \phi_n & = & \lambda_n \phi_n \nonumber \\
          \int dx \, \phi^{\dagger}_n \phi_m & = & \delta_{nm}.
\end{eqnarray}
Every function $\phi$ can then be expanded in this complete basis
$\phi(x) = \sum_n a_n \phi_n(x)$. Therefore, we define the measure of
the functional integral by
\beq
            [d\phi] = \prod_n d a_n \ .
\eeq
We do an infinitesimal transformation that may or may not leave the
action $S = \phi^{\dagger} D[A] \phi$ invariant. We only focus on a
possible Jacobian. Suppose that the transformation is given by
\beq
     \phi' = \phi  + \epsilon . M . \phi \ ,
\eeq
with an infinitesimal parameter $\epsilon$. $M$ is a field independent
matrix that determines the
transformation. In principle, it can be a differential operator too, but
for the sake of the argument we restrict ourselves to matrices. This case
includes for instance the important example of chiral rotations of
fermions, where $M=\gamma_5$. We can expand $\phi' = \sum_n a'_n \phi_n$,
with
\beq
     a'_n = a_n + \epsilon \sum_m M_{nm} a_m.
\eeq
The matrix $\cM$ is defined by its elements
\beq
     M_{nm} = \int dx \, \phi^{\dagger}_n  M \phi_m.
\eeq
It is then clear that
\begin{eqnarray}
   \prod_n d a'_n & = & \det [ \delta_{nm} + \epsilon M_{nm} ] \prod_n d
a_n \nonumber \\
       & \approx & e^{\epsilon \tr \cM}  \prod_n d a_n.
\end{eqnarray}
So far, we have only given a specification of the integration measure and
have determined what form the Jacobian then takes.

$\tr \cM$ is often ill-defined as an infinite sum.
K. Fujikawa proposed to replace it with the regularised expression
\beq
 \tr_{\alpha} \cM = \sum_n \int dx \,
           \phi^{\dagger}_n  M f(\lambda_n,\alpha) \phi_n \ ,
\eeq
where $f(\lambda_n,\alpha)$ is such that it suppresses the contributions
to the trace of the eigenfunctions associated with large eigenvalues
$\lambda_n$. $f(\lambda_n,\alpha=0) = 1$, so that the original,
divergent expression is reobtained in the limit $\alpha \rightarrow 0$.
We make the typical choice $f(\lambda_n,\alpha)= \exp(-\lambda_n^\beta
\alpha)$.
Since the $\phi_n$ are eigenfunctions of $D[A]$ with precisely the
eigenvalue $\lambda_n$, we find
\beq
   \tr \cM = \lim_{\alpha \rightarrow 0} \sum_n \int dx \,
           \phi^{\dagger}_n  M  \exp ( - \alpha D[A]^{\beta} ) \phi_n.
    \label{Fujfo}
\eeq
$\beta$ is chosen such that in the exponent there is a term of second order
in derivatives, leading to a Gaussian damping in momentum representation.
$D[A]^{\beta}$ is called the {\it regulator}. Using the methods of appendix
C, a regularised expression for the Jacobian can be obtained.

This method has however some shortcomings. First of all, there is the
arbitrariness in the choice $f(\lambda_n,\alpha)$. Certain choices give
an expression for the Jacobian that satisfies the consistency condition
(\ref{WZ}) while others do not \cite{Diaz}. Furthermore, there seems to
be no a priori reason for considering a complete orthonormal
set of eigenfunctions of the operator
$D[A]$ that determines the action of the theory. Clearly, the relation
between the regulator and action of the model under consideration is to
be put in by hand. This situation will be clarified in the next section.

Despite this criticism, we will always have a form like (\ref{Fujfo}) for
Jacobians, i.e.\ a trace over functional space of a matrix determining the
transformation times the exponent of a regulator that dampens the
contributions to this trace of eigenfunctions with a large eigenvalue.
There are other methods to regularise Jacobians, see e.g.
$\zeta$-regularisation \cite{Ramondboek}, but in the next section we will
take the different point of view that one does not need
regularise the determinant but the complete path integral. This then
implies regularised expressions for the Jacobians \cite{Diaz,WPT}.

\section{Pauli-Villars regularisation}

In this section, we describe a method \cite{Diaz,Hatsuda,WPT} to calculate
regularised expressions for Jacobians. The method starts by giving a
one-loop regularisation prescription for the complete functional integral.
This is in contrast with the method of Fujikawa which only regularises the
Jacobian determinant itself. Then we consider the
BRST transformation of this regularised functional integral. Remembering
the way the consistency condition was derived in the previous chapter, we
see that we are thus guaranteed to obtain a consistent anomaly (\ref{WZ}).

We introduce the regularisation method by means of a simple example that
nevertheless keeps many of the characteristic features. We
take as classical action the action for $W_2$ gravity (\ref{SOW2}):
\beq
   S_0 [\phi ,h] = \frac{1}{2\pi} \int d^2x \left[ \partial \phi \bar
\partial \phi - h (\partial \phi)^2 \right].
\eeq
The BRST transformation rules are given by
\begin{eqnarray}
   \label{BRSTex}
     \delta \phi & = & c \partial \phi  \nonumber \\
     \delta h & = & \bar \partial c - h. \partial c + \partial h . c \\
     \delta c & = & \partial c . c \;\; , \nonumber
\end{eqnarray}
and $\delta^2 = 0$ (the extended action (\ref{W2extac}) is linear in the
antifields). We define the {\it induced action} $\Gamma[h]$  by
\beq
    e^{- \Gamma[h] } = \int [d\phi] e^{- S_0 [\phi ,h]}.
    \label{indac}
\eeq
Notice that this is an example that satisfies the requirements that were
imposed to derive the consistency
condition (\ref{geinduc},\ref{WZ}). By a partial integration in both
terms of the classical action, we can bring $S_0$ in the form
\beq
  S_0[\phi ,h] = - \frac{1}{2\pi} \int d^2x \, \phi \partial \nabla \phi
   \; ,
\eeq
where $\nabla = \bar \partial - h \partial$.
As the integration over $\phi$ is Gaussian, we can formally write
\beq
     e^{- \Gamma[h] } = [ \det \partial \nabla ]^{-\frac{1}{2}} .
     \label{minhalf}
\eeq
The $\det A$ is formally the product of the eigenvalues of the
operator $A$.

\begin{figure}[t]

\begin{center}
\begin{picture}(28000,16000)
\put(8000,6000){\circle{2800}}
\pmidx = 7700 \pmidy = 5900
\put(\pmidx,\pmidy){$\phi$}
\pfrontx = 7000 \pfronty = 7000
\drawline\photon[\NW\REG](\pfrontx,\pfronty)[5]
\global\advance\pbackx by -650
\put(\pbackx,\pbacky){$h$}
\pfrontx = 7000 \pfronty = 5000
\drawline\photon[\SW\REG](\pfrontx,\pfronty)[5]
\global\advance\pbackx by -650
\put(\pbackx,\pbacky){$h$}
\pbackx = 9000 \pbacky = 7000
\drawline\photon[\NE\REG](\pbackx,\pbacky)[5]
\global\advance\pbackx by 650
\put(\pbackx,\pbacky){$h$}
\pbackx = 9000 \pbacky = 5000
\drawline\photon[\SE\REG](\pbackx,\pbacky)[5]
\global\advance\pbackx by 650
\put(\pbackx,\pbacky){$h$}
\end{picture}

\end{center}
\label{fig:lusprent}

\caption{An example of a one loop diagram contributing to the induced
action.}

\end{figure}
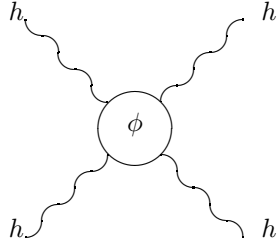

$\Gamma[h]$ is the generating function of the connected correlation
functions of an arbitrary number $n$ of
operators $T=\frac{1}{2\pi}(\partial \phi)^2$. To every such
correlation function contributes only one one-loop diagram.
For example, for $n=4$, we have the diagram Fig.\  12.1.

At the diagrammatic level, these diagrams can be regularised \`a la
Pauli-Villars \cite{PV} as follows. From every diagram\footnote{We use {\it
diagram} here as a metonymy for the {\it mathematical expression associated
with the diagram}.} we {\em subtract}
an identical one, with the only difference that instead of a massless,
bosonic particle $\phi$ in the loop,
we use a massive, bosonic particle $\chi$, with mass
$M$. These regulating diagrams disappear formally again upon taking the
limit
$M \rightarrow \infty $. Clearly, such diagrams can be obtained from the
action $S_{PV} + S_M$, with
\begin{eqnarray}
   \label{PVaction}
    S_{PV} & = & S_0[\chi ,h]  \nonumber \\
  & = & - \frac{1}{2\pi} \int d^2x \, \chi \partial \nabla \chi \nonumber
\\
    S_M & = & - \frac{1}{4\pi} \int d^2x \, M^2 \chi^2 .
\end{eqnarray}
The fact that the diagrams generated by this action have to be subtracted,
is implemented by the rule that every closed loop gets a minus sign. Owing
to this extra minus sign, we formally have that
\beq
   e^{- \Gamma_{PV}[h]} = \int [d\chi] e^{- S_{PV} - S_M } = [ \det
\partial \nabla + \frac{M^2}{2} ] ^{ + \frac{1}{2}}.
   \label{PVint}
\eeq
Notice that the power is now $+\frac{1}{2}$ in contrast to the integration
of the original $\phi$-fields (\ref{minhalf}). $\Gamma_{PV}[h]$ generates
all the regularising diagrams. The regularised expression for the induced
action is then by definition (we omit sometimes space-time
integrations and consider them understood):
\begin{eqnarray}
    \label{Greg}
       e^{- \Gamma_M [h]} & = & \int [d\phi] [d\chi] e^{- S_0[\phi,h]
       - S_0[\chi,h] + \frac{1}{4\pi} M^2 \chi^2} \nonumber \\
       & = & \left[ \frac{\det \partial \nabla + M^2/2}{\det \partial
\nabla} \right] ^{\frac{1}{2}}.
\end{eqnarray}

In a formal sense, the latter expression already shows that Pauli-Villars
regularisation removes the contributions to the induced action of the
large eigenvalues $\lambda_n$ of the operator $\partial \nabla$ and favours
the small eigenvalues. Indeed,
\beq
     e^{- \Gamma_M[h]} \sim \prod_n \left( \frac{\lambda_n +
M^2/2}{\lambda_n} \right) ^{\frac{1}{2}} ,
\eeq
where the RHS in fact defines the ratio of the two determinants in
(\ref{Greg}) in this regularisation scheme. For a fixed $M$, we see that
eigenvalues
$\lambda_n \gg M^2$ give a contribution $1$ to the product, while for
$\lambda_n \ll M^2$, the contribution is proportional to
$\frac{1}{\sqrt{\lambda_n}}$.

Instead of implementing the minus sign by hand with a Pauli-Villars (PV)
field
of the same (even) Grassmann parity as the original field, we could have
used a PV field $\psi$ of odd Grassmann parity. As with one such field we
would not be able to construct, for instance, a mass term ($\psi^2 = 0$),
we would
have to introduce two, $\psi_1$ and $\psi_2$. But then the PV path integral
with the action $S_{PV} + S_M \sim \psi_1 (\partial \nabla + M^2/2 )\psi_2$
would be proportional to $\det (\partial \nabla + M^2/2)$, i.e. half a
power too much. Therefore, a third, again bosonic, PV field would be needed
to generate an extra $\det (\partial \nabla + M_*^2/2)^{-1/2}$. Instead of
using this correct but cumbersome procedure, we will always use the
sleight-of-hand above, knowing that the procedure with three times as many
PV fields provides us with an ultimate justification if needed.

It is also important to notice that $S_{PV}[\chi,h]$ is BRST invariant if
we copy the BRST transformation of $\phi$ for the PV field
\beq
      \label{PVbrst}
      \delta \chi = c \partial \chi.
\eeq
The mass term $S_M$ is then the only term of the complete action
$S_R + S_M = S_0[\phi ,h] + S_0[\chi ,h] + S_M[\chi]$
that may not be BRST invariant. As a matter
of fact, this mass term is the only possible source of BRST non-invariance
of (\ref{Greg}). This is a consequence of the fact that the complete
measure of the fields $\phi$ and the PV fields $\chi$ is (BRST) invariant,
as the
Jacobian of the $\phi$ measure is compensated by the Jacobian of the $\chi$
measure\footnote{See the appendix of {\em this} chapter for more details.}
\cite{Fujikawa,Diaz}, owing to the extra minus sign for PV loops.

Taking this into account, we find that the BRST variation of the
regularised induced action is
\begin{eqnarray}
  \label{deltagamma}
    \delta e^{- \Gamma_M [h]} & = & - \int [d\phi][d\chi] e^{-S_R +
\frac{1}{4\pi} M^2 \chi^2}. \delta S_M \nonumber \\
      & = & - \int [d\phi][d\chi] e^{-S_R + \frac{1}{4\pi} M^2 \chi^2}.
           \frac{M^2}{4\pi}. \partial c. \chi^2.
\end{eqnarray}
We can now use (\ref{PVout}) of the appendix to integrate out the PV fields
\beq
   \label{delgamout}
   \delta \Gamma_M[h] = - \frac{M^2}{4} . \partial c. \tr[
\frac{1}{-\partial \nabla - M^2/2} ],
\eeq
where the factor $I$ on the RHS of (\ref{PVout}) allowed us to divide out a
factor $e^{- \Gamma_M [h]}$ on both sides. Notice that $\partial c$ can be
brought out of the trace over the function space, as it does not contain
derivatives
that act further nor matrix indices. As is also explained in the appendix
(\ref{DenReg}), we can now make contact with the expressions of Fujikawa
(\ref{Fujfo}), using \cite{Diaz,WPT}
\beq
\int_0^{\infty} d\lambda e^{-\lambda} e^{\lambda {\cal R}}
  = - \frac{1}{{\cal R} - \unity}.
  \label{DenReg2}
\eeq
$\cR$ is called the {\it regulator}. In our case, this gives,
\beq
    \delta \Gamma_M[h] = \frac{1}{2} \int d^2 x \; \partial c.
    \int_0^{\infty} d\lambda e^{-\lambda} . \tr \left[ \exp \left(
-\frac{\lambda}{M^2} 2 \partial \nabla \right) \right].
\eeq
Using (\ref{ResEx}), we can calculate the trace of the exponential of the
regulator in the limit $M \rightarrow \infty$. We find
\beq
    \label{Almost}
  \delta \Gamma_M[h] = \frac{1}{8\pi} \int d^2x \;\; \partial c .
    \int_0^{\infty} d\lambda e^{-\lambda} \left[ \frac{M^2}{\lambda} -
\frac{1}{3} \partial^2 h + \cO (\frac{1}{M^2}) \right].
\eeq

But now we see that we are in trouble. Even for a finite value of $M$, the
term of (\ref{Almost}) that is proportional to $M^2$ diverges owing to the
integration over $\lambda$: $\int_0^{\infty} d\lambda \; e^{-\lambda}
\lambda^{-1} = \Gamma(0)$ ( this is of course the gamma-function, not the
induced action !). The remedy is well-known (see e.g. chapter 7 of
\cite{ItzyZuber}) and goes as follows.
Instead of only one PV field, one introduces several copies $\chi_i$ with
a mass $M_i$. The precise
number may be determined from requirements to be specified below, but is
irrelevant. We take
\beq
   \label{PVcopies}
    S^i_{PV} + S^i_M =
 - \frac{1}{2\pi} \int d^2x \, \chi_i \partial \nabla \chi_i
 - \frac{1}{4\pi} \int d^2x \, M_i^2 \chi^2_i ,
\eeq
and $\delta \chi_i = c \partial \chi_i$.
The formal integration is defined by
\beq
    \label{PVC}
   \int [d\chi_i] e^{-S_{PV}^i - S_M^i} = [ \det \partial \nabla +
M^2_i/2]^{x_i/2}.
\eeq
All $M_i$ are taken to infinity in the end.
The precise values of the $x_i$ follow from some
relations which we will now specify. We should certainly have that
\beq
   \sum_i x_i = 1 \ ,
\eeq
in order to keep the complete measure of the original fields and all PV
fields BRST invariant (\ref{PVjacob}). When taking the BRST variation of
the regularised
action (\ref{deltagamma}), we now have $\sum_i \delta S^i_M$ instead of
$\delta S_M$. From (\ref{PVout}), we see that upon integration over the PV
fields, we get
\beq
  \delta \Gamma_M[h] = \frac{1}{8\pi} \int d^2x \;\; \partial c .
    \int_0^{\infty} d\lambda e^{-\lambda} \left[ \frac{\sum_i
x_i M_i^2}{\lambda} - \sum_i x_i \frac{1}{3} \partial^2 h + \sum_i x_i \cO
(\frac{1}{M_i^2}) \right].
\eeq
We now impose the extra condition
\beq
    \sum_i x_i M^2_i = 0,   \label{dees}
\eeq
to get rid of the diverging $\lambda$ integration. As is clear from the
formulas in the appendix (\ref{kernex}), when working in 4 or more
space-time dimensions, the expansion of the exponent of the
regulator starts with terms proportional to $M^4_i$ or more. Even more
conditions like
(\ref{dees}), with higher powers of $M_i$, are imposed then. After having
obtained the full set of such conditions, one can specify a minimal number
of PV fields, their $x_i$ and the relations between their masses.

We finally find that
\beq
  \label{cd3h}
  \cA \equiv \lim_{M \rightarrow \infty} \delta \Gamma_M[h] = \frac{1}{24
\pi} \int d^2x \, c \partial^3 h.
\eeq
It is easy to verify that $\delta \cA =0$, with the BRST transformations
(\ref{BRSTex}). It can be shown \cite{Polyakov1} that this expression for
$\delta \Gamma[h]$ is satisfied for
\beq
   \Gamma[h] = - \frac{1}{48\pi} \int d^2 x \, \partial^2 h
\frac{1}{\partial \nabla} \partial^2 h,
   \label{Polyakovaction}
\eeq
which is non-local. In the next section, we will derive this expression for
$\Gamma[h]$. No expression for $\Gamma[h]$, local in $h$, with $\delta
\Gamma[h] = \cA$ has been found.

Before turning to a further study of the use of the mass term in the PV
regularisation, let us briefly repeat the important steps of the above
procedure. We started by regularising the complete functional integral by
the introduction of a Pauli-Villars field $\chi$ for every field $\phi$.
$\chi$ has the same Grassmann parity as $\phi$, but with a closed
$\chi$-loop we associate an extra minus sign. The BRST transformation of
the PV fields and their action were obtained by direct substitution of
$\chi$
for $\phi$ in the original BRST transformation rules and action. As a
result of this construction, the measure of the regularised path
integral is BRST invariant. To complete the PV regularisation, one has to
choose a mass term $S_M$ for the PV fields. This mass term is the only
possibly BRST non-invariant factor in the regularised partition function.
In the PV scheme, anomalies come from the BRST variation of the mass term.
At first sight, one might think that one has no freedom in choosing the
mass term. However, the contrary is true. The possibilities this freedom
offers are explored in the next section. The regularised functional
integral gives rise to a regularised expression for Jacobians that is of
the same type as the regularised expressions proposed by Fujikawa
(\ref{Fujfo}). The heat kernel methods of the appendix C can then be used
to obtain an expression for the anomaly.

\section{Mass term dependence of the anomaly}

We now show how the freedom that one has in choosing a mass term for the PV
fields, can
be exploited \cite{ik1}. As it is this mass term that determines which
gauge
symmetries are anomalous (\ref{deltagamma}), one can try to keep preferred
symmetries anomaly free by a judicious choice of the mass term. If the mass
term is invariant under a certain symmetry, that symmetry will not become
anomalous. In other words,
we have yet another factor (the mass term) that determines the actual
expression of the anomaly. If the $\phi^A$ denote the original fields,
the most general choice of mass term that we can make is
$S_M = - \frac{1}{2} \chi^A T_{AB} (\phi) \chi^B M^2$. Here,
$T_{AB}$ is an invertible, field dependent matrix that satisfies $T_{AB} =
(-1)^{\gras{A}+\gras{B} +\gras{A}\gras{B}} T_{BA}$. Of course, in general
such mass terms are not fit for diagrammatic calculations, but they pose
no problem for the procedure sketched in the previous section.

Let us consider the $W_2$ example. A.M. Polyakov \cite{Polyakov1} found
that
a redefinition of the field --from $h$ to $f$-- makes the non-local induced
action (\ref{Polyakovaction}) local in this new variable $f$. The field $f$
is defined by
\beq
        h = \frac{\bar \partial f}{\partial f} \ .
        \label{repam}
\eeq
The BRST transformation of $f$ is $\delta f = c \partial f$. This
reproduces the BRST transformation of $h$ in (\ref{BRSTex}). Suppose now
that we take as mass term for the PV field
\beq
   \tilde S_M = - \frac{1}{4\pi} \int d^2x \, M^2 \chi^2 \partial f \ .
   \label{Minv}
\eeq
Although this is a local expression in $f$, it is non-local in the variable
$h$. Hence, the term {\it non-local regularisation}
was coined for this procedure. This mass term (\ref{Minv})
is clearly BRST invariant. Therefore, we find that (cfr.\
(\ref{deltagamma}))
\beq
     e^{- \Gamma_M}. \delta \Gamma_M = 0,
\eeq
and we can take $\Gamma[h] = 0$ in this regularisation scheme.

So far, we have in this chapter neglected the possibility to modify the
expression for the anomaly using a local counterterm $M_1$. More
specifically, the question arises whether one can compensate for the change
in the expression for the anomaly owing to a
different choice of local mass term by the addition of a {\it local}
quantum counterterm. It turns out that we can indeed do
that. A different mass term only leads to a different representant of the
same cohomology class for the anomaly.

Consider two induced actions, obtained by a regularised calculation
with two different mass terms. We allow for the presence of a counterterm
$M$, which only depends on the external gauge field $h$.
We have
\beq
    \label{gammanul}
    e^{- \Gamma^{(0)}[h]} = \int [d\phi][d\chi] e^{-S_0[\phi ,h] - S_0[\chi
, h] - S_M^{(0)} - M^{(0)}[h]},
\eeq
and
\beq
    \label{gammaeen}
    e^{- \Gamma^{(1)}[h]} = \int [d\phi][d\chi] e^{-S_0[\phi ,h] -
S_0[\chi , h] - S_M^{(1)} - M^{(1)}[h]}.
\eeq
We want to find the relation between the two counterterms, $X[h] = M^{(0)}
- M^{(1)}$, in order to have that the two induced actions are equal:
$\Gamma^{(0)}[h] =\Gamma^{(1)}[h]$.
We suppose that an {\it interpolating mass term}
$S_M^{(\alpha)}$ exists, i.e. a mass term depending on a parameter
$\alpha$, such that $S_M^{(\alpha = 1)} = S_M^{(1)}$ and
$S_M^{(\alpha = 0)} = S_M^{(0)}$. With this mass term, we define
\beq
   \cZ(\alpha) = \int [d\phi][d\chi] e^{-S_0[\phi ,h] - S_0[\chi , h]
   - S_M^{(\alpha)} + \alpha. X[h]} .
\eeq
{}From the definition of $X$, and from (\ref{gammanul}),(\ref{gammaeen}) it
follows that if we want $\Gamma^{(0)}=\Gamma^{(1)}$, we have to take $X$
such that
\begin{eqnarray}
  \label{condX}
 0 & = & \ln \cZ(1) - \ln \cZ(0) \nonumber \\
   & = & \int_0^1 \!\! d\alpha \; \frac{1}{\cZ(\alpha)} \frac{d
\cZ(\alpha)}{d\alpha} \ . \end{eqnarray}
Let us focus for a moment on $\frac{d \cZ(\alpha)}{d\alpha}$. We write
\beq
     S_M^{(\alpha)} = - \frac{1}{2} \chi^A T_{AB}(\alpha) \chi^{B},
\eeq
where $T_{AB}(\alpha)$ may depend on the original fields $\phi$.
Notice that we have also absorbed the mass parameter $M$ in $T_{AB}$.
Analogously, we define $\cO_{AB}$ by
\beq
S_0[\chi,h] =  \frac{1}{2} \chi^A \cO_{AB} \chi^B \  .
\eeq
Then we have that
\begin{eqnarray}
    \frac{d \cZ(\alpha)}{d\alpha} & = & \int [d\phi][d\chi] e^{-S_0[\phi ,h]
    - S_0[\chi , h] - S_M^{(\alpha)} + \alpha. X[h]}   \nonumber \\
 &  &   \times \left[ \frac{1}{2} \chi^A \frac{d T_{AB}(\alpha)}{d\alpha}
\chi^B + X[h] \right].
\end{eqnarray}
Integrating out the PV fields (\ref{PVout}), gives
\beq
   \frac{d \cZ(\alpha)}{d\alpha} = \cZ(\alpha) \left[ - \frac{1}{2} \tr
    \left( T^{-1}(\alpha) \frac{d T(\alpha)}{d \alpha} .
\frac{1}{T^{-1}(\alpha) \cO - 1} \right) + X[h] \right].
\eeq
Plugging this back in (\ref{condX}) we arrive at the most important result
of this section
\beq
  M^{(0)} -M^{(1)} = \frac{1}{2} \int_0^1 d\alpha \;  \tr
    \left( T^{-1}(\alpha) \frac{d T(\alpha)}{d \alpha} .
\frac{1}{T^{-1}(\alpha) \cO - 1} \right) \ .
   \label{deltaM}
\eeq
Of course, the limiting procedure $M \rightarrow \infty$ and the
introduction of copies of the PV field are understood. This relation
(\ref{deltaM}) was foreshadowed in \cite{peter2}, conjectured in \cite{WPT}
and proven in \cite{ik3}.

Let us now apply this result to our example \cite{ik1} and find the
counterterm
$M^{(0)}$, needed to obtain $\Gamma^{(0)}[h]=0$ with the PV mass term
$S_M^{(0)} = - \frac{1}{4\pi} M^2 \chi^2$.  When
calculated with the BRST invariant mass term
$S_M^{(1)} = - \frac{1}{4\pi} M^2 \chi^2 \partial f$ and with $M^{(1)} =
0$, we have the induced action $\Gamma^{(1)}[h] = 0$. Clearly we have
\begin{eqnarray}
T(\alpha ) &=& \frac{1}{2\pi} (\partial  f)^\alpha M^2 \nonumber \\
 \cO &=& - \frac{1}{\pi} \partial  \nabla \nonumber \\
T^{-1}(\alpha) \cO &=& \frac{\cR(\alpha)}{M^2} = - 2(\partial f)^{-\alpha
} \partial \nabla \frac{1}{M^2} \nonumber \\
T^{-1} \frac{d T}{d \alpha } &=& \ln (\partial f) \quad .
\end{eqnarray}
Using (\ref{DenReg2}), we have then
\beq
   M^{(0)} = \lim_{M \rightarrow \infty} - \frac{1}{2}\int_0^1 d\alpha
   \int_0^{\infty} \!\! d\lambda e^{-\lambda} \; \tr \left( \ln \partial f
. \exp[\frac{\lambda}{M^2} \cR(\alpha)] \right) \  .
\eeq
$\cR(\alpha)$ can again be rewritten as a covariant Laplacian
$\cR= \frac{1}{\sqrabsg} \partial_{\mu} \sqrabsg
g^{\mu \nu} \partial_{\nu}$ with
\beq
g^{\mu\nu}
=(\partial f)^{-\alpha} \left(\begin{array}{cc}
2h &  -1\\
-1 & 0
\end{array} \right)  \  .
\eeq
With the results listed in the appendix (\ref{gevalA}) (we need $E_2$ as
we work in two dimensions), we find that
\beq
    M^{(0)} = \frac{1}{48\pi} \int_0^1 d\alpha \int d^2 x \, \ln \partial f
. \sqrabsg R[g^{\mu \nu}] \ .
\eeq
The factor $\sqrabsg$ comes from $d vol(x)$ and $R[g^{\mu \nu}]$ is the
Riemann scalar of the metric $g^{\mu \nu}$.
We have taken the limit $M \rightarrow \infty$, and extra PV fields
were added so that we could integrate out $\lambda$ without running into
divergences. Observe that
\begin{equation}
\sqrabsg R[g^{\mu\nu}] = R[\tilde{g}^{\mu\nu}] + \frac{1}{2} \partial
_\mu \{\sqrabsg g^{\mu\nu} \partial _\nu \ln \vert g \vert \}
\label{er}
\end{equation}
for $\tilde{g}^{\mu\nu} = \sqrabsg g^{\mu \nu} = \left(\begin{array}{cc}
2h &  -1\\
-1 & 0
\end{array} \right)$. This gives us finally
\begin{eqnarray}
    M^{(0)} & = & \frac{1}{48\pi} \int_0^1 d\alpha \int d^2 x \, \ln
\partial f .  2 ( 1 -\alpha) \partial^2 h \nonumber \\
       & = & \frac{1}{48\pi} \int d^2 x \, \partial^2 h \ln \partial f \ .
\end{eqnarray}
It is clear that when written in terms of $f$, this is a local expression.
We can also reexpress it as a functional of $h$, using that $\partial
\nabla \ln \partial f = \partial^2 h$, which gives
\beq
 M^{(0)}=\frac{1}{48\pi}\int d^2 x \, \partial^2 h \frac{1}{\partial \nabla}
\partial^2 h.
\eeq
We find back the Polyakov-action (\ref{Polyakovaction}).
With this $M^{(0)}$, we have that
\beq
    1 = \int [d\phi][d\chi] e^{-S_0[\phi ,h] - S_0[\chi , h]
- S_M^{(0)} - M^{(0)}[h]},
\eeq
such that the induced action $\Gamma[h]$ defined in the previous section
(\ref{Greg}) is actually $- M^{(0)}$. It is easy to verify that indeed
\beq
      -  \delta M^{(0)} =  \frac{1}{24\pi} \int d^2x \, c \partial^3 h.
\eeq

Our method has hence become a way to calculate the induced action.
Moreover, it provides us with insight on why the induced action
can be written in a local way when the variable $f$ is used instead of $h$.
Indeed, we have obtained the Polyakov-action expressed in $f$, using a
procedure that is completely local in $f$.

We \cite{ik1} have also applied the same method to calculate the action
for gauge fields, induced by chiral fermions in two dimensions. There, the
reparametrisation corresponding to (\ref{repam}) is $A = \bar \partial g.
g^{-1}$ \cite{PolWieg}, i.e. the algebra element $A$ is written as a
function of a group element $g$. With this $g$, we can again construct an
invariant mass term for the PV regularisation fields. The induced action is
in that case, as is well-known, the Wess-Zumino-Witten model
\cite{WZcon,Witten3}. The $\alpha$ integral can not always be done
explicitly, it becomes the third dimension that is present in the
topological term of that model.

In this section, we have pointed out a third factor that influences the
actual expression for the anomaly, namely the mass term of the PV
regularisation. We have shown that the expressions for the anomaly obtained
with two different local mass terms are in general related by the BRST
variation of a local counterterm that can be determined from
(\ref{deltaM}). Hence,
whatever the chosen mass term, we obtain an anomaly of the same cohomology
class. As an application, we have seen how we can calculate the induced
action of $W_2$ gravity from a counterterm, since an invariant mass term
can be found. In our example, a redefinition of the field variables was
done from $h$ to $f$. The complete procedure, in particular the two
different mass terms and the counterterm, is local in the Polyakov variable
$f$. However, when everything is expressed in function of the original
variable $h$, non-local expressions appear.

\section*{Appendix}

The measure of the functional integral over the PV fields is defined
implicitly by the definition of the Gaussian integrals over the PV fields
(\ref{PVint},\ref{PVC}).
Hence, we can take the definition of these integrals as a starting point to
derive a consistent prescription for the Jacobian of the PV measure under
transformations.
Denote by $z^i$ a (finite) set of (bosonic) variables, and
define \beq
    \int [dz] \exp[ -\frac{1}{2} z^i D_{ij} z^j ] = [\det D]^{x/2}.
\eeq
Here, $x$ is a constant. In section 2 of this chapter, we have $x=1$ for
the PV fields and $x=-1$ for the original fields. Suppose that we want to
change integration variables to $y^i$:
\beq
     z^i = L^i{}_j y^j,   \label{trafo}
\eeq
with
\beq
    \int [dy] \exp[ -\frac{1}{2} y^i D_{ij} y^j ] = [\det D]^{x/2}.
\eeq
If we assume that $[dz]=[dy] . J$, we have that
\beq
  [\det D]^{x/2} = J [\det ( L^t D L) ]^{x/2}.
\eeq
{}From this we find that we should take for consistency
\beq
    \label{PVjacob}
    J = [\det L]^{- x}.
\eeq
This result generalises to superintegrations and superdeterminants.
Notice that $L^i{}_j$ may and will often depend on the fields $\phi$ if
the $z^i$ are the PV fields $\chi$.

\chapter{PV regularisation for BV}

We now use Pauli-Villars regularisation to construct a one-loop
regularised version of the BV scheme. In particular, we copy
the derivation of the quantum master equation based on canonical
transformations (\ref{deltachi}) using one-loop regularised path
integrals. This will lead us to a regularised $\cO(\hbar)$ master equation
and a regularised expression for $\sigma X$, also to $\cO(\hbar)$.
The basic features of the previous chapter will reappear.
Specifically, the $\Delta$-operators do not appear in the regularised
treatment, but are replaced by expressions of the Fujikawa
type (\ref{Fujfo}).
For instance, $\Delta S$ is replaced by an expression determined by
a regulator and by the antibracket of
the regularised extended action with the PV mass term. This statement is of
course the BV version of the result of the previous chapter
(\ref{deltagamma}).

This chapter is rather technical. Its purpose
is to translate the insights of the previous chapter in the BV language.
In section one\footnote{Some of the results in section one were developed
in discussions with R. Siebelink and W. Troost.}, we discuss the
construction of a one-loop regularised path integral for the expectation
value of an operator $X(\phi,\phi^*)$ in the BV scheme. It is shown in
section two what this one-loop regularised path integral looks like in a
second set of canonical coordinates, related to the first one by an
infinitesimal canonical transformation. By imposing that the two
expectation values are equal, i.e.\ by imposing gauge independence, we
derive a one-loop regularised expression for the one-loop master equation
and for the quantum BRST operator of the BV scheme. After a discussion of
the results in section three, we give an example in section four, where we
use our regularised expression for $\sigma X$ to calculate the Jacobian of
the measure under an infinitesimal canonical transformation.

\section{Setting up the PV regularisation}

In this section, we construct a one-loop regularised expression for the
expectation value of an operator $X(\phi,\phi^*)$ in the PV regularisation
scheme. We first describe the set-up of the PV regularisation scheme and
give some justification for this construction at the end of this section.

The $\phi^A$ denote, as usual, the complete set of fields and the
$\phi^*_A$ their
antifields. By the new notation $\phi^{\alpha}$, we denominate both
the $\phi^A$ {\em and} the $\phi^*_A$. In the same vein, we write the PV
fields as $\chi^{\alpha}= \{ \chi^A , \chi^*_A \}$. With every functional
depending on fields and antifields $A(\phi ,\phi^*)$ we associate
\beq
    A_{PV} = \frac{1}{2} \chi^{\alpha} \left[
    \frac{\dl}{\delta \phi^{\alpha}}
    \frac{\dr}{\delta \phi^{\beta}} A \right]  \chi^{\beta} \; .
    \label{APV}
\eeq
The regularised version of A is then taken to be $ A_R =  A +
A_{PV}$. One can straightforwardly prove that
\begin{eqnarray}
  \label{SBI}
  (A,B)_R  & = & (A,B) + (A,B)_{PV}  \nonumber \\
           & = &( A + A_{PV} , B + B_{PV} )_{\phi,\chi} + \cO(\chi^4)
           \nonumber \\
           & = & ( A_R,B_R)_{\phi ,\chi} + \cO(\chi^4).
\end{eqnarray}
By the notation $(A,B)_{\phi,\chi}$, we mean that the antibracket is to be
calculated
with respect to the fields {\em and} with respect to the PV fields. If no
such underscore is added to the antibracket, only derivatives with respect
to $\phi$ and $\phi^*$ are meant. We will
consistently drop all the terms of quartic or higher order in the PV
fields. When integrating over the PV fields, they lead to contributions of
higher orders in $\hbar$, or they disappear when putting the PV antifields
$\chi^*_A$ to zero after all antibrackets etc. have been evaluated.

As a first application of (\ref{SBI}), consider the extended action $S(\phi
,\phi^*)$, that satisfies the classical master equation $(S,S)=0$. If we
consider $S_R = S + S_{PV}$, we have that
\beq
   (S_R , S_R)_{\phi,\chi}= (S,S)_R + \cO(\chi^4) \approx 0 \ .
\eeq
This corresponds to our observation in the previous chapter that
$S_{PV}$ of $(\ref{PVaction})$ is BRST invariant if we take the BRST
transformation of the PV field as (\ref{PVbrst}). Here we see this as
follows. $S$ contains a term $\phi^* c \partial \phi$ (\ref{W2extac}), and
hence a term $\chi^* c \partial \chi$ is present in $S_{PV}$. The
antibracket $(\chi, S_{PV})_{\chi}$ then gives $\delta \chi = c \partial
\chi$.

To complete the regularisation scheme, we have to introduce a mass term
for the PV fields. We take as most general form for this mass term
\beq
   S_M = - \frac{1}{2} \chi^A T_{AB}(\phi ,\phi^*) \chi^B,
\eeq
where $T_{AB}$ also depends on the mass $M$. $T_{AB}$ is invertible and
satisfies $T_{AB} = (-1)^{\gras{A}+\gras{B} +\gras{A}\gras{B}} T_{BA}$.
Notice that this mass term is independent of the PV antifields,
but may depend on both the original fields $\phi$
and their antifields $\phi^*$. The {\it regularised extended action} is
then
\beq
   \cS = S + S_{PV} + S_{M} = S_R + S_M.
\eeq
If $S$ satisfies the classical master equation, we have that
\begin{eqnarray}
        \label{justabove}
   (\cS , \cS )_{\phi,\chi}& = & 2 (S_R,S_M)_{\phi,\chi} + \cO(\chi^4)
                      \nonumber \\
          & = & 2 (S,S_M)_{\phi} + 2(S_{PV},S_M)_{\chi} + \cO(\chi^4).
\end{eqnarray}
{\em The regularised extended action does not necessarily satisfy the
classical
master equation owing to the presence of the mass term}. Notice however
that the terms on the RHS of (\ref{justabove}) are at least quadratic in
the PV fields, and hence effectively of order $\hbar$.

With this regularised action, we can construct a regularised expression for
the expectation value of any operator $X(\phi ,\phi^*)$:
\beq
   \chi_R(\phi^*) = \int [d\phi][d\chi] \left[ X_R e^{\ihbar(\cS + \hbar
M_1)} \right]_{\chi^*=0} .
\eeq
That this is a one loop regularised expression follows from the general
principles of PV regularisation. In particular,
from every diagram with a
closed loop that contributes to the order $\hbar$ in perturbation theory to
the expectation value of $X$, a diagram is {\it subtracted} (remember the
extra minus sign for PV loops) corresponding to the expectation value of
the terms quadratic in the PV fields $\chi^A$ of $X_{PV}$. The terms in
$X_{PV}$ that contain one PV field and one PV antifield have expectation
value zero. The terms of $X_{PV}$ that are quadratic in the PV antifields
$\chi^*_A$ might cause new divergences owing to $\phi^A$-loops. To prevent
this, we construct the regularised path integral with $\chi^*_A = 0$.
At the diagrammatic level, we have for instance
\newline

\begin{picture}(18000,8000)

\put(500,4000){$\chi_R (\phi^*) \sim$}

\drawline\fermion[\E\REG](6700,4000)[2500]
\drawline\fermion[\NW\REG](\particlebackx,\particlebacky)[3500]
\drawline\fermion[\SW\REG](\particlefrontx,\particlefronty)[3500]
\global\advance\pfrontx by 2000
\put(\pfrontx,\pfronty){\circle{4000}}
\put(\pfrontx,\pfronty){$\phi$}
\global\advance\pfrontx by -2500
\global\advance\pfronty by  1500
\put(\pfrontx,\pfronty){$X$}

\global\advance\pfrontx by 6000
\put(\pfrontx,4000){$-$}

\global\advance\pfrontx by 2000

\drawline\fermion[\E\REG](\pfrontx,4000)[2500]
\drawline\fermion[\NW\REG](\particlebackx,\particlebacky)[3500]
\drawline\fermion[\SW\REG](\particlefrontx,\particlefronty)[3500]
\global\advance\pfrontx by 2000
\put(\pfrontx,\pfronty){\circle{4000}}
\put(\pfrontx,\pfronty){\circle{3700}}
\put(\pfrontx,\pfronty){$\chi$}
\global\advance\pfrontx by -2800
\global\advance\pfronty by  2000
\put(\pfrontx,\pfronty){$X_{PV}$}
\global\advance\pfrontx by  5500
\put(\pfrontx,4000){.}

\end{picture}

{}From the one-loop contribution (the $\phi$ loop is denoted by a single
circle) of the expectation value for the operator $X$ a one-loop diagram is
subtracted of the expectation value of $X_{PV}$ (the $\chi$ loop is denoted
by a double circle).

In the next section, we construct the regularised expectation value of
$X(\phi ,\phi^*)$ in a different set of coordinates, related to the first
one by an infinitesimal canonical transformation generated by $F = \unity +
f$. By demanding that the two expectation values are equal, we will
obtain a regularised version of the master equation and of $\sigma X$.
We copy the steps of section 2 of chapter 8 for one-loop regularised
path integrals.

\section{Regularised derivation of the master equation}

Under an infinitesimal canonical transformation generated by $F=\unity +
f$, a functional $A(\phi,\phi^*)$ transforms to $A'(\phi,\phi^*) = A -
(A,f)$, as was shown in chapter 8 (\ref{ClasTraf}). We can apply the recipe
of the previous section to construct a regularised operator $A'_R$ :
\begin{eqnarray}
   A'_R & = & A - (A,f) + A_{PV} - (A,f)_{PV} \nonumber \\
        & = & A_R - (A_R , f_R)_{\phi,\chi} + \cO(\chi^4).
\end{eqnarray}
We used again (\ref{SBI}). This result shows that we have in fact two
options
to obtain the regularised functional in the transformed coordinates. We can
first transform the unregularised operator and apply then the
regularisation prescription, or we can first regularise the operator and
then transform  this regularised expression $A_R$. The transformation rules
are
then to be derived from $f_R$. The PV fields need to be transformed too,
and the transformation rules of the original fields $\phi^{\alpha}$
may get corrections, quadratic in the PV fields
\cite{ToineString,TheBible}.

This result holds of course especially for $S_R$, which transforms to
$S'_R = S_R - (S_R , f_R)_{\phi,\chi}$. Having $S'_R$ is not sufficient to
construct a regularised path integral in the transformed coordinates. We
have to choose a mass term. We take $S'_M = S_M - (S_M,f_R)_{\phi,\chi}$.
This choice is made for two reasons. First of all, this choice makes a
derivation of the master equation copying the steps leading to
(\ref{deltachi}) possible. Secondly, we have seen in the previous
chapter that taking a different mass
term only results in the addition of an extra counterterm $M_1$. Hence, we
have that the regularised extended action transforms as
\beq
   \cS' = \cS - (\cS , f_R)_{\phi,\chi}  \ .
\eeq
The possible counterterm $\hbar M_1$ transforms as before, $M_1' = M_1 -
(M_1,f)$, but we can take there too $M_1' = M_1 - (M_1,f_R)_{\phi}$. The
extra terms are of order $\hbar\cO(\chi^2)$, and hence negligible.

When doing the canonical transformation on the fields in the unregularised
path integral, we had to take a Jacobian (\ref{deltaf}) into account. Like
in the previous chapter (section 2 and appendix),
we see that in the regularised functional integrals
the Jacobian from the fields $\phi$ is cancelled by the Jacobian of the PV
fields $\chi$. Indeed, we would now have $\ln J \sim \Delta_{\phi} f_R -
\Delta_{\chi} f_R$ instead of $\ln J \sim \Delta_{\phi} f$. The
relative minus sign is again a result of the implicit definition of the
path integral measure of the PV fields (\ref{PVjacob}). It is easy to see
that $\Delta_{\chi} f_{PV} = \Delta_{\phi} f$, such that we now have
$\ln J \sim \Delta_{\phi} f_{PV}$, which we can drop as this would
contribute a term of order $\hbar \cO(\chi^2)$ to the transformed quantum
extended action.

Finally, we have in the two sets of canonical coordinates the following two
regularised functional integrals for the expectation value of
$X(\phi,\phi^*)$:
\beq
   \chi_R(\phi^*) = \int [d\phi][d\chi] \left[ X_R e^{\ihbar(\cS + \hbar
M_1)} \right]_{\chi^*=0}
\eeq
and
\begin{eqnarray}
   \chi'_R(\phi^*) & = & \int [d\phi][d\chi] \left[  X_R -
(X_R,f_R)_{\phi,\chi} \right]_{\chi^*=0}  \nonumber \\
 &  & \times  \exp \left[ \ihbar \left( \cS + \hbar M_1
- (\cS + \hbar M_1, f_R)_{\phi,\chi} \right) \right]_{\chi^*=0} .
\end{eqnarray}
Subtracting the first from the second, we can again expand to linear order
in the infinitesimal fermion $f_R$ and find
\begin{eqnarray}
    \delta \chi_R(\phi^*) & = & \chi'_R(\phi^*) - \chi_R(\phi^*)  \\
  & = & \int [d\phi][d\chi] \left[ - X_R (\cW,f_R)_{\phi,\chi} e^{\cW}
  - (X_R , f_R)_{\phi,\chi} e^{\cW} \right]_{\chi^*=0} \ . \nonumber
\end{eqnarray}
We denoted $\cW = \ihbar ( \cS + \hbar M_1 )$. By writing out the
antibrackets, we can bring this in the form
\begin{eqnarray}
    \label{Intdelchi}
    \delta \chi_R(\phi^*) & = & \int [d\phi][d\chi] \left[ - \frac{\dr X_R
e^{\cW}}{\delta \phi^A} \frac{\dl f_R}{\delta \phi^*_A} - \frac{\dr X_R
e^{\cW}}{\delta \chi^A} \frac{\dl f_R}{\delta \chi^*_A} \right.
\nonumber \\
& & + X_R \frac{\dr \cW}{\delta \phi^*_A}.e^{\cW}.
\frac{\dl f_R}{\delta \phi^A}  + X_R \frac{\dr \cW}{\delta \chi^*_A}.e^{\cW}.
\frac{\dl f_R}{\delta \chi^A}  \nonumber \\
& & \left.
+ \frac{\dr X_R}{\delta \phi^*_A} \frac{\dl f_R}{\delta \phi^A} e^{\cW}
    + \frac{\dr X_R}{\delta \chi^*_A} \frac{\dl f_R}{\delta \chi^A}
e^{\cW} \right]_{\chi^*=0} \ .
\end{eqnarray}

In analogy with the unregularised derivation of the quantum master
equation, we would now like to do partial integrations to arrive at
the regularised version of (\ref{deltachi}). Owing to the
implicit definition of the
measure of the PV fields, it is unclear how the PV fields are to be
integrated by parts. Instead, we will use the following property:
\beq
  \label{lemmaaa}
  \int [d\phi][d\chi] \left[ \frac{\dr A_R}{\delta \phi^*_A} \frac{\dl
B(\phi ,\chi)}{\delta \phi^A} + \frac{\dr A_R}{\delta \chi^*_A} \frac{\dl
B(\phi ,\chi)}{\delta \chi^A} \right] = 0 \ ,
\eeq
for any $A(\phi,\phi^*)$ and $B(\phi,\phi^*,\chi,\chi^*)$.
The proof of this property, which is based on the freedom to redefine
integration variables, is given in the appendix of {\em this} chapter. As a
first consequence of this lemma, the first line of (\ref{Intdelchi}) is
seen to vanish identically $(A=f)$. Also, the third line of
(\ref{Intdelchi}) is equal to
\beq
 \label{reg1}
   \int [d\phi][d\chi] \left[ - \frac{\dr X_R}{\delta \phi^*_A}
   \frac{\dl \cW}{\delta \phi^A}. e^{\cW} .  f_R
    - \frac{\dr X_R}{\delta \chi^*_A} \frac{\dl \cW}{\delta \chi^A}.
e^{\cW}. f_R \right]_{\chi^*=0}   \ ,
\eeq
by taking $A_R = X_R$ and $B = f_R e^{\cW}$. As for the second line of
(\ref{Intdelchi}), instead of $A_R$ we now have $S_R + S_M + \hbar M_1$.
The extra contribution to the logarithm of the Jacobian is proportional
to $\Delta_{\phi} (S_M + \hbar M_1)$ and would lead to $\cO(\hbar^2)$
corrections to the quantum extended action, which we drop.
We find that the second line of (\ref{Intdelchi})
equals
\begin{eqnarray}
  \label{reg2}
   \int [d\phi][d\chi] &  & \left[ -  X_R \frac{\dr \cW}{\delta \phi^*_A}
   \frac{\dl \cW}{\delta \phi^A} . e^{\cW} . f_R
-  X_R \frac{\dr \cW}{\delta \chi^*_A}
   \frac{\dl \cW}{\delta \chi^A} . e^{\cW} . f_R  \right. \nonumber \\
   &  & \left. + \frac{\dr X_R}{\delta \phi^A}
   \frac{\dl \cW}{\delta \phi^*_A}. e^{\cW} .  f_R
    + \frac{\dr X_R}{\delta \chi^A} \frac{\dl \cW}{\delta \chi^*_A}.
e^{\cW}. f_R \right]_{\chi^*=0}  \  ,
\end{eqnarray}
if $A_R = \cW$ and $B= X_R . e^{\cW} . f_R$.

If we combine (\ref{reg1}) and (\ref{reg2}), we find the one
loop regularised version of (\ref{deltachi}):
\begin{eqnarray}
   \label{regdelchi}
   \delta \chi_R(\phi^*) & = & \int [d\phi][d\chi] \left[
   X_R \frac{1}{2} (\cW,\cW)_{\phi ,\chi} . e^{\cW} . f_R \right.
   \nonumber \\
& &\left. + (X_R ,\cW)_{\phi ,\chi} . e^{\cW} . f_R \right]_{\chi^*=0}.
\end{eqnarray}
In contrast to (\ref{deltachi}), we see that no $\Delta$ operators have
appeared, as the $\Delta_{\phi}$ were always cancelled by $\Delta_{\chi}$
terms. Notice that the first term of this very important result should be
interpreted as the regularised master equation, while the second term gives
a regularised expression for $\sigma X$. We discuss both expressions in
some detail in the next section.

\section{Discussion}

We first consider the (one loop) {\bf regularised master equation}. It is
contained in the first term of (\ref{regdelchi}):
\beq
  \label{regMas}
   \int [d\phi][d\chi] \left[
   X_R \frac{(-1)}{2\hbar^2} (\cS + \hbar M_1,\cS + \hbar M_1)
   _{\phi ,\chi} . e^{\ihbar(\cS + \hbar M_1)} . f_R
\right]_{\chi^*=0} = 0.
\eeq
As we have already pointed out (\ref{justabove}), $(\cS,\cS)_{\phi,\chi} =
2 (S,S_M)_{\phi} + 2 (S_{PV} , S_M)_{\chi} + \cO(\chi^4)$.
The only other contribution in (\ref{regMas}) that is
not of second or higher order in $\hbar$ with respect to $(S,S)_{\phi}$, is
$2(S,M_1)$. Taking these two remarks into account, we are led
to the {\em regularised one loop master} equation
\beq
  \label{1lusreg}
  \hbar (S,M_1)  I(PV)
  +  \int [d\chi] [ (S, S_M)_{\phi} + (S_{PV}, S_M)_{\chi} ]
e^{\ihbar(S_{PV} + S_M)}\vert_{\chi^*=0}  = 0.
\eeq
We denoted $I(PV) = \int [d\chi] \exp [\ihbar(S_{PV} +
S_M)]\vert_{\chi^*=0}$.
At this point, the PV fields can be integrated out (\ref{PVout}) in the
second term, and all steps following (\ref{deltagamma}) can be copied.
In particular, the limit $M \rightarrow \infty$ is understood, as is
the introduction of copies of the PV fields,
if needed (cfr. (\ref{PVcopies})). By comparing (\ref{1lusreg}) with the
formal, unregularised one loop master equation, $(S,M_1) - i \Delta S = 0$,
we see that the second term of the RHS is to be interpreted as a
regularised expression for $\Delta S$ \cite{WPT,TheBible}:
\beq
(\Delta S)_R = \frac{i}{I(PV) \hbar} \int [d\chi] [ (S_{PV} , S_M)_{\chi}
+ (S , S_M)_{\phi} ] e^{\ihbar(S_{PV} + S_M)} \vert_{\chi^*=0} .
\eeq
We denominate this regularised expression by $(\Delta S)_R$.

An important property of the formal, unregularised $\Delta S$ is that
$(S,\Delta S)=0$, if $S$ satisfies the classical master equation. This is
the translation in the BV language of
the Wess-Zumino consistency condition (see section 2 of chapter 11). As we
have duplicated the complete
structure of BV for the PV fields, we expect this property to be valid for
the regularised expression too. In the appendix of \cite{Stefan}, it is
explicitly proven that $((\Delta S)_R,S)=0$.

The second term of (\ref{regdelchi}) is a one loop {\bf regularised
expression
for $\sigma X$}, as follows from a comparison with (\ref{deltachi}). We
have \beq
   \label{sigmareg}
   (\sigma X)_R  I(PV) = \int [d\chi] \; (X_R , S_R + S_M + \hbar
M_1)_{\phi ,\chi} e^{\ihbar(S_{PV} + S_M)} \vert_{\chi^*=0}.
\eeq
Let us consider the integrand in some more detail:
\beq
  (X_R , S_R + S_M + \hbar M_1)_{\phi ,\chi} = (X,S)_R  + \hbar
(X,M_1)_{\phi} + (X_R , S_M)_{\phi ,\chi} + \cO(\chi^4) \ .
\eeq
The first two terms are the (regularised) order $\hbar$ contribution of the
antibracket part of $\sigma X$, $(X,S+\hbar M_1)$. The remaining term is
then interpreted as the regularised version of $-i\hbar\Delta X$:
\beq
  \label{DeltaXR}
(\Delta X)_R = \frac{i}{I(PV) \hbar} \int [d\chi] [ (X_{PV} , S_M)_{\chi}
+ (X , S_M)_{\phi} ] e^{\ihbar(S_{PV} + S_M)} \vert_{\chi^*=0} .
\eeq
For $X=S$, we find back the expression for $(\Delta S)_R$. Possible
applications of this
result are 1) infinitesimal canonical transformations, under which the
quantum extended action transforms by the addition of $\sigma f$
(\ref{Wtilde}) and 2) the study of the quantum cohomology at one loop.

Again, the properties derived for the formal, unregularised $\sigma X$ like
nilpotency etc. have to be verified for the regularised expression. This
remains to be done for the general result. In the next section, we will
give an example of a regularised calculation of $\sigma f$ for an
infinitesimal canonical transformation. In this example, it is easy to see
that $(\sigma (\sigma f)_R)_R = 0$.

\section{Example}

We consider an example in two dimensions. We sketch the first
step of the bosonisation procedure that is presented in \cite{DaNiS}, from
the BV point of view (see section 4 of chapter 8).
We start from an action\footnote{We take the following
conventions. The two Dirac $\gu{0}$ and $\gu{1}$ matrices satisfy the
anticommutation relation $\{\gu{\mu},\gu{\nu}\}=2 g^{\mu \nu}$. We define
$\gv =\gu{0} \gu{1}$ and $\gu{\mu} \gv = \epsilon^{\mu}{}_{\nu} \gu{\nu}$
with $\epsilon^{\mu}{}_{\nu}g^{\nu \sigma} = \epsilon^{\mu \sigma}$ the
antisymmetric 2x2 tensor with $\epsilon^{01} = -1$. We also have that
$\{\gu{\mu},\gv\}=0$.}
containing massless fermions, with
an external source $A_{\mu}$ coupled to their axial current:
\beq
    S_0 = i \bar \psi \dirac{\partial} \psi + A_{\mu} \bar\psi \gu{\mu}
    \gv \psi \; .
\eeq
An integration over two dimensional space-time is understood.
Including also a source term for the vector current, would only make the
algebra slightly more complicated.

We enlarge the set of fields with an extra scalar degree of freedom
$\alpha(x)$. As this field is not present in the classical action, $S_0$ is
invariant under arbitrary shifts of this scalar field. We introduce a ghost
$c$ for this shift symmetry and consider the extended action
\beq
    S = i \bar \psi \dirac{\partial} \psi + A_{\mu} \bar\psi \gu{\mu}
    \gv \psi + \alpha^* c \; .
\eeq
We now have a gauge symmetry that has to be gauge fixed. Therefore, we
consider the non-minimal solution of the master equation
\beq
     S_{n.m.} = S + b^* \lambda \; ,
\eeq
and do the canonical transformation generated by $F = \unity - b \alpha$.
We get the gauge fixed action
\beq
    S_{com} = i \bar \psi \dirac{\partial} \psi + A_{\mu} \bar\psi \gu{\mu}
    \gv \psi - bc - \alpha \lambda + \alpha^* c + b^* \lambda \; .
    \label{Esgef}
\eeq
Neither the ghost action $bc$ nor the term fixing $\alpha$ to zero lead to
propagating degrees of freedom, so that we only have to introduce PV fields
for the fermion fields $\psi$ and $\bar \psi$. We denote the PV fields
by $\chi$ and $\bar \chi$.

Let us now disguise the shift symmetry by doing an infinitesimal canonical
transformation generated by
\beq
   F = \unity - i \psi^{*'} \epsilon \alpha \gv \psi
              - i \bar \psi \epsilon \alpha \gv \bar \psi^{*'} \; .
\eeq
Here, $\epsilon$ is a global, infinitesimal parameter. This canonical
transformation is an infinitesimal chiral rotation,
as can be seen from the transformation rules it generates:
\begin{eqnarray}
     \psi & = & ( 1 + i\epsilon \alpha \gv )\psi' \nonumber \\
     \bar \psi & = & \bar \psi' ( 1 + i\epsilon \alpha \gv) \; .
\end{eqnarray}
However, also a transformation for $\alpha^*$ is generated:
\beq
   \alpha^* = \alpha^{*'} - i \psi^{*'} \epsilon \gv \psi'
              - i \bar \psi' \epsilon \gv \bar \psi^{*'}
\eeq
up to corrections of $\cO(\epsilon^2)$. With these infinitesimal
transformations, we find that
\beq
  (f,S) =  i \psi^* \epsilon \gv c \psi - i\bar\psi \epsilon \gv c \bar
\psi^* - \epsilon \partial_{\mu} \alpha \bar\psi \gu{\mu} \gv \psi \; ,
\eeq
with $f= - i \psi^* \epsilon \alpha \gv \psi - i \bar \psi^{*a}
\epsilon \alpha \gv_{ba} \bar \psi^b$.

We also have to take into account the Jacobian of this transformation, i.e.
we have to calculate $(\Delta f)_R$. As we
already pointed out, we only have PV fields for the fermions, so we have
\beq
  f_R = f - i \chi^* \epsilon \alpha \gv \chi
- i \bar \chi^{*a} \epsilon \alpha \gv_{ba} \bar \chi^b \ .
\eeq
The PV action for the fermions is given by
\beq
   S_{PV} = i \bar \chi \dirac{\partial} \chi + A_{\mu}\bar \chi \gu{\mu}
\gv \chi \  ,
\eeq
and we take the mass term that is invariant under ordinary phase rotations
$ S_M = - M \bar \chi \chi $.

We can now calculate $(\Delta f)_R$. $(f,S_{PV})_{\phi} = 0$, but we have a
contribution
\beq
    (f_{PV} , S_M )_{\chi} = -2i\epsilon \alpha M \bar \chi \gv \chi \  .
\eeq
Hence, we find that
\begin{eqnarray}
(\Delta f)_R & = & \frac{2\epsilon \alpha M}{\hbar I(PV)} \int [d\chi] \;\;
   \bar \chi \gv \chi \;\; e^{\ihbar(S_{PV} + S_M)} \nonumber \\
     & = & 2i\epsilon \alpha M \; \tr  \left[ \gv
\frac{1}{i\dirac{\partial} + \dirac{A} \gv - M} \right] \, ,
\end{eqnarray}
where in the last step we used again (\ref{PVout}) ($x=-2$). At this
stage, we can not yet use (\ref{DenReg2}) to obtain a Gaussian damping
regulator. If we symbolically write $\cO =i\dirac{\partial} + \dirac{A}
\gv$, we have (using (\ref{AddA})):
\beq
  M \tr [ \gv \frac{1}{\cO - M} ] =  M \tr [ \gv (\cO +M) \frac{1}{\cO^2 -
M^2} ]  \ .
\eeq
As we consider the limit $M \rightarrow \infty$, only the second term (with
$M^2$ in the numerator) survives and we get \cite{Diaz}
\beq
   (\Delta f)_R = 2i\epsilon \alpha  \tr \left[ \gv \frac{1}{\cO^2/M^2 -
1} \right]
\eeq
with $\cO^2 = - \Box + 2 (i A_{\nu} \epsilon^{\nu \mu})\partial_{\mu} - A^2
+ i \dirac{\partial} \dirac{A} \gv$. We denoted $\dirac{\partial} \dirac{A}
= \gu{\mu} \gu{\nu} \partial_{\mu} A_{\nu}$. We now go through the familiar
steps to obtain a Fujikawa type expression (\ref{DenReg2}) and use the
results of the appendix (\ref{gevalB}). As we work in two dimensions, and
provided additional PV fields are introduced to remove terms proportional
to $M^2$, we only need $E_2$ (\ref{gevalB}). We find that
$ \cO^2 = - (\partial_{\mu} + Y_{\mu})(\partial^{\mu} + Y^{\mu})
- E$ for $Y^{\mu} = i A_{\nu} \epsilon^{\nu \mu}$ and
$E = - \alpha^2 - \partial^{\mu} \alpha_{\mu} + A^2 -i \dirac{\partial}
\dirac{A} \gv$. Only the last term in $E$ contributes as $\tr \gv =0$.
Therefore, we have that
\begin{eqnarray}
    (\Delta f)_R & = & - 2 i \epsilon \alpha \tr [\gv E_2] \nonumber \\
         & = &  \frac{\epsilon}{2\pi} \partial^{\mu} \alpha . A_{\mu}.
\end{eqnarray}

The quantum extended action after canonical transformation is then seen to
be
\begin{eqnarray}
\tilde W & = & S_{com} + \sigma f \nonumber \\
         & = & S_{com} + i \psi^* \epsilon \gv c \psi
  - i\bar\psi \epsilon \gv c \bar \psi^*  \nonumber \\
   &  & - \epsilon \partial_{\mu} \alpha \bar\psi \gu{\mu} \gv \psi
   - \frac{i \hbar}{2 \pi} \epsilon \partial^{\mu} \alpha . A_{\mu} \  .
\end{eqnarray}
Let us mention two technical aspects of this result. The
transformation of $\alpha^*$ has generated $\psi^*$ and $\bar \psi^*$
terms. Canonical transformations produce the transformation rules in the
transformed coordinates. Furthermore, it is not difficult to see that
$(\sigma (\sigma f)_R)_R = 0$.

As for the physical point of view, we see that the Jacobian gives a term of
the form $\hbar \partial^{\mu} \alpha . A_{\mu}$. We find back the
typical bosonisation rule that the axial current of the fermions gets
replaced by $\partial^{\mu} \alpha$. If we also include an external source
$V^{\mu}$ for the vector current of the fermions, a term $ \epsilon_{\mu
\nu}V^{\mu} \partial^{\nu} \alpha $ appears. However, in $\tilde W$ a
term $\alpha \lambda$ is present that fixes $\alpha$ to zero, so that all
the $\alpha$-dependent terms are effectively zero.
The second step of the bosonisation procedure, as
described in \cite{DaNiS}, consists of changing the gauge fixing. Instead
of the gauge $\alpha=0$, a gauge is chosen where some fermion degrees
of freedom are fixed such that the fermion currents decouple from the
sources. This way, it is seen that 2D bosonisation is a matter of gauge
choice.

\section*{Appendix}

Here we give a proof of the property (\ref{lemmaaa}) used in the main text.
We start from an integral
\beq
      I = \int[d\phi][d\chi] B(\phi ,\chi) \; ,
\eeq
where the $\phi$ are ordinary fields and $\chi$ PV fields. The latter
statement
implies that $[d\chi]$ has the tranformation properties derived in the
previous chapter (\ref{PVjacob}). Consider any quantity $A_R = A + A_{PV}$,
and let us redefine the integration variables as
\begin{eqnarray}
     \phi^A & \rightarrow & \phi^A + \epsilon . \frac{\dr A_R}{\delta
\phi^*_A} \nonumber \\
     \chi^A & \rightarrow & \chi^A + \epsilon . \frac{\dr A_R}{\delta
\chi^*_A} \; ,
\end{eqnarray}
where $\epsilon$ is an infinitesimal parameter. As was already dicussed in
the main text, such redefinitions lead to a Jacobian $1$ (up to higher
orders in $\hbar$), owing to the definition of $A_R$ and of the measure
$[d\chi]$. We then have
\beq
   \int[d\phi][d\chi] B(\phi ,\chi) = \int[d\phi][d\chi]
   B(\phi+ \epsilon . \frac{\dr A_R}{\delta \phi^*_A}
,\chi +\epsilon . \frac{\dr A_R}{\delta \chi^*_A} ) \; ,
\eeq
which immediately implies the result
\beq
  \int [d\phi][d\chi] \left[ \frac{\dr A_R}{\delta \phi^*_A} \frac{\dl
B(\phi ,\chi)}{\delta \phi^A} + \frac{\dr A_R}{\delta \chi^*_A} \frac{\dl
B(\phi ,\chi)}{\delta \chi^A} \right] = 0.
\eeq

\chapter{Anomalous theories}

We have seen how we can detect genuine anomalies, at least to one-loop
order. A theory has a genuine, one loop-anomaly
if the one-loop master equation $(S,M_1) - i \Delta S = 0$ can not be
solved for a local $M_1$. As was already stressed a few times, we can not
derive the Ward identity $\langle \sigma X \rangle =0$ in that case, such
that the proofs of unitarity and renormalisability are jeopardized.
However, in the last decade, a large effort has been devoted to attempts to
proceed further with anomalous theories. These developments are
particularly stimulated by the interest in non-critical string theories,
i.e. string theories where one does {\em not} work in the specific
dimensions (26 for the bosonic string) required for the vanishing of the
anomaly. We will first briefly discuss the method of background charges in
conformal field theory. In the second section of this chapter, we discuss
in extenso the appearance of extra degrees of freedom in the case of an
anomalous theory \cite{ik3}.

\section{Background charges}

Instead of expanding the quantum master equation in powers of $\hbar$ as
Ansatz, we can also try $W = S + \sqrt{\hbar} M_{1/2} + \hbar M_1 + \ldots$
\cite{ik3}. When this expansion is plugged in in the quantum master
equation, we get the hierarchy of equations
\beq
  \begin{array}{lcl}
    \hbar^0 & \mbox{\hspace{1cm}} & (S,S)  =  0 \\
    \hbar^{1/2} &  &  (S, M_{1/2}) = 0 \\
    \hbar &   &  (S,M_1) + \frac{1}{2} ( M_{1/2} , M_{1/2} )=  i
\Delta S \\ \ldots
  \end{array}
\eeq
Let us again consider the example of $W_2$ gravity, where we take the
matter fields $\phi$ to be the only propagating quantum fields. The
extended action is given by (\ref{W2extac}):
\begin{eqnarray}
    \label{juistboven}
S & = & - \frac{1}{2\pi} \phi  \partial \nabla \phi
  + \phi^* c \partial \phi \nonumber \\
  &  & + h^* (\bar \partial c - h \partial c +
\partial h . c ) + c^* (\partial c) c .
\end{eqnarray}
It is easy to see that for
\beq
   M_{1/2} = a ( h \partial^2 \phi + \pi \phi^* \partial c ) \; ,
\eeq
with $a$ an arbitrary constant, we have that $(S,M_{1/2})=0$.
Furthermore,
\beq
    \frac{1}{2} (M_{1/2} , M_{1/2}) = - a^2 \pi . c \partial^3h \ .
\eeq
As $\Delta S$ is also proportional to $c \partial^3h$ for this model
(\ref{cd3h}), we can choose the constant $a$ such that
\beq
    \frac{1}{2} ( M_{1/2} , M_{1/2} ) -  i \Delta S = 0  \  .
\eeq
Therefore, no (non-local) counterterm $M_1$ is needed for the $\cO(\hbar)$
master equation to be satisfied. This is the translation in the BV language
of the method of {\it background charges} familiar in conformal field
theory (see e.g. \cite{Ginsparg})\footnote{We can interpret the terms of
$\sqrt{\hbar} M_{1/2}$ as follows. The term $a \sqrt{\hbar} \partial^2
\phi .h$ indicates that the energy-momentum tensor $T$ that is coupled to
$h$ gets a correction $\sim \sqrt{\hbar} \partial^2 \phi$. The
transformation rule for $\phi$ generated by $T$ also changes, it gets an
extra term $\sim \sqrt{\hbar} \partial c$. This is expressed by the extra
term $a \sqrt{\hbar} \pi \phi^* \partial c$ in $\sqrt{\hbar} M_{1/2}$.}.
For a more elaborate example, where background charges were
used to cancel the $W_3$ one loop-anomaly, we refer to
\cite{w3back} and to its translation in BV \cite{Stefan}.

\section{Hiding anomalies}

As we already heuristically argued above (\ref{exprop}), the most
interesting feature of gauge theories with genuine anomalies is that some
degrees of freedom that can be fixed to zero classically, start
propagating
at the quantum level. When faced with an anomalous theory, at least two
strategies have been followed in the literature up to now. The first
strategy (for examples, see e.g.\ \cite{Polyakov1,JacRaj}) consists in
making
a judicious choice of the order in which the functional integrals are done.
Consider for instance the $W_2$ example, with the classical action the
antifield independent part of (\ref{juistboven}).
Both the matter field $\phi$ and the field $h$ are considered to be
dynamical, i.e. are integrated over in the functional integral. If we
stipulate that one first has to do the $\phi$ integral, we see that $h$
gets a non-trivial action, the induced action $\Gamma[h]$
(\ref{Polyakovaction}). Thereafter, the integral over $h$ can be done. As
is generally the case, the induced action that one obtains, is non-local.
This need not worry us, precisely because we know this
non-local action is induced by a local quantum field theory. Moreover, it
has been shown that many of the (regularisation) procedures of local
field theory can be transplanted to non-local theories without any problem
\cite{Ruud}.

The second approach started with the work of L.D. Faddeev \cite{Faddeev}.
He argued that anomalies make the first class constraints of the gauge
symmetries in the Hamiltonian formalism second class and that they can be
made first class again by the introduction of extra degrees of freedom.
Later, it was recognised that in the functional integral approach to
quantisation, these extra degrees of freedom arise naturally. Indeed, the
integration over the volume of the gauge group does not factor out when
following the Faddeev-Popov \cite{FadPop} procedure for gauge fields
coupled to chiral fermions \cite{Shaposh}. The integration measure for the
fermions has an (anomalous) dependence on the gauge variables, effectively
producing the Wess-Zumino action for the extra variables.

In \cite{Jordi2}, the idea of adding extra degrees of freedom has been
implemented in the BV scheme. For every anomalous symmetry, an extra field
is introduced. The transformation rules for
these new fields under the original
symmetry are chosen such that a local $M_1$ can be constructed to solve the
quantum master equation to one loop. In other words, by adding extra
fields, and by choosing their BRST transformation rules, $\Delta S$ is made
BRST exact. This $M_1$ provides the dynamics
for these extra fields. In this approach, the anomaly has apparently
disappeared, where the price to pay is a minor change of the classical
theory.

We \cite{ik3} show here that these extra degrees of freedom can be
introduced
{\em without changing the classical theory} in the cohomological sense. The
choice that one makes for the transformation rules of the extra fields
under the original symmetries,
is determined by the condition that one can construct
a PV mass term, using these extra fields, that is invariant under these
symmetries. However, together
with the new fields, one has also introduced new symmetries (to keep the
classical cohomology unchanged) and the PV mass term is in general not
invariant under these new symmetries. As a consequence, the anomaly has not
disappeared, but is shifted to these extra symmetries. Since these extra
symmetries are often ignored in the literature, the anomaly is in fact
hidden in this way.

\subsection{Discussion of our method}

Suppose that we start from an extended action $S[\phi^A]$,
that is a proper solution of the classical master equation $(S,S)=0$.
After having set up the PV regularisation (see the two previous chapters),
we can calculate the operator $\Delta S$.
At least for theories with an irreducible gauge algebra, the result is
always of the form
\beq
(\Delta S)_{reg} = c^{\alpha} {\cal A}_{\alpha},
\eeq
at least up to antifield dependent terms.
Here, the values that $\alpha$ takes in the sum, depend on the
invariances of
the PV mass term. In other words, the mass term determines which symmetries
are anomalous. Let us assume that we have a genuine anomaly, i.e. that no
local functional $M_1$ exists such that $(S,M_1)=i (\Delta S)_{reg}$.

Now we propose to add trivial systems to $S$, one for every
anomalous gauge symmetry:
\beq
S \rightarrow S + \theta^{*}_{\alpha} d^{\alpha} \ .
\eeq
$\theta^{\alpha}$ has the Grassmann parity $\gras{\alpha}$
and $d^{\alpha}$ has the Grassmann parity $\gras{\alpha} + 1$.
Although these extra fields, and the entailing extra
symmetries, are completely trivial at this point,
regularisation in the quantum theory will interfere with this.
The new degrees of freedom clearly do not change the classical theory, as
they are cohomologically trivial for the (new) cohomology-operator
 $(S,\cdot)$.

At this point, we are still free to specify how these newly introduced
fields transform under the original symmetries related to the
$c-$ghosts. This choice can be encoded in the
extended action by doing a canonical transformation. Remember that
canonical transformations do not change the (classical) cohomology
(\ref{ClasTraf}). Our approach is
to choose this transformation such that a local PV-mass term can be
constructed that {\it is} invariant under the $c$-symmetries.

Suppose that one would like to have the transformation rules
\begin{equation}
 \delta_c \theta^{\alpha} = f^{\alpha}(\phi^A,\theta) \label{delalfa}
\end{equation}
for the $\theta^{\alpha}$ fields. This can be achieved
simply by taking as generating fermion for the canonical transformation
\beq
 F = \unity - d^{*'}_a f^a(\phi^A,\theta) \  .
\eeq
The fact that we used a canonical transformation guarantees that the
transformed extended action is still a solution of the classical
master equation, i.e. the new action is still BRST invariant with the
modified transformation rules.
 It is also important to remark that the extended action after the
transformation still contains terms with the $d$ ghosts. These are
necessary
to ensure the properness of the action. It is clear that, if we use the
extra $\theta$ fields to construct a
PV mass term that is invariant under the $c$-symmetries, it can not also be
invariant under the $d$-symmetries, which shift $\theta $. As a result
the anomalies will have been shifted to the new symmetries and one finds
\beq
(\Delta S)_{reg} = d^{b} {\cal B}_b   \ .
\eeq
It is only when one neglects the $d$-symmetries that one would conclude
 that there are no anomalies left.

To carry out calculations, one may still want to use the old
($c$ non-invariant) mass-term, for technical or other reasons.
The anomaly will still be left in the
$d$-symmetries, if a counterterm is added that matches
the interpolation between the two different regularisations, viz the two
mass terms. This has extensively been discussed above (\ref{deltaM}).
In practise, it is this counterterm that provides non-trivial
dynamics for the variables that were introduced as a trivial system
(Wess-Zumino term).

The next step is to integrate over these extra fields $\theta^{\alpha}$, as
they acquired non-trivial dynamics. However, at this moment, it is not
clear what measure one should take for these extra fields.
This is related to the fact that there may be different ways
to introduce extra fields (see the example in the next subsection).
Clearly, a guiding principle can be that one tries to construct
the complete theory to be invariant under the $c$-symmetries.

\subsection{Example}

We again use the $W_2$ gravity model as an example. Most of the technical
manipulations are the same as those discussed in the sections 2 and
3 of chapter 12, and will therefore not be repeated in detail. However,
there is an important difference. In chapter 12, we constructed a local
invariant mass term for the PV fields using the Polyakov variable $f$, that
is a reparametrisation of $h$. Here, we introduce an extra scalar
field in the theory to construct a local invariant mass term.

We start from the extended action for $W_2$ gravity (\ref{juistboven}), and
consider only the matter field $\phi$ as a quantum field. The PV field
for $\phi$ is denoted by $\chi$, and we have $S_{PV}= -1/(2\pi) \chi
\partial \nabla \chi + \chi^* c \partial \chi$ (\ref{APV}). To complete the
regularisation scheme, we try to construct a PV mass term that is invariant
under the $c$-symmetry, where we allow ourselves to introduce an extra
field $\theta$ with a transformation rule $\delta_c \theta$ that may be
chosen. It is easy to see that
\beq
     S_M = - \frac{1}{4\pi} . M^2. \chi^2. e^{\theta}
\eeq
is invariant under the $c$-symmetry if we take $\delta_c \theta = c
\partial \theta + \partial c$. We also could have taken
\beq
     S_M = - \frac{1}{4\pi} . M^2. \chi^2. \xi
\eeq
with $\delta_c \xi = \partial (c \xi)$. It is clear that introducing a
flat measure in the functional integral for $\xi$ differs from the
introduction of a flat measure for $\theta$ by a Jacobian of the
redefinition by $\xi = e^{\theta}$.

Let us now calculate the anomaly, using the first invariant mass term.
Define $ S_M^{(\alpha)} = - 1/(4\pi) . M^2. \chi^2. e^{\alpha \theta}$. For
$\alpha=0$, we have the $c$ non-invariant mass term while for $\alpha=1$ we
have the conformal $c$ invariant mass term, if $\delta_c \theta = c
\partial
\theta + \partial c$. We add a trivial system and consider the extended
action $\tilde S = S + \theta^* d$. The field $d$ is the ghost field for
the shift symmetry of the extra field $\theta$. In order to have a term
$\theta^* \delta_c\theta$ in the extended action, we do a canonical
transformation generated by
\beq
    F = \unity - d^{*'} ( \partial c + c. \partial \theta ).
\eeq
After the canonical transformation we have
\beq
    \tilde S' =  S + \theta^* ( d + \partial c + c \partial \theta ) + d^*.
\partial d.c \;.
\eeq
This action is the same as in \cite{Jordi2}, except for the terms
containing the $d$-ghost. These terms are needed to assure that we have a
proper solution of the classical master equation. They also guarantee that
the classical cohomology of the original theory is the same as the
classical cohomology of the theory with the extra field $\theta$.

We now calculate the anomaly due to the matter fields\footnote{We work
again in Euclidean space and $\hbar=1$.}. We use
(\ref{DeltaXR}), for $X=S$. The first ingredient is
\beq
   ( \tilde S'_R , S_M^{(\alpha)})_{\phi ,\chi} = \frac{M^2}{4\pi} [
\partial c (1 - \alpha) - \alpha d] . e^{\alpha \theta}.\chi^2 \; .
\eeq
Copying the steps of section 2 of chapter 12, we find
\beq
   (\Delta S)_R (\alpha) = - \frac{1}{24\pi} \int d^2 x \;
  [ \partial c (1 - \alpha) - \alpha d] [ \partial^2 h -\alpha \partial
\nabla \theta ] \; .
\eeq
For $\alpha =0$, we find back (\ref{cd3h}), but for $\alpha = 1$, we have
\beq
   (\Delta S)_R (\alpha = 1) =  \frac{1}{24\pi} \int d^2 x \;\;
   d . [ \partial^2 h - \partial \nabla \theta ] \; .
\eeq
The $c$-symmetry is seen to be anomaly free, the anomaly has been shifted
to the extra symmetry that comes with the extra field.

Instead of using the $c$ invariant mass terms ($\alpha=1$) one might prefer
to use the $c$ non-invariant mass term ($\alpha=0$) in practical
calculations. The anomaly will still be proportional to the $d$-ghost, if
we add a counterterm that compensates the change of mass term. The way to
calculate such counterterms is described in section 3 of chapter 12. This
counterterm provides non-trivial dynamics for the extra field $\theta$.
We find
\beq
    M_1 = \frac{1}{24\pi} \int d^2x \;\; \left[ -\frac{1}{2} \theta
\partial \nabla  \theta + h \partial^2 \theta \right] \; .
\eeq
Such counterterms are named {\it Wess-Zumino terms}. In the literature, one
often introduces extra fields and adds a Wess-Zumino term to make
the anomaly BRST exact. As we have shown, the anomaly has not disappeared,
it has been shifted to extra symmetries that come with the extra
fields and that are needed to keep the classical theories cohomologically
equivalent. Moreover, the presence of the Wess-Zumino term
involves a choice: one works with the $c$ non-invariant mass term and the
counterterm is added to shift the symmetry to the $d$ symmetries.

\chapter*{Conclusions}

After this extensive discussion of the {\bf Batalin-Vilkovisky scheme}, we
can  list some of its {\bf assets}, as was promised 14 chapters ago in the
introduction.

\begin{itemize}

\item
First of all, whatever the properties of the gauge algebra are, be it an
open or a closed, a reducible or an irreducible algebra, in the BV scheme
one has to solve the classical
master equation $(S,S)=0$ as the first step in the quantisation process
(chapters 5 and 6). This way, the BV scheme provides a unified
approach for all types of gauge theories.

\item
To formulate the classical master equation, we have already used the most
remarkable feature of the BV scheme. The fields and antifields are
canonically conjugated with respect to a Grassmann odd symplectic bracket,
the antibraket (chapters 5 and 8).
All manipulations that constitute the BRST quantisation
formalism (chapter 3) can be rephrased using antibrackets {\em and} their
canonical transformations (chapter 8): gauge fixing, taking a different set
of gauge generators, the cohomology of gauge independent operators, Ward
identities, ...

\item
The antifields serve many purposes in the BV scheme. One is that they act
as sources for the BRST transformation of their associated field. This
allows, for instance, for a natural derivation of the Zinn-Justin equation
for the generating function of connected diagrams in the BV
scheme (chapter 8). The equation holds not only
for Yang-Mills theory, its original application area, but for all types of
gauge theories, provided that the theory is anomaly free.

\item
Moreover, owing to the antifields, one does not have to choose a
gauge. If one keeps track of all antifield dependence in practical
calculations, one can afterwards always transform the result obtained in
one gauge (one set of canonical coordinates) to
other gauges by canonical transformations. For one thing, this prevents the
accidental vanishing of anomalies that may be caused by gauge fixing
symmetries that are anomalous. In the BV scheme, the anomaly will then be
function of the antifields.

\item
Finally, both the classical and the quantum cohomology contain respectively
the classical and quantum equations of motion.

\end{itemize}

Given all these assets of the BV scheme, we found it worthwhile to
give a derivation of the main features of the scheme with a strong emphasis
on their relation with the analogous constructions in the BRST
quantisation prescriptions. This way it is clear that the BV scheme is an
elegant, unified reformulation of previously known (BRST) quantisation
recipes. Here, the BV scheme is constructed by
imposing that the BRST symmetry of a gauge theory be enlarged such that the
Schwinger-Dyson equations for all fields can be obtained as Ward identities
of this enlarged BRST symmetry. This requirement can be implemented using a
collective field formalism. Owing to a doubling of the configuration space
in that formalism, a new gauge symmetry --the
Schwinger-Dyson (SD) shift symmetry-- is present,
and the antighosts introduced to gauge fix this symmetry become the
antifields of the BV scheme.
Starting from the BRST quantisation prescription, we have derived both for
closed and open algebras the classical and quantum master equation of the
BV scheme. Especially for open algebras, the collective fields are seen to
play a crucial role in the construction of a BRST invariant, gauge fixed
action.

Using a slightly modified collective field formalism, we have constructed
an antifield scheme for BRST--anti-BRST invariant quantisation. Instead of
one, one has three antifields for every field in that case.

We have also shown how the Lagrangian Batalin-Vilkovisky scheme follows
from the Hamiltonian description of gauge theories. Here too, our guiding
principle has been that the BRST symmetry of the theory has to be enlarged
to include the SD shift symmetry. After integration over the momenta of the
Hamiltonian formalism, a Lagrangian extended action is obtained that
satisfies the quantum master equation of the BV scheme, owing to the gauge
independence of the Hamiltonian path integral.

In a third part of this dissertation, some aspects of a one-loop
regularised study of anomalies are presented.
For that purpose, we have used a Pauli-Villars (PV)
regularisation scheme. In particular, we have studied the effect of the
mass term of the regulating PV fields on the actual expression of the gauge
anomaly. We have shown how in some examples, preferred symmetries can be
kept anomaly free by constructing a mass term
that is invariant under these symmetries. This can be done
by using either reparametrisations of fields that are already present in
the field content of the theory or by
introducing extra fields. We have also used PV regularisation to give a new
derivation of the one-loop regularised quantum master equation and of the
one-loop regularised quantum BRST operator of the BV scheme.

Some results where obtained in the study of concrete models as well, using
the general methods that are described. We have reexamined the
construction of four-dimensional topological Yang-Mills theory,
taking advantage of the many uses of the canonical transformations of the
BV scheme. Prompted by this example, we have derived a general recipe
for a classical and quantum BRST invariant energy-momentum tensor in the BV
scheme. We have studied induced $W_2$ gravity.
Using the Polyakov variable, we can construct an invariant mass term
for the regularisation of this model, which was exploited to
calculate the induced action.
This method provides more insight on why the Polyakov reparametrisation
leads to a local expression for the induced action.

\part{Appendices}

\parskip 0truemm
\parindent 1 truecm

\appendix

\chapter{Grassmannology}

With every field $\phi^A$ of the configuration space, we associate a
Grassmann parity, which we denote by $\gras{A}$. $\gras{A}$ is an element
of $\mbox{Z} \!\! \mbox{Z}_2$.
By definition, $\gras{A}=0$ for a bosonic degree of freedom, and
$\gras{A}=1$ for a fermionic one. The antifield of
$\phi^A$, the field $\phi^*_A$, has Grassmann parity $\gras{A}+1$. Although
our notation seems to allow for the possibility that the Grassmann parity
of a field depends on the space-time point, this will never be the case.

The Grassmann parity of a product of two or more fields is the sum of their
Grassman parities:
\beq
    \gras{\prod_i A_i} = \sum_i \gras{A_i}.
\eeq
The Grassmann parities are mainly introduced to keep track of the
signs which appear when exchanging places between two fields (or monomials
of fields):
\beq
     X.Y = (-1)^{\gras{X}.\gras{Y}} Y.X,
\eeq
so only when both $X$ and $Y$ are fermionic, an extra minus sign has to
be taken into account.
The relation between the right derivative $\frac{\dr X}{\delta \phi^A}$ and
the left derivative $\frac{\dl X}{\delta \phi^A}$ is
\beq
   \frac{\dr X}{\delta \phi^A}= (-1)^{(\gras{X}+1)\gras{A}}
                        \frac{\dl X}{\delta \phi^A}.
\eeq
In the Leibnitz rule for directional derivatives, the signs are as follows:
\begin{eqnarray}
       \frac{\dr}{\delta \phi^A} (FG) & = & F. \frac{\dr G}{\delta \phi^A}
           + (-1)^{\gras{G}\gras{A}} \frac{\dr F}{\delta \phi^A} . G
           \nonumber \\
       \frac{\dl}{\delta \phi^A} (FG) & = &
            \frac{\dl F}{\delta \phi^A} . G + (-1)^{\gras{F}\gras{A}}
        F. \frac{\dl G}{\delta \phi^A} .
\end{eqnarray}
Two directional derivatives both acting from the left or from the right do
not commute. We have for instance,
\beq
    \frac{\dl}{\delta \phi^A} \frac{\dl}{\delta \phi^B} X =
    (-1)^{\gras{A}\gras{B}}
    \frac{\dl}{\delta \phi^B} \frac{\dl}{\delta \phi^A} X.
\eeq
However,
\beq
    \frac{\dl}{\delta \phi^A} \frac{\dr}{\delta \phi^B} X =
    \frac{\dr}{\delta \phi^B} \frac{\dl}{\delta \phi^A} X.
\eeq
Finally, $\gras{\frac{\delta X}{\delta \phi^A}} = \gras{X} + \gras{A}$,
whatever the derivative.

\chapter{Properties of the antibracket and the $\Delta$-operator}

As defined in the main text (\ref{antibracket}), the {\bf antibracket} of
two function(al)s $F$ and $G$ of arbitrary Grassmann parity is given by
\beq
   (F , G) = \frac{\dr F}{\delta \phi^A} \frac{\dl G}{\delta \phi^*_A} -
 \frac{\dr F}{\delta \phi^*_A} \frac{\dl G}{\delta \phi^A}.
\eeq
It is then clear that $\gras{(F,G)} = \gras{F}+\gras{G}+1$.
Another obvious property of the antibracket is that
$ \gh{(F,G)} = \gh{F} + \gh{G} + 1 $, since the sum of
the ghostnumber of a field with the ghostnumber of its antifield is $-1$.

{}From the properties of directional derivatives listed in the previous
appendix, we can derive that
\beq
     ( G , F ) = (-1)^{\gras{F}\gras{G} + \gras{F} +\gras{G}} (F,G).
\eeq
It follows that for any function(al) $F$ of odd Grassmann parity
$(F,F) = 0$ trivially. In contrast to the Poisson brackets of classical
mechanics, the antibracket of a bosonic $B$ function(al) --like the
extended action-- with itself is not necessarily zero. Indeed,
\beq
    ( B , B )  = 2 \frac{\dr B}{\delta \phi^A} \frac{\dl B}{\delta
\phi^*_A}.
\eeq
{}From the Leibnitz rule for directional derivatives, one easily arrives at
\beq
   ( F , GH) = (F,G)H + (-1)^{(\gras{F} + 1)\gras{G}} G ( F , H ).
\eeq
The antibracket version of the Jacobi identity is
\beq
   (F,(G,H)) = ((F,G),H) + (-1)^{(\gras{F} +1)(\gras{G}+1)} (G,(F,H)).
\eeq

The {\bf delta-operator} is defined by
\beq
   \Delta X = (-1)^{\gras{A}+1} \frac{\dr}{\delta \phi^*_A}
\frac{\dr}{\delta \phi^A} X  = (-1)^{\gras{X}}
   (-1)^{\gras{A}} \frac{\dl}{\delta \phi^*_A}
\frac{\dl}{\delta \phi^A} X                   ,
\eeq
and it emerged naturally by integrating out the fields in the collective
field formalism\footnote{Notice that in
\cite{TheBible} a slightly different definition of
the delta-operator is used:
\beq
\Delta X =
   (-1)^{\gras{A}} \frac{\dl}{\delta \phi^*_A}
\frac{\dl}{\delta \phi^A} X                   .
\eeq
It is clear that this only differs by a minus sign when the quantity $X$ is
fermionic.} (\ref{DELTA}). $\Delta$ is a fermionic operator, $\gras{\Delta
X} = \gras{X}+1$.
The basic properties are
\beq
   \begin{array}{lrcl}
   1. & \Delta^2 & = & 0 \\
   2. & \Delta (FG) & = & F \Delta G + (-1)^{\gras{G}} \Delta F.G +
   (-1)^{\gras{G}} (F,G) \\
   3. & \Delta (F,G) & = & (F,\Delta G) - (-1)^{\gras{G}} (\Delta F,G).
   \end{array}
\eeq

Using all these results, we can derive the following property of the
quantum BRST operator $\sigma X = (X,W) - i\hbar \Delta X$:
\beq
  \sigma [ A.B ] = A . \sigma B + (-1)^{\gras{B}} \sigma A . B - i\hbar
(-1)^{\gras{B}} (A,B) .
\eeq


\chapter{Technical details of the anomaly calculation}

This rather technical appendix contains a detailed description of
the steps needed to calculate an expression for the regularised
Jacobian, once the regulator is chosen. First, we show how the
Pauli-Villars regularisation leads to Fujikawa-type expressions
\cite{Diaz}.
The mathematical results needed to evaluate these expressions are then
presented. Thereafter, we treat a special case in great detail, which
allows an understanding from a low-brow point of view of many of the
characteristic features of the mathematical results.

\section{From Pauli-Villars regularisation to a consistent Fujikawa
regulator}

Pauli-Villars regularisation typically leads to the evaluation
of the following type of functional integral over PV fields (see e.g.
(\ref{deltagamma})):
\beq
   I[K] = \int [d\chi] \;\; \chi^C K_{CD} \chi^D \; e^{- \frac{1}{2}
    \chi^A D_{AB} \chi^B}.
\eeq
Here, the supermatrices of the bosonic type $K_{CD}$ and
$D_{AB}$\footnote{This means that $\chi^A K_{AB} \chi^B$ has even
Grassmann parity, and hence $\gras{K_{AB}} = \gras{A} + \gras{B}$}
are considered to be of the form
\begin{eqnarray}   \label{diffop}
      K_{CD} & = & K_{kl}(x).\delta(x-y) \nonumber \\
      D_{AB} & = & D_{ij}(x).\delta(x-y) ,
\end{eqnarray}
where $K_{kl}(x)$ and $D_{ij}(x)$ are differential operators in the
variable $x$ that act on the $\delta$-functions.
We have split up the capital indices $A$,$B$,\ldots in
space-time indices $x$,$y$,\ldots and internal indices $i$,$j$,\ldots
Remember that $\frac{1}{2} \chi^A D_{AB}
\chi^{B} = S_{PV} + S_{M}$. Hence, $D_{AB}$ is by construction
supersymmetric, which means $D_{BA} = D_{AB} (-1)^{\gras{A} + \gras{B}
+ \gras{A}\gras{B}}$.  On the other hand, $K_{CD}$ has in general no
such property.

In order to have a consistent evaluation of the integral $I[K]$, we can
derive it from
\beq
    I = \int [ d\chi] e^{-\frac{1}{2} \chi^A D_{AB} \chi^B}
      = \left( \sdet D \right)^{x/2},
\eeq
which is defined in the regularisation procedure (\ref{PVC}). Here, $x$ is
the {\it PV weight} of the field when several copies of PV fields
are introduced. $I[K]$ is seen to be
\begin{eqnarray}
    \label{PVout}
   I[K] & = & -2 K_{CD} \frac{\dl I}{\delta D_{CD}} \nonumber \\
        & = & -2 K_{CD} \frac{\dl \left( \sdet D \right)^{x/
           2}}{\delta D_{CD}} \nonumber \\
        & = & - x.I.\str \left( K \frac{1}{D} \right).
\end{eqnarray}
In the last step, we used that $\delta \sdet D = \sdet D . \str \left(
D^{-1} \delta D \right)$ \cite{superBryce}. Notice that the {\it PV weight}
multiplies the whole expression.
Using the explicit form (\ref{diffop}) of the two operators in the
supertrace, we can rewrite this as
\beq
     I[K] = -x.I.\str \int dz \int d{z'} \; K(z). \delta(z-z'). \frac{1}
{D(z)}.\delta(z-z'),
\eeq
where the $\str$ now only runs over the internal indices.

In order to show how this procedure leads to Fujikawa regulators
(\ref{Fujfo}) \cite{Diaz}, we take $D_{lk}$ to be of the
form (see e.g.\ (\ref{delgamout})):
\beq
   D_{lk}(z) = {\cal O}_{lk}(z) - T_{lk}(z) M^2.
\eeq
The first term ${\cal O}$ contains all the differential operators, and
is determined by $S_{PV}$. The second term $T$ is the
matrix determining the mass term of the PV fields.
Using the cyclicity of the supertrace, we can rewrite
\beq
   I[K] = - x. I. \str \int dz \int dz' \; J(z) .\delta(z-z'). \frac{1}
{\frac{{\cal R}}{M^2} - \unity} . \delta(z-z').
\eeq
Here, $J = T^{-1} {\cal K}$ and ${\cal R} = T^{-1} {\cal O}$, where
${\cal K}$ is defined by
$K = M^2 {\cal K}$.
The operator ${\cal R}$ is the regulator in the Fujikawa approach,
as can be seen by using
\beq
 \label{DenReg}
\int_0^{\infty} d\lambda e^{-\lambda} e^{\lambda {\cal R}}
  = - \frac{1}{{\cal R} - \unity},
\eeq
which leads to
\beq
  \label{Gaussklok}
  I[K] = x. I .\str \int dz \int dz' \int_{0}^{\infty} d\lambda
   e^{-\lambda} J(z).\delta(z-z') e^{-\lambda \frac{{\cal R}}{M^2}}.
   \delta(z-z').
\eeq
In the next section we discuss some mathematical results useful for
the evaluation of this supertrace in the limit $M \rightarrow \infty$, for
the case of a regulator
${\cal R}$ that is an elliptic second order differential operator. If this
is not the case, see for instance the example in section 4 of chapter 13,
one can use the equality
\beq
    \label{AddA}
     \str \left( K \frac{1}{D} \right) = \str \left( K A \frac{1}
{(DA)} \right) ,
\eeq
to get a second order differential operator in the denominator, by a
judicious choice of the operator $A$.

\section{Mathematical results for calculating the supertrace}

We first present the main results of \cite{Gilkey},
which allow us to evaluate the supertrace (\ref{Gaussklok}).
These results belong to the branch of differential geometry where the
properties of elliptic operators are studied.
In the mathematical literature the names of Seeley and Gilkey are
associated with this study. Physicists have time after time
rederived these results from a more algebraic point of view,
unaware of the complete results obtained by the mathematicians.
One such derivation can be found in the third section of this
appendix. In the physics literature, these results are associated
with Schwinger, DeWitt, 't Hooft and Veltman, Avramidi \ldots
and they are often denoted by the name {\it heat kernel
expansion}\footnote{For a sketch of the field, we refer
to \cite{VilHeat}.}.

The results sketched below are derived for bosonic matrices, but
they can be generalised for supermatrices.
We start from a second order differential operator
\beq
    D = - \left( h^{ij} \frac{d^2}{dx_i dx_j} + a_i \frac{d}{dx_i}
   + b \right).
\eeq
Here, $a_i$ and $b$ can be arbitrary square matrices of dimension $r$
and  $h^{ij} = g^{ij} \unity_r$, with $\unity_r$
the unit matrix of dimension $r$.
With the operator $e^{-tD}$, where $t$ is a positive parameter, we can
associate an integral kernel, which describes how it acts on
an arbitrary function $f(x)$:
\beq
    e^{-tD} f(x) = \int d vol(y) \; K(t,x,y) f(y).
\eeq
$d vol(y)$ is the Riemannian volume element. In the limit
$t \stackrel{ > }{\rightarrow} 0$, and for equal points $x=y$, this
kernel has the following expansion:
\beq
     K(t,x,x) = \sum_{n=0} E_n(x) t^{\frac{n-d}{2}}.
\label{kernex}
\eeq
Here, $d$ is the dimension of space-time. Notice that
the expansion starts with terms of negative power in $t$, depending
on the dimension $d$. The origin of these terms will become clear
in the explicit calculation in the next section. In fact, we are not
interested in terms with a strict positive power of $t$, as they
disappear in the limit $ t \rightarrow 0$, which corresponds to taking
the mass $M$ of the  PV fields to infinity.

We now rewrite the operator by defining a metric and
gauge connections. The quantities $E_n$ can then be expressed
using the invariant geometrical objects that can be constructed
from these ingredients. This includes the Riemann- and Ricci tensor,
the curvature of the gauge fields, covariant derivatives.
Although results are available for operators where both a gauge and
a metric connection is needed, we will restrict ourselves to
the case where we have either a non-flat metric and no gauge connection,
or a flat metric and a gauge connection, as these are the only cases
encountered in the main text. The explicit example of the third
section belongs to the first category.

First, suppose that no gauge connection has to be introduced
to rewrite $D$. Then, we define $g^{ij}$ and the
(matrix) $E$ by:
\beq
    D = - \frac{1}{\sqrt{g}} \partial_{i} \sqrt{g} g^{i j}
     \partial_{j} \unity - E.
\eeq
The first term is the covariant Laplacian associated with the
metric $g^{i j}$. All the derivatives are explicit, meaning that $E$
does not contain free derivative operators.
If we denote the Riemann curvature tensor associated
with this metric by $R_{ijkl}$, then the first three non-zero
contributions to (\ref{kernex}) are determined by:
\begin{eqnarray}
      \label{gevalA}
     E_0 & = & \frac{1}{ (4\pi)^{d/2}} I \nonumber , \\
     E_2 & = & \frac{1}{ (4\pi)^{d/2}}  \left( E - \frac{1}{6}
        R_{ijij} \right)  , \\
     E_4 & = & \frac{1}{ (4\pi)^{d/2}}  \left( -\frac{1}{30}
   R_{ijij;kk} + \frac{1}{72} R_{ijij} R_{kmkm} \right. \nonumber \\
& &   \left. - \frac{1}{180}
   R_{ijik}R_{njnk} + \frac{1}{180} R_{ijkn}R_{ijkn} -\frac{1}{6}
   R_{ijij} E + \frac{1}{2} E^2 + \frac{1}{6} E_{;kk}   \right). \nonumber
\end{eqnarray}
The covariant derivative of the metric connection is denoted by ;
and repeated indices are summed over.

Secondly, suppose that we can rewrite $D$ using a flat metric
$\eta^{ij}$, a gauge connection $A_i$ and a matrix $E$ defined
by:
\beq
   D = -(\partial_i \unity + A_i) \eta^{ij} (\partial_j \unity + A_j)
  - E.
\eeq
Again, all derivatives are explicit. $E$ and $A_i$ do not contain
free derivative operators.
Define now a covariant derivative $\nabla_i X = \partial_i X
+ [ A_i , X]$, and the curvature of the gauge field by $W_{ij} =
\partial_i A_j - \partial_j A_i + [ A_i , A_j ]$.
We then have:
\begin{eqnarray}
   \label{gevalB}
     E_0 & = & \frac{1}{ (4\pi)^{d/2}} I \nonumber , \\
    E_2 & = & \frac{1}{ (4\pi)^{d/2}} E \nonumber , \\
    E_4 & = & \frac{1}{ (4\pi)^{d/2}}  \left( \frac{1}{12}
     W_{ij} W^{ij} + \frac{1}{2} E^2 + \frac{1}{6} \nabla_i
    \eta^{ij} \nabla_j E \right) .
\end{eqnarray}

We will now show how these results can be used to calculate the
the supertrace derived in the first section of this appendix.
First of all, identifying $t = \frac{\lambda}{M^2}$, we see
that we indeed are studying the limit $ t \rightarrow 0 $ if $M
\rightarrow \infty$. Introducing the integral kernel $K(t,x,y)$ for the
operator $ e^{-t {\cal R}} $, we can derive
\begin{eqnarray}
      &  & \int dz \int dz' J(z).\delta(z-z') e^{-t {\cal R}}.
        \delta(z-z') \nonumber \\
      & = & \int dz \int dz' \int d vol(y) J(z). \delta(z-z')
          K(t,z,y) \delta(y-z') \nonumber \\
      & = & \int dz \int d vol(y) J(z).\delta(z-y) K(t,z,y).
\end{eqnarray}
If we now make the extra restriction that $J(z)$ does not contain
differential operators acting on the $\delta$-function, then the
integral over $z$ can also be done, leading to
\beq
   I[K] = x.I. \str \int_0^{\infty} d \lambda e^{-\lambda} \int d vol(y)
  J(y)   K(\frac{\lambda}{M^2}, y, y).
\eeq
Here it is clear that indeed the expansion (\ref{kernex}) above can be
used.
Thereafter, calculating the supertrace over the internal
degrees of freedom yields the final result.
Notice that the terms with a strict negative power of $t$ lead
to ill-defined integrals for $\lambda$. These have to be removed by
the regularisation or renormalisation, for which we refer to the main text.

We close this section with a comment on the restriction
that $J(z)$ does not contain derivatives. If it does, one of
the two following tricks may be useful. Using the fact that
$D$ is a supersymmetric matrix, and that the supertrace is
invariant under supertransposition, the supertrace can be replaced
by a supertrace over a symmetrised Jacobian. See \cite{ik2},
where this trick was used to remove linear derivatives from $J$.
Another possibility, which works if $J$ is up to second order
in the derivatives, is to study $\str( e^{-B+ \alpha J})$ instead
of $\str( J e^{-B})$, as the latter can be obtained from the
former by a derivation with respect to $\alpha$.

\section{An explicit example}

In this section, we will calculate one explicit example.
This will enable us to understand some of the features
of the previous section from an algebraic point of view.
In the mean time, it will become clear that a non-negligible
amount of work is saved by using the results of \cite{Gilkey} as
they are listed in the previous section.
As anounced, we will study the case where there are no
gauge fields, only a metric. $J(z)$ will be taken to contain
no derivatives, and $E=0$. Thus we study
\beq
   I = \int d^2z \int d^2z' \int_0^{\infty} d\lambda e^{-\lambda}
   J(z) \delta(z-z') e^{\lambda t {\cal R}(z)} . \delta(z-z'),
\eeq
with ${\cal R}= \frac{1}{\sqrt{g}} \partial_{\mu} \sqrt{g}
g^{\mu \nu} \partial_{\nu}$ the covariant Laplacian, and
with a two-dimensional space-time. The limit $t \rightarrow 0$
is of course understood. We work in two space-time dimensions.

The first thing to do is to write both the $\delta$-functions
as Fourier-integrals:
\beq
   I = \int d^2z \int d^2z' \int_0^{\infty} d\lambda e^{-\lambda}
      \int \frac{d^2p}{(2\pi)^2} \int \frac{d^2q}{(2\pi)^2}
        J(z) e^{i p(z-z')} e^{a {\cal R}(z)} e^{i q(z-z')}.
\eeq
We introduced $a=\lambda t$. Now pull the $e^{i q(z-z')}$ through
the operator to the left. For $e^{-i qz'}$ this is trivial, as
${\cal R}$ acts on $z$, not on $z'$. Integrating over $z'$ then
leads to $ (2\pi)^2 \delta(p+q) $. Some more effort is required
to pull through the $e^{i qz}$. First observe that $\partial_{\mu}
( e^{i qz} g(z) ) = e^{i qz} [ \partial_{\mu} + iq_{\mu} ] g(z)$,
so that
\beq
    {\cal R}(z) \left[ e^{i qz} g(z) \right] = e^{i qz}
    \frac{1}{\sqrt{g}}
    (\partial_{\mu} + i q_{\mu}) \sqrt{g} g^{\mu \nu}
    (\partial_{\nu} + i q_{\nu}) g(z) \stackrel{def.}{=}
     e^{i qz} {\cal R}_q(z) g(z).
\eeq
This defines ${\cal R}_q(z)$.
It straightforwardly follows that
\beq
    I = \int d^2z \int \frac{d^2 q}{(2\pi)^2} \int d\lambda
    e^{-\lambda} J(z) e^{a {\cal R}_q(z)} .1  ,
\eeq
where we explicitly wrote the $1$ on which the operator is acting.

We now regroup the terms in ${\cal R}_q(z)$ according to their
power in the momentum $q$. ${\cal R}_q(z) = - A + B + C$, where
\begin{eqnarray}
     -A & = & - q_{\mu} q_{\nu} g^{\mu \nu} , \\
     B & = & i q_{\nu} \left[ 2 g^{\mu \nu} \partial_{\mu} +
      \frac{1}{2} \partial_{\mu} \ln g . g^{\mu \nu} +
      \partial_{\mu} g^{\mu \nu} \right],\\
     C & = & \frac{1}{\sqrt{g}} \partial_{\mu} \sqrt{g} g^{\mu \nu}
     \partial_{\nu}.
\end{eqnarray}
By doing the rescaling of the integrationvariable $q' = \sqrt{a} q = b q$,
we get, dropping the primes:
\beq
  I =
   \int d^2 z \int \frac{d^2q}{(2\pi)^2} \int d\lambda e^{-\lambda} \;
      \frac{1}{b^2} e^{-A + b B + b^2 C}.1  .
\eeq
Notice that a factor $t^{-1}$ has appeared (in the $b$). For
arbitrary dimension of space-time this
generalises to $t^{-\frac{d}{2}}$,
which is exactly the power of $t$ of the first term in (\ref{kernex}).

Let us now concentrate on the operator $\frac{1}{b^2} e^{-A + b B
+ b^2 C} = \frac{1}{b^2} e^{-A} F$, which defines $F$.
$F$ is determined using the Baker-Campbell-Hausdorf formula.
Defining $F(x)= e^{xA} e^{x(-A + b B + b^2 C)}$, we can write
\beq
    F = F(1) = \sum_{n=0}  \frac{1}{n!} F^{(n)}(0) .
\eeq
We have that
\beq
   \frac{d F}{d x} = \left( b B + b^2 C + x b [ A , B ] + x b^2
    [ A , C ] + \frac{x^2}{2} b^2 [ A , [ A , C ] ] \right) F(x) \; ,
\eeq
owing to the fact that $A$ is a space-time
dependent scalar, $B$ a linear differential operator and
$C$ is a second order differential operator, which makes all
possible higher order commutators vanish.
As we take the limit $b \rightarrow 0$, we are only interested in
the terms of the Taylorseries up to quadratic degree in $b$.
Therefore, the $F^{(4)}(0)$ is the last term we need, and we find
\begin{eqnarray}
   F & = & 1 + b B + b^2 C + \frac{1}{2} b [ A , B ] +\frac{1}{2}
  b^2 [ A , C ] + \frac{1}{2} b^2 B^2  \\
   &  & + \frac{1}{6} b^2 [ A , [ A , C]] + \frac{1}{3} b^2
   [ A , B ] B + \frac{1}{6} b^2 B [ A , B ] + \frac{1}{8}
   b^2 [ A , B ]^2 \nonumber.
\end{eqnarray}
This operator acts on $1$, so we can drop the term $b^2 C$. Also,
the two terms linear in the operator $B$, $b B$ and $\frac{1}{2}
b [ A , B]$ disappear upon integration over $q$, as $B$ is linear in
q.

We now take our metric to be of the form discussed in the main
text (\ref{Almost}). That is (${\mu}$,${\nu}$ run over $z$,$\bar{z}$):
\beq
   g^{\mu \nu} = \left(  \begin{array}{cc}
                   2h(z,\bar{z}) & -1 \\
                    -1 & 0          \end{array} \right).
\eeq
Constructing the operators $A$,$B$ and $C$ using this form
for the metric and calculating the terms in $F.1$, one ends up
with momentumintegrals of the form
\beq
\int d^2 q f(q,\bar{q}) e^{-A} ,
\eeq
where we denoted $q_z$ by $q$ and $q_{\bar{z}}$ by $\bar{q}$.
$f(q,\bar{q})$ can be of the form $q\bar{q}$,$q^2\bar{q}^2$,$q^2$,
$q^4$,$q^3\bar{q}$,$q^6$,$q^4\bar{q}^2$ and $q^5\bar{q}$. Only
the first two give a non-zero contribution, because
\begin{eqnarray}
         \int d^2 q e^{-A} & = & \pi \sqrt{g},\nonumber \\
         \int d^2 q q_{\mu} q_{\nu} e^{-A} & = & \frac{1}{2} \pi \sqrt{g}
          g_{\mu \nu} , \nonumber \\
      \int d^2 q q_{\mu} q_{\nu} q_{\rho} q_{\sigma} e^{-A} & = &
            \frac{1}{4} \pi \sqrt{g} \left( g_{\mu \nu} g_{\rho
           \sigma} + g_{\mu \rho} g_{\nu \sigma} + g_{\mu \sigma}
            g_{\nu \rho} \right),
\end{eqnarray}
and because $g_{zz}=0$. The two remaining terms come from $B^2 .1$,
which contributes a $4 q\bar{q} \partial^2 h$, and from $B[A,B].1$,
giving a contribution $8q^2\bar{q}^2 \partial^2 h$. Putting it all
together, the final result is:
\beq
   \label{ResEx}
   I = \frac{1}{4\pi} \int d^2 z \int d\lambda e^{-\lambda} J(z)
   \frac{1}{b^2} \left( 1 - \frac{1}{6} 2\partial^2 h b^2 \right)
    .
\eeq
Notice that the scalar curvature associated with the chosen metric
is indeed $2\partial^2 h$, so that our result agrees with the $E_0$
and $E_2$ of (\ref{gevalA}).


\bibliographystyle{unsrt}
\bibliography{phd}

\end{document}